\documentclass[submission, PhysLectNotes]{SciPost}

\hypersetup{
    colorlinks,
    linkcolor={red!50!black},
    citecolor={blue!50!black},
    urlcolor={blue!80!black}
}

\usepackage[bitstream-charter]{mathdesign}
\DeclareSymbolFont{usualmathcal}{OMS}{cmsy}{m}{n}
\DeclareSymbolFontAlphabet{\mathcal}{usualmathcal}

\numberwithin{equation}{section}

\usepackage{graphicx}
\usepackage{epstopdf} 
\usepackage{pdfsync}
\usepackage[normalem]{ulem}
\usepackage{hyperref}
\usepackage{slashed}
\usepackage{dsfont}
\usepackage{upgreek}
\usepackage{tikz}
\usetikzlibrary{decorations.markings}
\usetikzlibrary{positioning,arrows}
\usetikzlibrary{decorations.pathmorphing}

\def\-{\sout}

\newcommand{\be}[1]{ \begin{eqnarray} \mbox{$\label{#1}$} }
\newcommand{\ee}{\end{eqnarray}}

\newcommand\ie {{\it i.e. }}
\newcommand\eg {{\it e.g. }}
\newcommand\etc{{\it etc. }}

\newcommand\etcc{{\it etc., }}
\newcommand\cf {{\it cf.  }}
\newcommand\etal{{\it et al.}}

\newcommand\half{\frac 1 2 }

\newcommand\ket [1] {|#1 \rangle }

\newcommand\tr [1]{\mathrm{tr} [#1]}
\newcommand\Tr{\mathrm {Tr}}
\newcommand\bra [1] {\langle #1 |}

\newcommand{\av}[1]{\langle #1\rangle}

\newcommand\noi{\noindent}
\newcommand{\p}{\partial}
\newcommand{\gamd}{\gamma^{D+1}}

\newcommand{\rmr}{{\mathrm r}}

\newcommand{\mbr} {\mathbf{r}}
\newcommand{\onam}{\operatorname}

\newcommand{\psib}{\bar\psi}
\newcommand{\etab}{\bar\eta}

\def\ov{\overline}

\def\Z{\mathbb{Z}} 
 
\def\RR{\mathbb{R}} 

\def\nl{\nonumber \\}

\def\de{\partial}

\def\Tr{{\rm Tr}} 
\def\dag{\dagger}

\def\CR{\color{red}}

\def\a{\alpha} 
\def\b{\beta} 
\def\g{\gamma} 
\def\G{\Gamma} 
\def\D{\Delta} 
\def\d{\delta} 
\def\e{\epsilon} 
\def\eps{\varepsilon} 
 
\def\h{\eta} 
\def\k{\kappa} 
\def\l{\lambda} 
\def\L{\Lambda} 
\def\m{\mu} 
\def\n{\nu}

\def\p{\pi} 
\def\P{\Pi} 
\def\r{\rho} 
\def\s{\sigma} 
\def\S{\Sigma} 
\def\t{\tau} 
\def\vf{\varphi} 
\def\c{\chi} 
\def\w{\omega} 
\def\W{\Omega} 
\def\Th{\Theta} 
\def\th{\theta}

\begin{document}

\begin{center}
  {\Large \textbf{ Quantum Field Theory Anomalies\\
      in Condensed Matter  Physics${}^\star$ }}
\end{center}

\begin{center}
R. Arouca\textsuperscript{1,2,3},
Andrea Cappelli\textsuperscript{4} and
T. H. Hansson\textsuperscript{5}
\end{center}

\begin{center}
{\bf 1} Department of Physics and Astronomy, Uppsala University,
  Uppsala, Sweden
\\
{\bf 2} Institute for Theoretical Physics, Utrecht University,
the Netherlands
\\
{\bf 3} Instituto de F\'isica, Universidade Federal do Rio de Janeiro,
Brazil
\\
{\bf 4} INFN, Sezione di Firenze, Italy
\\
{\bf 5} Fysikum, Stockholm University, Stockholm, Sweden
\\
\end{center}

\begin{center}
\today
\end{center}

\section*{Abstract}
{\bf
We give a pedagogical introduction to  quantum anomalies,
how they are calculated
using various methods, and why they are important in condensed matter
theory. We discuss axial, chiral, and gravitational anomalies as well
as  global anomalies. We illustrate the theory with examples
such as quantum Hall liquids, Fermi liquids, Weyl semi-metals,
topological insulators and topological superconductors. The required
background is basic knowledge of quantum field theory, including
fermions and gauge fields, and some familiarity with path integral and
functional methods. Some knowledge of topological phases of matter is
helpful, but not necessary.
}

\vfill \noi
$^\star$Partially based on lectures given at the Spring 2020 Delta ITP
Advanced Topics Course at Utrecht University
and at the PhD school of the University of Florence. \\
email: rodrigo.arouca@physics.uu.se, andrea.cappelli@fi.infn.it, hansson@fysik.su.se

\vspace{10pt}
\noindent\rule{\textwidth}{1pt}
\tableofcontents\thispagestyle{fancy}
\noindent\rule{\textwidth}{1pt}
\vspace{10pt}

\section{Introduction} 
\label{sec:intro}

\subsection{Overview}

The history of anomalies in quantum field theories (QFT) is long and
interesting. In this introductory section we shall mention some of the
most important concepts and give a brief outline of the lectures. The
account is far from complete and will zoom in on the parts that are of
most importance for condensed matter physics.

A symmetry in field theory is said to be anomalous if one cannot
quantize the theory without violating it. If a continuous symmetry has
an anomaly the corresponding current is no longer conserved
at the quantum level.
Such non-conservation can have important consequences
in that physical processes that are forbidden by selection rules
derived from the symmetry in question, are in fact allowed because of
an anomaly.
Historically, the first instance was the axial anomaly,
  whose existence explains the rapid decay of the
neutral pi-meson into two photons, $\pi^0\rightarrow
\gamma\gamma$, in agreement with experiments.
In this and other cases, anomalies are important to understand the
physics described by well-defined QFTs, as they leave characteristic
fingerprints in the form of exact and universal effects.

However, if an anomaly violates the conservation of a current that is
coupled to a dynamical gauge field, it destroys the gauge invariance,
and usually makes the whole theory inconsistent.  This is the case of
the chiral gauge coupling of weak interactions in the Standard Model
of elementary particles, where anomaly cancellation gives important
constraints on the particle content.

Anomalies often occur in QFTs with fermions, and in a strict sense
only in continuum theories with massless (gapless)
particles. However, if a gapless system has an anomaly, this affects
the physics also when the system is perturbed to develop a gap.
A continuum QFT has an infinite number of degrees of freedom even in a
finite volume.  As a consequence, when calculating physical
quantities, typically in perturbation theory, one often encounters
infinities that must be regulated by introducing a high-energy 
(short-distance) cut-off.  This can be done in various ways, such as
dimensional regularization, 
introducing Pauli-Villars regulators or formulating the theory on a
lattice.  In important cases, there are symmetries that are violated
by any regularization -- they are anomalous.

This clash between symmetry and regularization can be seen in the
perturbative loop expansion of QFTs involving fermions coupled to
external gauge fields. Returning to the important case of the pion
decay as an example, the axial current, $J_5^\mu$, which is directly
related to the $\pi^0$ field, is no longer conserved but satisfies
$\partial_\mu J_5^\mu = c\, e^2 \vec E \cdot\vec B$
where $\vec E$
and $\vec B$ are the electromagnetic fields related to the emitted
photons.  
The anomalies are, however, very peculiar loop effects. Although a
regularization is needed for the calculations, the results are finite
at one loop level and do not get corrections from higher loops.
In the example we just gave, this amounts to the constant $c$ having the
exact value $1/(2 \pi^2)$.
 
Anomalies are usually associated with relativistic fermions, as in the
case of the decay of the $\pi^0$ meson, but Lorentz invariance is not
a prerequisite in condensed matter applications, although in many
cases a relativistic dispersion relation provides a simple model that
captures the essence of the system in question.  One example
where anomalies are important is the Luttinger liquid which
describes electrons in various one-dimensional systems such as
nanowires.  Other examples, of great current interest, are the Weyl
semi-metals, which are three-dimensional materials where bands cross to
form ``Dirac cones''.

Later we will discuss both these examples, but it is now fair
to ask why anomalies are at all relevant in these systems which have a
finite (although very large) number of degrees of freedom?  The quick
answer is that in order to give a consistent description of the
original many-body quantum mechanical system, the effective continuum
theory must be such that the anomalies cancel.  More precisely, it is
only the total anomalies in conserved currents that have to
vanish, meaning that contributions resulting from
different fields should cancel.
For example, in the case of the
Luttinger liquid there are different currents describing right and left
moving electrons, and only when taken together do the anomaly in the
electric current vanishes.

In condensed matter physics, many experimental techniques are based on
subjecting a system to classical electric and magnetic fields and measuring the
response.  An important theoretical tool to describe this is the
effective response action, $S_{\rm eff}[A]$, which is obtained by
coupling the system to a background gauge potential $A_\mu$ and then
integrating out the matter fields. Various response functions can then
be determined from $S_{\rm eff}[A]$.  It turns out that in several important
cases, the low-energy response is robust, and in a certain sense
``geometric'', \ie independent of the detailed dynamics.  The deep
reason for this is that the response is determined by anomalies which
are related to topological invariants that, by nature, are
insensitive to small deformations of the Hamiltonian including the
addition of disorder.  Indeed, certain {\it global} quantities,
like conductances, are topological and can only take discrete values.
The integrated anomaly in four dimensions   
$\int \vec{E}\cdot \vec{B}$ is an example of topological quantity.

Another concept that you will encounter in these lectures is that of
an effective low energy theory. These are dynamical QFTs that encode
not only the response to external fields, but also the effective low
energy dynamics. An early example from high energy physics, is the
chiral pion theory that captures the low energy phenomenology of quantum
chromodynamics. An important example in condensed matter physics that
we shall discuss in some detail is the hydrodynamic theory of quantum
Hall liquids, where the effective degrees of freedom are gauge
fields. We shall also encounter a special kind of effective field
theories that only encode topological response, these are the
topological quantum field theories (TQFT). Anomalies in a microscopic
theory, put severe restrictions on the low energy theories, in that
these have to reproduce the anomalies at the low energy scale -- these
are the t' Hooft anomaly matching conditions. 
  
Most condensed matter systems studied in the laboratory have
boundaries,\footnote{Exceptions do occur in one and two dimensions
  where one can have closed loops and surfaces.} and anomalies are
often crucial in order to understand the connection between bulk and
boundary. It turns out that the boundary might be
described by an anomalous theory, which by itself would be
inconsistent, typically by not conserving the electrical current. This
is, however, compensated by a flow of current from the bulk. The bulk
theory is not anomalous, but has an action that violates gauge
invariance, and thus current conservation, at the boundary.  This
cancellation of a quantum anomaly by a classical current in one extra
dimension is called ``anomaly inflow'' or the ``Callan-Harvey
mechanism''.

The first, and most well known, example in which anomaly inflow
  takes place is the quantum Hall effect: The system is gapped in the
  two-dimensional bulk but hosts low-energy boundary excitations that are
  massless one-dimensional fermions.  
This system also provides a paradigm for understanding other topological
phases of matter, in different dimensions and possessing different
symmetries.  Their gapless boundary excitations are again
characterized by specific forms of anomalies and corresponding
topological invariants.  

As a consequence, anomalies provide a classification of topological
phases of matter.  This can be compared to the ``ten-fold way''
classification based on the properties of free fermions in band
systems, and in many cases, as for topological insulators, the two
approaches match.  In others, as for topological superconductors, they
differ in the presence of fermion interactions. Since anomalies can be
exactly determined even in this case, they provide the correct
characterization, and thus are very important for getting a complete
understanding of topological phases of matter.

There are other ways, beside perturbation theory, to calculate
anomalies and some will be presented in these lectures.  A very
important method is based on path integrals, where the anomalies
appear since the integration measure cannot be regulated in a way
that respects all symmetries of the classical Lagrangian.  The
simplest, and most intuitive, approach to anomalies is
 by regularizing
the infinite Dirac sea of relativistic fermions, and this gives a direct
way to understand why anomalies involve simple numerical factors like the
$1/(2 \pi^2)$ mentioned earlier. This is much harder to comprehend in
the context of perturbation theory.  A closely related, but more
general, method is via the study of the spectral flow of
eigenvalues of the Dirac operator: the related index theorems
reveal a connection to certain topological invariant quantities
made out of the gauge field, as well as the background metric if the
systems are put on a curved manifold.
  
Concerning the latter, one would think that curved spaces would only be of interest in high energy physics or string theory, but they are also
relevant in condensed matter physics. As originally shown by
Luttinger, thermal transport can be related to 
gravitational response.  This is an important tool for
characterizing topological phases of matter, especially in cases where
there is no conserved $U(1)$ current, as for superconductors.

The anomalies we talked about so far are all about non-conservation of
currents related by Noether's theorem to continuous symmetries. In
addition to these, there are global anomalies that can violate
discrete symmetries like parity or time reversal.  In these cases, the
partition function is  not invariant under the discrete symmetries,
and also not under  ``large'' gauge transformations.
Important examples are topological insulators
and superconductors in three dimensions. We shall discuss global
anomalies in the last three sections.

\medskip

\subsection{Outline of the lectures and reading instructions}

The chiral anomaly is introduced in Sect. \ref{sec:II}, through the
discussion of $(1+1)$-dimensional fermions and the Luttinger model --
how it originates from the regularization of the Dirac sea, its
relation to the monopole charge and the associated flow of energy
eigenvalues.

Sect. \ref{sec:QHE} shows anomalies at work in the
fractional quantum Hall effect. We discuss
the bulk effective field theory of Chern-Simons type, and derive
the conformal field theory of edge excitations, starting from
free fermions, and then including interactions with the help of 
bosonic currents (chiral Luttinger liquid).
We explain how the anomaly  characterizes the Hall current,
and thus make a case for the  exactness of the Hall conductance. 
Next we  show that a gravitational anomaly  can occur in the presence of
a metric background. This implies a nonconservation of the matter
current that corresponds to heat flow, and an exact result
for the thermal Hall conductivity.

In Sect. \ref{sec:pert} we derive anomalies by a perturbative
expansion in the gauge coupling. This is the historical approach in
particle physics, and we briefly discuss the application to $\pi^0$
decay. The spectral flow argument is extended to higher dimension, and
in Sect. \ref{sec:pathint} we introduce the Fujikawa method for
calculating anomalies in the path-integral formulation. The section is
concluded with some general comments on the importance of anomalies.

The use of anomalies in three-dimensional condensed matter systems is
reviewed in Sect. \ref{sec:WSM}, which describes anomaly governed
low-energy properties of Dirac and Weyl materials such as the
chiral magnetic effect and other electromagnetic responses.

In Sect. \ref{sec:indexth}, we describe general mathematical aspects
of gauge and gravitational anomalies in even dimensions by explaining
the relevant topologically invariant quantities that are involved.
These results are later used in Sect. \ref{sec:global}.
Sect. \ref{sec:heattrans} provides an extended analysis of condensed
matter systems on curved backgrounds, and in particular the role of
gravitational anomalies in the fractional quantum Hall effect.
Sect. \ref{sec:indexth} and \ref{sec:heattrans} are more technical and
require some familiarity with curves space calculus, which is briefly
reviewed in App. \ref{app:calculus}, as a help for the reader.

In Sect. \ref{sec:spt} we discuss effective theories for
symmetry protected topological states, dwelling in some detail on
topological insulators and topological superconductors in three space
dimensions, and explaining the nature of their fermionic boundary excitations.

The study of these and other topological states requires addressing
global anomalies, a concept that is explained in
Sect. \ref{sec:global}, in terms of the non-invariance of the
partition function under ``large'' gauge transformations and discrete
symmetry transformations. We analyze several examples
involving topological insulators and superconductors in various
dimensions.

Sect. \ref{sec:framing} explains the anomalous origin of
the gravitational Chern-Simon action that has physical applications
in the quantum Hall effect and other systems.
Sect. \ref{sec:global} and \ref{sec:framing} are the most
  technical parts of the lectures, but we tried to simplify them as much as
  possible, limiting the presentation to the key physical ideas.

Finally, the concluding section
gives a list of the most important points made in the lectures.

A number of appendices provide technical discussions that
complement the analysis in the text: App. \ref{app:notations}
summarizes our spinor conventions; App. \ref{app:jain} describes the
effective theory of general fractional quantum Hall states;
App. \ref{app:infinitehotel} details some steps in the derivation of
the three-dimensional spectral flow.  App. \ref{app:mixed} spells out
the two definitions of anomalous currents that are possible in the
presence of both vector and axial vector gauge fields. This rather
technical aspect has nonetheless important physical applications.  The
last, App. \ref{app:calculus} provides an overview of differential
geometry and curved space formulas that are used in the text,
complemented with brief discussions of the main concepts behind them.

We emphasize that this is not a review paper, but a set of lecture
notes covering a wide range of topics, with anomalies as the
overreaching theme. Thus we have not attempted to provide a comprehensive
bibliography. In addition to fundamental works, we have
referenced accessible review papers and books, and also some recent work
that points students to new and exciting directions.

This rather long exposition can be read at different levels of
ambition and with different focus.  A first reading can be limited to
Sect. \ref{sec:II}-\ref{sec:WSM}. Starting from the simple case of
$(1+1)$-dimensional fermions and the application to edge excitations
in the quantum Hall effect, one can progress to perturbative and
path-integral analyses of the anomaly.  Then, the application to Weyl
semi-metals is rather straightforward.  In the table below, we show how
the different sections are interconnected to help the readers to select
the parts of most interest to them.

\begin{center}
\begin{tabular}{c|c|c|c|c|c|c|c|c|c|c}
\hline\hline
Section & 2 & 3& 4& 5& 6& 7 & 8 & 9 & 10 & 11 \\
\hline
Needs sections & & 2 & 2 & 2, 4 & 2, 4& 4, 5& 2, 3& 2, 3& 2, 3, 4, 7, 9
                                         &2, 3, 8, 10 \\
     \hline\hline
  \end{tabular}
  \end{center}

\subsection{List of acronyms}

\begin{tabular}{ll}
  APS & Atiyah-Patodi-Singer (theorem) \\
  BdG & Bogoliubov-de Gennes (Hamiltonian)\\
  C & Charge conjugation \\
  CFT & Conformal (invariant) Field Theory \\
  CS & Chern-Simons (action)\\
  D & Spacetime dimension\\
  d & Space dimension \\
 $(3+1)$ & Three space dimensions and time (Minkowskian)\\
  FQH, FQHE & Fractional quantum Hall (effect)\\
  P & Parity (symmetry)\\
  PCAC & Partially conserved axial current \\
  QCD & Quantum Chromodynamics \\
  QFT & Quantum field theory \\
  QH, QHE & Quantum Hall (effect)\\
  SPT & Symmetry protected topological (phase)\\
  TI & Topological insulator \\
  T, TR & Time reversal (symmetry) \\
  TO & Topological ordered (phase)\\
  TQFT & Topological quantum field theory \\
  TSC & Topological superconductor \\
  WSM & Weyl semi-metal\\
\end{tabular}

\section{The axial anomaly in (1+1) dimensions and the
  infinite hotel} \label{sec:II}

In this section we start by discussing the field theory for fermions
in (1+1) dimensions and showing how to regularize the infinities related to
the filled Dirac sea. We then derive the axial anomaly, first by a
direct calculation and then by a simple and intuitive method based on
spectral flow. Next we apply what we learned to an important physical
system, namely the Luttinger liquid, and we end the section with a
first look at the topological aspect of anomalies, to which we shall
return in greater detail later.

\subsection{Two dimensional fermions} \label{sec:2dfermions}
Our first, and simplest, example of an anomalous theory is given by free
relativistic fermions in $(1+1)$ dimensional Minkowski space,
\be{2dferm}
{\cal L}_f = \psib (x)  i \slashed  D  \psi (x)  =
\psib (x)\g^\mu (i \partial_\mu + e A_\mu) \psi (x) \, ,
\ee
where $x^\mu=(t,x)$, $\mu=0,1$,  is the  space-time coordinate,
$A_\mu=(A_0,A_x)$ the electromagnetic potential and we set $\hbar=c=1$.
Throughout these lectures the covariant derivative $D_\mu$ is set for the
electron charge $-e$ and the positive constant $e$ is incorporated
in the definition of $A_\mu$. Our conventions for the
Dirac equation are summarized in App. \ref{app:notations}.

This Lagrangian is invariant both under the vector gauge transformation,
\be{gtr}
\psi (x) \rightarrow e^{i\lambda(x)} \psi (x) \ , \qquad\qquad
A_\mu \rightarrow A_\mu + \partial_\mu \lambda (x) \; ,
\ee
and the global axial  transformation,
\be{agtr}
\psi (x) \rightarrow e^{i\gamma^3 \xi} \psi\; ,
\ee
where $\gamma^3=  \gamma^0\gamma^1$, obeying $\g^{3\dag}=\g^3$ and
$(\g^3)^2=1$.
By Noether's theorem, it follows that both the vector current 
$j_\mu = \psib\gamma_\mu\psi$ and the axial current
$j^5_\mu = \psib \gamma_\mu \gamma^3\psi$ are conserved.\footnote{
We shall use the notation $j^5_\mu = \psib \gamma^\mu \gamma^{2n+1}\psi$ 
for the axial current in any even spacetime dimension.}
Since $\{\gamma^3,\gamma^\mu\} = 0$ it follows that if $\psi$ solves
$i\slashed D \psi =0$, so does $\gamma^3\psi$. We can thus label the
eigenfunction by their eigenvalue of $\gamma^3$ which is just a sign,
\ie $\gamma^3 \psi_\pm = \pm \psi_\pm$, with,
\be{projop}
\psi_\pm = P_\pm \psi= \half (1 \pm \gamma^3)\psi_\pm \; , 
\ee
where $P_\pm$ are projectors.
Written in these components, the Lagrangian decouples into two independent
expressions,
\be{2dferm2}
{\cal L}_f = \psi_+^\dagger (iD_0 + iD_x ) \psi_+ +
\psi_-^\dagger (iD_0  - iD_x ) \psi_-  \, ,
\ee
which associates $\psi_+$ and $\psi_-$ to right and
left moving excitations, respectively. These components are called
``chiral'', borrowing the name from the 4D case, where they actually
correspond to the different property of ``handedness'' of particles,
\ie whether the spin in  parallel or antiparallel to the momentum
\cite{nairbook}.

To quantize, we choose the gauge condition $A_0=0$ and
read off the equal time anticommutation relations
from the time-derivative terms, 
\be{2dcomr}
\{\psi_+ (x,t), \psi^\dagger_+ (y,t) \} = \{\psi_- (x,t),
\psi^\dagger_- (y,t) \} = \delta(x-y) \, ,
\ee
and the corresponding Hamiltonian is, 
\be{2dham}
H = \int dx\,\left[- \psi^\dagger_+ ( i\partial_x  + A_x) \psi_- +
  \psi^\dagger_- ( i\partial_x  + A_x) \psi_- \right] .
\ee

Now define the theory on a circle with circumference  $L = 2\pi R$,
and impose antiperiodic and periodic boundary conditions on
the fermions and the gauge field, respectively,\footnote{
    Antiperiodic (Neveu-Schwartz) boundary conditions are standard for
    fermions,  since the minus sign follows by bringing the fermion
around the circle. Periodic (Ramond) conditions can also be chosen,
they introduce additional features to be discussed in Sect. 9.3. }
\be{2dbc}
\psi_\pm (0, t) = - \psi_\pm (L, t) \ , \qquad\qquad
A_x(0, t) = A_x(L,t) \, .
\ee

\begin{figure}
  \centering
\includegraphics[width=\linewidth]{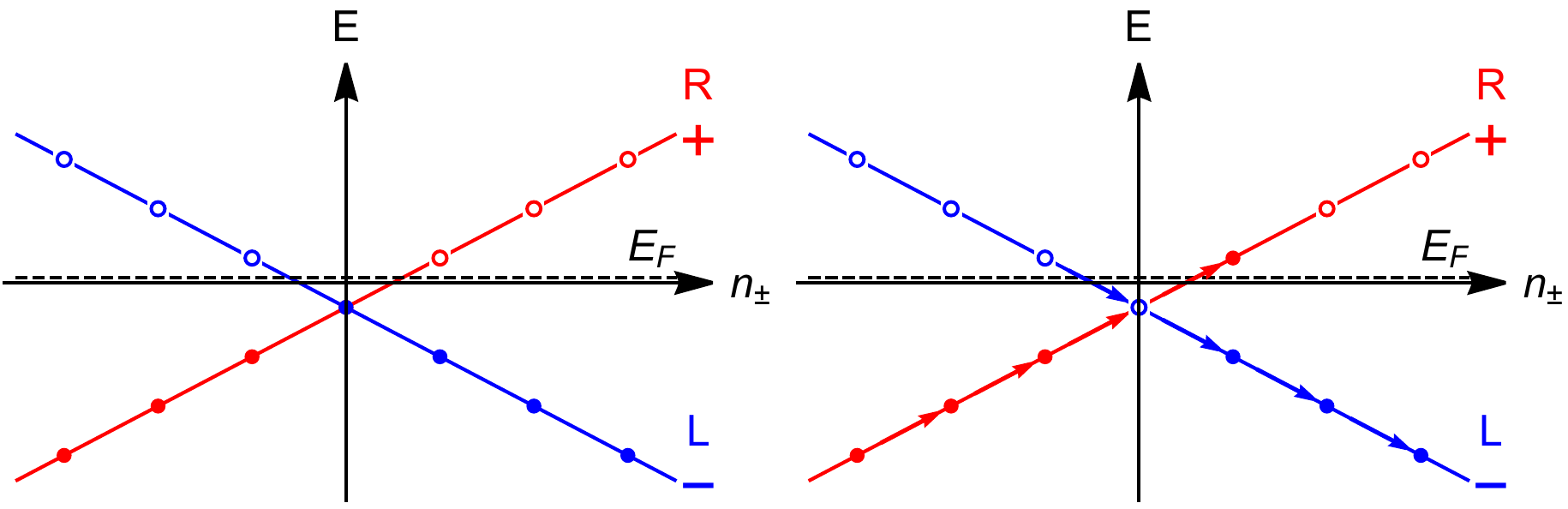}  
\caption{Energy levels of the Dirac fermion and Fermi energy $E_F$.
  Left panel: Ground state filling. Right panel: Filling after adiabatic
  insertion of one flux quantum, with arrows showing the spectral flow.}
\label{fig:1dhotel}
\end{figure}

Since there is no magnetic field in one space dimension, any spatial
dependence in $A_x(t,x)$ can be removed by a suitable gauge
transformation \cite{SHIFMAN1991341}.\footnote{
  This finite gauge transformation is $U(x)=\exp(-i \int^x dx' A_x(x'))$.}
Note also that a spatially constant field $A_x(t)={\rm const.}$ does 
have physical significance since in a gauge $A_0=0$,
the electric field is simply $E^x=F_{0x}=\partial_t A_x$.
From \eqref{2dham} we have ${\cal E}^{(\pm)} = \pm (p -  A_x)$,
and since $p$ is quantized on the circle, we get the spectrum,
\be{2dspec}
{\cal E}_{n_+}^{(+)} =  (n_+ -\half)\frac 1 R - A_x \ , \qquad\quad
\ {\cal E}_{n_-}^{(-)}=- ( n_- + \half) \frac 1 R + A_x \ ,
\qquad\quad n_{\pm}
\in \Z \, ,
\ee
where the sign of the half unit offset is chosen for future convenience.
To define the ground state we must as usual fill up the Dirac sea to some
chemical potential (or Fermi energy) as shown in the left panel of
Fig. \ref{fig:1dhotel} (assuming $A_x(t)=0$ at $t=0$).

Let us define the charge operators for left and right fermions, or
  alternatively the vector and axial charges,
\be{chdef}
\hat Q_\pm= \int_0^L dx \,  \bar\psi_\pm \gamma^0 \psi_\pm, \qquad
\hat Q = \int_0^L dx\,  \bar\psi \gamma^0\psi ,\qquad
\hat Q_5 = \int_0^L dx\,  \bar\psi \gamma^0\gamma^3\psi \, ,
\ee
where  $\hat Q = \hat Q_+ + \hat Q_-$ and
$\hat Q_5 =\hat  Q_+ - \hat Q_-$. 

\subsection{Regularization by point-splitting} \label{sec:pointsplitt}

We expand the fields $\psi_\pm$ in plane wave eigenmodes
as follows,
\be{f-modes}
&&\psi_\pm(x,t)=\sum_{n\in \Z} \vf_{\pm,n} (x,t)\; \hat c_{\pm,n},
\\
&&\vf_{\pm,n}(x,t)= \frac{1}{\sqrt{L}}
\exp\left(\frac{i}{R}(n\mp\frac{1}{2})(x\mp t)\pm iA_xt\right),\ \ 
\ee
where $\hat c_{\pm,n}$ are the usual anticommuting
annihilation operators. Putting the Fermi energy to zero, $E_F=0$,
we get the following expressions for  the  expectation values of
the right and left moving charges in the ground state $\ket{\Omega}$:
\be{formc}
&&Q_ + = \langle\W|\hat Q_+|\W\rangle =
\sum_{n = -\infty}^{0} \int_0^L dx\,
\vf_{+,n}^\dagger (x) \vf_{+,n} (x) , \nl
&& Q_- = \langle\W|\hat Q_-|\W\rangle =
\sum_{n=0}^\infty \int_0^L dx\, \vf_{-,n}^\dagger (x) \vf_{-,n} (x)
\, .
\ee
These are formal expressions, since, as written, they are infinite
sums of ones. A first try to make sense of this would be to bound the value of
$n$, to regularize the sums. This does not work: 
a cutoff on the momentum $p_x$, a gauge non-invariant quantity,
would violate gauge symmetry, or
equivalently, vector charge conservation.  However, since the
divergence at high momenta is due to having two quantum fields at the
same point, we can regularize the sums by a gauge invariant
point-splitting. Let us verify that the following
prescription does the job,
\be{psp}
\psi^\dagger_\pm (x, t) \psi_\pm (x,t)\ \rightarrow\ \psi^\dagger_\pm (x, t)
\exp\left(-i\int_x^{x+\varepsilon} dx' A_x(x')\right)
\psi_\pm (x+\varepsilon,t) \; .
\ee 
For a spatially constant potential, the phase factor is simply
$\exp(i\varepsilon A_x)$, and we get the regularized versions of \eqref{formc},
\be{formc2}
Q^r_+ = \sum_{n = 0}^{\infty}
\exp\left[ -i(n+\half)\frac \varepsilon R - i\varepsilon A_x \right]
\ , \qquad\quad
Q^r_- = \sum_{n=0}^\infty
\exp\left[i(n+\half)\frac \varepsilon R - i \varepsilon A_x \right]  \, .
\ee
Note that the expressions in the exponents are just the energies of
the eigenmodes, and thus gauge invariant.  Summing the geometric
series, and extracting the leading terms in the limit
$\varepsilon \rightarrow 0$, we get,\footnote{ Strictly speaking the
  series does not converge, but you could add a small imaginary part
  to $\epsilon$ to remedy this.  Alternatively we can just cut off the
  series with the gauge invariant factor $\exp(-\tau |(p_n + eA_x)|)$,
  since any gauge invariant regularization of the sum should be
  equally good. This last prescription is similar to the ``heat kernel''
  regularization \cite{nairbook}.}
\be{psp2}
Q^r_+ =  \frac R {i\varepsilon} - R A_x  \ , \quad\qquad\qquad
Q^r_- = -  \frac R {i\varepsilon} + R A_x \, ,
\ee
giving $Q=0$ and $Q_5=2R/i\eps -2RA_x$.

After subtracting the divergent term,
\footnote{A detailed analysis of the subtraction procedure, unambiguously
determining the finite terms, will be presented in section 3.3.
  }
we can check whether
  the two charges are conserved at the quantum level in the presence
  of the background gauge field, 
  \be{locan3}
  \dot Q(t)=0, \qquad\qquad
\dot  Q_5(t) = - 2R  \dot A_x = - \frac 1 \pi \oint dx\, E^x(x,t) \, .
\ee 
Note that the last expression on the right-hand side has been rewritten
in gauge invariant form, since a non-vanishing contribution
$-\partial_x A_0$ to $E^x$ as well as a gauge transform of $A_x$
would vanish when integrated around the circle with  periodic boundary
conditions.
Using translational invariance, \eqref{locan3} can be written in the 
following Lorentz invariant form, using $E^x=F_{0x}$ and
$\eps_{0x}=-\eps^{0x}=1$,
\be{locan2}
\partial_\mu J^\mu =0, \qquad\qquad
\partial_\mu {J^\mu_5} = \frac{1}{2 \pi} \eps^{\mu\nu}F_{\mu\nu} \, .
\ee
These expressions show that the regularization of the infinite Dirac sea
leads to the violation of a classical conservation law, that of the
axial current. This is the axial anomaly in $(1+1)$ dimensions.

A natural question to ask at this point is how much this result depends
on the regularization method. Let us consider another possible choice, 
where the fermion is subjected to an axial gauge field background
$A_{5\mu}$, with the coupling
$J_5^\mu A_{5\mu}$.  Then, taking the same steps as above
but with the point-splitting,
 \be{psp3}
 \psi^\dagger(x, t) \psi(\rmr)\ \rightarrow\ \psi^\dagger(x, t)
\exp\left(-i \gamma^3 \int_x^{x+\varepsilon} dx' A_{5x}(x')\right)
 \psi(x+\varepsilon,t) \, ,
\ee
which respects the axial gauge symmetry \eqref{agtr}, we get,
  \be{psp4}
Q^r_+ =  \frac {R} {i\varepsilon} - R A_{5x} , \quad\qquad\qquad  
Q^r_- = -  \frac {R} {i\varepsilon} - R A_{5x} \, .
\ee
So with this regularization, $Q_{5}$ is independent of $A_5$
but $Q$ is not; the axial current is conserved but not the vector charge,
\be{locan4}
\partial_\mu J^\mu = \frac{1}{2\pi} \eps^{\mu\nu}F_{5\mu\nu}, \qquad\qquad
\partial_\mu J^\mu_5 = 0\, .
\ee

In conclusion we have found that a regularization preserving one
symmetry leads to violation of another.   How to choose one among
two incompatible symmetries is dictated by the physical problem
and amounts to carefully fixing the renormalization
conditions.  Once this is done, the anomaly is unambiguously
determined. That anomalies can be
shuffled around by using different cutoff procedures is something we
shall encounter several times in the following. 

You might, and with all rights, wonder if there could not be some
regularization that would completely remove all anomalies. That this
is not possible is an important result in QFT, which in its most
elegant formulation uses deep mathematical results in differential
geometry. We will touch on this important subject in
Sect. \ref{sec:indexth}.

Finally, a remark about the physical interpretation of the vector
fields $A$, and $A_5$: In condensed matter systems you should think of
$A$ as the usual electromagnetic potential which couples to the
electric current which is always conserved.  The axial field $A_5$, on
the contrary, is a non-dynamical ``emergent gauge potential'', and
there is no fundamental principle that requires the axial current to
be conserved.  We shall see in Sec. \ref{sec:WSM} 
that $A_{5}$ plays an important role in the
physics of Weyl semimetals and the chiral magnetic effect,
while in App.  \ref{app:mixed} we discuss
aspects of the mixed anomalies where both vector and axial
(background) fields are present.  The full effective theory for $A$
also includes a kinetic term for the vector field $A$ which eventually
derives from the microscopic Maxwell theory but is modified by
material effects such as electric polarizability.  For many phenomena
discussed in these lectures, however, we can neglect the photons, and
consider $A$ as a background field just as $A_5$.  In a couple of
cases, such as in the Luttinger model in Sect. \ref{sec:luttinger}, we
will include the electromagnetic interaction in the guise of an
instantaneous Coulomb potential.

\subsection{Spectral flow and the infinite hotel}
\label{sec:infinitehotel}

From the previous section one would naturally conclude that anomalies
are ultraviolet effects since they depend on how you regularize the
high-energy end of the spectrum.  There is, however, another approach
which reveals the infrared nature of the anomalies. In
condensed matter this is all important, since all the relevant field
theories are effective low energy descriptions that
should not much care about the ``ultraviolet completion''. The key to
understanding the infrared face of anomalies is the concept of
spectral flow.

We start by noting that on a circle, even a spatially constant $A_x$ 
has a physical significance since it can provide a nontrivial gauge invariant 
phase factor, 
\be{phasefactor}
W[A_x] = e^{i\oint dx\, A_x} \, ,
\ee
to an electron that makes a full turn around the circle. 

Our next observation is that there is a concrete way to change
  $A_x$.  Let us suppose that the $(1+1)$-dimensional theory is a part
  of a higher-dimensional system: The integral ${i\oint dx\, A_x}$ can
  be changed by inserting a long solenoid inside the circle and
  varying the magnetic flux in time, for $-T<t<T$ to insert one flux 
  (here and in the following, this  means one quantum of flux), 
   so called ``flux insertion''. If $T$ is large enough,
the process is adiabatic and the system at all times remains
in the ground state.
The field, 
\be{fluxins}
A_x = -\frac{f(t)}{R}, \qquad \mathrm{with} \qquad f(-T)=0 \ ,\qquad
f(T)=1 \; ,  
\ee
is obtained by adding a single flux through a surface
$D_2$ bounded by the circle $S_1$, 
 \be{fluxin}
\Phi(T)-\Phi(-T)=\int_{-T}^T \!\!\! dt
\int_{D_2} \partial_t B_z =
-  \int_{-T}^T dt\int_{ S_1} E^x =
  2\pi\int_{-T}^T dt \partial_t f(t) = \Phi_0\, ,
  \ee
  where $\Phi_0=h/ec$ is the flux quantum (after restoring the
  electric charge and all units). 

Note that the entire process, going from $t=-T$ to $t=T$, amounts to
a gauge transformation, $A_x\to A_x+\de_x\l$,
with $\lambda =-x/ R $, that is not itself periodic in space
but keeps $W[A_x]$ periodic, since $A_x$ appears at the exponent.
This is a so-called ``large'' gauge transformation that {\it does}
change the physics, in the following way.

Being a gauge transformation respecting the boundary conditions,
the spectrum of the $(1+1)$-dimensional theory at $t=-T$ and $t=T$
are equal.
On the other hand, upon integrating the axial anomaly \eqref{locan3},
  we find,
\be{eq-hotel}
\D Q_5=\int_{-T}^T dt\, \dot Q_5 =
2\left(f(T)-f(-T) \right)=2,
\qquad \qquad \D Q=0.
\ee

This means that the ground state changes in the process, in that a
right-moving fermion is appearing from the Fermi sea while a
left-moving one is diving down, as illustrated in
Fig. \ref{fig:1dhotel}.  Actually, each level in the right-moving
(left-moving) Dirac spectrum is mapped into the next (previous) level,
carrying the electron with itself, if present: this is the 
so-called spectral flow.  In conclusion, the flux insertion has
caused an anomalous non-conservation of the chiral charges!\footnote{
  Note that this is not in contradiction with the Byers-Yang
  theorem that states that if a system consists entirely of particles
  with charge $e$, there is no way to detect whether there are an integer
  number of fluxes through a puncture in the
  plane \cite{imry}. }

A popular analogy is that of the ``infinite
hotel''\footnote{ This name seems to have been coined by H.B. Nielsen
  in reference to Hilbert's infinite hotel paradox
  \cite{nielsen1991} \cite{nielsen1983}. }:
In a hotel with an infinite number of rooms, there is always space for
a new guest even if all rooms are taken. The manager just moves the
guest in room one to room two, the person in room two moves to room
three, and so on. Now, room one is free for the new guest.  This
relabeling amounts to the process we just
described. Obviously this trick does not work in a finite hotel ...

The spectral flow is an adiabatic process that, as a whole, amounts to
a gauge transformation, but where the intermediate states are
physically distinct.  This argument shows that, from a physics point of
view, the axial anomaly is an infrared effect; it is about what is
happening at the Fermi surface.  Thus, the axial anomaly has a Janus
face.  The physical aspect, \emph{viz.} the non-conservation of a
current, is the infrared face, while the need to regularize in a way
consistent with the symmetries is the ultraviolet one. We will next
explain that this double nature of the anomaly is deeply rooted in its
topological origin.

\subsection{Anomalies and topology -- a first look}\label{sec:antop}

A recurrent theme of these lectures will be the close connection
between anomalies and various topological invariants. This is a very
deep and interesting topic by itself, but it is also of paramount
importance to understand topological phases of matter which is at the
frontline of current research in condensed matter physics.
\footnote{For a general review on topological aspects in condensed matter
  physics, see \cite{Abanov}.}

We now  return to the (1+1)D fermions, and rephrase our previous  
discussion in a mathematically more precise way, which will be of use 
later when we discuss anomalies in higher dimensions.
For this we shall look more carefully at a gauge transformation 
that gives rise to a spectral flow ($A_\mu=(A_t,A_x)$),
\be{eq-gauge}
&&A_\mu(t=-T) =(0,0) , \qquad\qquad
A_\mu(t=T) =\left( 0, -\frac{n}{R} \right),\qquad \mu=0,x,
\nl
&&A_\mu(T) - A_\mu(-T) =-\de_\mu \l,\qquad
\l(x)= n \frac{x}{R}\; .
\ee
We already argued that this is an acceptable gauge transformation
since it leaves the spectrum invariant, but another, and more 
fundamental, observation is that the group element
$u(x)=\exp( i \l(x))\in U(1)$  is periodic on the circle. 
From a mathematical point view, $u(x)$ is a map from 
the spatial circle to the group manifold of $U(1)$, that is the manifold
of phase factors, which is also a circle. 
Such functions from $x\in S_1$ to $u(x)\in U(1)\sim S_1$
are loops that wind in the ``target space'' and can be divided into
classes that cannot be continuously deformed into each other: those that
do not wind, that wind once in one direction, twice, and so on.
These classes form a group by addition that is called the first
homotopy group $\pi_1(U(1))=\Z$. The example given
by \eqref{eq-gauge} has winding number $n$.

To proceed, we return to the adiabatic flux insertion \eqref{fluxins}.
For large $T$ the ramp function $f(t)$ 
can be chosen to vary very slowly near the extremes $t=\pm T$, so that
$F_{\mu\nu}$ effectively vanishes. We can then identify the field 
configurations at $t = \pm T$, so that the spacetime manifold becomes
$S_1 \times R \to S_1 \times S_1$ \ie a torus.
Integrating  the gauge invariant form
\eqref{locan2} of the anomaly over this torus, we get,
 \be{eq-topo}
\D Q_5 =\int_{-T}^{T} dt \int_0^L dx \, \de_\mu \ J_5^{\mu}=
\frac{1}{2\pi}\int_{S_1\times S_1}\!\!\!\!\!\!\! d^2x\;
\eps^{\mu\nu}F_{\mu\nu} =2 I[A]=2n  \, .		
\ee
The quantity $I[A]$ is quantized as an integer,
\be{eq-mono}
I[A]=\frac{1}{4\pi}\int_{S^1\times S^1}\!\!\!\!\!\!\! d^2x\;
\eps^{\mu\nu}F_{\mu\nu}= n\in\Z \; , 
\ee
and is usually referred to as the monopole charge, while it is called
the first Chern class in the mathematical literature.

$I[A]$ is a very special type of functional of the gauge field. It is a
topological invariant, \ie it is independent of any
smooth deformation $A_\mu(x)\to A_\mu(x)+\d A_\mu(x)$.
These cannot alter the integer value
which only depends on the global properties of the field configuration.
Let us understand how the quantization comes about.
Since the integrand is a total derivative, a naive use of Stokes
theorem would imply that the integral \eqref{eq-mono} over a compact
surface vanishes. However, the gauge field $A_\mu$ is not well defined
over the entire surface, and the theorem must be applied with
care. Remembering the discussion of the gauge transformation
\eqref{eq-gauge}, $A_\mu$ is double valued at $t=\pm T$, the
difference being a non-trivial gauge transformation $u(x)\in U(1)$
with winding number $n\in\Z$.
The use of Stokes' theorem then gives,
\be{eq-stokes}
\!\!\!\!\! I=\frac{1}{4\pi}\int_{S_1\times S_1} \!\!\!\!\!\!\!\!\!\!\!\!
d^2 x\;\eps^{\mu\nu}F_{\mu\nu}=
\frac{1}{2\pi} \int_{S_1} \!\!\!\!\!  dx^\mu\; (A_\mu(-T)-A_\mu(T))=
\frac{1}{2\pi} (\l(2\pi R)-\l(0))= n \, .
\ee
Note that, any 
local deformations of the gauge field $A_\mu(x,t)$, for
$-T<t<T$, do not affect its value as anticipated. 

In conclusion, we have shown that the integrated form of the anomaly
is related to a topological invariant of the gauge field,
the first Chern class, that takes integer values, owing to the
homotopy group $\pi_1(U(1))=\Z$. The spectral flow with $n$ right-moving
electrons pumped out of the Dirac sea (and left-moving holes pumped in)
is realized
precisely when the gauge field configuration is that of a charge $n$
monopole inside the space-time cylinder (compactified to a torus).
It is apparent that the anomaly is only sensitive to the global infrared
properties of the underlying theory.

Finally we should comment on the connection between the adiabatic 
flux insertion \eqref{fluxin} of the previous section,
and the purely $(1+1)$-dimensional approach just discussed, 
leading to \eqref{eq-topo}. First you should notice that topology
is important also in the flux insertion argument, since the winding number 
is a topological invariant, just as the Chern class.
Secondly, the Yang-Byers quantization of magnetic flux in three dimensions
\cite{imry} is the counterpart
to the topological electric flux quantization in the (1+1) dimensional case. 
For physical phenomena that takes place at boundaries, 
it is rather natural to think in terms of embeddings in a higher dimensional
space, but in the following discussion of other systems 
we shall explore both approaches. 

\subsection{Connection to the  Luttinger model} \label{sec:luttinger}

In condensed matter physics, fermion systems in one dimension can be
realized in various ways, such as quantum wires, carbon nano-tubes or
edge states in a quantum Hall droplet.  All these systems have a rich
and fascinating phenomenology, but here we shall just mention one
aspect of the theory which is intimately connected to the anomaly we
derived in the previous sections.

\begin{figure}
  \centering
\includegraphics[width=6in]{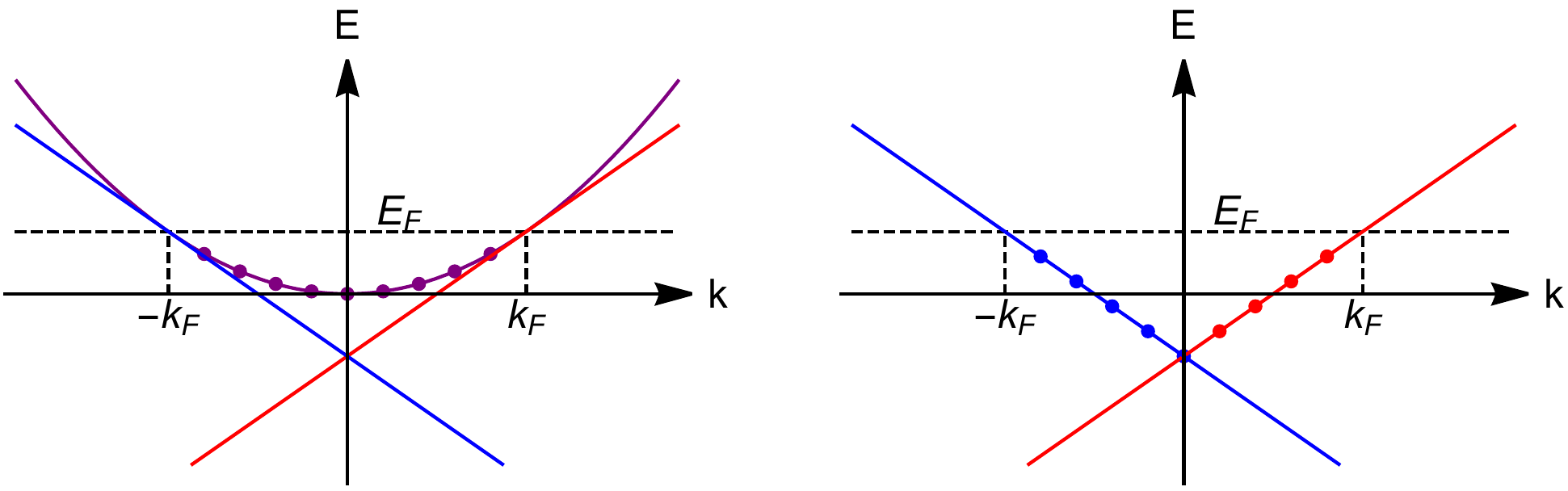}
\caption{Left panel: Dispersion relation for a parabolic band, with
  the Fermi energy $E_F$ and momenta $\pm k_F$.  The dots indicate
  filled levels in the ground state. The red and blue lines are
  linear approximations near Fermi points. Right panel: Linear
  dispersion relation for the Luttinger model, with filled levels.}
 \label{fig:luttinger}
\end{figure}

The way relativistic fermions emerge in a condensed matter framework
is illustrated in Fig. \ref{fig:luttinger}.  The left, and more
realistic picture, represents a single parabolic band in a one-dimensional
crystal. It should however be clear that the low-lying excitations in
this model are the same as those of the $(1+1)$ dimensional Dirac theory
\eqref{2dferm}, shown on the right part of the figure. That the
former has a finite number of electrons filling the band, and the
other an infinite Dirac sea should not change the physics close to
the Fermi points. The second model, which was originally studied by
Luttinger, is given by the Tomonaga-Luttinger Hamiltonian,
\be{lutt}
H=\sum_{\alpha=\pm 1} v_F \int dx\, \psi^\dagger_\alpha
(i\alpha\partial_x - k_F)\psi_\alpha - \half \int dx dx'\, \rho (x)
V(x-x') \rho (x')\; ,
\ee
where $\alpha$ labels the two Fermi points,
$v_F$ is the Fermi velocity, and $\rho$ the total density, 
$\rho = \\ \rho_+ + \rho_- = \psi^\dagger_+ \psi_+ + \psi^\dagger_-\psi_-$.
In addition to the kinetic term, there is a density-density
interaction.  This model can be studied using perturbation theory,
renormalization group methods, and most importantly with the powerful
method of bosonization, to be described in the next section.

Let us anticipate that the Fourier components for the chiral density
operators $\r_{\pm,n}$, with $k=(n+1/2)/R$ on a circle of length
$2\pi R$, can be shown to satisfy,
\be{kacm}
[\rho_{\pm, n}, \rho_{\pm , n'} ] = \pm n \delta_{n+n',0}\; ,
\label{eq:KacMoody}
\ee
which is known as current algebra  ($U(1)$ Kac-Moody algebra \cite{Ginsparg}).
Just as the axial anomaly, this is a quite surprising result. Normally
we think that the density should commute with itself! Indeed, upon
expressing $\rho$ in terms of fermionic creation and annihilation
operators $c_n$ and $c^\dagger_n$, we would naively find that the
commutator vanishes.  However, a finite term is generated when the UV
regularization of the densities is taken into account.\footnote{ See
  \eg \cite{Haldane_1981 }. }  As will be described in
Sect. \ref{sec:curr}, this is a manifestation of the anomaly.

We already stressed that the anomaly can also be 
understood as an infrared effect, so we expect that 
 the anomalous commutator could also be derived is a similar manner.
 To do this, we start from the one-dimensional
(non-chiral) density and current operators in first quantization,
$\rho(x) = \sum_{n=1}^N \delta(x - x_n)$ and
$J(x)=\sum_{n=1}^N [p_n \delta(x - x_n) + \delta(x-x_n) p_n]/2m $.
Using the quantum mechanical commutators,
$[x_n,p_m] = i\hbar \delta_{mn}$, we  get the
non-relativistic current algebra relations,
\be{dencurrcom}
[\rho(x),J(x')] = -i \frac{\hbar}{m} \rho(x) \partial_x \delta (x - x')\; ,
\ee
and $[\rho(x),\rho(x')] = [J(x), J(x')] =0$. Now assume that there are
only small excitations around both the left-moving and right-moving
Fermi points, and write $\rho=\rho_0 + \rho_+ +\rho_-$ and
$J = v_F (\rho_+ - \rho_-)$. Here
$\rho_0$ is a constant background density which is
much larger than the perturbations $\rho_+$ and $\rho_-$.  With this
we have,
\be{KacMoody2}
[\rho_+(x),\rho_+(x')] &=&
\frac{1}{4} \left[\rho(x)+\frac{J(x)}{v_F}, \rho(x') +
\frac{J(x')}{v_F} \right]  \nonumber \\
	&=& -i \frac{\hbar}{2 v_F m} \rho(x') \partial_x \delta(x - x') \\
	&=& -i \frac{1}{2 \pi} \partial_x \delta(x - x')\, .  \nonumber
\ee 
In going to the last line we have substituted $\rho_0$ for
$\rho(x)$ on the right, since it is assumed to be much larger than
$\rho_+$ and $\rho_-$.  We also used the values
$\rho_0 = k_F/\pi = m v_F /(\pi \hbar)$ for spinless
one-dimensional fermions. 
Finally, Fourier transforming \eqref{KacMoody2} gives
\eqref{eq:KacMoody}.

Note that just as in the
 ``infinite hotel'' derivation of the spectral flow,
we only considered effects close to the Fermi surface. To
show the connection to the chiral anomaly, we couple the right moving
charge to electromagnetism.  In the axial gauge $A_x = 0$, this
coupling is $H= -\int dx\, A_0\, \rho_+ $, and we can use Heisenberg's
equation to get,
\be{anomalyrel}
 \partial_t Q_+
 &=& -\frac 1 i \int dxdy\, [\rho_+(x), A_0(y) \rho_+(y)]
 \nl
 &= & \frac{1}{2\pi} \int dxdy\, \partial_x \delta(x-y) A_0(y)
      = - \frac {1} {2\pi} \int dx\, E_x \; .
\ee
Therefore, the charge commutator \eqref{KacMoody2}
implies the chiral anomaly \eqref{anomalyrel}.

\section{ Anomalies and effective field theories of 
the Quantum Hall \\ Effect} \label{sec:QHE}

\subsection{Introduction}\label{sec:LL}
 
The best known example of a topological phase of matter is the quantum
Hall effect \cite{Prange}.  A two-dimensional electron system is
realized in a layered semiconductor, \eg Gallium Arsenide, and
subjected to an orthogonal magnetic field $B$.

\begin{figure}
    \centering
    \includegraphics[scale=0.3]{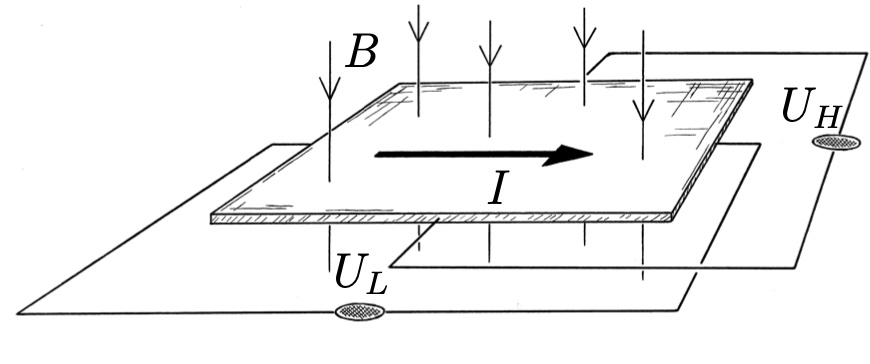}
    \caption{Quantum Hall bar with circuits to measure the longitudinal
      and transverse voltage drops (Figure thanks to S\"oren Holt).}
    \label{fig:qhexp}
\end{figure}

A typical experiment, which is schematically shown in
Fig. \ref{fig:qhexp}, is performed in very clean samples at a very low
temperatures ($T \sim 10 \text{ mK} $) and high magnetic field
($B \sim 10 \text{ Tesla} $). The transverse (Hall) resistance shows
extremely stable plateaus when $B$ is varied over a sizeable range. To
analyze this remarkable phenomenon, one must first understand that in
in 2d, the Hall resistance $R_H = U_H/I$ (using the notation of the
figure) equals the Hall resistivity $\rho_H$ which is a material
property independent of the geometry \cite{Prange}. On the plateaus,
the corresponding Hall conductivity takes the values,
\be{sigma-H}
 J_i=\s_H \eps_{ij} E^j,\qquad\qquad
 \s_H=\n \frac{e^2}{h}\; ,\qquad\qquad
 \nu=1,2,\ldots,\frac{1}{3},\frac{2}{5},\ldots\, .
\ee
In this equation, $J_i,E^i$, $i=1,2,$ are the in-plane current and
electric field, respectively, $h$ is the Planck constant; $\nu$ is the
so-called filling fraction, the density of electrons in units of a
filled Landau level as explained below.
At the same time, the longitudinal conductance vanishes, indicating a
gap in the spectrum.
The prominent plateaus occur for integer filling and for the
fractional values of the Laughlin series, $\nu=1/p$, with $p=3,5,\dots$.
Experiments show remarkably precise values, 
$\Delta\s_H/\s_H\sim 10^{-9}$, that are independent of the
sample details. As will be explained below, this 
is due to the invariance of $\s_H$ under both continuous deformations
of the sample, and the interaction between the electrons;
$\sigma_H$ is a topological quantum number.

The main physical picture for the quantum Hall effect, due to
Laughlin \cite{Laughlin2} \cite{Prange},
is that the electrons form an incompressible quantum fluid,
characterized by a constant density in the bulk and a gap that forbids
density waves.  This type of ground state can most easily  be  understood for
non-interacting electrons completely filling one or more Landau
levels.  In this case, the Coulomb interaction can be neglected and
the gap is  the Landau level spacing, \ie the
cyclotron energy $\hbar\w=\hbar eB/m_e c\gg k_B T$
($m_e$ the mass of the electron).

Let us recall some properties of one-particle states in the Landau levels
\cite{CTZ}. For polarized electrons, the Zeeman term can be
disregarded,\footnote{In relevant materials, the magnetic dipole moment
  is high enough for the Zeeman gap to be much larger than the cyclotron gap.}
and the Hamiltonian is, 
\be{LandHam}
H=\frac{1}{2m_e}\left(\mathbf{p}+\mathbf{A}\right)^2,
\ee 
within our conventions for the electron charge.
The planar momentum is $\mathbf{p}=(p_1,p_2)$ and the
downward magnetic field in Fig. \ref{fig:qhexp}
is realized by the vector potential
$\mathbf{A}=B/2(x_2,-x_1)$, in the so-called symmetric gauge.
We  introduce  two independent pairs of creation-annihilation operators,
\be{ab-oper}
&&a=\frac{z}{2}+ \ov{\de}, \qquad\quad a^\dag=\frac{\ov{z}}{2} -\de, \qquad
\quad\left[a,a^\dag\right]=1,
\nl
&&b=\frac{\ov{z}}{2}+ \de, \qquad\quad b^\dag=\frac{z}{2} -\ov{\de}, \qquad\ 
\quad\left[b,b^\dag\right]=1,
\ee
expressed in the complex coordinate, $z=x_1+ix_2$ and derivative
$\de=\de/\de z$. In these, and many of  the following expressions,
we set the magnetic length,\footnote{
  Another common convention is
  $\tilde \ell_B=\sqrt{\hbar c/eB} \to 1$, as in Ref. \cite{jainbook}.}
$\ell_B=\sqrt{2\hbar c/eB} $, as well as
$c$, $e$, $\hbar$ and $m_e$, equal to one, so $\w=B=2$.
The Hamiltonian and angular momentum $M$ can then be rewritten as,
\be{HandJ}
H=\w\left(a^{\dag}a+\frac{1}{2}\right),\qquad
M=\left(b^\dag b-a^\dag a\right) .
\ee

The single particle wave functions $\psi_{n,m}(z, \ov{z})$ are labeled 
by a Landau level index $n=0,1,\dots$ and an angular momentum $m=-n,-n+1,\dots$.
The energies are degenerate with respect to $m$ as a consequence of
the translation invariance of the cyclotron orbits.
In the following  we shall limit ourselves to the first level,
where the wavefunctions are, 
\be{Psinm}
\psi_{0,m}\left(z,\ov{z}\right)=e^{-z\ov{z}/2}\frac{z^m}{\sqrt{\pi m!}}\,  .
\ee
One can easily check that these states are localized on the classical
cyclotron orbits of radius $r_{m}=\ell_B\sqrt{m}$, in that the area bounded by
the $m$-th orbit encloses $m$ fluxes $\Phi_0$ (equal to $2\pi$ in our units).

Let us consider the Hall system in the geometry of a disk of radius $R$.
Then the angular momentum has an upper bound, approximatively given by the
largest orbit $m_{\rm Max}=R^2$, so the Landau degeneracy 
is equal to the number of enclosed fluxes, ${\cal D}=BA/\Phi_0$. Filling all
the available levels determines the average density $\r_0={\cal D}/A $ and the
Hall conductivity $\s_H=\r_0 e c/B=e^2/h $, corresponding to filling $\nu=1$
in \eqref{sigma-H}.

\begin{figure}
    \centering
    \includegraphics[scale=0.2]{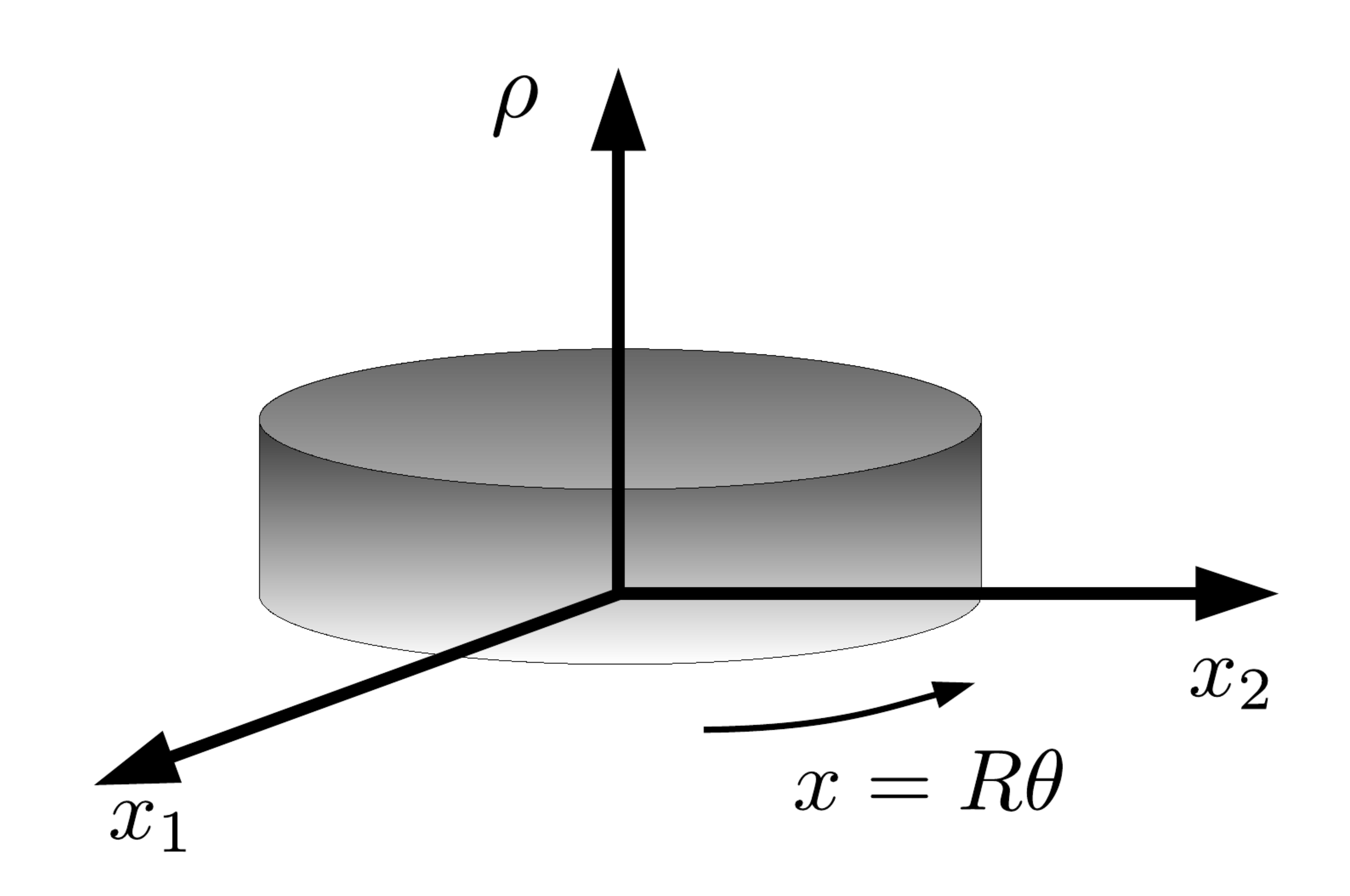}
    \caption{Ground-state electron density \eqref{rhoOmega}
      in the geometry of the disk with radius $R$; the coordinate along the
    edge is $x=R\th$.}
    \label{fig:density}
\end{figure}

The $\nu=1$ ground state wave function $\psi$  is the Slater determinant of the
single particle states \eqref{Psinm}, which can be written as
the  Vandermonde determinant \cite{jainbook}, 
\be{Laugh-wf}
\psi(z_1,\cdots, z_N)=e^{-\sum_{i=1}^{N}|z_i|^2/2}
\prod_{i<j=1}^{N}\left(z_i-z_j\right),
\ee
where $z_i$ are the positions of the $N$ electrons.

The profile of the ground state density is most easily obtained using
second-quantized expressions. The non-relativistic field operator is,
\be{field-op}
\hat\Psi\left(\vec{x},t\right)=
\sum_{m=0}^{\infty}\psi_{0,m}\left(\vec{x}\right)\hat{c}_m \, ,
\ee
where $\vec x=(x_1,x_2)$,
$\hat{c}_m$ is the fermionic annihilation operator and the ground state
energy $\w/2$ has been discarded.
The  $\nu=1$ ground state $\ket{\Omega}$ is built over the Fock vacuum
$\ket{0}$, as follows,
\be{fermVacuum}
\ket{\Omega}=\hat{c}_0^\dag \hat{c}_1^\dag \ldots\hat{c}_{N-1}^\dag\ket{0}.
\ee
The expectation value of the
density $\hat{\r}=\hat{\Psi}^\dag\hat{\Psi}$ then becomes,
\be{rhoOmega}
\bra{\Omega}\hat{\r}\left(r\right)\ket{\Omega}=\sum_{m=0}^{N-1}
|\psi_{0,m}\left(\vec{x}\right)|^2=\frac{1}{\pi}e^{-r^2}\sum_{m=0}^{N-1}
\frac{r^{2m}}{m!}\, , \qquad\qquad r=|\vec{x}|,
\ee
as shown in Fig. \ref{fig:density}. This  describes a droplet of
fluid with constant density $\r_o=1/\pi$ in the interior, which is
rapidly falling  to zero at the edge within a few magnetic lengths.

\subsection{Edge excitations}
\label{sec:edge}

\begin{figure}[t]
\begin{center}
\includegraphics[width=0.6\textwidth]{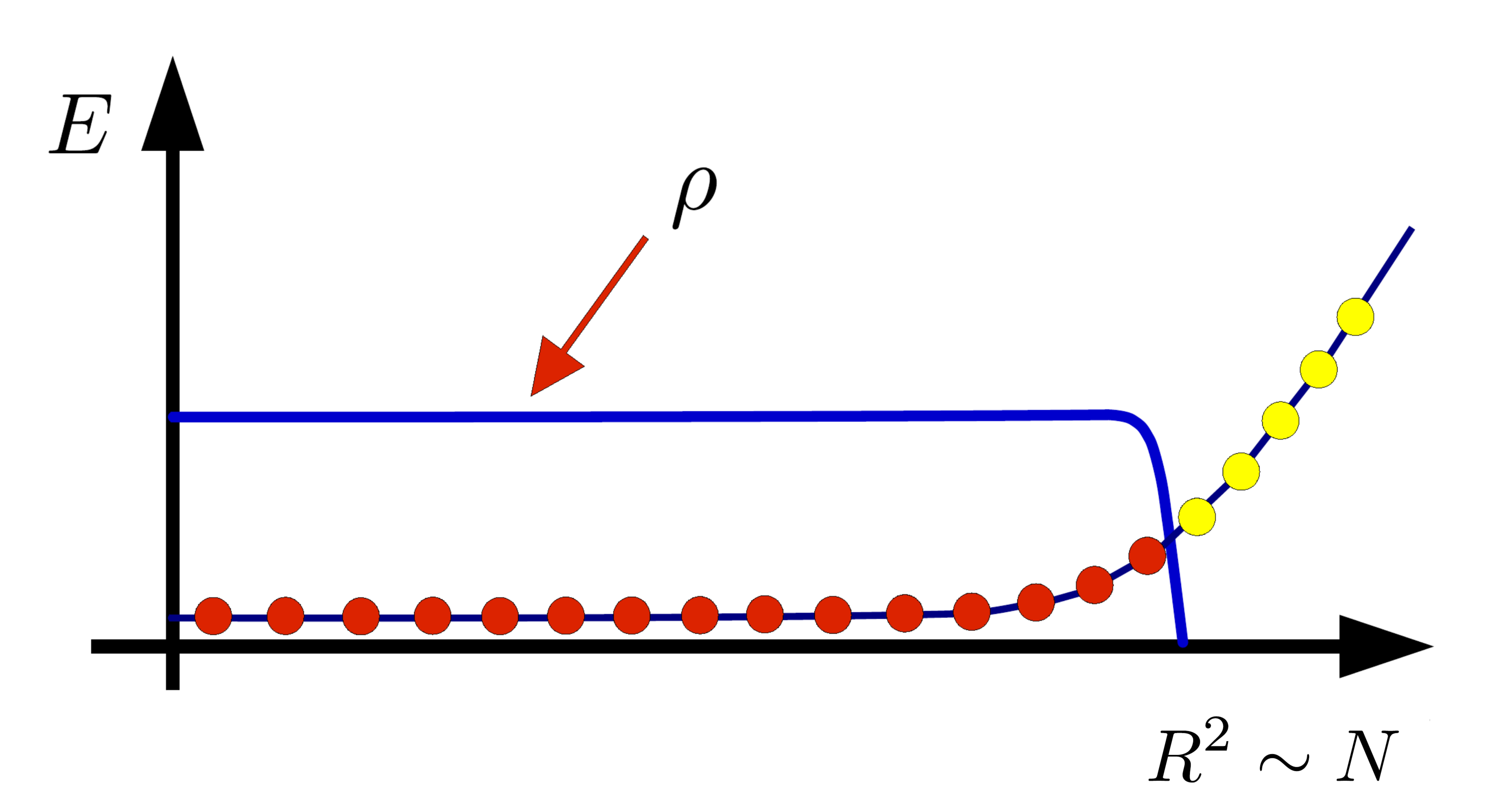}
\caption{\small{Energy levels of the first filled Landau level as a function
    of angular momentum $m\propto r^2$ (red filled, yellow empty levels).
    Near the edge, $m\sim N$, $r^2=R^2\sim N$, the confining potential breaks
    the degeneracy leading to a kind of Fermi surface
    in coordinate space. The radial shape of the density $\r(r^2)$
    is also shown.}}
\label{fig:edge}
\end{center}
\end{figure}

Fig. \ref{fig:edge} shows a radial profile of the density, where the
one-particle states are represented by dots.  A confining potential at
the edge of the system prevents the electrons from escaping.  Such a
potential breaks the Landau level degeneracy so the states will be
filled up to a Fermi energy and empty above it.

Thus, the quantum Hall droplet has a Fermi surface, or rather Fermi point,
but in coordinate space!  Following the steps of Sect. \ref{sec:luttinger},
we can linearize the energy of the
one-particle states around the ``Fermi angular momentum'' $m_F\sim R^2$,
\be{spec}
\eps(m)\sim\frac{v_F}{R}(m-R^2-\mu), \qquad \qquad R\to\infty,
\ee
where $v_F$ is the Fermi velocity  and
$\mu$ is the chemical potential to be determined later.
This corresponds to a relativistic dispersion relation,
$\eps(k)=v_F (k-k_F)$, where $k=m/R$ is the
one-dimensional momentum of waves propagating along the boundary circle.
The resulting $(1+1)$D chiral, relativistic particle is 
called a Weyl fermion. This theory is  a chiral version of the Luttinger
model discussed in Sect. \ref{sec:luttinger}.

Let us verify that the Landau electrons near the edge indeed
behave as a relativistic fermions.
For this we consider the Landau orbitals \eqref{Psinm} in the
combined limit \cite{CM2018}, 
\be{edge-lim}
m=R^2+m', \quad |m'|<R, \qquad\quad r=R+r', \quad |r'|=O(1),
\qquad\quad R \to\infty .
\ee
This corresponds to looking at a finite region of a few magnetic lengths
around the boundary in the limit of large droplet sizes.
In this limit, the angular momentum range is in the linear part of the
spectrum \eqref{spec}.
Using the Stirling approximation in the Landau wavefunctions
\eqref{Psinm}, we obtain,
\be{psi-edge}
\psi_{0, R^2+m'}\left(r,\theta\right)= {\cal N}
\frac{e^{i\left(R^2+m'\right)\theta}}{\sqrt{2\pi R}}\ 
e^{-\left(r'-\frac{m'}{2R}\right)^{2}}
\left(1+O\left(\frac{1}{R}\right)\right)\, ,
\ee
where the normalization constant is ${\cal N}=(2/\pi)^{1/4}$.
In this expression, one recognizes
the wave function for the $(1+1)$-dimensional Weyl fermion,
$\psi_{m'}(x)=e^{ik x}/\sqrt{2\pi R}$, with $k=m'/R$ and $x=R\, \th $.
Note also that edge states with $m'$ in the
range \eqref{edge-lim} are superposed within a distance of one magnetic length,
up to $1/R$ corrections. Thus the radial dependence is inessential and
can be neglected.

Therefore, in a suitable kinematic range,  the Landau level field operator
$\hat{\Psi}$ \eqref{field-op}, becomes,
\be{Weylfield}
&&\hat\Psi(R\th,t)\sim e^{i(R^2+\mu)\th} \hat\psi_-(x,t) ,
\nl
&&\hat \psi_-(x,t)=\sum_{m=-R^2}^{\infty}\frac{1}{\sqrt{2\pi R}}
\exp{\left[\frac{i}{R}\left(m-\mu\right)(x-v_F t)\right]}\;
\hat c_{m}.
\ee
In this expression, the energy dependence \eqref{spec} has been included and
the indexing of the Fock operators has been redefined by
$\hat c_{R^2+m} \to \hat c_m$.
So, if we set $v_F=1$ and $\mu=1/2$ (antiperiodic boundary conditions)
we recover, up to an overall phase, the expression for the
relativistic fermion field \eqref{f-modes} discussed in Sect.
\ref{sec:pointsplitt}

\subsection{Current algebras and conformal field theory -- a primer}
\label{sec:curr}

The Weyl fermion theory provides a simple example for introducing
$(1+1)$-dimensional conformal field theory (CFT) \cite{Mussardo}, and, 
before proceeding, we provide some background material on  conformal symmetry.

Massless theories, such as the Weyl fermion, do not involve any
dimensionful parameter and therefore the action \eqref{2dferm} is
invariant under scale transformations, or dilatations,  $x^\mu\to \l x^\mu$ and
$\psi\to\l^{-1/2}\psi$.
Conformal invariance amounts to a local extension of scale invariance 
where infinitesimal dilatations are independently chosen at any point.
In $(1+1)$ dimensions, this generalization is possible under the generic
assumptions of unitarity, local degrees of freedom and a local Hamiltonian.
Conformal transformations of the metric are given by local scale
factors $g_{\mu\nu}\to\l(x)g_{\mu\nu}$ which can be realized as follows:
In Euclidean flat coordinates, the line element can be written,\footnote{
  See App. \ref{app:calculus} for an introduction.}
\be{line-el}
ds^2=dx_0^2+dx_1^2=dz d\bar{z}, \qquad\qquad z=x_0+i x_1 \, ,
\ee
in terms of complex coordinates.
Thus, analytic coordinate changes, $z\to w$,
\be{gzz}
z=z(w), \qquad\qquad ds^2=
\left\vert\frac{dz}{dw}\right\vert^2 dw d{\bar w}
=2g_{w\bar w}\; dw d{\bar w},
\ee
 realize the conformal transformation of the metric  with $\l(x)=|dz/dw|^2$.
Note that the conformal symmetry is  much more powerful than
just scale invariance because it involves an infinite number of parameters
that characterize the analytic function $z=z(w)$.
Note also the factorization of the transformation in $z,\bar z$,
which, after continuation
to Minkowski space,  are the natural coordinates to describe 
right- and left-moving particles, respectively.

Since conformal transformations are  
analytic reparameterizations, for infinitesimal changes,
\be{conf-z}
z\to z+\eps(z), \qquad\qquad \eps(z)=\sum_{n=-\infty}^\infty \eps_n z^{n+1},
\ee
the transformation of functions reads,
\be{conf-f}
f(z)\to f(z+\eps(z))=
\left[1+\sum_{n=-\infty}^\infty \eps_n z^{n+1}\de_z\right]f(z) \, ,
\ee
which amounts to having the following (classical) generators and algebra,
\be{L-class}
L_n=- z^{n+1}\de_z, \qquad\qquad \left[L_n,L_m \right]=(n-m)L_{n+m} \, ,
\ee
which is the famous Virasoro algebra \cite{Mussardo}.

The stress-energy tensor $T_{\mu\nu}$ plays a central role in conformal
field theories. It expresses the response of
the theory to a varying background metric $g_{\mu\nu}$, 
\be{t-def}
\d S[\psi, g_{\mu\nu}]= -\frac{1}{2}\int d^2x\sqrt{g}\, T_{\mu\nu}\d g^{\mu\nu},
\ee
similarly to currents being responses to background gauge fields.

Coordinate reparameterizations,
  $x^\mu\to x^\mu+ \eps^\mu(x)$, correspond to the metric variations
  $\d g^{\mu\nu} = D^\mu \eps^\nu+D^\nu \eps^\mu $,
  where $D_\mu$ is the covariant derivative with respect to the metric, while
conformal mappings \eqref{gzz} are instead given by
$ \d g^{\mu\nu}=-\d \l(x)g^{\mu\nu}$.
Using \eqref{t-def}, we see that invariance of the action with
respect to these transformations implies the following conditions,
\be{class-conf}
D_\mu T^{\mu\nu}(x)=0, \qquad\qquad\qquad T_\mu^\mu(x)=0,
\ee
expressing (covariant) momentum conservation
and conformal invariance, respectively.

Again there are anomalies that violate these symmetries at the
quantum level. The lack of conservation is captured by the gravitational
anomaly \cite{Witten:GravAnom},
\be{a-grav}
D^\mu T_{\mu\nu}= \frac{c}{48\pi}\de_\nu {\cal R}, \qquad
 \qquad
T_\mu^\mu=0 \,,
\ee
and the non-vanishing trace by the conformal  (or
trace) anomaly,
\be{a-conf}
D^\mu T_{\mu\nu}=0, \qquad \qquad
T_\mu^\mu=-\frac{c}{24\pi} {\cal R} \; ,
\ee
where ${\cal R}$ is the scalar curvature of the background
metric. The coupling to gravity and its anomalies will be further
discussed in Sect. \ref{sec:heattrans}, and  our conventions for
curved-space calculus are summarized  in App. \ref{app:calculus}.
In the following we mention some basic facts.

As in earlier cases, the two expressions \eqref{a-grav} and \eqref{a-conf}
are related; by using different regularizations, \ie adding counterterms
to the effective action, the anomaly can be moved
from one law to the other, but cannot be eliminated completely.
The dimensionless universal number $c$ is the ``central
charge'' which characterizes the conformal theory.
In the case of chiral theories, there are two independent central charges,
$(c,{\bar c})$, one for each chiral component.\footnote{
  The expressions of anomalies given in \eqref{a-grav} and \eqref{a-conf}
  are only valid for $c=\bar c$. The general case will be discussed in
Sect. \ref{sec:heattrans}.}
This is due to the factorization of conformal transformations \eqref{gzz}
in independent analytic and anti-analytic reparameterizations,
$z=z(w)$ and $\bar z=\bar z(\bar w)$.
For example, the Weyl and Dirac fermions have
$(c,{\bar c})=(1,0)$ and $(c,{\bar c})=(1,1)$, respectively.  

Of course, many more things could be said about CFTs and the interested
reader can find extensive reviews in the literature, at different
level of sophistication: for the beginner, we already suggested
\cite{Mussardo}, a  modern standard references is \cite{Ginsparg},
and a comprehensive account is given in  \cite{DF}.

The CFT approach is algebraic, \ie  it is based on obtaining
correlation functions and observable quantities from the study of
representations of the Virasoro Algebra.
Deep mathematical results on the  representations
for interacting theories have made it possible to find 
 exact solutions of models with very interesting dynamics.

 In these lectures we shall, however, only consider the Luttinger
 model, also called the compactified boson in the CFT literature,
 which is a free theory where all results can be obtained rather
 easily, and which also describes interacting fermions through the
 method of bosonization. Below we shall explain how this can be used
 to describe the fractional charge and statistics of the edge
 excitations in the Laughlin quantum Hall states.

To continue the analysis of the $\nu=1$ edge excitations 
we define the Fourier modes of density and energy (hereafter,
we suppress hats on operators) as,
\be{rho-def}
\r_n &=&\int_0^{2\pi R} dx\; e^{-ixn/R}\; \psi^\dag_-(x,t)\psi_-(x,t)=
\sum_{k=-\infty}^{\infty} : c^\dag _{k-n} c_k : ,
\\\label{L-def}
L_n &=&R\int_0^{2\pi R} dx\; e^{-ixn/R}\; \psi^\dag_-(x,t)(-i\de_x)\psi_-(x,t)
\nonumber \\
&=& \sum_{k=-\infty}^{\infty} \left( k -\frac{n+1}{2}\right)
: c^\dag _{k-n} c_k : \, ,  
\ee
where we again put  $\mu=1/2$.
The stress tensor is a two-dimensional symmetric and traceless matrix
and therefore possesses two independent components
that correspond to the two chiral terms in the fermionic Hamiltonian
\eqref{2dham}.
One chiral component appear  in \eqref{L-def}, so the $L_n$ operators
are the stress-tensor Fourier modes which generate the conformal
transformations.
They will turn out to be  the quantum version of the classical
generators \eqref{L-class}.

In Sect. \ref{sec:pointsplitt} we showed that bilinears of the fields require
a regularization in the presence of an infinite Dirac sea,
and used the point-splitting method.
In this section, we shall use another technique called
normal ordering $:\dots:$, that
works as follows: We fill the Fermi sea up to an edge, which corresponds
to the $k=0$ state after the momentum relabeling \eqref{edge-lim}. 
This ground state satisfies the conditions,
\be{f-gs}
&&c_k\vert \W \rangle=0,\qquad k>0,
\nl
&&c_k^\dag\vert \W \rangle=0,\qquad k \le 0 \, .
\ee
Normal ordering amounts to putting $ c_k$ to the right of $c^\dag_k$
for $k>0$, and the other way around for $k\le 0$,
\be{norm-ord}
&&: c^\dag_k c_k : = - : c_k c^\dag_k : =c^\dag_k c_k,\qquad\ \ k>0,
\nl
&&: c^\dag_k c_k : = - : c_k c^\dag_k : =- c_k c^\dag_k,\qquad k \le 0,
\ee
while no conditions are imposed for unequal indices.
Once implemented in the definition of $\r_0$ and $L_0$ in \eqref{rho-def} and
\eqref{L-def}, this prescription removes the infinities in the
ground-state values, therefore, 
\be{gs-cond}
\r_0\vert \W \rangle= L_0\vert \W \rangle=0 \, .
\ee
Furthermore, we have:
\be{hws}
\r_n\vert \W \rangle= L_n\vert \W \rangle=0, \qquad n>0 \, ,
\ee
while for $n<0$  these operators create particle-hole excitations
by pulling electrons out of the filled Dirac sea. 

In Sect. \ref{sec:luttinger}, we have already derived the commutator
between the $\r_n$'s by analyzing what happens close to the Fermi
points. We now complement this by deriving the full algebra of the
$\r_n$'s and $L_n$'s using a ultraviolet regularization. These two methods give
the same result which demonstrates that the anomalous commutators have
both an infrared and an ultraviolet aspect. 

Using Fock space anticommutators, we find, 
\be{rho-alg}
[\r_n,\r_m]=\sum_{k=-\infty}^\infty c^\dag_{k-n}c_{k+m}-
\sum_{k=-\infty}^\infty c^\dag_{k-n-m}c_{k}=n\d_{n+m,0} \, .
\ee
The two sums in the r.h.s. cancel each other after shifting the summation
variable in the second term. However, for $m= -n$, the Fock operators have
the same index so we need the normal ordering \eqref{norm-ord}, which is
not translation invariant in $k$. Upon enforcing it, there remains the factor
$\delta_{n+m,0}\left(\sum_{k\le n} 1 -\sum_{k\le 0}1\right)=n\d_{n+m,0} $; 
thus, the non-vanishing commutator originates from a finite offset of
normal orderings. The same argument can be applied to the commutators involving
$L_n$. The resulting algebras are found to be \cite{CDTZ},
\be{curr-alg}
&& \left[ \rho_n,\rho_m \right]=n\delta_{n+m,0}\,,
\\\label{rhocom}
& &\left[ L_n,\rho_m \right]=-m \rho_{n+m}\,, 
\\\label{rholcom}
  & &\left[ L_n,L_m \right]=(n-m)L_{n+m}+\frac{c}{12}
       n(n^2-1)\delta_{n+m,0}\, ,\qquad\quad c=1\, .
\ee
The first relation is the current algebra  for the
generators $\r_n$, already introduced in Sect. \ref{sec:luttinger};
the third expression is the Virasoro algebra  \eqref{L-class} realized in
the fermionic Fock space. Note the additional $c$-number term in the
right-hand side
of the algebra, which is the so-called central extension.
This is a quantum addition to the algebra, that follows from
the conformal anomaly; actually, the coefficient $c$, called
central charge, is the same as the parameter 
in Eqs. \eqref{a-conf} (see App. \ref{app:calculus}
for the proof of this statement).
Note that in the Dirac theory, there is a corresponding set of
commutation relations  for the other chirality,
with central charge $\bar c=1$.

Let us also mention a third derivation of the commutation relations
\eqref{curr-alg} that, like the  one in Sect. \ref{sec:luttinger}, 
does not rely on shifts of summation indices in \eqref{rho-alg}.
Consider the expectation
value of the density commutators, and impose the ground state conditions
\eqref{hws} as well as their Hermitian conjugates,
$\langle\W|\r_n=0$, $n\le 0$, (recall $\r_n^\dag=\r_{-n}$), to obtain,
\be{rho-norm}
\langle \W |[ \r_n,\r_m]|\W\rangle=
\d_{n+m,0}\langle \W | \r_n\r_{-n}|\W\rangle=
\d_{n+m,0}n \langle \W |\W\rangle, \qquad n>0 \, . 
\ee
In these equations, we used the definition \eqref{rho-def} and the Fock-space
conditions \eqref{f-gs}.
This calculation can be extended to $\bra{\W}L_n L_{-n}\ket{\W}$ and 
$\bra{\W}L_n\r_{-n}\ket{\W}$, $n>0$, showing that the anomalies are uniquely
determined once
the ground-state conditions are fixed; they depend not only on
the algebra but also on the norm of the Hilbert space.

To summarize this section: The conformal field theory of the Weyl fermion
is characterized by central charges $(c, {\bar c})=(1,0)$.
The states in the theory can be organized into representations of
the algebras \eqref{curr-alg} -- \eqref{rholcom}, whose
central terms are  due to chiral and conformal anomalies.
The ground state $|\W\rangle$ is the bottom state of one
representation, characterized by the conditions
\eqref{gs-cond}, \eqref{hws}. Other states are
the particle-hole excitations,
\be{tower}
|\{n_i\},\W\rangle=\r_{-n_k}\cdots\r_{-n_2}\r_{-n_1}|\W\rangle,
\qquad n_k\ge\cdots n_2\ge n_1>0,
\ee
that  form an infinite tower of states.
Charged excitations, with some net number of particles or holes,
correspond to other representations. 
For example, for charge $Q=n>0$ 
the bottom state is $| n \rangle= c^\dag_1c^\dag_2\cdots c^\dag_n |\W\rangle$,
satisfying,
\be{hn-cond}
&&\r_0|n\rangle= n| n \rangle, \qquad L_0|n \rangle=\frac{n^2}{2}|n \rangle,
\ee
together with the conditions \eqref{hws}, as readily obtained from
the Fock space expressions \eqref{f-gs}. Note that the $L_0$ eigenvalue
 by definition gives the energy of the charged excitation within the
linear relativistic spectrum.
The tower of particle-hole excitations has energies $n^2/2+\sum_k n_k$.
From the Hamiltonian  \eqref{L-def} it follows that this is the
original Landau level angular momentum, \ie
$z\de/\de z\to -i \de/\de\th$.  
Therefore, $L_0$ also measures the angular momentum of the edge excitations.

\subsection{Hall current and the chiral anomaly}
\label{sec:hallcurr}

\begin{figure}[t]
\begin{center}
\includegraphics[width=0.6\textwidth]{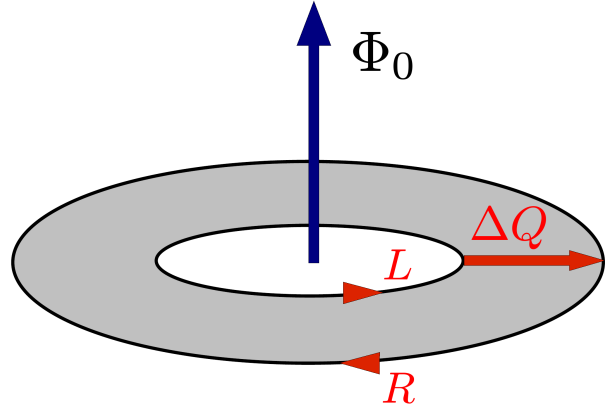}
\caption{\small{Laughlin flux argument: the adiabatic insertion of a
    flux quantum $\Phi_0$ inside the annulus moves a charge $\D Q=1$
    from the inner to the outer edge.}}
\label{fig:flux}
\end{center}
\end{figure}

Having established that the edge excitations in the quantum Hall
state are described by the chiral $(1+1)$-dimensional fermion, let us
explain the consequences of the chiral anomaly in this context.
Consider the annulus shown in Fig. \ref{fig:flux}, which is a
convenient geometry for describing the Hall currents. The outer edge
harbors a chiral fermion, and the inner boundary hosts one of opposite
chirality.  Together they can be viewed as the two components of a
Dirac fermion, although they are spatially separated.

Let us recall the spectral flow in the formulation in
Sect. \ref{sec:infinitehotel}.  Upon adiabatically inserting one unit
of flux in the hole of the annulus the spectrum returns to itself, but
one electron is removed from the left-moving Dirac sea and one is
added to the right-moving sea, so effectively one electron has moved
from the inner to the outer edge,
\be{QHE-flow}
\D Q= \D Q_R= -\D Q_L= \frac{\Phi(T)-\Phi(-T)}{\Phi_0}=1 \, .
\ee
Physically, the  magnetic flux insertion induces an
azimuthal electric field $E_x$, which in turn generates a radial Hall current
$I=2\pi\s_H d\Phi/dt$.
Therefore, we have shown that the spectral flow of edge states
corresponds to the charge transport by the $\nu=1$ Hall current!  From
the perspective of spectral flow, the bulk Hall current appears as a
collective radial motion of electrons, each one moving from one
orbital to the next -- the only difference with respect to the
relativistic case is that the two Dirac seas are joined at the bottom
into a single Fermi sea.

This reformulation in terms of the anomaly brings in an extra bonus.
As stressed in Sect. \ref{sec:antop},
the integrated anomaly is proportional to a topological
 number, the first Chern class, which takes  integer values.
 This correspondence provides a  profound explanation for the
 robustness and universality  of the quantization of
the Hall conductance. For example, it is independent of small
local fluctuations of the density and/or  the magnetic field,
impurities, lattice defects, \etcc
which explains the remarkable experimental precision.   

Let us remark that the original Laughlin argument \cite{Laughlin} for
the exactness of the integer Hall conductance is basically the same as
the spectral flow just described, but is formulated in the context of
the non-relativistic Landau problem, and supplemented by an important
discussion of the effect of impurities that is crucial to understand
real quantum Hall devices \cite{Prange}.

We remark that there is no anomaly in the whole system: the charge
non-conservation at the two edges of the annulus is compensated by the
classical Hall current flowing radially. The apparent violation only
appears in the effective relativistic theory that considers just a
single edge.
The general mechanism of a quantum anomaly (edge) canceled by a
classical current in one dimension higher (bulk) is called ``anomaly
inflow'' \cite{callan1985anomalies}, and we shall return to this
phenomenon at several occasions in the context of the so called
bulk-boundary correspondence in topological states of
matter.\footnote{
  See section 6.2 and appendix D.4, in particular.}

In the pivotal work of Niu \etal \cite{niu1985quantized}, 
the Hall conductivity was also  related to a topological
quantity characterizing the non-relativistic Landau levels,
namely the first Chern class of the Berry phase acquired by
wave functions during the adiabatic flux insertion.
The two arguments of the anomaly in the edge theory and of the Berry phase
of wavefunctions are equivalent for the integer Hall effect.
However, for interacting electrons, as in the fractional Hall effect,
the anomaly argument generalizes most easily.\footnote{For a
  discussion of the Berry phase argument, see \cite{niu1985quantized}
  and \cite{kvorning2018topological}.}  It will be our first example
of how to use anomalies to characterize topological phases of
interacting matter.

\subsection{Fractional Hall states and Chern-Simons effective
  theory}\label{sec:fracQHE}

We now discuss some features of the fractional Hall effect at
the Laughlin filling factors $\nu=1/k \\=1/3,1/5,\dots$.
Before deriving the conformal field theory of the edge excitations,
we first briefly review the theoretical description of the bulk system in terms
of wave functions and effective low-energy field theory.

\subsubsection{The Laughlin wave function} \label{sect:laughlinwf}
For fractional fillings, the free electron system in the lowest
Landau level is highly
degenerate, so the existence of a gap is a non-perturbative
effect due to the Coulomb repulsion among the electrons.
Laughlin argued that the ground state is again an incompressible
fluid with constant density and gapful density wave excitations
\cite{Prange}.
He proposed the following ansatz wave function,
\be{Laughwf}
\psi(z_1,\cdots,z_N)=e^{-\sum_{i=1}^{N}|z_i|^2/2}
\prod_{i,j=1,i<j }^{N}\left(z_i-z_j\right)^k\ ,
\qquad\qquad \nu=\frac{1}{k} \; .
\ee
This wave function  minimizes the kinetic energy by  being entirely in
the  lowest Landau  level,  and  has low  Coulomb  energy  due to  the
$k^{th}$ power  of the  Vandermonde determinant  \eqref{Laugh-wf} that
effectively  keeps  the   electrons  apart;  since   $k$  is  odd,
\eqref{Laughwf} is fully antisymmetric.

The Laughlin wave function has several ``magic'' properties that are
difficult to explain in terms of one-particle states, because it
is an involved superposition of Slater determinants \cite{Dunne}.
Thus other methods must be used, and an important technique,
again due to Laughlin, is the mapping to
a two-dimensional plasma\cite{Prange}.
Here the idea is  that the norm of the wave function,
\be{plasma}
\Vert\psi\Vert^2 &=& \int \prod_{i=1}^N d^2z_i\; \exp\left(
  -\b\; \cal H_{\rm plasma} \right)\ ,
\nl
{\cal H_{\rm plasma}}&=&k^2\sum_{i<j}^N\log |z_i-z_j|^2 -
k \sum_i^N |z_i|^2, \qquad\quad \b =\frac{1}{k}\; ,
\ee
can be thought of as  the partition function of a classical two-dimensional
Coulomb plasma of particles with charge $k$  subject to
a uniform neutralizing background.
(Verify this for yourself using that the two-dimensional
Laplace operator can be written $\nabla^2 = 4\partial_z\partial_{\bar z}$.).

It is known that when the ``temperature'' of this ``analogous'' plasma
is low, $k\ll 70$, it is in a ``screening phase'' where the Coulomb
potential is dynamically screened over a range $O(\ell)$.  In the
original electronic system, this corresponds to the electrons forming
an incompressible liquid with density $\r_0=\nu/\pi=1/{\pi k}$, and
with a  gap to density (or sound) wave excitations. The lowest charged
excitations are vortices in the fluid which correspond to having terms
$k^{-1}\sum_i^N (z_i-\eta)$ in ${\cal H_{\rm plasma}}$ where $\eta$ is
the position of the vortex. This term depletes the plasma at $\eta$,
and describes a particle with fractional charge $e/k$. With a bit of
work one can also establish that the wavefunction acquires a
non-trivial phase $\exp(i\th)$, with $\theta=\pi/k$, when two such
vortices are exchanged, thus demonstrating that they obey fractional
exchange statistics \cite{Leinaas}. Such particles are
called anyons \cite{Wilczek-frac}. These and other properties of the
Laughlin wave function have been confirmed both numerically and
experimentally, as described \eg in \cite{jainbook,DasSarma-Pinczuk}.

\subsubsection{The hydrodynamic Chern-Simons theory}
\label{sec:CS}

The experimentally observed precision and universality in quantum Hall
systems, suggest the existence of an effective field theory, just as
in the description of critical phenomena \cite{Mussardo}.

Since the quantum Hall liquids are topological states of matter with a
bulk gap, the low energy theory should lack dynamical bulk degrees of
freedom, while still coding for quantized conductance, fractionalized
excitations and gapless edge modes. Such a ``topological'' field
theory can be derived, or rather guessed, from these properties
together with the relevant symmetries \cite{Wenbook}.

To do so, we first express the conserved   matter
current $J^\mu$, which describes excitations of the incompressible fluid,
in terms of a ``dual'' vector potential  $a_\mu$,
\be{CScurrent}
J^\mu = \frac{1}{2\pi}\eps^{\mu\nu\rho}\partial_\nu a_{\rho},
\ee
which by construction satisfies $\partial_\mu J^\mu=0$,
and is invariant under the gauge transformation, 
$a_\mu\to\a_\mu+\de_\mu \vf$.
Since  $a_\m$  parameterizes a current it is referred to as a
hydrodynamic gauge field, and is not to be confused with the
background electromagnetic field $A_\mu$. 

We now claim that the effective hydrodynamic action is, 
\be{Seff}
S_{\rm eff}[a;A]=-\frac{k}{4\pi}\int d^3x\,
\eps^{\m\n\r}a_\m\de_\n a_\r + \frac{1}{2\pi} \int d^3x\, 
A_\m\eps^{\m\n\r}\de_\n a_\r + \int d^3x\, {\mathcal J}_\mu a^\mu \; ,
\ee
and go on to motivate the individual terms. The first is the
celebrated Chern-Simons action,
\be{CSaction}
\mathrm S_{CS}[a] = \frac{1}{4\pi}\int d^3x\,
\eps^{\m\n\r}a_\m\de_\n a_\r \, ,
\ee 
which is gauge invariant and dominates
at low energy because it has one derivative less than the Maxwell term. 
The coupling constant $k$ is called the ``level'',
and should take integer values for the proper definition of the theory.
\footnote{This quantization is related to that of
  the monopole charge \eqref{eq-mono}, see Sect. 2.3 in \cite{Witten-lect}.}  

Fixing the gauge  $a_0=0$, the action
reduces to a symplectic form, $S\sim\int a_1\dot a_2 \sim \int p \dot q $,
and the Hamiltonian vanishes.
There are no propagating ``photons'' associated to
this gauge theory: the Chern-Simons action only describes global effects
that occur in the Hall system at energies below the bulk gap. 
Furthermore, it violates time
reversal (TR) and parity (P) invariance, as is pertinent for describing
quantum Hall systems. The properties of this theory will be discussed
in more detail later.

The second term in \eqref{Seff} is just the electromagnetic potential
coupling to the current \eqref{CScurrent}, and since the action is
quadratic in $a$ it can be integrated out using the equations of
motion to get the response action,
\be{S-ind}
S_{\rm eff}[A]=\frac{1}{4\pi k}\int d^3 x\,\eps^{\mu\nu\rho}
A_{\mu}\partial_{\nu} A_{\rho}\; ,
\ee
having neglected the last term in \eqref{Seff}. 
From \eqref{S-ind} we can calculate the electromagnetic charge and current by, 
\be{Jind}
J^{\mu}_{\rm ind}=\frac{\d S_{\rm eff}[A]}{\d A_{\m}}=
\frac{1}{2\pi k}\eps^{\mu\nu\rho}\partial_{\nu}A_{\rho},
\ee
that are induced by background variations and correspond to
  the response of the system.
Written in components, they read,
\be{CS-sigma}
\r_0= \sigma_H B\ , \qquad \qquad
J^i_{\rm ind}=\s_H \eps^{ij} E_j\ , \qquad\qquad
\sigma_{H}=\frac{1}{2\pi k }  \, .
\ee 
These expressions reproduce the desired results for charge density
and conductance if we identify the level number $k$ of the
Chern-Simons action with the $k$ in the Laughlin wave function.

To understand the meaning of the last term in \eqref{Seff}, consider
the equation of motion in the presence of a static source $\mathcal{J}^0$,
\be{CS-eq0}
\frac {1} {2\pi} b=\frac{1}{2\pi k} B +
\frac 1{k} {\cal J}^0 \equiv \rho_{em} \, ,
\ee
where $b (B)$ is the magnetic field of $a (A)$ field
and we used \eqref{CS-sigma}. Equation \eqref{CS-eq0}
shows that the charge of the current $\mathcal J$ is $1/k$ times
the electric charge, so we can identify it with the quasiparticle
current.

We also learn from \eqref{CS-eq0}
that a static quasihole with
fractional charge $Q=n/k$, represented by the source
$\mathcal{J}_0(x)=n \d^{(2)}(x)$,
generates a flux $b=n/k$ of the field $a_\mu$. 
Thus, when another quasi-hole with
charge $m/k$ winds around it we expect an Aharonov-Bohm phase,
and a calculation gives the braiding phase,\footnote{
Some care is needed not to get a factor of 2 wrong when doing this.}
\be{AB-phase}
2\th=\frac{2\pi n m}{k} \; ,
\ee
which by definition is twice the  ``exchange'' phase that determines
the quantum statistics \cite{Wilczek}.

These results show that the Chern-Simons theory \eqref{CS-eq0} also
describes excitations with fractional charge and fractional exchange
statistics, in agreement with the Laughlin theory.  Among the
excitations, the electron has $Q=1$ and statistics $\theta/\pi=k$, so
the level number $k$ is odd consistent with Fermi statistics.

A final important remark: The Chern-Simons action \eqref{Seff} does
not change in curved space, \ie in the presence of a background metric
$g_{\mu\nu}$, since the usual $\sqrt{g}$ factor in the measure of
integration cancels against an inverse factor coming from the
covariant form of the antisymmetric tensor $\eps^{\mu\nu\r}/\sqrt{g}$.
Since the action does not couple to the metric, it is invariant under
smooth changes of the geometry. In physical terms, this means
independence on lattice defects, strain, and impurities.

\subsection{Chiral Luttinger theory of edge excitations}
\label{sec:chilut}

In this section we show how the Chern-Simons effective action
allows us to derive the $(1+1)$-dimensional theory of edge
excitations also for fractional fillings. We shall then canonically quantize
this theory and explore the consequences of chiral and conformal anomalies
\cite{CDTZ}.

In the $a_0=0$ gauge, and in absence of $A_\mu$ background and sources,
the equations of motion of the Chern-Simons action \eqref{CSaction}
imply $a_i=\de_i \vf$.
In a disk ${\cal D}$, the action reduces to a
boundary term (${\cal C}=\de{\cal D}$), 
\be{S-edge0}
S_{\rm edge}=-\frac{k}{4\pi}\int_{\mathcal{C}} d^2 x\,\de_x\vf\de_t\vf \,
,\ee
where $d^2x\equiv dt dx$, with $x=r\th$ the circle coordinate. This is
just a symplectic form, that  defines the canonical brackets as discussed later.
The Hamiltonian is zero as expected, since
\eqref{S-edge0} derives from the topological Chern-Simons action.

We should  introduce a dynamics for $\vf$, and 
the simplest choice is to add a term quadratic in the momenta,
$H\sim \de_x \vf \de_x \vf$ to get the ``chiral Luttinger liquid''
\cite{Wenbook},
\be{S-edge}
S_{\rm edge}=-\frac{k}{4\pi}\int_{C} d^2 x\,\de_x\vf
\left(\de_t\vf+v_F\de_x\vf\right)\, ,
\ee
which is called ``chiral boson'' in the conformal
field theory literature \cite{Floreanini:1987}.
A simple physical way to derive the Hamiltonian in \eqref{S-edge} is to assume a
constant electric field $E_y = E$ transverse to the edge. For an edge
excitation with profile $h(x)$, the density is
$\rho(x) = \rho_0 h(x)$, with $\rho_0 = B/(2\pi k) $,
and the electrostatic energy density $\cal E$ is,
  \be{elstat}
{\cal  E}(x) =  \int_0^{h(x)} \!\! \!\! dy\, \rho_0 E y =  E \rho_0 
 \half h(x)^2= \frac {E}{ B} \pi k\; \rho(x)^2 = \pi k v_F  \rho(x)^2 =
  \frac {kv_F} {4\pi} (\partial_x\varphi)^2 \; ,
\ee 
where we wrote the electrostatic potential  $Ey$, the
drift velocity at the edge $v_F=E/B$ and expressed the density
in terms of the scalar field (see \eqref{bulk-q} below),
\be{edge-j}
\r(x)=-\frac{1}{2\pi} \de_x\vf\, .
\ee
It also follows that the conserved electromagnetic current is,
\be{emcurrent}
\qquad\qquad\qquad\qquad\qquad\qquad
J^\mu = -\frac 1 {2\pi} \epsilon^{\mu\nu} \partial_\nu \varphi \,
\qquad\qquad (\mu,\nu=t,x)\;. 
\ee

The equation of motion of the edge theory \eqref{S-edge} is,
\be{edge-eq}
\left(\de_{t}+v_F\de_x\right)\de_x\varphi=0,
\ee
which shows that excitations are indeed gapless and chiral
(we set the velocity $v_F=1$ in the following).
Note the residual gauge symmetry $\vf\to\vf +{\rm const.}$.
The Hamiltonian does not include a potential term because
it would gap the edge excitations.

We can summarize the  discussion so far by saying that the bulk gauge
degrees of freedom have become dynamical at the edge.  In the
following, we canonically quantize the theory \eqref{S-edge} and
derive the axial and conformal anomalies (later we shall understand
why this is possible in a bosonic theory).  We also clarify the issue of gauge
invariance of the whole system, made by bulk and edge.

As mentioned earlier, one important property of the scalar field
$\varphi$ is that it is periodic, being the phase of the $U(1)$ gauge field,
\be{g-u}
u\in U\left(1\right)\,, \qquad u=e^{i\varphi}\, ,
\ee
the period of $\varphi$ being $2\pi$. 
We shall temporarily consider the more
general period $2\pi r$, where $r$ is the so-called compactification radius
\cite{Ginsparg},
\be{compa}
\vf (x,t) \equiv \vf (x,t)+2 \p n r,\qquad\quad n\in\mathbb{Z} \, .
\ee
Let us rescale the coordinates $t\rightarrow R t$ and write $x=R\th$,
for simplicity. The field $\vf(\theta)$ 
maps the edge circle $\th\in (0,2\pi)$ into another circle
$\vf\in(0, 2\pi r)$. 
Using the solutions of the field equations we can expand it as,
\be{modes}
\vf(\th, t)=\vf_0 - \a_0 \left( \th -t \right) +
i \sum_{m\neq 0}\frac{\a_m}{m} \text{exp}\left( i m\left(\th -t \right)
\right),
\ee
where $\a_m^*=\a_{-m}$. Moreover, $\vf_0\equiv \vf_0+2 \p r$ in order to
satisfy \eqref{compa}.
Note that the field expansion contains both chiral waves and
``solitonic'' modes ($\vf_0, \a_0$) that are nontrivial for compact fields.

The action \eqref{S-edge} is first order in time-derivatives,
and thus already in Hamiltonian form, $S\sim\int p \dot q - H(p,q)$,
with the first term being the symplectic form. The way to quantize such a
  theory differs from the usual canonical procedure: Rather than
  dividing the phase space into coordinates and momenta, and imposing
  canonical quantization conditions, the commutators are obtained
  directly from the symplectic form. In our case, the detailed recipe,
  that is given in \cite{Floreanini:1987},\footnote{This method is
    very useful also for quantizing fermion and Chern-Simons actions
    which are also first order in time derivatives.} yields,
\be{canrel}
\left[\vf(\th,t),\partial_x\vf(\th',t)\right]=\frac{2\pi i}{k} \d(\th-\th')\; .
\ee 

Upon substituting the expansion \eqref{modes}, we obtain the following
commutation relations among the modes,
\be{comm-modes}
[\vf_0,\a_0]=\frac{i}{k}, \qquad\qquad [\a_n, \a_m]=\frac{n}{k} \d_{n+m,0}\, .
\ee
The quantum operators $\vf_0$, $\a_0$ and $\a_n$ act
in a bosonic Fock space, whose ground state $\ket{\Omega}$ is defined by,
\be{b-gs}
\a_n \ket{\Omega}=0, \qquad\qquad n \ge 0.
\ee

In the previous section, we saw that the Weyl fermion corresponds to a
conformal theory with central charge $c=1$. We shall find that
the chiral boson also has $c=1$ and, furthermore, that it can
describe the edge states at both integer and fractional fillings,
for different values of the parameter $k$. We first observe that
the $\a_n$ commutation relation \eqref{comm-modes} are the  same as the
current algebra relations \eqref{curr-alg}  obeyed by the modes $\r_n$
of the edge current in the Weyl fermion theory,
for $k=1$.
Our definition of the edge charge is also consistent with the
bulk expression \eqref{CScurrent} as follows,
\be{bulk-q}
Q=\int_{\cal D}\! \! \! dx^1dx^2\,J^0=-\frac{1}{2\p}\int_{\cal C}d\th\,a_\th=
-\frac{1}{2\pi}\left(\vf\left(2\pi\right)-\vf\left(0\right)\right)=\a_0 \, ,
\ee
so we can identify $\a_n=\r_n$ for all $n$.

The Virasoro generators are again introduced as the modes of the
Hamiltonian and are therefore quadratic in the $\a_n$,
\be{L-bose}
L_n = \frac{k}{2}\sum_{l=-\infty}^{\infty}:\a_{n-l}\a_l:\ .
\ee
This expression requires normal ordering, consistent with the
ground-state conditions \eqref{b-gs}, that amounts to setting the
$\a_n$ with positive index to the right of those with negative index.
This procedure only affects $L_0$, that becomes,
\be{L0-bose}
L_0 =\frac{k}{2}\a_0^2+k\sum_{l=1}^{\infty}\a_{-l}\a_{l}\,.
\ee
Using the commutation relations \eqref{comm-modes} and
paying attention to normal ordering, we can obtain the rest of
the current algebra relations \cite{CDTZ},
\begin{align}
    &  [L_n,\a_m]=-m\a_{n+m}, \\
    \label{Vir-alg2}
    &[L_n,L_m]=(n-m)L_{n+m} + \frac{c}{12}n(n^2-1)\d_{n+m,0},
      \qquad\quad c=1.
\end{align}
It is apparent that these have the same form as in the Weyl fermion,
thus showing that the chiral boson is a conformal field theory with
central charge $c=1$ and current-algebra symmetry. It is, however,
more general than the fermion theory, owing to the freedom to choose $k \ne 1$.

\subsection{Spectral flow and fractional Hall current}
\label{sec:specflow}

The minimal coupling of the chiral boson to the electromagnetic field
amounts to adding the following term to the action \eqref{S-edge},
\be{edg-int}
S_{int}= -\int_{\de \mathcal{C}}d^2x\,J^\mu A_\mu =
\frac{1}{2\p}\int_{\de \mathcal{C}}d^2x\,
\left(A_t\de_x\vf - A_x\de_t\vf\right)\, ,
\ee
where we used the expression \eqref{emcurrent} for the current and
added a minus sign for the electron charge. 
The equation of motion becomes, 
\be{eq-E}
\left(\de_t+\de_x\right)\de_x \vf=-\frac{1}{k}E^x \, .
\ee
Integration over space, $x\in [0,2\pi R]$, leads to charge non-conservation,
\ie to the chiral anomaly, 
\be{d-Q}
\de_t Q=-\frac{1}{2\pi k}\int_{\mathcal C} dx E^x \, .
\ee
We can now repeat the spectral flow argument and obtain, after the
insertion of $n$ flux quanta $\Phi_0$,
\be{frac-flow}
\D Q=\frac{1}{4\pi k}\int_{S^1\times S^1} \!\!\!\!\!\!\! d^2x\,
\eps^{\mu\nu} F_{\mu\nu}=\frac{n}{k}\, .
\ee
This charge variation corresponds to a Hall conductivity
$\s_H=1/(2\pi k)$, as seen from the perspective of edge physics.  The
result is again expressed in terms of the topological quantity given
by the first Chern class of the gauge field \eqref{eq-mono}, and is
exact in the low-energy limit in which the effective field theory was
derived.  Note that this limit is fully justified since the adiabatic
process occurs in the presence of a bulk gap. Furthermore, the charge
non-conservation at the edge is compensated by the bulk radial current
related to the anomaly inflow mechanism already explained for integer
fillings in Sect. \ref{sec:hallcurr}.

In conclusion, the chiral boson theory displays the chiral anomaly and
generalizes the anomaly inflow mechanism to fractional fillings.  It
is rather remarkable that the exact quantization of the Hall current
can be proven in interacting systems by using the effective field
theory approach and anomalies. Further exact results due to the
gravitational anomaly will be discussed later.
Finally note that restoring $\hbar$ in \eqref{frac-flow} we get
$\Delta Q \sim \hbar$ as expected.

That the bosonic low-energy theory, for $k=1$, reproduces the anomaly
present in the original fermionic theory is an example of the 't Hooft
anomaly matching conditions \cite{Hooft1980, coleman1982t}.  These put
important constraints on low-energy theories by requiring them to have
the same anomalies as the microscopic theories from which they derive.

\subsection{Excitation spectrum and bosonization of
  $(1+1)$-dimensional fermions}
\label{sec:bose}

To determine the spectrum of allowed charged excitations
in the bosonic theory, we must understand the quantization
of the solitonic modes in \eqref{comm-modes}.
Since $\vf_0$ is  $2\p r$ periodic, and $\vf_0$ and $k\a_0$ are
canonically conjugate, we have the quantization, 
$ k\a_0=n/r$, with $n\in\mathbb{Z}$.
Thus, there are two periodicities,
\be{period2}
\vf (2 \p,t)=\vf\left(0,t\right)+2 \pi r m  -2\pi \frac{n}{kr},
\qquad\qquad n, m \in \mathbb{Z}\; .
\ee
We require the two periods to be commensurable, then
$kr^2$ takes rational values, $kr^2=p/q$, with $p$ and $q$ coprime
integers \cite{CDTZ}. In the conformal field theory literature,
this choice is called ``rational compactification'',
because the spectrum of conformal dimensions contains
a finite number of fractional values, in agreement with the observed
critical exponents of two-dimensional statistical models \cite{DF}. Indeed,
from the expressions \eqref{L0-bose} and \eqref{b-gs}, we have
$L_0=k \a_0^2/2 = n^2/(2kr^2)=n^2q/(2p)$.
In the application to the fractional Hall effect,
we know that $r=1$ and $k$ is integer, so we should identify $p=k$ and $q=1$.
Thus, the spectrum of charged excitations is given by,
\be{alpha0}
\a_0=\frac{n}{k}, \qquad\qquad n\in \Z \; .
\ee

We conclude that the fractional values \eqref{alpha0} obtained by canonical
quantization of the bosonic edge theory match the charges
$Q=n/k$ found in the bulk Chern-Simons theory. Note that
an anyonic bulk charge corresponds to a solitonic mode
in the edge theory, which is another instance of the 
bulk-boundary correspondence.

As already explained in the case of the Weyl fermion, excitations
of the conformal field theory form infinite towers of states.
Each of them is characterized by a bottom state, called highest-weight state
in the CFT literature \cite{DF},
that actually is the ground state in the solitonic sector $\a_0$, 
\be{a0eig}
\a_0\left\vert\frac{n}{k}\right\rangle=\frac{n}{k}
\left\vert\frac{n}{k}\right\rangle ,
\qquad\qquad n\in\mathbb{Z} \, .
\ee
This state obeys $\a_n\ket{\a_0}=0$, for $n>0$, while its infinite
tower of particle-hole excitations are obtained by repeatedly acting with
$\a_{n}$,  $n<0$.
The bottom state is also characterized by a value of $L_0$,
\be{L0eig}
L_0\left\vert\frac{n}{k}\right\rangle=
\frac{n^2}{2k} \left\vert\frac{n}{k}\right\rangle \, .
\ee

Let us now verify that these angular momentum values reflect 
the anyonic statistics of the edge excitations, which is in agreement with the 
results for the bulk Chern-Simons theory.
We first need the field that creates a charged excitation
at a point on the edge. This is the so-called vertex operator,
which is  well known in conformal field theory \cite{Ginsparg},
\be{vertex}
V_n\left(z\right)= :\exp\left(in\vf\left(z\right)\right): ,
\ee
Note that powers of the field $\vf$ are well defined thanks to the normal
ordering prescription for  the $\a_n$ modes
in \eqref{b-gs}. The  commutation relation,
\be{charge-vert}
[\a_0, V_{n}(z)]=\frac{n}{k} \ V_n(z), 
\ee
is easily derived since 
\eqref{b-gs} implies $\a_0\sim(-i/k)\de/\de\vf_0$, and
shows that  the vertex operators indeed
describe the insertion of a fractional charge $Q=\a_0=n/k$
at the point $z=Re^{i\th}$ on the boundary.

Next we compute the two point function of vertex operators.
Using the mode expansion
\eqref{modes}, and the ground state conditions \eqref{b-gs},
we  get the correlator of two currents,
\be{j-corr}
\langle\W| \de_\th \vf(\th,0)\; \de_\chi \vf(\chi,0)|\W\rangle=
\frac{z w}{(z-w)^2}, \qquad\qquad z=Re^{i\th}, \ w=Re^{i\chi}\, ,
\ee
and upon integration in $z$ and $w$ we obtain the field correlator,
$\langle \vf(\th) \vf(\chi)\rangle= - \log(z-w)$. We now have the elements
to compute the vertex operator two-point function. 
By expanding the exponentials in  power series and using
the $\langle\vf\vf\rangle$ correlator for any pair of fields
\cite{Mussardo}, we get,\footnote{
  The experts of conformal field theory should excuse us for the little
  abuse of notation at this point.}
\be{V-corr}
\langle V_{n}(z)V_{n}(w) \rangle= (z-w)^{n^2/k} .
\ee
Upon setting the two points opposite to each other on the edge,
\ie $z=w e^{i\pi}$, and then moving
both of them around by an angle $\pi$, we can realize the exchange
of the two excitations in the plane.
In this process, the edge correlator \eqref{V-corr} acquires the 
phase $\exp(i\pi n^2/k)$, corresponding to the fractional statistics 
already found in the Chern-Simons theory. 

Another fundamental property implied by the expression \eqref{V-corr} is that
the vertex operators with $n=\pm 1$, for $k=1$,
represent Weyl fermion fields, as follows,
\be{bf-corr}
\psi\left(\th,t\right)\equiv V_{-1}=
\, :\exp\left(-i\vf\left(\th,t\right)\right):\,,
\qquad\quad
\psi^\dag\left(\th,t\right)\equiv V_{1}=
\, :\exp\left(i\vf\left(\th,t\right)\right):\,.
\ee
This is the bosonization of fermions in $(1+1)$ dimensions,
an exact map between the two theories,
that actually are two descriptions of the same Hilbert space of
states \cite{Ginsparg}.  We already saw that they
 have the same conformal charge $c=1$ and realize the same
 chiral algebra relations (cf. \eqref{curr-alg} -- \eqref{rholcom} and
 \eqref{comm-modes}, \eqref{Vir-alg2}).
 Also, the two theories (for $k=1$) share
 the same set of representations of this algebra, and the
partition functions and multi-point correlators are  also found to be equal.

It has been shown that the correspondence
extends to bosons at $k\neq 1$ and fermions with
current-current interaction (four-fermions coupling) \cite{Tsvelik}.
This relation is rather natural from the point of view of bulk Hall
physics, since the Laughlin state describes interacting fermions, and
the interaction extends to the edge \cite{Wenbook}.
The bosonic theory described in these sections has been
extensively applied to the dynamics of edge excitations and
has been important for interpreting many experiments \cite{DasSarma-Pinczuk}.
In particular, the fractional charges (first observed in Refs.
\cite{dePicciotto:1997, Saminadayar:1997}), and the resonant tunneling
and fractional statistics recently observed \cite{Nakamura2020}.

We close this section with a couple of remarks.
\begin{itemize}
  \item
The bosonization of fermions in $(1+1)$ dimensions
makes it clear how the chiral anomaly is manifested  in a bosonic theory. 
\item
  In conformal theories involving both chiralities, \ie $c=\bar c$,
  the correlation functions of chiral vertex operators generalize to,
  \be{V-corr2}
\langle V_{n}(z)V_{n}(w) \rangle= (z-w)^{2 h} (\bar{z}-\bar{w})^{2 \bar h}\, ,
\ee
where $(h, \bar h)$ are the eigenvalues of $(L_0,\bar L_0)$.
The sum $\D=h+\bar h$ is called the conformal (or scaling) dimension,
because it gives the power-law behavior of the field under scale
transformations. The difference, $s=h-\bar h$, is an angular momentum
that is referred to as ``conformal spin'' in the CFT literature, or
``orbital spin'' in the context of quantum Hall physics.
\end{itemize}

\subsection{Thermal currents and the gravitational anomaly}
\label{sec:thermgrav}

We now discuss the coupling of the conformal theory of edge
excitations to a metric background.  While a general analysis of the
field theory response to deformations of the geometry will be
presented in Sect. \ref{sec:heattrans}, here we shall discuss a simple
physical consequence of the trace and gravitational anomalies
introduced in \eqref{a-conf} and \eqref{a-grav}.  The notations for
curved space calculus are summarized in App. \ref{app:calculus}.

We consider theories that have both chiral and antichiral
components and corresponding central charges $(c,\bar c)$:
The gravitational anomaly generalizing \eqref{a-grav} is \cite{heat2},
\be{chi-grav}
D^z T_{zz}= \frac{c}{48\pi}\de_z {\cal R}, \qquad\quad
D^{\bar z} T_{\bar z\bar z} =\frac{\bar c}{48\pi}\de_{\bar z} {\cal R}, 
\qquad\quad T_{z \bar z}=T_{\bar z z}=0 \, ,
\ee
where we use the complex Euclidean coordinates
$z=x_0+ix_1$, $\bar z =x_0-ix_1$ that allow for the factorization of
conformal transformations as explained in Sect. \ref{sec:curr}
(\emph{cf}. \eqref{line-el}  and \eqref{gzz}); the corresponding
vector indices are raised and lowered with the 
metric, $g_{z \bar z}=g_{\bar z z}=1/(g^{z\bar z})$ and
$g_{zz}=g_{\bar z\bar z}=0$ \cite{Ginsparg}.

The non-conservation laws \eqref{chi-grav} imply that the stress
tensor does not transform covariantly under conformal transformations,
but acquires an additive term proportional to the central charge
\cite{Ginsparg}. When mapping from the plane
to another geometry, like the cylinder having one periodic coordinate,
this additive term causes a non-vanishing ground state value,
that is  the source of the 1d Casimir effect
\cite{Mussardo}.

Next consider the edge theory at finite temperature $T$, by taking the
Euclidean time  periodic, with period $\b=1/(k_BT)$.
In the limit where the spatial period $R$ is much larger than $\b$,
the theory is effectively defined on a ``time cylinder''.
In this geometry, the ground-state expectation value of the stress tensor gives the energy density,
\be{Casimir}
{\cal E}(x)=\langle \W|T_{00}(x)|\W\rangle =\frac{ \pi c}{12 v_F\b^2}\, .
\ee
This result is very general as it holds for any chiral
conformal theory with central charges
$(c,0)$, $v_F$ being the velocity of edge excitations.\footnote{
  The proof of this result is rather technical and is given in
  App. \ref{app:calculus}.}
In a chiral theory, the energy density ${\cal E}$ is proportional to
the momentum
density, ${\cal P}$, by ${\cal P}=v_F{\cal E}$. Furthermore, 
in the presence of antichiral excitations there is another
contribution corresponding to
\eqref{chi-grav}, but parameterized by $\bar c$, that adds to the
energy and subtracts from the momentum,
\be{E-P}
{\cal E}(x) =\frac{ \pi k_B^2 T^2}{12 v_F} (c+\bar c),\qquad\qquad
{\cal P}(x) =\frac{ \pi k_B^2 T^2}{12 }(c-\bar c)\; .
\ee
In applications of conformal theories to statistical models, one generally
deals with parity invariant theories where $c=\bar c$; in this
case, ${\cal P}(x)$ vanishes and  ${\cal E}(x)$ determines the specific heat
by $c_V=\de{\cal E}/\de T$.
In the quantum Hall effect, it is rather natural to have edge
excitations with $c\neq \bar c$, implying a nonvanishing momentum
density that is a matter, or thermal, edge current
$J_T\equiv{\cal P}$, as we now explain.\footnote{
  The specific heat of the edge excitations is actually
  negligible in comparison with the lattice phonon contribution.}

While the Laughlin states, are purely chiral, with $\bar c=0$,
other prominent plateaus belonging  to the Jain series with filling fractions,
\be{Jain-series}
\nu=\frac{n}{2p n\pm 1}, \qquad\quad n,p=1,2,\dots ,
\ee
have, in general, both chiralities. 
As explained in App. \ref{app:jain}, these  states can be described
as  $n \ge 1$ filled ``effective Landau levels''.
 The edge excitations correspondingly possess $n$ branches:
for the $(+)$ sign in \eqref{Jain-series},
all branches are chiral, thus making
a conformal theory with $(c,\bar c)=(n,0)$.
For the $(-)$ sign, there is one chiral branch and $n-1$ antichiral branches
of neutral excitations, corresponding to central charges
$(c,\bar c)=(1,n-1)$.

\begin{figure}
  \begin{center}
\includegraphics[height=4.2cm]{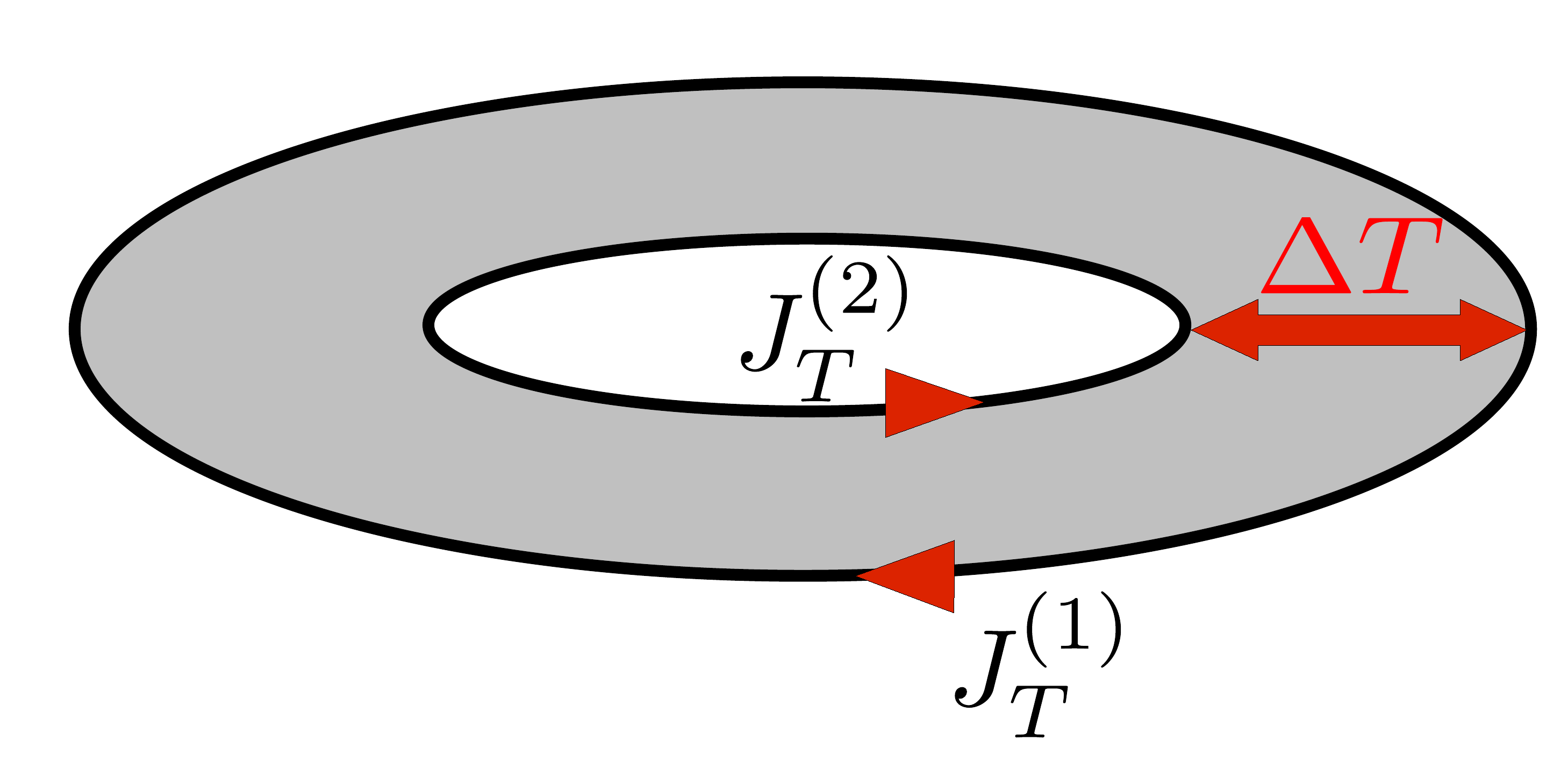}
\caption{Annular geometry with the two edges at different temperatures and
  their thermal currents.}
\label{fig:thermal}
\end{center}
\end{figure}

Let us consider the annular geometry of Fig. \ref{fig:thermal}, and assume
that the two edges are in thermal equilibrium. The expression \eqref{E-P} 
refers to one edge, say the outer one.
At equilibrium the currents on the two edges add up to zero,
$J_T^{(1)}(T)+J_T^{(2)}(T)=0$. If we  heat up one edge, there is a
disequilibrium and a net heat flow,
\be{delta-j}
\D J_T=J_T^{(1)}(T+\D T)+J_T^{(2)}(T)\sim \frac{\de J^{(1)}_T}{\de T}\D T=
\k_H \D T \, .
\ee
Using \eqref{E-P}, we find that the thermal Hall conductance $\k_H$
is \cite{heat1}\cite{heat2},
\be{kappa-H}
\k_H=\frac{\pi k_B^2 T}{6}(c- \bar c) \, .
\ee
Note that the thermal current is orthogonal to the
temperature gradient, just as the Hall current is orthogonal to
the potential gradient. 

We have shown that $\kappa_H$ is proportional to the gravitational
anomaly \eqref{a-conf}; this is another example of how anomalies
determine universal transport properties, thus measuring
in a unique way a fundamental property of
the underlying effective theory.
The result is very general and applies to any conformal theory
of quantum Hall edge excitations with specific central charges $(c,\bar c)$.

{\bf Experimental confirmations.}  The thermal current has been
measured and the results agree fairly well with the 
prediction \eqref{kappa-H}.
The experimental probe used in Ref. \cite{heat-exp} has a geometry
with four arms and corresponding edge currents,
that can be switched on and off.
One measures the power  $\D P$ dissipated by the thermal currents
flowing in $N$ edges,
\be{delta-pow}
\D P(N) = N \frac{\kappa_H T}{2}= N \frac{\kappa_0 T^2}{2} |c-\bar c|,
\qquad\qquad \kappa_0=\frac{\pi k_B^2}{6}\; ,
\ee
as well as the difference of powers $\l$ between measures with
different number of active edges $N_1$ and $N_2$, suitably normalized,
\be{delta-dp}
\frac{\l}{\D N}=\frac{\D P(N_2)-\D P(N_1)}{\D N \kappa_0/2}=
T^2 |c-\bar c|, \qquad\qquad \D N= N_1-N_2 \; .
\ee
This quantity is useful because the constant contribution by lattice phonons
cancels out. Note that the experiment cannot determine the sign of the
thermal current, \ie of $(c-\bar c)$.

\begin{figure}
  \begin{center}
\includegraphics[height=6.4cm]{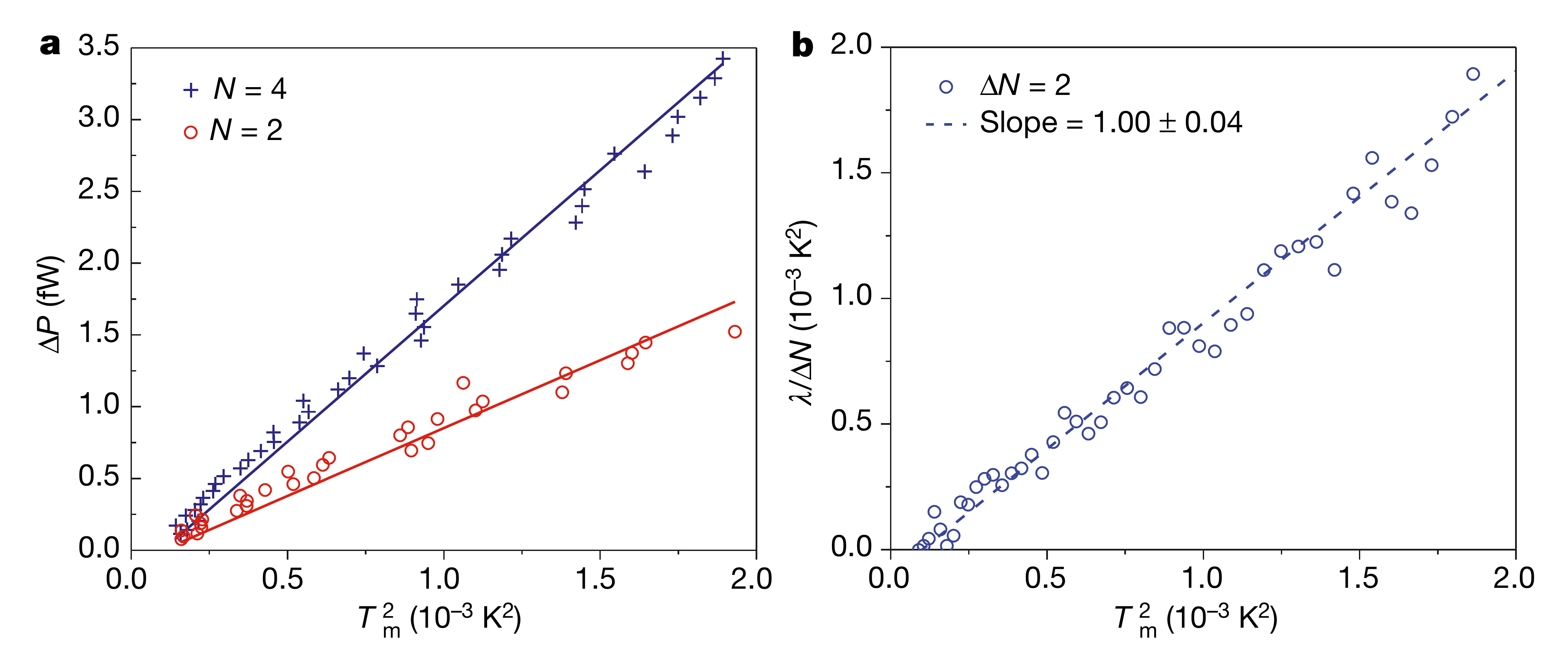}
\caption{Measurement of the thermal current for $\nu=1/3$
  \cite{heat-exp}: (a) Power dissipation $\D P$ as a function of $T^2$
  for $N=2,4$ active edges (currents), compared with the theoretical
  prediction \eqref{delta-pow} for one chiral mode;
  (b) Difference of power dissipations
  for $N=4$ and $N=2$ edges, and linear fit
  for measuring the slope $|c- \bar c|$ in \eqref{delta-dp}.}
\label{fig:heat1}
\end{center}

\end{figure}

\begin{figure}
  \begin{center}
  \includegraphics[height=6cm]{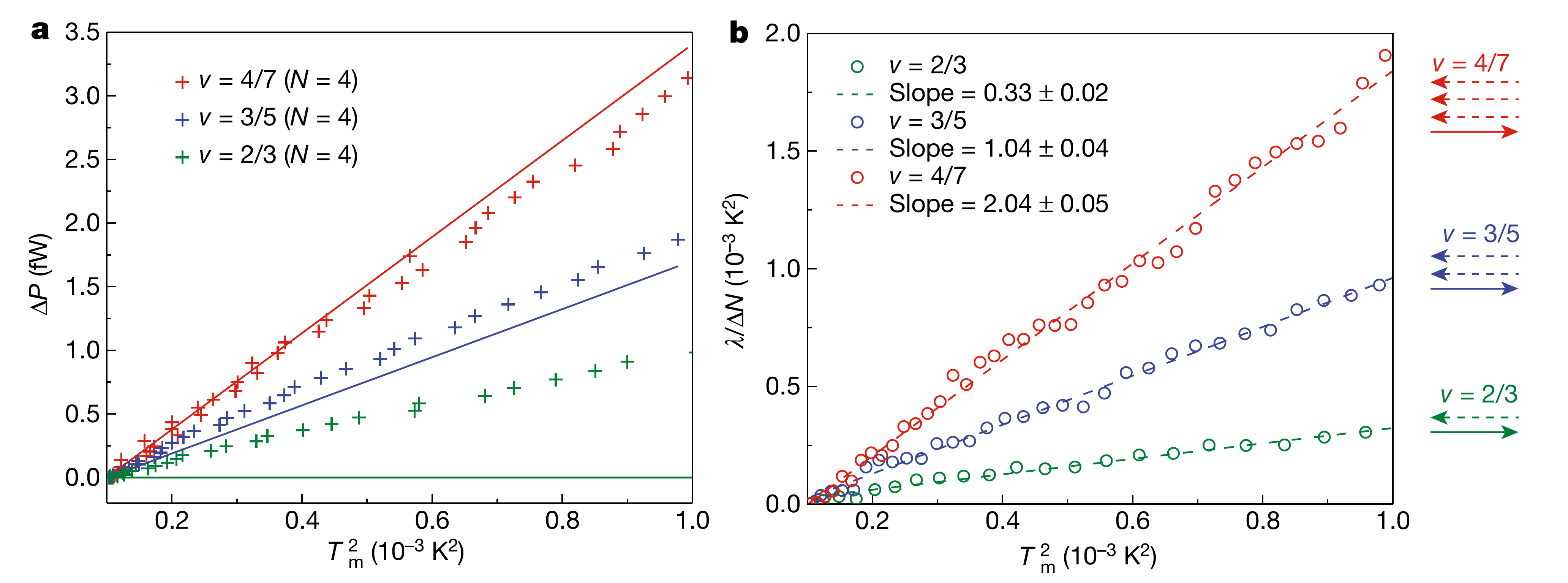}
  \caption{ Measurement of thermal current for Jain states with both chiral
and antichiral edge excitations at $\nu=2/3,3/5,4/7$ \cite{heat-exp}:
    (a) Power dissipation $\D P$ as a function of $T^2$ for $N=4$
    edges versus \eqref{delta-pow} with $|c-\bar c|$ predicted values for
    the up- and down-stream modes shown on the right; (b) Difference of power
    dissipations between $N=4$ and $N=2$ edges and linear fits determining
    the experimental $|c-\bar c|$ values in \eqref{delta-dp}.}
 \label{fig:heat2}
\end{center}
\end{figure}

The results for the Laughlin state at $\nu=1/3$ are shown in Fig.
\ref{fig:heat1}, where one sees that the experiment measures
$|c-\bar c| =1.00\pm 0.04$, in good agreement with the chiral Luttinger theory.
The measures for filling fractions belonging to the Jain series
\eqref{Jain-series}
are reported in Fig. \ref{fig:heat2}. In these cases, there can be 
negative (upstream) contributions to the current by the antichiral
modes. For $\nu=3/5$, for example, the theory predicts 
one chiral and two antichiral modes leading to $|c-\bar{c}|=|1-2|=1$;
the experimental result $1.04\pm 0.04$ confirms the expectations.
Other cases are also fairly well verified.
For $\nu=2/3$, there is a discrepancy between the predicted
$c-\bar{c}=1-1=0$ and measured $0.33\pm 0.02$ values, which might be
explained by the lack of thermal equilibration between the up- and
down-stream modes \cite{feldman}.


Let us finally add some remarks.
\begin{itemize}
\item
  The determination of the thermal Hall conductance by the
  gravitational anomaly is a genuine result of the conformal theory of
  edge excitations. In Laughlin and Jain states, each edge mode
  contributes $\pm 1$ to the central charge, so the expression
  \eqref{kappa-H} reduces to a mode counting. However, non-trivial
  fractional values are found in edge theories involving excitations
  with non-Abelian fractional statistics, for example,
  $(c, \bar c)=(3/2,0)$ for the Pfaffian state at $\nu=5/2$, including a
  bosonic edge mode and a neutral Majorana fermion \cite{Pfaffian}.
\item
  The measurement of the thermal current is particularly
  interesting in cases where there are several competing theories for
  the same filling fraction, as for example at $\nu=5/2$ where
  two other theories have been proposed beside the Pfaffian
  state.  Unfortunately, the measures of the thermal current are not
  yet definitive \cite{feldman}.  The determination of the
  correct theory could confirm the presence of non-Abelian anyons.
  It has been proposed that using them, one could model qubits that
  are decoherent-free due to their topological nature and realize
  a platform for quantum computations \cite{TQC}.
\item
  The anomaly inflow,  \ie the compensation of the edge anomaly
  by a bulk current, also takes place in the gravitational case as
  discussed in Sect. \ref{sec:heattrans}.
\item
  The spacetime integral of the trace anomaly \eqref{a-conf} is
  again related to a topological invariant quantity: Actually,
  the integral of the scalar curvature ${\cal R}$ gives the so-called
  Euler characteristic,
\be{chi}
\chi=\frac{1}{4\pi}\int_{\cal M} d^2x \sqrt{g}{\cal R}=2-2g-b= n_F-n_E+n_V.
\ee
This integer counts the number of handles $g$ and boundaries $b$ of
the two-dimensional surface ${\cal M}$, as well as the alternating sum
of faces (F), edges (E) and vertices (V) in a triangulation of the
surface \cite{Nakahara:Geometry}.
\end{itemize}

\section{Anomalies from perturbation theory} \label{sec:pert}

Although the method to calculate the axial anomaly described in
Sect. \ref{sec:II} has a great pedagogical value, and can with some
extra effort be generalized to higher spacetime
dimensions\cite{nielsen1991}, it cannot in any obvious way be used to
calculate non-Abelian or gravitational anomalies.  We now describe how
to compute anomalies from perturbation theory using Feynman diagrams.
Later we will return to the infinite hotel in the 4D.

\subsection{ 2D anomalies from perturbation theory}
\label{sec:perturb}

We shall make the calculations in Euclidean space-time,
and for this we need some notation and conventions ($D=d+1$),
\be{conv}
x\equiv x_E^\mu  = (\vec x, ix^0) = (\vec x, x^D) \ ,\qquad\qquad
  p\equiv p_\mu = (\vec p, -i p_0) = (\vec p, p_D)=
-i( \vec\partial, \partial_D) \, .
\ee
With the conventions given in App. \ref{app:notations}
  we write the Euclidean action of a Dirac fermion in the presence of
  the gauge backgrounds $A_\mu$ and $A_{5\mu}$ as follows,
\be{euclact}
S_E =  \int dx_E \, \bar\psi (i\slashed D_E  + im ) \psi  \, ,
\ee
where $
i\slashed D_E = \gamma^\mu_E (i\partial_\mu + A_\mu +A_{5\mu} \g^{D+1}) 
$
is an Hermitian operator, and $dx_E$ is  the D-dimensional Euclidean measure
(when there is no risk of confusion we will just write $dx$).
We also included the mass for future reference.\footnote{
  Note that $(i\slashed\partial - im)(i\slashed\partial + im)
  = -\partial^2 + m^2 = p^2 + m^2$ which is positive definite.} 

The quantization of the fermion field is defined by the path integral,
\be{partf2}
Z[A,A_5] = e^{-\G[A,A_5]  }   =
\mathcal N \int {\mathcal D}\psib {\mathcal D}
\psi \, e^{-S_E [\psi,\psib,A,A_5] },
\ee
where the normalization constant is given by,
$\mathcal {N} = 1/Z[0,0]$.
The quantity $\Gamma[A,A_5]$ gives the response of the quantum system
to varying backgrounds.  It is the same as $S_{\rm eff}[A]$
\eqref{S-ind} introduced in the previous section; both are 
called ``effective actions'', but condensed matter and particle physics
papers often use different notations. In the following, we shall
use both $S_{\rm eff}[A]$ and $\Gamma[A]$
for this response action, adapting to the standard conventions of the
topic discussed.

\begin{figure}  \centering
\includegraphics[height=2.8cm]{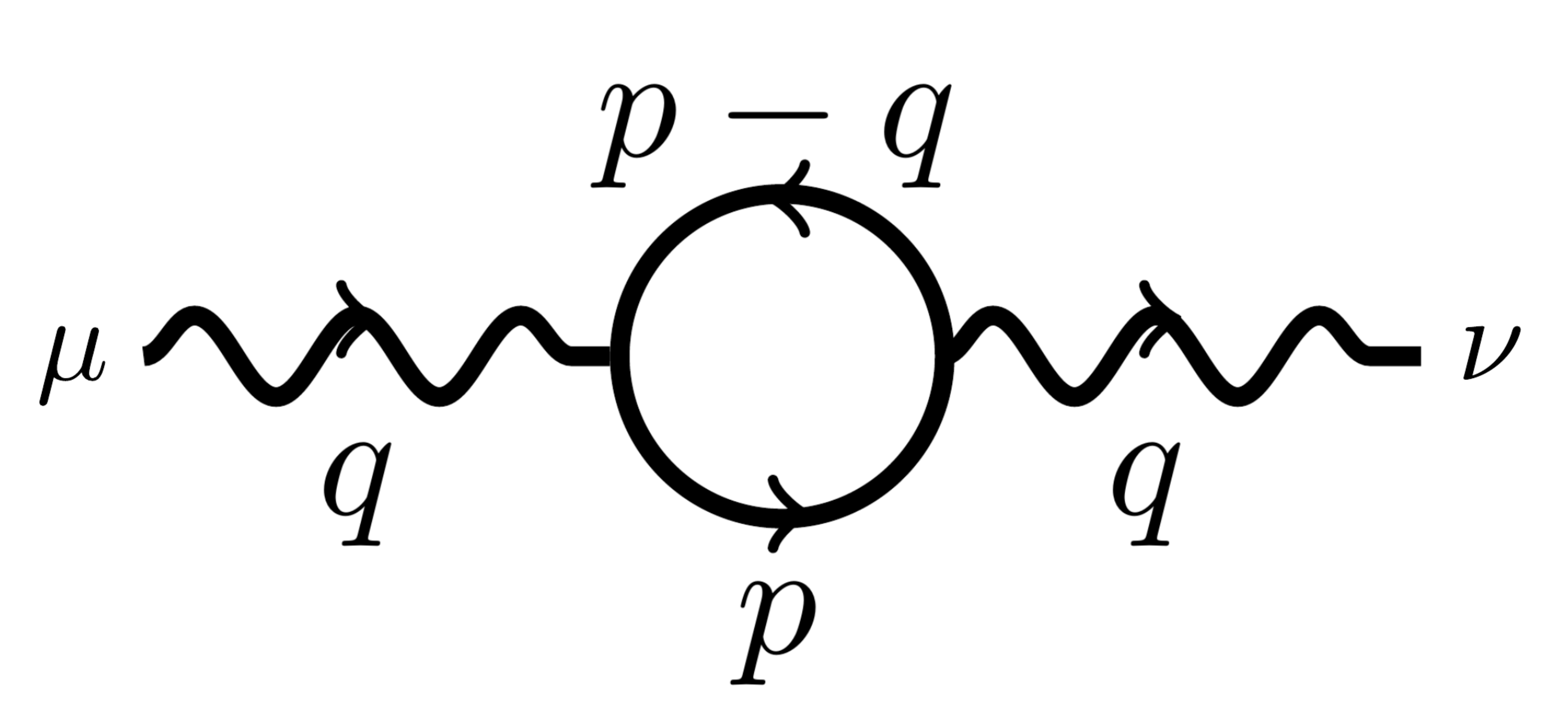}  
\caption{Feynman diagram for the 2D polarization tensor $\Pi_{\mu\nu}(q)$.}
\label{fig:polten}
\end{figure}

We now proceed to calculate the effective action
$\Gamma[A, A_5] $ in two dimensions to quadratic order in the gauge fields.
This is obtained by expanding the logarithm of the fermionic determinant,
as follows,
\be{2deffact}
\Gamma[A,A_5 ] &=& \Tr\left[- \ln\left(i\slashed\partial - \slashed {
  A} -\slashed {A}_5 \g^3\right) + \Tr\ln( i\slashed\partial) \right]=
\Tr \left[\ln \left( 1 - \frac 1 {i\slashed \partial}
  ( {\slashed A}_\mu +\slashed {A}_{5\mu} \g^3)\right)\right]
\nonumber \\
&=& \half \int \frac {d^2q}{(2\pi)^2}
\left( A_\mu (q) +\tilde A_{5\mu}(q)\right)
\Pi^{\mu\nu}(q)
\left( {A}_\nu (-q) +\tilde A_{5\nu}(-q)\right) + \dots\; ,
\ee
where the term
linear in the potentials vanished because of the gamma matrix trace.
We also used the two-dimensional form of the Euclidean gamma matrices, \ie
the Pauli matrices, for expressing
$\g^3\g^\mu=i\eps^{\mu\nu}\g_\nu$, such that the axial field is
expressed in terms of $\tilde A_5^\mu=i\eps^{\mu\nu}A_{5\nu}$.
In momentum space the polarization tensor is given by the Feynman
integral ($m=0$),
\be{polten}
\Pi^{\mu\nu} (q) = \int\frac {d^2 p} {(2\pi)^2}
\frac{\mathrm{tr} [\gamma^\mu \slashed p \gamma^\nu (\slashed q +
  \slashed p)]} {p^2 (p -q)^2}\; ,
\ee
as illustrated in Fig. \ref{fig:polten}.
The loop integral is logarithmically divergent in the ultraviolet,
and must be regulated. 
The normal procedure is to add a local counter term to $\G[A,A_5]$
that removes the singularity, that is actually independent of the external
momentum, for dimensional reasons. One finds,
\be{polten2}
\Pi^{\mu\nu} (q) = \frac 1 {\pi} \left( c \d^{\mu\nu} -
  \frac {q^\mu q^\nu} {q^2} \right) .
\ee
In this expression, the constant $c$ expresses the freedom to
adjust the subtraction by a finite amount,\footnote{Not to be confused with the
  central charge of the previous section.} after removing the infinite part,
it is a so-called choice of renormalization conditions that fully
specify the renormalized quantities \cite{nairbook}. In general, this is
dictated by some physical requirement.
It is apparent that the effective action $\G$
\eqref{2deffact} cannot be simultaneously gauge invariant
with respect to both the vector and axial fields.
Vector gauge symmetry would require $c=1$, \ie a transverse polarization,
while axial gauge invariance needs $c=0$
(Note the two-dimensional identity
$\eps_{\mu\a}p^\a\eps_{\nu\b}p^\b=p^2\d_{\mu\nu}- p_\mu p_\nu$).
We thus have the same two options already encountered
in the derivation of the anomaly by point-splitting of currents
in Sect. \ref{sec:pointsplitt}. 

The axial and vector currents are obtained by variation of the
effective potential, respectively $J^\mu=\d \G[A,A_5]/\d A_\mu$ and
$J^5_\mu=\d \G[A,A_5]/\d A_{5\mu}$.
Next we compute the divergence of these currents, adding a factor of $(i)$ 
for switching back to Minkowskian notation, as explained in
App. \ref{app:notations}.
With vector gauge invariance, choosing $c=1$ and $A_{5\mu}=0$, we find,
  \be{2dpertanom0}
\quad  \de_\mu J^\mu = 0 \, , \qquad \qquad\quad\ 
  \de_\mu J_5^\mu = \frac{1}{\pi}F \, , 
    \ee
 and with axial gauge invariance, choosing $c=0$ and $A_{\mu}=0$,
  \be{2dpertanom1}
  \de_\mu J^\mu =   \frac{1}{\pi}F_5 \, , \qquad\qquad
  \de_\mu J_5^\mu = 0 ,
  \ee
where $F=\eps^{\a\b}\de_\a A_\b \equiv E_x$  and
$F_5=\eps^{\a\b}\de_\a A_{5\b} \equiv E_{5x}$. (Recall that the electric
and axial charges are included in the definition of gauge fields).

In conclusion, the perturbative calculation reproduces the earlier
results for the 2D anomaly, \eqref{locan2} and
\eqref{locan4}, clearly showing that the anomaly depends on the choice
of renormalization conditions, and that we can preserve at most one
symmetry out of the two. In Sect. \ref{sec:mixed} we shall also discuss
the case of ``mixed anomalies'' when both vector and axial
fields are present.

\subsection{ 4D anomalies from perturbation theory}
\label{sec:4Dpert}

We now very briefly outline how the 4D anomalies come about in
the perturbative expansion.\footnote{Details of this analysis
  can be found in \cite{srednicki}. A comprehensive discussion
  of 4D Dirac theory is given in \cite{IZ}.}
A more detailed derivation will be given in the following section
using path integrals.

\begin{figure} \centering
  \includegraphics[height=5cm]{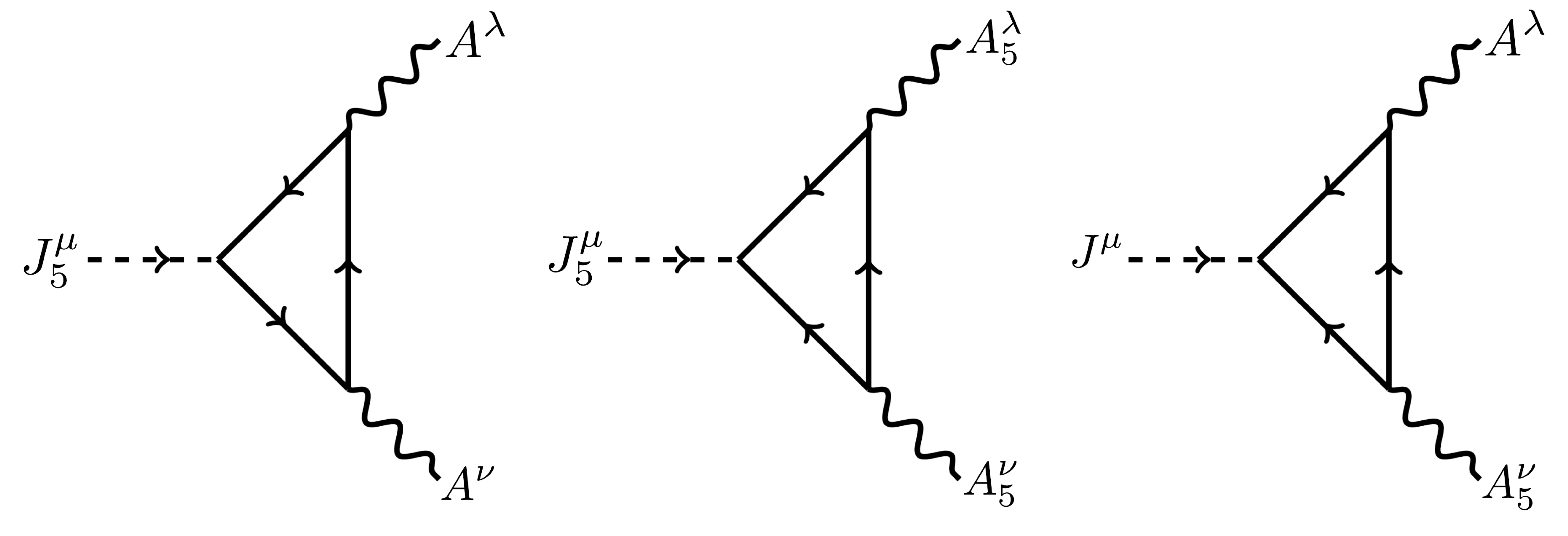}
  \caption{The two leftmost diagrams contribute to the axial current,
    while the third one is used to calculate the vector current. There
    are three more diagrams where the external $A$-lines are
    exchanged.}\label{fig:triangles}
\end{figure}

Let us first anticipate which Feynman diagrams could contribute to the
axial anomaly. As in two-dimensions, we want to calculate the vector and 
axial currents as functions of the gauge fields $A$ and $A_5$. Since
$\partial_\mu J_5^\mu$ is a pseudoscalar quantity, we need the
Levi-Civita tensor $\epsilon^{\mu\nu\sigma\lambda}$ for parity to work
out: its indices should be contracted with gauge invariant quantities,
so we expect the anomaly to be
$\sim \epsilon^{\mu\nu\sigma\lambda} F_{\mu\nu} F_{\sigma\lambda}.$
From this follows that the relevant diagrams must have one vertex with
the axial or vector current insertion, and two vertices with
insertions of $A$ and/or $A_5$.
They correspond to the third order expansion of the fermionic determinant
\eqref{2deffact}. The contributing ``triangle
diagrams'' are shown in Fig. \ref{fig:triangles} .

Note that diagrams with an even number of $\gamma^5$ insertions vanish
because of time reversal invariance, as dictated by Furry's theorem
(you should consult your QFT textbook for this!). Thus, as shown in
Fig. \ref{fig:triangles}, all diagrams have either a single or three
insertions of $\gamma^5$.

The Feynman amplitudes take the form,
\be{spinortrace}
\Pi^{\mu\nu\l}(q,k,p)\propto\int \frac{d^4r}{(2\pi)^4}\frac{ \tr{
    \Gamma^\mu (\slashed r + \slashed p ) \Gamma^\lambda \slashed r
    \Gamma^\nu (\slashed r - \slashed k) }} {(r+p)^2 r^2 (r-k)^2}\, ,
\ee
where $\Gamma^\nu$ is either $\gamma^\mu$ or
$\gamma^\mu\gamma^5$, for vector or axial-vector insertions, respectively.
By commuting two $\gamma^5$ matrices and
using $(\gamma^5)^2 =1$ we conclude that all diagrams give the
same contribution, although with different statistical factors.
The actual evaluation of the integral is a bit tricky since
it involves cancellation between linearly diverging terms.
In fact, there is no unique answer -- although the divergences cancel,
the finite result depends on the details of the regularization employed.
As in two dimensions, there is a freedom of finite counterterms
in the effective action, that are polynomial in the momenta.
A direct calculation using Minkowski signature gives\cite{srednicki},
\be{gtensor}
q_\mu \Pi^{\mu\nu\lambda}  &=&   \frac{-i}
{8\pi^2} (2c) \epsilon^{\nu  \lambda \alpha \beta} k_\alpha p_\beta\ ,
\nonumber\\
k_\nu \Pi^{\mu\nu\lambda} &=&  \frac{-i}
{8\pi^2}  (1-c) \epsilon^{\lambda \mu \alpha \beta} p_\alpha q_\beta\ ,
\nonumber  \\ 
p_\lambda\Pi^{\mu\nu\lambda}  &=&   \frac{-i}
{8\pi^2} (1-c)\epsilon^{ \mu \nu  \alpha \beta}
q_\alpha k_\beta \, ,
\ee
where the momentum $q$ flows into the vertex with the $\gamma^5$,
and $c$ is the free parameter. The choice $c=1$ amounts
to conserving the vector currents at the other vortices,
thus enforcing vector gauge invariance. Setting $A_5=0$ we get a
contribution from the first diagram in Fig. \ref{fig:triangles},
and using $q_\mu \rightarrow -i\partial_\mu$ \etc to go back to coordinate space one finally gets 
\be{anompert}
\partial_\mu J ^\mu = 0, \qquad\qquad
\partial_\mu J_5 ^\mu = \frac{1}{16\pi^2}\eps^{ \mu \nu  \alpha \beta}
F_{\mu\nu}F_{\a\b} = \frac {1} {8\pi^2} F_{\mu\nu}\tilde F^{\mu\nu}, 
\ee
where $\tilde F_{\mu\nu}=\eps_{ \mu \nu  \a \b}F^{\a\b}/2$ is
the dual field strength.
The result for both $A, A_5\neq 0$, with contributions from all 
three diagrams, will be discussed in Sect. \ref{sec:mixed}
and App. \ref{app:mixed}: This is the case of so-called mixed anomalies,
that requires a further analysis of the anomalous currents, and 
will be applied to the physics of Weyl semi-metals.

\subsection{ Decay of the neutral pi meson}

Here we shall give a short account of how a quantum field theory
anomaly for the first time was evoked to explain an observed
phenomenon, the decay of the neutral pion $\pi^0$ into two photons.
In this historical
excursus we shall assume some knowledge of strong interaction physics.

Before the advent of Quantum Chromodynamics (QCD), the non-Abelian gauge
theory of strong interactions, theoretical approaches used
effective field theories based on approximate symmetries and
phenomenological parameters. The first discovered of these symmetries
was the isospin $SU(2)_V\times SU(2)_A$ vector-axial symmetry,
in which up and down quarks $u$ and $d$ form a isodoublet.
The vector part is responsible for 
isospin conservation, while the axial symmetry is 
spontaneously broken. The corresponding three Goldstone bosons
are identified as the pseudoscalar pions, $\pi^a=(\pi^+,\pi^0,\pi^-)$,
in the account for their 
masses being much smaller than those of other mesons like $\rho$ and 
$\omega$. The vector and axial currents are exactly conserved
in the limit of vanishing masses, $m_u=m_d=0$ and $m_{\pi^a}=0$,
called ``chiral limit''.

The spontaneous symmetry breaking, means that the axial current
$\tilde J_{5\mu}^a$, $a=1,2,3$, which is the generator of $SU(2)_A$
transformations, does not leave the ground state invariant, but creates
a Goldstone boson, leading to the expression \cite{IZ},
\be{ssb} \langle \W|\tilde J_{5\mu}^a (p) |\pi^b \rangle
=-i f_\pi \d^{ab} p_\mu\, ,
\ee
where $\ket{\pi^a}$ is the pion state,
 $f_\pi \approx 93$ MeV is a phenomenological
parameter, and the momentum factor $p_\mu$ signals the Goldstone
nature of pions. Taking the derivative of \eqref{ssb} we can write a
operator relation between the pion field
and the divergence of the axial current,
\be{pcac}
\de_\mu \tilde J^{\mu a}_5(x) = f_\pi m_\pi^2 \pi^a(x) \; ,\qquad\qquad
(p^2=m^2_\pi),
\ee
where $\langle\W|\pi^a(p)|\pi^b\rangle=\d^{ab}$ is the pion field normalization.
Equation \eqref{pcac} is known as the 
partially conserved axial current (PCAC) relation,  since 
the current is only conserved in the chiral limit.

However, this spontaneous symmetry breaking argument overlooked the presence
of the axial anomaly, and for this reason we used the notation
$\tilde J^a_{5\mu}$ for the anomaly-free current. 
The PCAC relation should be modified\footnote{
  We use the $U(1)$ current for the neutral
  component of the pion triplet.} \cite{PhysRev.177.2426} as follows,
\be{pcac-anom}
\de_\mu \tilde J^{\mu}_5(x) = f_\pi m_\pi^2 \pi^0(x) +
\frac{e^2}{8\pi^2} F_{\mu\nu} \tilde F^{\mu\nu}
\ee
by including the contribution of the anomalous Feynman diagram on the
left in Fig. \ref{fig:triangles}, according to \eqref{gtensor} and
\eqref{anompert}.

The old PCAC relation \eqref{pcac} would imply the vanishing of
$\pi^0$ decay, $\G(\pi_0\to \g\g)=0$, in the chiral limit ($m_\pi=0$),
and even taking into account the non-vanishing mass, the value of $\G$
was three orders of magnitude smaller than the observed
value.\footnote{Weinberg gives a detailed discussion of this estimate
  in \cite{weinberg1995quantum}. }

The modified PCAC relation \eqref{pcac-anom} leads to the following
matrix element,
\be{pidecay}
\langle \g \g | \pi^0(p)|\W\rangle =\frac{1}{m_\pi^2 f_\pi}
\left\langle \g\g \right\vert\left(
ip^\mu \tilde J_{5\mu}(p) -\frac{Q^2}{8\pi^2} F\tilde F(p)\right)
\left\vert\W\right\rangle\; ,
  \ee
  where the anomaly term creates the photons and the contribution
  of $\tilde J^5_\mu$ is now negligible ($Q$ is the charge of the
  particle circulating the loop).
  A standard field theory calculation leads to the decay
  rate \cite{nairbook,weinberg1995quantum},
\be{pidecayrate}
\G(\pi_0\to \g\g) = S^2\frac{\a^2}{64\pi^3} \frac{m_\pi^3}{f_\pi^2}
\approx 7.63 \mathrm{eV}\; ,
\ee
where $\a=e^2/4\pi$. Note the factor $S=3[(2/3)^2-(1/3)^2]=1$ involving
the charges $2e/3$ and $-e/3$ of the up and down quarks circulating
the loop, the multiplicity $3$ due to color of $SU(3)$
Quantum Chromodynamics and a minus sign due to the isospin wave function
of $\pi^0$.

The good agreement with with the experimental value of
$(7.7 \pm 0.6)\ eV$, clearly demonstrated that anomalies are in fact
essential for understanding how symmetries are manifested and broken
in quantum field theories.

\bigskip
\emph{Historical note}: Tracing the early history of quantum
anomalies, the first paper we know of that clearly describe the axial
anomaly in 2D is by Ken Johnson \cite{JOHNSON1963253}. The first, and
arguably most important, papers with direct application to the $\pi^0$
decay were the ones by Bell and Jackiw \cite{belljackiw}, and
Adler \cite{PhysRev.177.2426}.
It is interesting to mention that the very first derivation
was made by Steinberger \cite{PhysRev.76.1180} long
before the discovery of quarks and QCD. He calculated a triangle graph
with the proton in the loop, and got the same result as from
\eqref{pidecay}. The reason is that the anomaly does not depend on the
mass of the fermion in the loop, and the factor 
is $S=1$ in \eqref{pidecayrate} for a single charge $e$ proton, just
as for the sum of the colored quarks.  If, however, quarks circulate
in the loop, the needed factor of $3$ is an experimental confirmation
of QCD having three colors.
Another early calculation was done by H. Fukuda and Y. Miyamoto,
  \cite{Fukuda}. These old papers preceded the
development of renormalization in quantum field theory and did not
offer a proper understanding of  the axial anomaly.

\subsection{The three-dimensional infinite hotel} \label{sec:3dhotel}

\begin{figure} \centering
  \includegraphics[width=\linewidth]{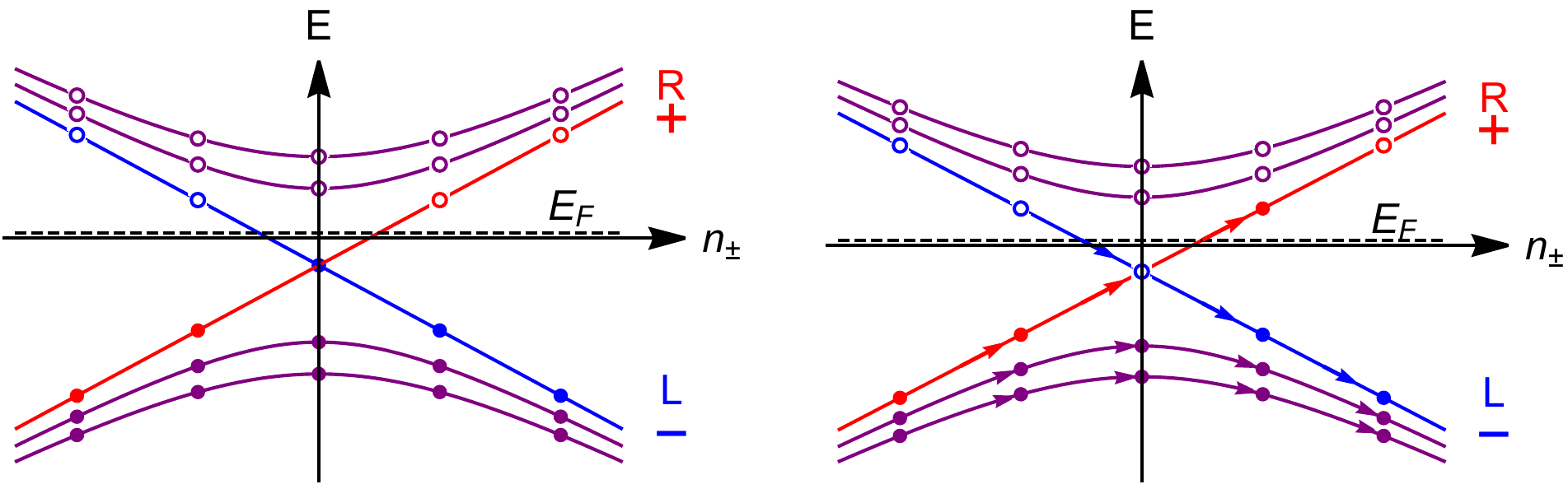}
  \caption{Left panel: Energy spectrum of the $3d$ infinite hotel with
    ground state filling. Right panel: Filling after
    insertion of one flux quantum with spectral flow shown by arrows.}
\label{fig:hotel}
\end{figure}

As promised, we will rederive the 4D axial anomaly using a spectral
flow argument. The
identity $F_{\mu\nu} \tilde F^{\mu\nu} = 4 \vec E \cdot \vec B$ 
tells us to look at the response of fermions for
parallel electric and magnetic fields.
More precisely, by assuming a constant magnetic field,
$\vec B = -B \hat z$ and then adiabatically turning on an electric
field in the same direction, we will have a situation similar to
that of the 1d spectral flow in Sect. \ref{sec:infinitehotel}.
We need the spectrum for the
Dirac particle in a constant magnetic field. To proceed, we recall
that the massless Dirac field can be separated into chiral components,
$\psi=(1+\g^5)\psi/2+(1-\g^5)\psi/2=\psi_L+\psi_R$,
by using the Weyl representation for the (Minkowski space) gamma
matrices,

\be{weylgamma}
\gamma^0 = 
\begin{pmatrix}
0 & -1 \\ -1 & 0
\end{pmatrix} \; ,\qquad
\gamma^i = 
\begin{pmatrix}
0 & \sigma^i \\ -\sigma^i & 0
\end{pmatrix} \; , \qquad 
\gamma^5 = 
\begin{pmatrix}
1 & 0 \\ 0 & -1
\end{pmatrix} \, .
\ee
Since we are interested in the spectrum, we use the matrices
$\alpha^i = \gamma^0 \gamma^i$, and write
the Dirac Hamiltonian as,
$
H_{Dirac} = \vec\alpha\cdot (\vec p + \vec A)
$.
Specializing to the case of a magnetic field in the $z$-direction,
and doing some manipulations of the Pauli matrices
as explained in App. \ref{app:infinitehotel}, we obtain the following
expression for the square of the Hamiltonian,
\be{hwsquare}
H_{Dirac}^2 = 
\begin{pmatrix}
  p_z^2 + (\vec q + \vec A)^2  - B\sigma^z & 0 \\
  0 & p_z^2 + (\vec q + \vec A)^2 -  B\sigma^z
\end{pmatrix} \, ,
\ee
where $\vec q$ is the momentum in the $xy$-plane and $\vec A$
the two-dimensional vector potential describing the magnetic field. 
Next note that $ (\vec q + \vec A)^2$ is the Hamiltonian for a 2d
non-relativistic particle with mass $m=1/2$ moving in a magnetic field
that gives the Landau levels discussed in Sect. \ref{sec:LL}, with
energies $\eps_n = \omega_c(n + \half)  = (2n + 1) B$. Thus the
energy eigenvalues of \eqref{hwsquare} are,
\be{lleigen}
E_{p_z, n, s} = \pm \sqrt{ p_z^2 + (2n + 1) B + sB },\qquad\quad
n=0,1,\dots,\ \ s=\pm 1,
\ee
where $s$ is the eigenvalue of $\vec \sigma\cdot \vec B /|\vec B| = s \sigma^z$.
This spectrum is shown in Fig. \ref{fig:hotel}. 
From \eqref{hwsquare} follows that the energy levels are doubly
degenerate since $2n - 1 = 2(n-1) + 1$, with the exception of the
$n=0, s=-1$ case, corresponding to the lowest Landau level.

The spectral flow has significant effects only for the eigenvalues
that cross the Fermi level $E_F$ under the adiabatic change of the
potential; thus we only have to consider the linearly dispersing modes
corresponding to $n=0$ and $s=-1$, as illustrated in the right panel
of Fig. \ref{fig:hotel}.
The mode with $E=p_z$ is right-moving, while the one at $E=-p_z$ is
left-moving. For antiperiodic
boundary conditions along the $z$-axis, the momentum is quantized as
$p_z=(n_z-1/2)/R$.
So for these modes, the situation is identical to the one-dimensional
spectral flow in Sect. \ref{sec:infinitehotel}, with charge density
$\rho_0 = B/2\pi$ of the lowest Landau level. Thus, introducing the
spatially constant potential $A_z(t)$, we get, in complete analogy
with \eqref{locan3} and \eqref{eq-hotel},
\be{3dhotel}
\dot Q_\pm = \mp R {\cal A}_{xy} \rho_0 \dot A_z(t) =
\mp \frac {Vol} {2\pi} \rho_0 E^z \; ,
\ee
where ${\cal A}_{xy}$ is the area in the $xy$-plane, and $Vol$ the volume.
Since the system is homogeneous we can write the chiral currents
(recalling that $B^z=-B$),
\be{3dhotel2}
\de_\mu J_\pm^\mu = \pm \frac 1 {4\pi^2} E^z B^z =
\pm \frac 1 {4\pi^2} \vec E \cdot \vec B =
\pm \frac 1 {16\pi^2} F \tilde F \, ,
\ee
whose sum and differences reproduce the anomaly \eqref{anompert}.

The spectral flow is moving an integer number of electrons in agreement with 
flux quantization. Let us verify that the normalizations are correct,
the total charge displacement in the adiabatic process
being, for each chirality,
\be{4dquant}
\D Q_\pm=\int d^4 x \de_\mu J^\mu_\pm=\pm \frac{1}{32\pi^2}\int_{\cal M} d^4x
\; \eps^{\mu\nu\a\b}F_{\mu\nu}F_{\a\b} = n \in \Z \; .
\ee
This completes the infrared derivation of the anomaly for 4D fermions. 

As in the two-dimensional case, the integrated anomaly in the r.h.s.
is expressed by a topological invariant quantity, a gauge-invariant
functional of the $A$ field taking integer values.  This
is called the second Chern class in the mathematical literature.  While
more technical aspects will be dealt with in Sect. \ref{sec:indexth}, let us
momentarily check its normalization.
The integration is over a compact space-time manifold $\cal M$,
the four-dimensional torus $S^1\times S^1\times S^1\times S^1$, 
because we took periodic boundary conditions in space and in time,
by identifying the points $t=-T$ and $t=T$, as explained in Sect.
\ref{sec:antop}.
The four-torus is actually the product of two two-torii, such that
the integration in \eqref{4dquant} splits in two orthogonal fluxes,
up to a factor of two for exchange symmetry. One then recovers the
square of the monopole charge (first Chern class) \eqref{eq-mono},
taking square integer values, as \eg one.
Therefore, the insertion of one flux quantum in each torus leads
to the spectral flow of one electron per chirality, as required.

\subsection{When are anomalies dangerous, and why?}

Before moving on, we make some
general comments about gauge theories.  We already briefly discussed
the physical meaning of vector and axial gauge fields, and mentioned
that anomalies might make theories inconsistent. Here we shall expand
on these points.

The notion of a local gauge symmetry is in fact misleading. A real
symmetry operation, like a space rotation, takes you from one
realization of a system to another, as \eg when rotating a sphere.  A
gauge transformation does not change the state of the system but only
changes the description. That there are different descriptions of the
same state means that certain apparent degrees of freedom are
redundant and thus not physical, such as the longitudinal and
time-like components of photons. If one starts from a gauge theory and
then adds terms that do not respect the local symmetry, such as
$A_\mu A^\mu$ in quantum electrodynamics, one completely changes the
theory. In this case the longitudinal photons become propagating, \ie
physical, and the ultraviolet behavior changes, which generically
makes the theory non-renormalizable.

``Spontaneously broken'' gauge symmetry is yet another concept prone to
be misunderstood.  Here the longitudinal gauge bosons indeed become
physical by the Higgs mechanism, but no new degrees or freedom
appear. Rather they are ``repackaged'', so that a degree of freedom in
a scalar field appear in guise of a longitudinal photon.
 
Here it is important to understand that gauge theories can be realized
in different phases. The electrodynamics in insulators is a gauge
theory realized in the Coulomb phase, with propagating transverse
photons. In a superconductor, electrodynamics is in a Higgs phase,
meaning that there are no gapless photons at low energy.\footnote{In
  non-Abelian theories, and also in ``compact'' electrodynamics, there
  is also the possibility of a confining phase.}

There is an important distinction between ``dynamical'' gauge fields
and ``external'' gauge fields.  We have discussed the chiral anomaly of
$(1+1)$-dimensional fermions and found that it has interesting
physical effects.  The gauge fields were, by necessity, treated as
classical background fields, because there are no photons in $(1+1)$
dimensions.  In higher dimensions, on the contrary, we have the option
to quantize the gauge fields or not. Let us discuss the consequences
of having anomalies in the two cases.

Classical, or ``external'', gauge field do not describe physical
degrees of freedom.  Sometimes they are just auxiliary quantities that
are used to calculate various responses.  For example, the metric
background can mimic the thermal response as well as the mechanical
stress on materials. Background gauge fields are widely employed in
condensed matter system.
In these cases, anomalies cause no problems, but actually characterize
the behavior of the system.  Several examples of this have already been
discussed.  Needless to say, any realistic theory of nature must
respect the experimentally verified conservation laws. For condensed
matter physics this means that we must regularize in a way that leaves
the electric current non-anomalous, even if we choose not to include
photons in our theory.

Dynamical gauge fields in $d \ge 1$ have a kinetic term in the action
and describe propagating degrees of freedom. The archetypical example
is the electromagnetic field, but we also have gluon fields and fields
related to the weak interaction. In condensed matter physics, it can
also happen that ``emergent'' gauge fields describing low energy
effects acquire dynamics.
Anomalies in currents coupled to dynamical gauge fields destroy the
gauge invariance, and, just as adding explicitly non-invariant terms to
the action, drastically change the theory. 

 In the Standard Model of fundamental interactions, chiral anomalies
 of currents coupled to dynamical gauge fields do vanish. This
 cancellation occurs between the contributions of several particles
 with different (in general non-Abelian) charges.  For an account of
 this see \eg \cite{nairbook}. In more general theories, such as
 string theory, anomaly cancellations provide important consistency
 checks. In condensed matter systems, the only fundamental 
 gauge symmetry is that of electromagnetism. There are however examples
 of ``emergent gauge symmetries'' which describe effective low energy 
 degrees of freedom. In this case, anomalies have to be absent, or cancel, 
 just as in the Standard Model.

 In condensed matter systems there are relevant examples of theories
 with unavoidable chiral anomalies, such as a single right-moving
 fermion at the $(1+1)$ dimensional edge of the quantum Hall state,
 as discussed in Sect. \ref{sec:chilut}. 
 In this and other examples to be discussed in following chapters,
 charge conservation is preserved due to the presence of a
 compensating current flow from an attached bulk.  This is the general
 mechanism that takes place at the boundary of various topological
 states of matter.

\section{Anomalies from path integrals}
\label{sec:pathint}

\subsection{The Schwinger model, and a path integral
   riddle} \label{sec:schwinger}

We shall start this chapter by presenting a paradox.  Let us consider
the Schwinger model, which is the one-dimensional Dirac theory in
\eqref{2dferm} coupled to dynamic electromagnetism,
 \be{schmod}
{\cal L} = \psib (i \slashed  \partial + e\slashed A)\psi -
\frac 1 4 F^2  \, ,
\ee
where $F^2 = F_{\mu\nu} F^{\mu\nu} = 2 E^2_x$. Without
transverse dimensions, there are no photons, only  electrons
moving on a line interacting
via a linear Coulomb potential (the solution of the one-dimensional Laplace
equation). This means that any state with non-zero charge has an
infinite energy, so that electrons cannot, even approximately, exist
as widely separated particles. The analogy with quark confinement
in 4D QCD is obvious, and the Schwinger model has been
used to gain insights about this problem.

In 2D we can always decompose the vector potential as,
\be{decomp}
eA_\mu = \partial_\mu \lambda + \epsilon_{\mu\nu} \partial^\nu \xi \, .
\ee
The first term is just a gauge transformation while the second encodes
the electric field, $E_x= -\partial^2 \xi$.\footnote{
  This is a special case of the Hodge decomposition which in general
  also includes gauge transformations that cannot be continuously
  deformed to the identity. These ``large gauge transformations'' are
  topological in nature, and in our case are precisely the ones
  discussed in Sect. \ref{sec:infinitehotel} when the theory is
  defined on a circle.}

We pick the gauge  $\partial_\mu A^\mu = 0$  by taking $\lambda = 0$
and  perform the local axial transformation, 
\be{schtran}
\psi \rightarrow e^{-i \xi (x)\gamma^3} \psi \, ,
\ee
under which,
\be{x}
\bar\psi i\slashed\partial \psi \rightarrow  \bar\psi i\slashed\partial \psi
+  (\partial_\mu\xi) \bar\psi \gamma^\mu\gamma^3 \psi = 
\bar\psi i\slashed\partial \psi  - \epsilon^{\mu\nu} (\partial_\nu \xi)
\psib \gamma_\mu \psi   =\bar\psi i\slashed\partial \psi -
\psib  e \slashed A \psi \, .
\ee
Substituting this in \eqref{schmod}, it looks like the Coulomb force
suddenly disappeared! Here you will presumably protest and say that
this is a purely classical consideration and that one has to be much
more careful in quantum field theory. The problem, however, apparently
remains if one uses the path-integral approach which tells us that all
correlation functions can be obtained from the expression,
\be{partf}
Z[J,\eta,\etab] = \mathcal {N}  \int \mathcal {D} A
\mathcal {D}\psi \mathcal {D}\psib
\,e^{-S_E + \etab\psi + \psib\eta + J^\mu A_\mu} \, .
\ee
Since this expression only involves classical variables,
it seems that our problem remains. 

The solution to this apparent paradox was given by
Fujikawa \cite{fujikawa1980} who showed that the measure
$d\mu=\mathcal {D}\psi \mathcal {D}\psib$ is not invariant under the
chiral transformation\eqref{schtran}, but picks up a dependence on the gauge
field.

In the next section we shall derive this result, and its
counterpart in 4D, and show how it is related to the
anomaly derived in the previous section. But first we shall finish the
story about the Schwinger model. To calculate how the finite
axial transformation \eqref{schtran} affects the measure, we refer
to the work \cite{roskies1981}. Here we just give the result,
\be{schmass}
d\mu \rightarrow e^{-\frac {e^2} {2 \pi} \int d^2x\, A_\mu A^\mu  } d\mu \, ,
\ee
where we restored the charge $e^2$ which in 2D has the dimension of mass.
Thus, picking the Lorenz gauge $\partial_\mu A^\mu = 0$, the
action for the gauge field becomes,
\be{schwact}
{\mathcal L}_{Sch} = - \frac 1 2  {A}_\mu \left( \partial_\mu \partial^\mu
  - \frac {e^2} \pi \right) A^\mu \; ,
\ee
which means that longitudinal photon, which was originally just a
gauge freedom, is reborn as a free boson with the ``Schwinger mass''
$m^2 = \frac {e^2} \pi $.
In the fermion theory, this boson is actually the particle-hole
excitation discussed in the context of bosonization in Sect. \ref{sec:bose}. 

Also,  by adding a fermion mass term to \eqref{schmod} we
get the ``massive Schwinger model'' which has very interesting
topological features \cite{COLEMAN1975267,COLEMAN1976239}.

\subsection{Ward-Takahashi identities}
\label{sec:wtident}

Before moving to anomalies, we shall recall some
general results about symmetries in quantum field theories.
The symmetries manifest themselves in the so called Ward-Takahashi
identities between Green's functions (Ward identities in short).
The original one that you
presumably learned about in QED, is a relation between the vertex
function and the inverse propagator. Path integrals provide a simple
and transparent way to derive these identities, and also show how they
are modified by anomalies.

The starting point is the partition function \eqref{partf2}, again
with a background gauge field $A$ and sources for the fermions,
keeping $A_5 =0$, 
\be{partf5}
Z= e^{-\G[A,\eta,\etab]} =  \mathcal N \int  {\mathcal D}\psib {\mathcal D}
\psi \, e^{-S_E [A,\psi,\psib] +\int dx\, (\etab\psi + \psib\eta  )} \, .
\ee

The Euclidean  action was given in \eqref{euclact}.
That a QFT has a symmetry means that $W$ is invariant under the
corresponding symmetry transformations, \ie $\delta W = 0$, that
is implemented by a change of integration variables.  Under an
infinitesimal vector and axial transformation of the fermion
fields,
$\delta \psi = i(\lambda + \xi \g^{D+1})\psi$,
$\delta \psib = i\psib (-\lambda + \xi \g^{D+1})$,
we obtain the following expression inside the path integral,
\be{varw}
\lambda(x) \left[- \partial_\mu J^\mu +i(\bar\eta\psi-\psib\eta)\right] +
\xi(x) \left[- \partial_\mu J_5^\mu + 2 m \psib \g^{D+1} \psi
+i(\bar\eta\g^{D+1}\psi +\psib \g^{D+1}\eta) \right]\; .
\ee
Since $\l(x)$ and $\xi(x)$ are independent, the two corresponding local
expressions in square brackets should vanish independently,
as expectation values in the presence of the sources $\eta,\bar\eta$.
Taking functional derivatives with respect
to $\eta(x)$ and $\etab(x)$ of the path integral including \eqref{varw},
gives the Ward identities between multipoint
correlators. The simplest ones, with no additional fermion fields,
are obtained by setting the sources to zero, and read,
\be{wident}
\partial_\mu J^\mu = 0 \qquad \mathrm{and} \qquad \partial_\mu
J_5^\mu = 2m \av{ \psib \gamma^{D+1}\psi}\; .
\ee

Clearly there is something wrong with this argument since it appears
that for $m=0$ the axial current is conserved, and we know that this
is not the case because of the anomaly. We now explain how this 
is resolved by carefully defining the integration measure.

\subsection{Anomalies from the path integral measure}
\label{sec:anpath}

We  learned in Sect. \ref{sec:schwinger} that the
path integral measure must be treated with care, and to be more
precise, it must be properly defined. The problem is that the
(continuum) path integral, even in a finite volume, is over an
infinite number of degrees of freedom, so there are potential
ultraviolet divergences. We shall closely follow the original
treatment by Fujikawa \cite{fujikawa1980}.

Fujikawa's central insight was that in the above derivation of the Ward
identities, we assumed that the functional integration measure was
invariant under the symmetry transformations. To show that this is not
the case, we note that in Euclidean space $i\slashed D$ is an
Hermitian operator (see App. \ref{app:notations}),
so the eigenvalues defined by,
\be{direig}
i\slashed D \varphi_n (x) = \lambda_n \varphi_n (x)\; ,
\ee
are all real. Now expand the spinors in terms of Euclidean 4D eigenfunctions,
as follows,
\be{spinexp}
\psi(x) = \sum_n \varphi_n (x) c_n \ , \qquad\qquad
\psib(x) = \sum_n \varphi^\dagger_n (x) \bar c_n\; ,
\ee
obeying,
\be{spinexp2}
\int dx\, \varphi_n^\dagger(x) \varphi_m(x) = \delta_{mn}\ ,\qquad\qquad
\sum_n\varphi_n^\dagger(x) \varphi_m (y)=\langle x| y\rangle,
\nonumber
\ee
and define the fermionic path integral measure as,
\be{defmes}
d\mu = {\mathcal D}\psib(x) {\mathcal D}\psi(x) = \prod_n d\bar c_n dc_n \; .
\ee
In any even spacetime dimension D, we can define the axial transformation,
\be{inax} \psi \rightarrow \psi' = e^{i \xi(x) \gamma^{D+1} } \psi =
\sum_n c_n\, e^{i \xi(x) \gamma^{D+1}} \varphi_n (x) = \sum_n c'_n\
\varphi (x) \, ,
\ee
so we can solve for $c_n'$ by using the
orthogonality of the $\varphi_n$,
\be{newcoeff} c'_m = \sum_n \int
dx\, \varphi_m^\dagger e^{i \xi(x) \gamma^{D+1} } \varphi_n c_n \equiv
\sum_n T_{mn} c_n \, .
\ee
Recalling that the $c$ variables are
Grassmann numbers, so that under a transformation $c\rightarrow a c$,
the measure transforms as $dc \rightarrow a^{-1} dc$, we get,
\be{meastrans}
\prod_n dc'_n = \left[\mathrm {det} T_{mn} \right]^{-1}
\prod_n dc_n \, .
\ee
For infinitesimal $\xi$,
\be{infT}
T_{mn} =\delta_{mn} + i \int dx\, \xi(x) \varphi^\dagger_m\gamma^{D+1}
\varphi_n \; ,
\ee
giving,
\be{Tdet}
\left[{\mathrm{det}} T_{mn}
\right]^{-1} = e^{-\mathrm {Tr} \ln T_{mn} } = e^{-i\sum_n \int dx\,
  \xi(x)\varphi^\dagger_n\gamma^{D+1} \varphi_n } \, .
\ee
Combining this with an \emph{equal} result from
the antichiral part, we get,
\be{meastrans2}
d\mu \rightarrow d\mu\, e^{-2i\int dx\, \xi(x)
  \mathcal{A}(x) } \, ,
\ee
where $\mathcal A$ is given by,
\be{anomaly}
\mathcal{A} (x) = \sum_n \varphi_n^\dagger(x) \gamma^{D+1}
\varphi_n(x) \; .
\ee

This quantity involves an infinite sum and requires  regularization.
We can use the eigenvalues $\lambda_n$ of the Hermitian operator
$i\slashed D$ to introduce a gauge invariant Gaussian cutoff as
follows,\footnote{See App. \ref{app:calculus} and
  Sect. 5 of \cite{Bertlmann} for details.}
\be{acutoff}
{\mathcal A}(x) &=& \lim_{M\rightarrow \infty} \sum_n
\varphi^\dagger_n\gamma^{D+1} e^{- \left(\frac {\lambda_n} M
  \right)^2} \varphi_n =
\lim_{M\rightarrow \infty} \sum_n \varphi^\dagger_n\gamma^{D+1}
e^{- \left(\frac {i\slashed D} M \right)^2} \varphi_n \\
&=& \lim_{M\rightarrow \infty} \langle x|{tr}\left[ \gamma^{D+1} e^{\frac
    {\slashed D^2} {M^2}} \right]|x\rangle =
\lim_{M\rightarrow \infty} \mathrm \int
\frac{d^D k} {(2\pi)^D} e^{ikx}\;
{tr}\left[  \gamma^{D+1} e^{\frac {\slashed D^2}
  {M^2}} \right]e^{-ikx} \nonumber \, ,
\ee
 where tr means summing over the spinor indices.
The last identity follows by inserting a complete set of states, 
which we choose as the plane
waves which are  eigenfunctions  of $i\slashed\partial$. To evaluate
the integral, we rewrite $\slashed D^2$ as,
\be{rewr}
\slashed D^2 = \left[ \half \{ \gamma^\mu , \gamma^\nu \} + \half [
  \gamma^\mu , \gamma^\nu ] \right] D_\mu D_\nu = D^2 +\half
\gamma^\mu\gamma^\nu [D_\mu , D_\nu] = D^2 - \frac i 2\gamma^\mu
\gamma^\nu F_{\mu\nu}\; ,
\ee
and then expand the exponential and pick up the leading term in an
expansion in $1/M^2$. In doing this we can also replace $D^2$ with
$\partial ^2$, which dominates the ultraviolet behavior.
This is how far we can proceed in a general even
space-time dimension, and we now specialize to D=2 and D=4.

For D=2, $\tr{\gamma^3 \gamma^\mu\gamma^\nu} = 2 i\epsilon^{\mu\nu} $,
and the momentum integral gives a factor $M^2/4\pi$. Thus, expanding
the exponent to order $1/M^2$ and using \eqref{specrel},
we get $\mathcal{A}(x)_{2D} =  F(x)/2\pi$,
with $F=\epsilon^{\mu\nu}\partial_\mu A_\nu$,
and finally using \eqref{meastrans2},
\be{2dtr}
d\mu \rightarrow d\mu\exp\left(-\frac i \pi \int dx\, \xi (x) F(x) \right)
\ , \qquad\qquad  \mathrm {D}=2\, .
\ee

For D=4, $\tr{ \gamma^5\gamma^\mu\gamma^\nu\gamma^\sigma\gamma^\lambda}
=- 4 \epsilon^{\mu\nu\sigma\lambda}$ so
we must expand the exponent to second order in $F_{\mu\nu}$
for the trace not to vanish. In this case the momentum integral
gives $M^4/16\pi^2$ and the anomaly,
\be{anexp4D}
\mathcal{A}(x)_{4D} =  \frac 1{16\pi^2}  F(x)_{\mu\nu}\tilde F^{\mu\nu}(x) \; ,
\ee  
and finally  for the change in the  measure,
\be{4dtr}
d\mu \rightarrow d\mu\,  \exp\left(-\frac i {8\pi^2} \int dx\,
  \xi (x) F(x)_{\mu\nu}\tilde F(x)^{\mu\nu} \right) ,
\qquad\qquad \mathrm {D}=4\, .
\ee
(recalling that $\tilde F_{\mu\nu}=\eps_{\mu\nu\a\b} F^{\a\b}/2$).
A very important aspect of this result is that the anomalous
contribution to the measure is a pure phase, both in Minkowski and
Euclidean space.

The change in the measure under the axial transformation will
give an ``anomalous'' contribution to the WI \eqref{wident}.
Considering only the axial transformation, it  becomes,
\be{varw2}
\partial_\mu J_5^\mu = 2m \av{\psib \gamd \psi}  - 2 i \mathcal {A} \, ,
\ee
giving, after putting $m=0$ and going back to Minkowski space,
the axial anomalies,
\be{anomfin}
D=2 &:&\qquad  \partial_\mu J_5^\mu = 2\mathcal{A}(x) =
\frac 1 {2\pi}\eps^{\mu\nu} F_{\mu\nu}\; ,
\nl
D=4 &:&\qquad \partial_\mu J_5^\mu= 2\mathcal{A}(x) =
\frac 1 {8\pi^2} F_{\mu\nu}\tilde F^{\mu\nu} \, .
\ee
These findings reproduce the earlier results \eqref{locan2} and
\eqref{anompert}.

We close this part with a comment on the nature of the  anomaly
relations. In the above derivation they appear as  relations between vacuum
expectation values of derivatives of currents, and external classical
gauge field. However, from the Ward identity \eqref{wident}
we learn that we can trade an insertion of a divergence of the
axial current for an insertion of the anomaly $\mathcal A$ also in a
general amplitude. This means that the anomaly relations
\eqref{anomfin} \etcc also hold as operator equations.
There is a more direct way to reach this conclusion by using
point-splitting \cite{SHIFMAN1991341}.  In this connection it should
also be mentioned that there is also a purely Hamiltonian approach
to anomalies \cite{nelsonalvarez85}.

\section{4D anomalies at work -- Dirac and Weyl materials  }
\label{sec:WSM}

At least since the discovery  of graphene, condensed matter physicists
have been interested in finding systems in more than 1d where fermions
have relativistic dispersion, which happens when the Fermi level is at
a band crossing.\footnote{ The history is in fact longer and is
  briefly outlined in \cite{ashvinview}.}
For a review of such ``Dirac materials'', see
\cite{diracmaterials}.  Graphene, for example, has a band
structure with two such ``Dirac points'', where the dispersion
relation takes the form of ``Dirac cones''.  When the Fermi energy is
tuned to these Dirac points, the density of states vanishes, and then the
system is called a  semi-metal. At low energies this is described by a
Dirac equation, thus quasiparticles are
gapless (or massless, if we use the high energy  language), but
by adding symmetry breaking terms, we can open up a gap. The gapless
Dirac equation has modes of both chiralities, but there are also
materials where the chiralities are separated so that a Dirac cone
splits in two ``Weyl cones'', and these Weyl semi-metals, have
many intriguing properties. The chiral and axial anomalies we
discussed so far are important for understanding the physics of these
systems, and in this section we shall discuss a couple of the many
interesting phenomena, namely the chiral magnetic effect, and
anomalous transport.  But before doing so, we should get some
understanding  on the mixed anomalies obtained by including
both nonvanishing vector $A$ and axial $A_5$ gauge backgrounds.

\subsection{Mixed anomalies and consistent -- covariant currents}
\label{sec:mixed}

In the absence of a mass term, the two chiralities are decoupled, and
the path integral calculations of previous sections can be
extended to give the anomalies in the presence of both $A$ and $A_5$
fields, or equivalently chiral and antichiral components
$A_\pm=A\pm A_5$.
One finds \cite{PhysRevD.29.285}: 
\be{covchir}
D=2 &:& \qquad 
\de_\mu J_\pm=\pm \frac{1}{4\pi}\eps^{\mu\nu}F_{\mu\nu}\ ,
\nl
D=4 &:&\qquad 
\de_\mu J_\pm^\mu=\pm  \frac { 1} {16\pi^2}  F_\pm\tilde F_\pm \, . 
\ee
Rewritten in terms of vector and axial currents and field strengths,
they read:
\be{covanom}
D=2 &:&\qquad 
\de_\mu J^\mu=\frac{1}{2\pi}\eps^{\mu\nu}F_{5\mu\nu}\; , \qquad\qquad
\partial_\mu J_5^\mu = \frac 1 {2\pi}\eps^{\mu\nu} F_{\mu\nu}\; ,
\nl
D=4 &:& \qquad
\partial_\mu J^\mu =   \frac 1 {4\pi^2}  F_5\tilde F \, ,
\qquad \qquad\qquad
\partial_\mu J_5 ^\mu = \frac 1 {8\pi^2} ( F\tilde F + F_5\tilde F_5 )\; ,
\ee
(with \eg $F\tilde F\equiv F_{\mu\nu}\tilde F^{\mu\nu}$).
That the electric charge is not conserved when both
 backgrounds are present is troublesome, 
and we shall see how this issue can be resolved in various physical
settings.

In the next section, we discuss the realization of the mixed anomalies
in the quantum spin Hall effect, that can be thought of a time-reversal
invariant generalization of the quantum Hall effect.  In this case,
the non-conservation of currents, both vector and axial, at the edge
is compensated by the anomaly inflow from the bulk of the system, thus
causing no problem at all. This model will give us some intuition
about the physical meaning of $A_5$ backgrounds.

The Weyl semi-metals will then be analyzed in the following section.
In these 3d systems, there is no compensating bulk in higher
dimension and vector current conservation should be verified.  In such
a case, another kind of anomalous currents can be defined, that do not
obey the inflow relation. They are called ``consistent'' because they
can be obtained by the variation of the 3d effective action
$\G[A,A_5]$. The anomalies discussed so far in these lectures are
instead ``covariant'', because they have the correct gauge
transformations, (are invariant in the Abelian case), and in fact are
related by inflow to bulk currents that are themselves
covariant. This terminology is conventionally used in the literature,
especially in discussing non-Abelian gauge theories
\cite{BARDEEN1984421}.

The difference between covariant and consistent currents has both
physical and geometrical meaning and it is unrelated to the freedom of
adding local terms to the effective action.  In some physical
settings, like the spectral flow and index theorem, the covariant
anomalies should be used, while in other situations, such as for the
response of isolated chiral systems, the consistent ones are
appropriate.  The discussion of properties of the two kinds of
currents is rather technical and is presented in App.
\ref{app:mixed}.  In the following we simply list the consistent
anomalies and compare them to the covariant ones already given in
\eqref{covanom}.

Consistent currents are denoted by $j_\mu$ and $j_{5\mu}$ and read,
\be{2Djcons}
D=2 &:&\qquad j^\mu =  J^\mu +\D J^\mu,\qquad\qquad
\D J^\mu = - \frac{1}{\pi} \eps^{\mu\nu}A_{5\nu},
\nl
&&\qquad  j^\mu_5=J^\mu_5,
\\\label{4Djcons}
D=4 &:&\qquad j^\mu =  J^\mu +\D J^\mu,\qquad\qquad
\D J^\mu = - \frac{1}{4\pi^2} \eps^{\mu\nu\a\b} A_{5\nu} F_{\a\b},
\nl
&&\qquad j^\mu_5 =  J^\mu_5 +\D J^\mu_5,\qquad\qquad
\D J^\mu_5 = - \frac{1}{12\pi^2} \eps^{\mu\nu\a\b} A_{5\nu} F_{5\a\b},
\qquad\qquad\qquad
\ee
where the pieces $\D J_\mu, \D J_{5\mu}$ being added to the
covariant currents $J_\mu, J_{5\mu}$ are called Bardeen-Zumino terms
\cite{BARDEEN1984421}. It is apparent that these modifications cannot be
realized by adding local terms to $\G[A,A_5]$, as anticipated.

The corresponding consistent anomalies read, using \eqref{covanom},
\eqref{2Djcons} and \eqref{4Djcons}:
\be{2Dcons_an}
 D=2 &:&\qquad \de_\mu j^\mu=0,\qquad\qquad \de_\mu  j^\mu_5=
\frac 1 {2\pi}\eps^{\mu\nu} F_{\mu\nu},
\\ \label{4Dcons_an}
 D=4 &:&\qquad \de_\mu j^\mu=0,\qquad\qquad \de_\mu  j^\mu_5=
\frac{1}{8\pi^2}\left(F\tilde F + \frac{1}{3}F_5\tilde F_5  \right) .
\ee

Note the relevant feature that
the consistent vector current is conserved in the mixed case too.
The comparison with \eqref{covanom} shows that the two kinds of 
anomalies coincide for vanishing $A_5$, thus explaining why
the consistent currents were not introduced in the earlier analyses.

\subsection{Spectral flow in the
  quantum spin Hall effect} \label{sec:SQHE}

Topological insulators in $(2+1)$ dimensions are topological phases of
matter sharing properties with the quantum Hall effect, in that they
have a gapped bulk and gapless edge excitations.  However, they
respect time reversal (TR) symmetry. Later we shall discuss them in more
detail, but in this section we shall consider a special case which is
easy to analyze given what we already know. This is the quantized spin
Hall effect which can be modeled by two independent quantum Hall
liquids with opposite chirality and also opposite spin
polarizations. TR invariance is achieved since the two component are
mapped into each other by time-reversal, which flips both chirality
and spin.

\begin{figure}[!htb] \centering
\includegraphics[width=0.58\linewidth]{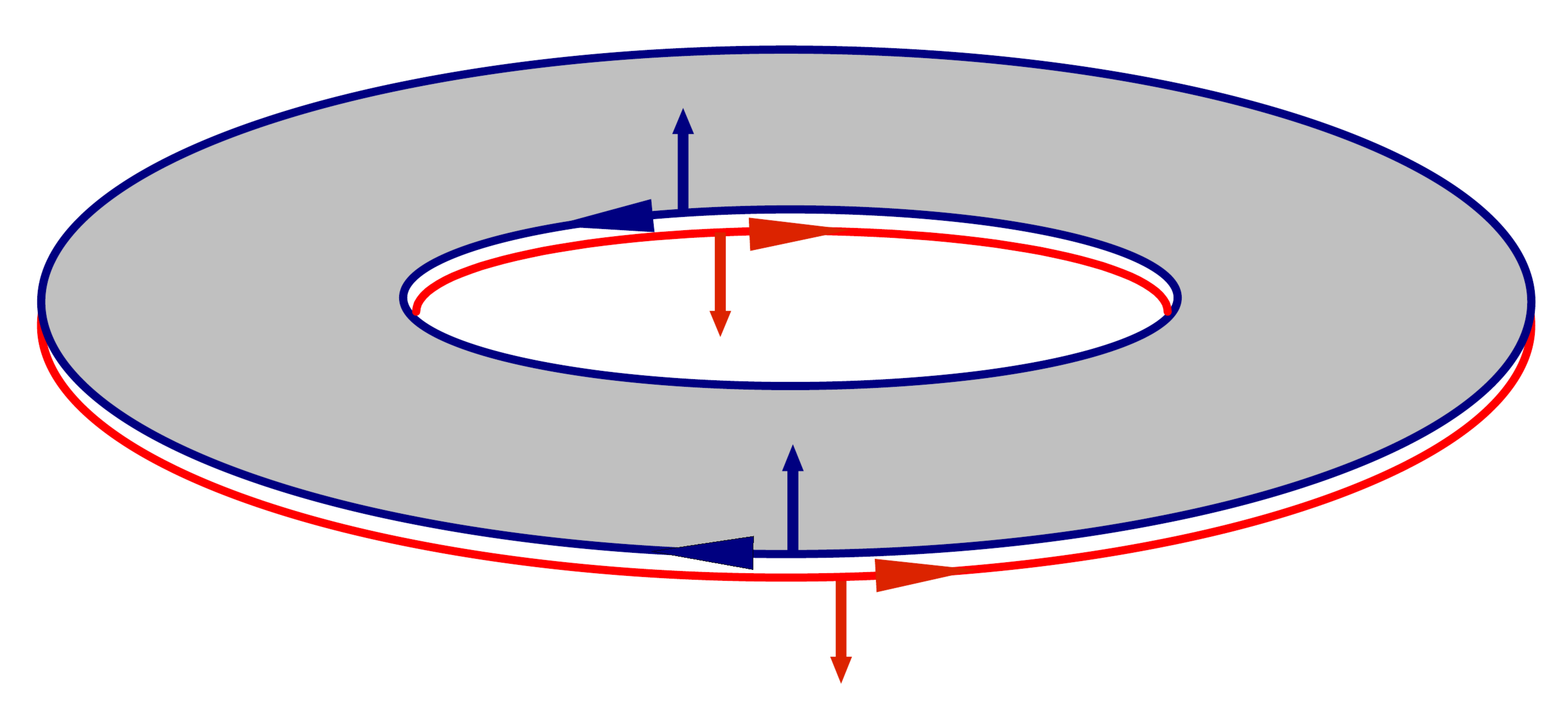}
\caption{Edge excitations in the quantum spin Hall
  effect. On the outer edge, right-moving spin-down modes are drawn in red,
left-moving spin-up ones in blue. Velocities are reversed in the inner edge.}
\label{fig:ti}
\end{figure}

On the boundary there are two massless Weyl fermions, a chiral one
with spin-up and an antichiral with spin-down
along some direction, \eg orthogonal to the plane.
As a result there are no charge currents, but the two edges carry net
spin currents with opposite chirality, as illustrated in
Fig. \ref{fig:ti}.  Together the chiral modes form a
$(1+1)$-dimensional Dirac fermion on each edge, and the system
possesses a $U(1)_Q \times U(1)_S$ symmetry, where the second, axial,
part corresponds to the conserved spin current, and is responsible for
the quantized spin Hall effect.

Before studying the spectral flow in this system, let us recall
that of the quantum Hall effect in Sect. \ref{sec:hallcurr}.
There we discussed the Laughlin flux argument in the geometry of an annulus
with two edges and a radial Hall current, exemplifying
the anomaly inflow mechanism.
Upon adiabatically adding a flux quantum, an electron is pulled out
of the Fermi sea, say on the inner edge, and another 
is pushed down into the sea on the outer edge, altogether corresponding
to the displacement of one charge by the bulk  $\nu=1$ Hall current
(see \eqref{QHE-flow}).

On each edge, the anomalous charge non-conservation is compensated by
the flux of the bulk current $J_r$ obtained from the Chern-Simons
action $S_{CS}$ \eqref{S-ind}. In formulas,
\be{inflow}
\mathrm {Anomaly:} \qquad \dot Q_L&=&
\frac{1}{2\pi}\oint\eps^{\mu\nu}\de_\mu A_\nu, \qquad\qquad
\mu,\nu=1,2\ ,
\nl
\mathrm {Inflow:} \qquad \dot Q_L&=& \oint J_r=\oint
\frac{\d S_{CS}}{\d A_r}=\frac{1}{2\pi}\oint \eps^{r\mu\nu}\de_\mu A_\nu\; ,
\ee 
where the integrals are over the spatial boundary. The first equation
follows from the $(1+1)$-dimensional theory, the second  expresses
the continuity of the bulk current as it reaches the boundary.

Referring back to our study of the fermion spectrum in
Sect. \ref{sec:2dfermions}, we recall that the induced boundary electric
field $E_x$ is due to the constant potential $A_x(t)$, which acts as a
chemical potential and shifts the energy levels in opposite ways on
the two edges/chiralities; $\mu_L=A_x$, $\mu_R=-A_x$. It follows that
the spectral flow can also be seen as due to an imbalance of the Fermi
level across the system.

The vector and axial backgrounds, $A=(A_L+A_R)/2$ and
$A_5=(A_L-A_R)/2$, act on the two chiralities/edges with different signs.
For example, inserting a $A_5$ flux in the quantum Hall effect would
lead to the same charge variation on both edges, leading to no net
Hall current. Equivalently, one can see this as the
simultaneous lowering/raising of the chemical potential on both edges
$\mu_L=\mu_R=A_{5x}$.

\begin{figure}[!htb] \centering
\includegraphics[width=0.9\linewidth]{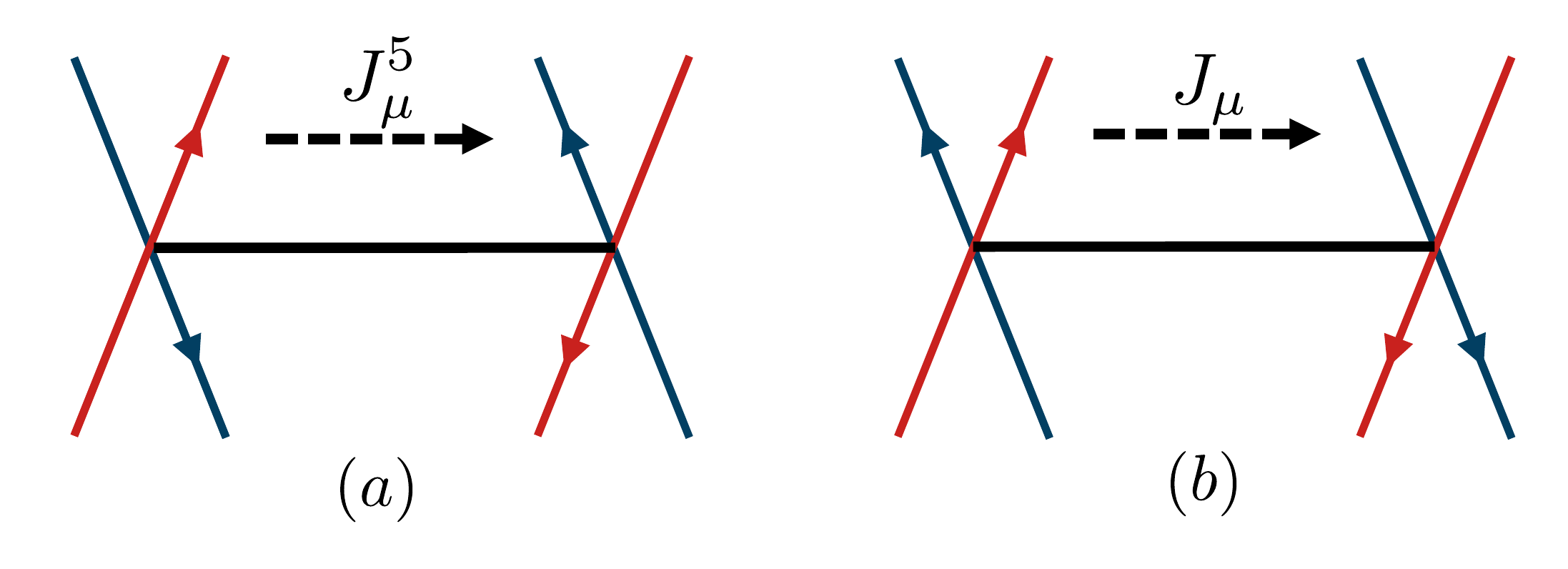}
\caption{Radial view of the annulus geometry (black line) with energy spectra
  of right-moving (red) and left-moving (blue) edge modes in the quantum
  spin Hall effect. The arrows show the spectral flow under
  adiabatic changes of: $(a)$ the $A$ field; $(b)$ the $A_5$ field.
  The compensating bulk currents are indicated.}
\label{fig:mix}
\end{figure}

We now turn to the quantum spin Hall system, described as a
pair of Hall states (see Fig. \ref{fig:ti}). 
One copy, say that of spin-up electrons (in blue), is subject to the
usual spectral flow  when its background field $A_+$ is adiabatically
switched on. For the  other copy, with
 spin-down electrons (in red), the  corresponding flow is associated to $A_-$.
Turning to the  bulk, the  Chern-Simons theory is
now,\footnote{The action is expressed in terms of differential forms, see
  App. \ref{app:calculus}.}
\be{BFtheory}
S_{TI}=\frac{1}{4\pi}\int A_+dA_+ - A_-dA_-= \frac{1}{\pi}\int Ad A_5\; ,
\ee
where we recall that $A_\pm=A\pm A_5$.
This action is time-reversal invariant by mapping
$A_+\leftrightarrow A_-$,
or equivalently by the corresponding transformations of vector and axial fields.
The  anomaly inflows are given by  the radial currents, 
\be{BFflow}
J_r=\frac{1}{\pi}\eps^{r\mu\nu}\de_\mu A_{5\nu},\qquad\qquad
J_{5r}=\frac{1}{\pi}\eps^{r\mu\nu}\de_\mu A_\nu \ .
\ee
These currents cancel the mixed anomalies of the edge fermions 
given by the \eqref{covanom}, using the inflow relations
\eqref{inflow}. Specifically (see Fig. \ref{fig:mix}):
\begin{itemize}
\item
  The $A$ field creates a chiral anomaly in the Dirac theory at
  each edge, opposite on the two edges, corresponding to
  non-conservation of axial charge, $\D Q_{5L}=-\D Q_{5R}$ (\ie a
  relative imbalance of chemical potentials).  The bulk axial (or
  spin) current $J_{5r}$ compensates for this effect between the two
  edges.
  \item
  The $A_5$ field creates a vector Hall current $J_r$ that balances the
  charge non-conservation since $\D Q_L=-\D Q_R$. Actually, electrons of
  both chiralities are pushed down the Fermi sea at one edge and pulled
  up at the other one.
  \item
    The value $\nu=2$ of the conductance  is correct since
    these adiabatic processes always transport pairs of electrons.
\end{itemize}

In conclusion, the spectral flow in the quantized spin Hall system
exemplifies the action of mixed anomalies, in the presence of both vector
and axial backgrounds.

There is a close similarity between the quantized spin
Hall system and the $ 2d$ topological insulator. In fact, often the
distinction is not made, since in real materials, the spin is not a
good quantum number, and the $U(1)_Q\times U(1)_S$ symmetry is
explicitly broken to $U(1)_Q\times \Z_2$ by time-reversal invariant
interactions that do not conserve spin $S_z$, such as the Rashba term,
and disorder \cite{PhysRevLett.95.146802}. So rather than having a
$\Z$ valued topological number, which you can think of as number of
filled Landau levels, or layers, there is only a $\pm 1$ number related
to $\Z_2$, which corresponds to the spin parity $(-1)^{2S}$. That
this symmetry is present even when no component of the spin is
conserved is highly non-trivial and at the heart of the discovery of
topological insulators. We shall come back to this in later sections.

\subsection{Mixed anomalies and  Weyl semi-metals}
\label{sec:WSManom}

In this section we shall learn how chiral anomalies are crucial for
understanding the electromagnetic response of Weyl semi-metals (WSM).
This is not the place for any deep explanation of the physics and
material science of WSMs, but we still have to set the stage by
providing some basics of these very interesting systems.

\subsubsection{Weyl semi-metals -- a primer}
\label{sec:WSMprimer}

The WSMs are 3d cousins of graphene, but with distinct and interesting
new properties. The first prediction of a WSM was in certain iridate
materials with rather complicated lattice structure, but 
it was subsequently found experimentally in clean crystals of TaAs
\cite{ashvinview}, which is a simpler compound. Still the band
structure is involved, and several steps are needed to extract an
effective low energy model simple enough to be described here. Details
about how that is done can be found \eg in
\cite{PhysRevX.5.011029}, where you can learn
under which conditions a Dirac cone can split into Weyl cones, each
having only one chirality. An important point is that you cannot have
both TR and inversion symmetry, but if you keep one of
these symmetries the Weyl points come in pairs related by symmetry.
 
\begin{figure}[!htb] \centering
  \includegraphics[width=0.6\linewidth]{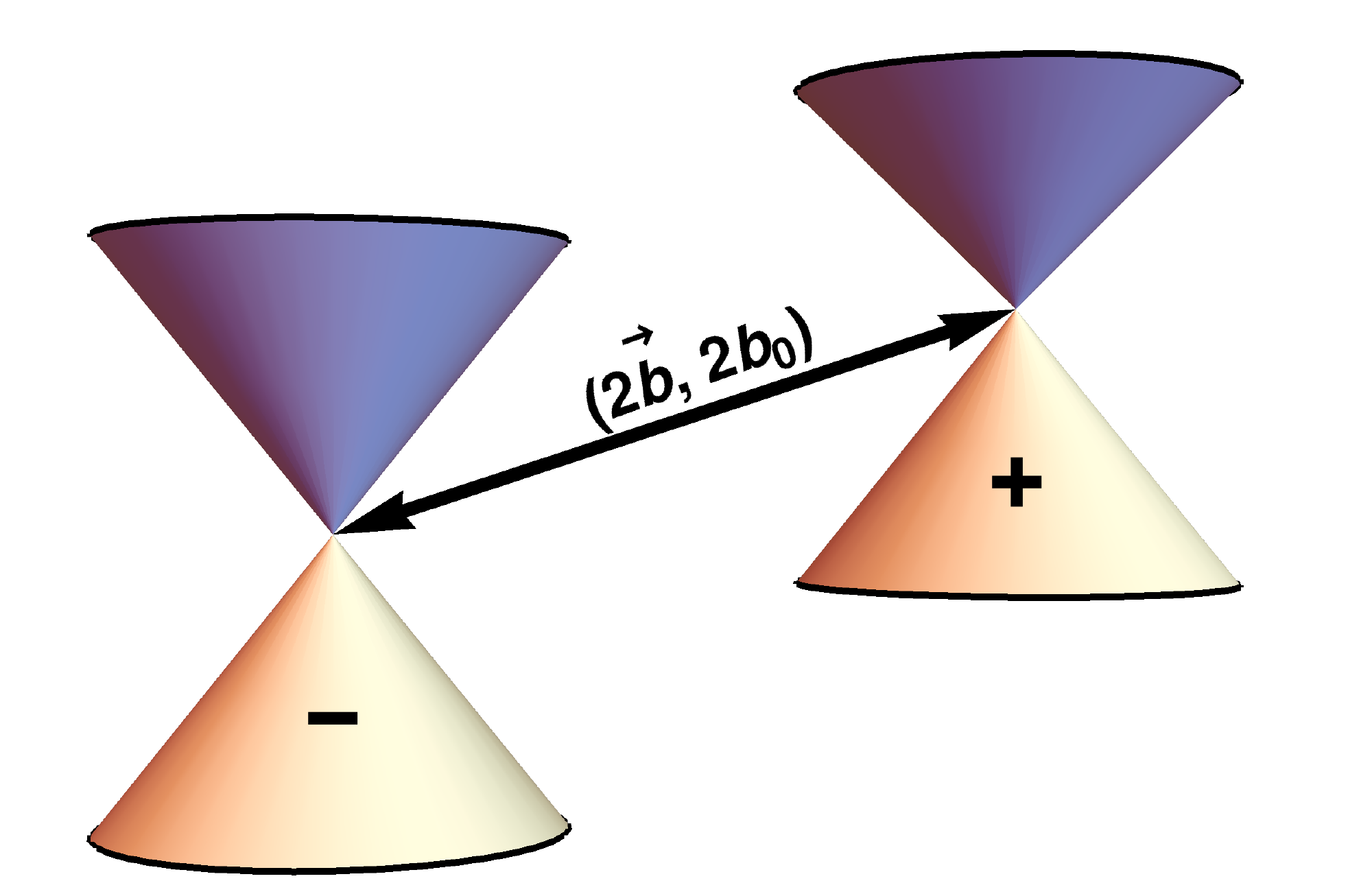}
  \caption{Dirac cones around Weyl points in a Weyl semi-metal.}
\label{fig:weyl}
\end{figure}

A simple and instructive model, that will suffice to illustrate the
importance of anomalies, was given in
\cite{PhysRevLett.111.027201}. They started from a model on a
cubic lattice for a 3d topological insulator and added symmetry
breaking terms to get non-degenerate Weyl points, that are shifted
both in energy and momentum. Assuming isotropy, the effective theory
for the modes close to Weyl points located at $\vec p = \pm \vec b$,
and shifted in energy by $\pm  b_0$, is,
\be{wham}
 H_W =\pm  b_0 \pm \mathrm{v} \vec\sigma\cdot (\vec p \mp \vec b)\ ,
\ee
where $\mathrm v$ a characteristic velocity, and the
sign denotes the helicity of the Weyl point.\footnote{ Since in real
  materials, there are typically several Weyl points, we have a bona
  fide semi metal only when $b_0 + \mu = 0$ for all Weyl points where
  $\mu$ is the relevant chemical potential.} The Dirac cones in these
Weyl points are depicted in Fig.~\ref{fig:weyl}.

WSMs, together with the quantum Hall and spin Hall effects discussed
earlier, are examples of topological phases of matter, to be discussed
in more detail in Sect. \ref{sec:spt}. A very useful concept to
understand these phases is that of Berry phase which we already
alluded to in Sect. \ref{sec:hallcurr}. In general, during a slow,
adiabatic variation, an eigenstate of a quantum system acquires a
phase that consists of two components \cite{berryphase}. One is a
dynamical contribution, that depends on the energy spectrum, the other
just depends on the trajectory defined by the Hamiltonian in the
parameter space. Because the latter does not depend on the details of
the evolution, it is a geometric or even topological quantity. This
phase is expressed in terms of the integral of a gauge connection, the
Berry connection ${\cal A}$, that is originating from the parameter
variation of one-particle states during evolution. It is like a
particle moving in an external electromagnetic field, but this is
generated inside the quantum problem.  As we will see, non-vanishing
Berry phase and Berry curvature often characterizes the presence of a
topological phases of matter.

We now explain why the WSM is a topological state of matter, by
analogy with the more familiar case of the integer QHE. In
that case, the topology is coded in the map from the Brillouin zone to
the space of one particle states,
$\{\vec k\in BZ\}\to \{\psi_{\vec k}\in U\}$,
where $U$ is one patch of the Hilbert space.
More precisely, one first defines the Berry potential by,
\be{berrypot}
\mathcal{A}_{k_i}(\vec{k})=-i\bra{\psi_{\vec{k}} }
\partial_{k_{i}} \ket{\psi_{\vec{k}} }\ ,
\ee
and then the Berry field strength by,
$\mathcal {B}^i = \epsilon^{ijn} \partial_{k_j} \mathcal{A}_{k_n} $. The
first Chern number is given by (cf. \eqref{eq-mono}),
\be{chernnum}
I =\frac 1 {2\pi} \int_{BZ} d\s^i \mathcal{B}^i =
\frac 1 {2\pi} \int_{BZ} d^2k\, \mathcal{B}^3 \, ,
\ee
and since the Brillouin zone is a closed surface (a torus) we can, in
analogy with an ordinary magnetic field $\vec B$ in the presence of
monopoles, conclude that the integral must be an integer number.
As originally shown in \cite{tknn}, the
Hall conductance is given by $\sigma_H = I\; e^2/h$.
This way of defining topology is based on having a gapped system, and
cannot be directly applied to a WSM. What we can do, however, is to
consider an integral not over the full zone, but over a 2d surface
enclosing a Weyl point. Since the energy is finite everywhere except
at the Weyl points this is a well defined object, which is actually
easy to calculate by the following analogy: The Hamiltonian $H_W$ you
can think of as a momentum space version of the Hamiltonian
$H_B = -\mu \vec S \cdot \vec B$ for a particle with magnetic moment
$\mu$ and spin $\vec S$ in a magnetic field. Changing the direction of
$\vec B$ adiabatically provides the archetypical example of a
geometrical phase, and Berry showed that this phase only depends on
the value $m$ of the spin projection on $\vec B$ and equals $m\Omega$,
where $\Omega$ is the solid angle subtended by the loop traced by
$\vec B$ \cite{berryphase}. For a full sphere, $\Omega = 4\pi$ so for
a spin half particle corresponding to the two level Hamiltonian $H_W$,
the integrated Berry flux is $\pm 2\pi$, where the sign is given by
the helicity. Thus the total flux of a pair of Weyl points with
different helicity vanish, meaning that they can annihilate into a
trivial state if they are made to collide.

\subsubsection{Electromagnetic response}  \label{sec:WSMresp}

We are now prepared to study the response of a WSM to electromagnetic
fields, and let us consider the simplest case with only two Weyl
points. Using the Weyl representation
\eqref{weylgamma} for the Dirac matrices, the two cones can be
described by the Dirac action,
\be{wdirac}
S =\int dx \bar\psi (i\slashed\de + \slashed A
+ \slashed A_5 \gamma^5)\psi \ ,
\ee
where $A_{5\mu} = (b_0, \vec b)$, and we take
$\mathrm{v} =1$ and $e = 1$, as usual.
The shift in energy and momentum between the
Weyl points enters the Dirac theory as the axial gauge potential $A_5$.
More importantly, by applying magnetic fields and/or
strain it is possible to have a space dependent axial potential, and
thus non zero axial field strengths. We shall not dwell here on how
this can come about, but details and more references can be found \eg
in \cite{Landsteiner_2016}.

Now consider a situation where $A_5=0 $, but with non-vanishing and
parallel electric and magnetic fields, so $\vec E \cdot\vec B \neq
0$. Because of the chiral anomaly \eqref{covchir} electric charge will
not be conserved in the separate chiral channels, but taken together,
the electric charge is conserved. A more interesting situation occurs
when there is a non vanishing axial magnetic field, $\vec B_5$
and a parallel electric field $\vec{E}$. In this case,
the result \eqref{covanom} and a corresponding variant of the
``infinite hotel'' calculation in Sect. \ref{sec:3dhotel}, seem to
show that the electric charge is not conserved,
\be{wanom}
\partial_\mu J^\mu = \frac 1 {2\pi^2} \vec E\cdot\vec B_5 \, ,
\ee
which is troublesome.

Let us see how the issue is resolved in this   context.
An axial gauge potential $A_5$ is certainly
present since without it there are no Weyl cones with non degenerate
spectra. But to get an axial magnetic field, $\vec B_5$
the potential must be space dependent. This can happen if the stress
in the material is inhomogeneous, but we shall look at a simpler
realization which is a boundary between a Weyl metal and a Dirac metal
where two Weyl cones have merged \cite{Landsteiner_2016}.
Consider a slab of a WSM with sharp boundaries, $x\in (-a,
a)$. Further assume that
$\vec A_5 = b[\Theta(x+a)-\Theta(x-a)] \hat y$, where $\Theta$ is
the Heaviside step function, and a constant electric field
$\vec E =E_z\hat z$. The axial magnetic field becomes,
\be{axmag}
B_5 = b \left( \delta(x+a)-\delta(x-a) \right) \hat z\ .
\ee
Using this in \eqref{wanom} we get the explicit expression,
\be{wsman}
\partial_\mu J^\mu = \frac{E_z b} {2\pi^2}
\left( \delta (x-a) - \delta (x+a) \right),
\ee
which shows that the electric charge is not conserved at the boundaries.
The total  electric charge is conserved, but that is 
not good enough, since electric charge must be conserved locally.

The situation is somehow similar to that of the spin Hall effect in
Fig. \ref{fig:mix} (b) (see Sect. \ref{sec:SQHE}), with a charge
unbalance between the two cones. Actually, there is a current flowing
within the system itself, not through an external bulk, because deep
inside the Fermi sea the bands of the two cones reconnect, similar to
the case of the Luttinger model in Fig. \ref{fig:luttinger}.  How this
ultra-violet feature can be modeled in our effective theory which is
valid near the Fermi surface? The answer is found by using the
consistent currents \eqref{4Djcons} introduced in
Sect. \ref{sec:mixed}, instead of covariant ones in
\eqref{wanom}. Here are their important features:
\begin{itemize}
  \item
They describe isolated systems without inflow from outside, as described
    in App. \ref{app:mixed}.
  \item
    The vector current is conserved also in the case of mixed anomalies,
    see \eqref{4Dcons_an}.
  \item
    The Bardeen-Zumino term modifying the vector current
    in \eqref{4Djcons} re-establishes local charge conservation.
\end{itemize}
  
Let us evaluate this Bardeen-Zumino term $\D J^\mu$ in our geometry.
Its value is non vanishing inside the slab,
\be{deltaj}
\D J^x = - \frac 1 {2\pi^2} A_{5y} E_z , \qquad\qquad x\in (-a, a)\; ,
\ee
and zero outside, and the resulting divergence
$\partial_x \D J^x$ indeed cancels the anomaly \eqref{wsman}.
Therefore, this current correctly describes the charge compensation
between the two Weyl cones.
It is an interesting possibility that the difference between covariant
and consistent currents \eqref{deltaj} might originate 
by a shift of normal-ordering conditions for the ground state
in the region $x\in (-a,a)$.

There is lot more to be said about the role of the mixed anomalies in
the study of transport properties of WSMs.  An important issue is to
what extent the simple effective models just described reflect the
physics of real materials associated to concrete lattice models. An
example of this are the recent studies
\cite{Grushin,PhysRevB.96.125123,PhysRevLett.118.127601,PhysRevB.99.140201},
which also include an $E_5$ field and thus can probe the full
expression of the anomaly \eqref{4Dcons_an}.  Another example is
\cite{PhysRevX.6.041021}, which presents a model calculation where
torsion gives rise to a uniform axial magnetic field in a WSM wire,
and goes on to studying detail how the effect of the bulk anomaly in
the presence of an electric field is canceled by modes concentrated on
the surface. Refs. \cite{PhysRevB.102.235163,
PhysRevResearch.2.033269} contains later work on the effects of
torsion.

\subsection{The chiral magnetic effect}
\label{sec:chieff}

The chiral magnetic effect was originally proposed in the context of
ultra-relativistic heavy ion collisions \cite{PhysRevD.78.074033}, but
later also generalized to Dirac materials and then experimentally
discovered \cite{cmedm}. It predicts the appearance of an electric
current when there is an imbalance between the chemical potential of
the right and left moving modes. Such an imbalance can can occur if there is
non-zero axial potential $\mu_R \neq \mu_L $,
a possibility that was discussed in Sect. \ref{sec:SQHE}. 

In  equilibrium, the current 
vanishes,\footnote{This follows from a theorems originally attributed  to
  Bloch \cite{Bohm} and recently generalized generalized by Yamamoto
  \cite{Yamamoto}. } 
but in the presence of both electric and magnetic fields there can
be a steady non-equilibrium current. To show this, first assume that
we subject a material with non-zero $A_{5\, 0}$ 
to a constant magnetic field $\vec B$, and also to a constant axial
electric field $\vec E_5$.  The mixed anomaly of the covariant vector current
\eqref{covanom} is again,
\be{xx}
\dot Q = \frac{\mathrm {Vol}} {4\pi^2} \vec B \cdot \vec E_5 \, ,
\ee
and note that the magnetic field is defined in the frame where there
is originally no current flowing. We learned in
Sect. \ref{sec:mixed} that this unacceptable anomaly can be eliminated
by the Bardeen-Zumino term $\D J^\mu$ in \eqref{4Djcons},
which in this frame  provides the full current
(in the limit $\vec E_5=0$):
\be{BZcme}
j^i \equiv \D J^i = \frac 1 {2\pi^2} \epsilon^{0ijk} A_{5\, 0}
\partial_j A_k =\frac {\mu_5}{2\pi^2} B^i \; .
\ee
Note that to leading order in $\vec B$ we used equilibrium relation
$A_{5\, 0} = \mu_5$ which is maintained by inter-valley scattering
between states at the two Weyl Fermi surfaces that transfer
particles between them.

For a detailed explanation for why a non-zero $\mu_5$ can occur in a
quark gluon plasma, or in a Dirac material, we refer to the original
papers, and here we shall just outline the argument given in \eg
\cite{cmedm}. Assuming that we apply a weak electric field,
$\vec E$, again in the same reference frame as before, we get the
axial anomaly \eqref{4Dcons_an},
\be{drude}
\dot \rho_5 =  \frac 1 {4\pi^2} \vec E\cdot \vec B -
\frac {\rho_5} \tau\; ,
\ee
where we also added a Drude-like term, with a time constant $\tau$ to
describe the relaxation  due to inter-valley scattering.
Solving \eqref{drude} gives the steady state value
$\rho_5 = \frac 1 {4\pi^2} \vec E \cdot \vec B\, \tau$, and
combining this with the thermodynamic relation between $\rho_5$
and $\mu_5$, gives the chiral magnetic linear response,
\be{cmresp}
j_{CME}^i \sim \tau B^i \vec B \cdot \vec E \, .
\ee
We remark that this expression does not depend on $\vec E_5$, which
should be thought of as a probe field that is taken to zero at the end
of the calculation.\footnote{ An alternative derivation of this
  relation is given by \cite{PhysRevD.78.074033}. }
For $\vec E\parallel \vec B$ we get an enhanced longitudinal
conductivity $\sim B^2$. This contribution is in addition to the usual
Ohmic term, and the characteristic quadratic dependence on the
magnetic field was experimental observed in \cite{cmedm}.

\section{Anomalies and index theorems } \label{sec:indexth}

In the perturbative derivation of the anomalies, we glossed over an
important question: Are there corrections due to higher order
diagrams? From the path integral derivation one would be tempted to
say no, since adding interactions will not change the fermionic
measure. The infinite hotel method, using spectral flow and
point-splitting did involve the Hamiltonian, so there it is not clear
how anomalies could be changed by interactions. Not long after the
discovery of the axial anomaly, it was proven that
the only effect of interactions is to replace bare charges and masses
with renormalized ones, a result know as the Adler-Bardeen theorem
\cite{PhysRev.182.1517} .

There is, however, a much more general and elegant way to show that
the anomalies we have discussed, as well as their generalization to
non-Abelian gauge groups, and curved geometries, are not changed by
interactions. It follows from deep results in algebraic
topology, but unfortunately the original literature is not very
accessible to non mathematicians. The important results for us are the
index theorems of Atiyah and Singer and Atiyah, Patodi and
Singer. These theorems link properties of differential operators to
topological properties of gauge fields, a relation anticipated in
Sect. \ref{sec:antop}. As we now show, we have already indirectly
derived a special case of one of these theorems, and after explaining
this point we shall just state some general results.

\subsection{Definition of the index of the Dirac operator}
\label{sec:dirind}

First we recall the definition of chiral projectors and the
Dirac operator in even dimensional spacetime,
\be{projectors}
P_\pm = \half (1 \pm \gamma^{D+1})  \  ; \qquad\qquad
\slashed D = \gamma^\mu (\partial_{\mu}  - i A_\mu) \, ,
\ee
where we use Euclidean signature, and the same gamma matrices
$\gamma^\mu_E$ as in Sect. \ref{sec:perturb}, but will not write 
the subscript $E$ in what follows.  Next define,
\be{chiraldop}
D = \slashed D P_+  \ , \qquad\qquad  D^\dagger = -P_+ \slashed D =
- \slashed D P_-  \, ,
\ee
where we recalled that $\slashed D$ is anti-Hermitian.
Since $P_\pm$ are projection operators they define two subspaces,
$\mathcal {H}_\pm$ so that for  D=4,
\be{chir-dec}
\psi_\pm \in \mathcal {H}_\pm, \qquad\qquad \mathrm{obeying} \qquad\qquad
P_\pm\psi_\pm=\psi_\pm, \qquad
\gamma^5 \psi_\pm = \pm \psi_\pm \, .
\ee
With these definitions we get, 
\be{chirmap}
\gamma^5 D^\dagger \psi_- = -\gamma^5  \slashed D  P_- \psi_- =
 \slashed D \gamma^5 P_- \psi_-
=  -\slashed D P_-\psi_- =  D^\dagger \psi_- \; ,
\ee
and similarly $\gamma^5 D \psi_+ = -D\psi_+$, so the operators $D$
and $D^\dagger$ gives a one-to-one mapping
between the states in two subspaces,   $\mathcal {H}_\pm$, except for
the zero modes of any of these operators. The index of the Dirac 
operator is defined \cite{nairbook} as,\footnote{
  In mathematics the number of zero modes of an operator defines
  the kernel of that operator, so in this language the 
  definition of the index is:
  $\mathrm{ind} (i\slashed D) =  \mathrm{dimension\ of \ kernel \ of }
  \ D -  \mathrm{dimenson\ of \ kernel\ of}\  D^\dagger$.
}
\be{defindex}
\mathrm{ind} ( \slashed D )= \#\ \mathrm{zero\ modes\ of } \ D -
\# \ \mathrm{zero\ modes\ of }\  D^\dagger \, .
\ee
In order to calculate the index, it will be useful to work with
the Hermitian operators $D^\dagger D $ and 
$D  D^\dagger $. Obviously any zero mode of $D$ is also a zero mode
of $D^\dagger D$ and the same for the pair $D^\dagger$ and $DD^\dagger$.
It is also not too hard to  prove that the index can also be re-expressed
in terms of the quadratic operators so we have the alternative definition,
\be{defindex2}
\mathrm{ind} ( \slashed D )= \#\ \mathrm{zero\ modes\ of } \ D^\dagger D -
\# \ \mathrm{zero\ modes\ of }\  D  D^\dagger  \, .
\ee

\subsection{Calculation of the index}
Since $D^\dagger D$ is a Hermitian operator on $\mathcal H_+$ it can
be diagonalized with real eigenvalues, $\lambda^+_n$, and similarly
for $D D^\dagger$ on $\mathcal H_-$. Since the non-zero modes
$\lambda^\pm \neq 0$ are matched in pair, we get the following formal
expression for the index,
\be{indexformal}
\mathrm{ind} ( \slashed D ) = \sum_{n\in \mathcal{H}_+} 1 -
\sum_{n\in \mathcal{H_- }}1 =
\int dx  \left(  \sum_{n\in \mathcal{H}_+}   \psi^{+\star}_n (x) \psi^+_n (x)
  -  \sum_{n\in \mathcal{H}_-}   \psi^{-\star}_n(x) \psi^-_n (x) \right) \, ,
\ee
whose proof is left as an exercise.
In this expression, $\psi^\pm_n$ are the eigenfunctions associated
  to $\l^\pm_n$, similarly to those in Sect. \ref{sec:anpath}.
Now  we  regularize the above expressions as follows, 
\be{indexformal2}
\mathrm{ind} ( \slashed D ) &=& 
\int dx  \left(  \sum_{n\in \mathcal{H}_+}   \psi^{+\dagger}_n(x) \gamma^5
  e^{-\frac {D^\dagger D} {M^2}}   \psi^+_n (x)  + 
  \sum_{n\in \mathcal{H}_-}   \psi^{-\dagger}_n(x)\gamma^5
  e^{-\frac {D D^\dagger } {M^2}}  \psi^-_n (x) \right)
\nonumber \\
&=& \mathrm{ Tr} \left[  \gamma^5 e^{-\frac {(i\slashed D)^2} {M^2}}
\right]_{\mathcal H}        \, ,
\ee 
where we used $D^\dagger D \psi^+_n = - \slashed D \slashed D  \psi^+_n$ and
$D D^\dagger \psi^-_n =  -\slashed D \slashed D  \psi^-_n$ to get the last equality.
The trace is both over the gamma matrices, and the full Hilbert space
$\mathcal {H} = \mathcal{H}_+ \oplus \mathcal{H}_-$.
As in Sect. \ref{sec:anpath} we used a gauge invariant cutoff,
and note that since the nonzero eigenvalues of $(i\slashed D)^2$ comes
in pairs, the expression is independent of $M^2$. This means that we
can evaluate the regularized sum for any $M^2$, and in particular for
the limit $M^2 \rightarrow \infty$.  But this is precisely the
calculation we already did in Sect. \ref{sec:anpath}, the trace
in \eqref{indexformal2} is given by the integrated anomaly \eqref{acutoff},
so we immediately get the result,
\be{asth4}
\mathrm{ind} (  \slashed D )_{D=4} = \frac 1 {16\pi^2} \int dx\,
F^{\mu\nu}(x) \tilde F_{\mu\nu}(x)  \, .
\ee
This is a special case of the Atiyah-Singer theorem.
There are in fact a number of index theorems, but this is the only one that
will concern us here. Similarly we can extract the index in D=2 from
\eqref{2dtr} ,
\be{asth2}
\mathrm{ind} (  \slashed D )_{D=2} = \frac 1 {4\pi} \int dx\,
\epsilon^{\mu\nu} F_{\mu\nu} \, .
\ee

Now comes an important point. The index is by definition an integer,
but it is not obvious that this is the case for the right-hand sides
of \eqref{asth4} and \eqref{asth2}. We have however not been precise
enough in defining the eigenvalue problem for the operators. To do so,
we assume that the field strength vanishes on a circle at infinity,
which, after identifying this circle as a point, is the same as
defining the theory on a sphere. Thus, the index is given by the flux
of the magnetic field through a closed surface, which is a
topological quantity, as discussed in Sect. \ref{sec:antop}.

Another way to understand \eqref{asth2} is to recall that  a
non-zero magnetic flux through a closed surface $\Sigma$ is obtained by
enclosing a number of Dirac monopoles, each attached to a flux tube
of strength $2\pi$, since it should be invisible for a particle of unit
charge \cite{nairbook}.
Assume that there is only one monopole, and integrate over an infinitesimal
circle $S_\epsilon$ around the Dirac string, such
that $\Sigma$ covers the whole sphere in the limit when the radius
of the  $S_\epsilon$ goes to zero.
Note that this does not depend on the shape of the surface, which proves
that the index \eqref{asth2} is a topological quantity.

\subsection{The Atiyah-Singer theorem for the Dirac operator
  in general dimensions} \label{sec:astheorem}

We will now state the general result for the index of the Dirac
operator defined on an even dimensional closed (and orientable) curved
manifold, and for this we need some notation. In case you are not
familiar with field theory on curved backgrounds and some basics
notions of differential geometry, you can skip this section and just
    note the final result, \eqref{genasth4} for the 4D Dirac index.
    App. \ref{app:calculus} may nonetheless help you with some basic
    notions and formulas.

The Dirac operator on a closed curved manifold coupled to a
non-Abelian vector field $A_\mu$, is  defined by,
\be{gendirac}
\slashed D = e_a^\mu \gamma^a  ( \partial_\mu  +
i\omega^{ab}_\mu \s_{ab} - i A_\mu) \; ,
\ee
where $e_a^\mu$ is the vierbein field $\omega^{ab}_\mu $
the spin connection, $\s_{ab} = i[\gamma_a,\gamma_b]/4$ and 
$A_\mu \equiv A^\alpha_\mu T^\alpha$ with $T^\alpha$  generators of
the appropriate representation of the gauge group.\footnote{
  Our notations for  differential forms and calculus
  on curved surfaces are summarized in App. \ref{app:calculus}.}

The Chern characters, $Ch_n(F)$ are defined by the generating function
\cite{Bertlmann}:
\be{chchar}
Ch(F) = \mathrm{Tr}\, e^{  F/2\pi} =
\sum_{n=0} \frac 1 {n!}  \mathrm{Tr}\left(  \frac F {2\pi} \right)^n\ ,
\ee
where $F$ is the field strength two form
$F =  \half F_{\mu\nu} dx^\mu \wedge dx^\nu$ and the trace is on the gauge
indices.

For a curved space  the  $\hat A$-genus is defined by,
\be{agenus}
\hat A = 1 + \frac 1 {48\pi^2} \mathrm{Tr}\,
R \wedge R + \dots \, , 
\ee 
where $R$ is  the Riemann curvature two form
$R^a_b =  \half R_{\mu\nu,}{}^a_{\ b} dx^\mu \wedge dx^\nu$, that is obtained from
the spin connection $\omega^a_{\ b}$ and is also a matrix valued two form.
In this case the trace acts on the Lorentz indices.
The dots in \eqref{agenus} are other pseudoscalar quantities that
are higher-order polynomials of the curvature and its covariant derivatives.

The Atiyah-Singer theorem gives a formula for the index of the Dirac operator,
\be{asing}
\mathrm{ind} (  \slashed D ) = \int_{{\mathcal M}_D} \hat A(R)
\wedge Ch(F) = \int_{{\mathcal M}_D} \Omega_D\; , \qquad\qquad
(D\ {\rm even}),
\ee
where $\mathcal M$ is D dimensional closed manifold with D even,
and the last equality defines the 
index density $\Omega_D$. Using this together with \eqref{chchar}  gives the
2D index (prove this!), and also using \eqref{agenus}, and writing out
the indices, we have,
\be{genasth4}
\mathrm{ind} (  \slashed D )_{D=4} = \int dx\,   \epsilon^{\mu\nu\sigma\lambda}
\left(  \frac {1} {32\pi^2}  \mathrm{Tr} F_{\mu\nu} F_{\sigma\lambda} +
  \frac {1} {192 \pi^2}
  \mathrm{Tr}\, R_{\mu\nu} R_{\sigma\lambda}\right) = P - \hat A \, , 
\ee
where we specialized to a $U(1)$ connection, and introduced the
notation $P$ for the first term which is (also) called the instanton number.
The term $\hat A$ is the corresponding gravitational instanton number.
We should mention that this result can be obtained by
more pedestrian methods. Just as the electromagnetic vector potential
couples to the electromagnetic current, the metric couples to the
energy-momentum tensor. Thus the second term in \eqref{genasth4} can
be obtained from a triangle graph with two insertions of $T^{\mu\nu}$
\cite{Kimura},
and it can also be obtained using Fujikawa's method as shown in his
original paper \cite{fujikawa1980}.

Using the theory of connections on fiber bundles, and characteristic
classes, one can prove that both terms in this expression are
topological, \ie they cannot be changed by small deformation of the
gauge field or the metric, and that they are properly normalized to be
integers. This generalizes the result for the 2D index that we
discussed above, and amounts to a general proof that the anomaly in
any even dimension is a topological object.\footnote{
    Note, incidentally, that there is no gravitational contribution in
  2D (see \eqref{asth2}) because the Riemann tensor has
  one independent scalar component only, and there is no counterpart
  to a pseudoscalar quantity.}
The observant reader might
have spotted a loop hole in this argument -- we have not proven that
we cannot add a local counter term that could change the anomaly. At a
pedestrian level, this can be excluded just by enumerating all
operators with the right dimensionality. This method is very
cumbersome in the non-Abelian case, and there is a more sophisticated
and general way to reach the same conclusion introduced by 
Ref. \cite{WITTEN1983422} (for a detailed explanation, see
\cite{nairbook}).

As a final comment, we mention that in higher dimensions there are
also mixed anomalies which contain both gauge fields and
gravity. Expanding \eqref{asing} to next order we get the 6D index
density,
\be{D6anom}
\Omega_6 =  \frac 1 {48\pi^3} \mathrm{Tr}\, F^3   +
\frac 1 {384 \pi^3} \mathrm{Tr}\, F\; \mathrm{Tr}\, R^2\; .
\ee 
We can now give the result for the axial anomaly in a general even dimension
and on a curved space. Recalling the formulas   \eqref{meastrans2}
and \eqref{anomfin} which  \emph {mutatis mutandis}
holds in any even dimension, we get,
\be{anomfingen}
\partial_\mu J_5^\mu = 2 \mathcal{A}(x) = 2 \Omega_D  \, \qquad\qquad\  \ 
D=2,4,\dots.
\ee

\section{ Gravitational anomalies and responses}
\label{sec:heattrans}

Introducing a coupling to a background metric $g_{\mu\nu}$ gives
another way to probe quantum systems and to calculate important
response functions. In this section we shall explore this with an
emphasis on the consequences of gravitational anomalies.

The most direct way to introduce a metric is to consider an
essentially two-dimensional material on a curved surface. This could
be achieved by growing films on top of curved substrates, see \eg
Refs. \cite{QHKS} and \cite{QHW}.  Another possibility is to deform the
lattice geometry by applying strain or inserting
dislocations. Furthermore, as already discussed in Sect. \ref{sec:thermgrav},
gravitational fields can mimic thermal gradients and allow us to
calculate thermal response.

The defining feature of a gravitational background is that it couples
to all form of matter; it is not specific to fermions, nor does it
distinguish chiralities.  This is a great advantage when studying the
response of neutral matter, and it reveals new aspects of the violation
of classical conservation laws.  The gravitational \eqref{a-grav} and
trace \eqref{a-conf} anomalies can again be understood, using the
perturbative and path-integral methods of the previous sections, as a
consequence of the renormalization of loops or the integration
measure.  Also here, one anomaly can be traded for the other by adding
finite counterterms to the effective action.

For example, in the case of non-chiral theories like Dirac fermions
($c=\bar c=1$), one can choose to cancel the gravitational anomaly and
be left only with the trace anomaly \eqref{a-conf}. Integrating it
again gives a topological invariant, the Euler characteristic
\eqref{chi}, that generalizes to any even dimension
\cite{Nakahara:Geometry}. In chiral theories with $c\neq \bar c$, the
standard choice in conformal field theory is to cancel the trace, see
\eqref{chi-grav}.  As discussed in Sect. \ref{sec:thermgrav}, this
non-conserved stress tensor acquires an anomalous transformation law
which is responsible for the thermal Hall current.

The coupling to gravity is more technically involved than that to
(Abelian) gauge fields, and here we we shall only consider 2d
theories. Higher dimensions are discussed in the context of
index theorems in Sect. \ref{sec:indexth} and global anomalies in
Sect. \ref{sec:global}.
The physical consequences of gravitational anomalies in
higher-dimensional topological states of matter is a topic of active
research, and we refer to the more advanced review \cite{Vozmediano}
for recent achievements.

In what follows, we first explain in more detail how heat flow can be
related to gravity, and then analyze the gravitational bulk-boundary
correspondence by extending the results in
Sect. \ref{sec:thermgrav}. We shall also use a metric to introduce a
time dependent strain in a quantum Hall system, as a mean to calculate
the so-called Hall viscosity. Our notations and conventions for curved
space calculus are given in App. \ref{app:calculus}.

\subsection{Heat transport from gravitational response}
\label{sec:lutgrav}

\begin{figure} \centering
\includegraphics[height=5.5cm]{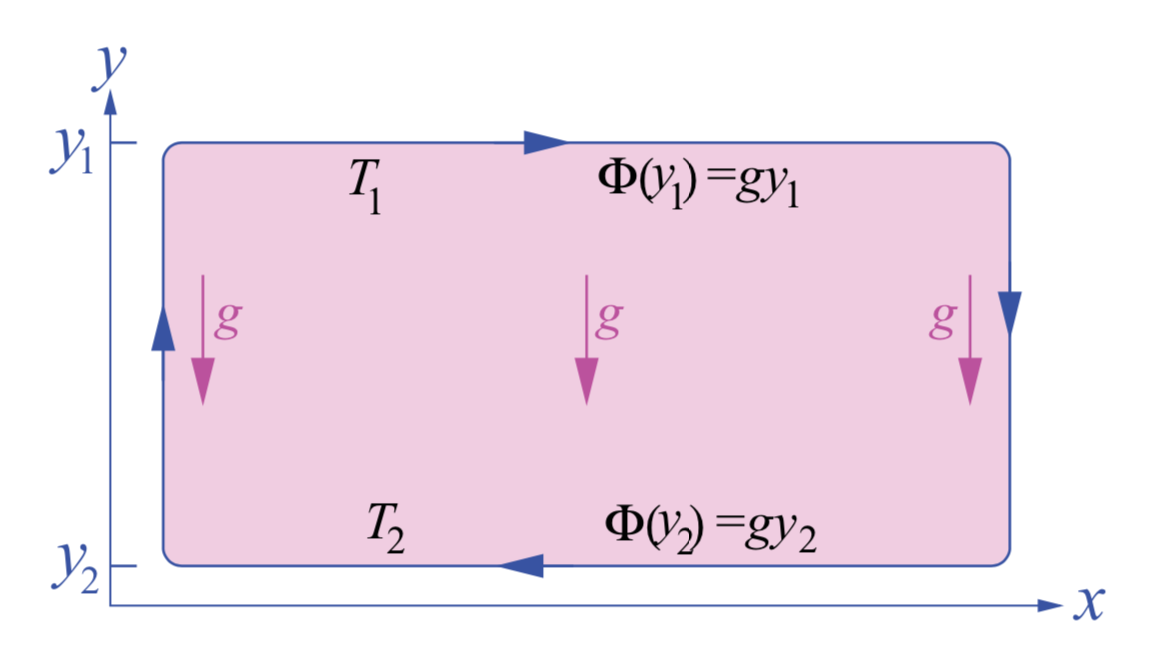}
\caption{Quantum Hall bar placed in a gravitational field along the
  $y$ direction, with upper edge kept at temperature $T_1$ and lower
  one at $T_2$ ($T_1<T_2$). The chiral edge modes are indicated by
  blue arrows.  figure from \cite{PhysRevB.85.184503}. }
 \label{fig:hallbar}
\end{figure}

We start by giving a general argument, due to Luttinger, which
connects temperature gradients with gravitational potentials
\cite{PhysRev.135.A1505}.
For this, consider a bar of material subject to both a temperature gradient
and a gravitational field in the $y$-direction, as illustrated in
Fig. \ref{fig:hallbar}, and let us first think of this as an
insulating crystal where the only excitations are phonons, and the
only transport is that of heat. The bottom is at
temperature $T_2$, and the top at $T_1$, with $T_1 =T_2+\D T< T_2$.
Now consider a fluctuation that transfers the heat
$\delta Q = \delta E$ from the hot to the cold region. The entropy
changes in the two regions are $\delta S_1 = \frac {\delta E} {T_1}$
and $\delta S_2 = -\frac {\delta E} {T_2}$, respectively. This gives a
change in the free energy,
\be{deltaf}
\delta_T F \sim -T (\delta S_1  + \delta S_2) \sim
\frac {T_1 - T_2} T \delta E =   \frac {\Delta T  } T \delta E   \, ,
\ee
where we assumed that the temperature gradient is small, so
$|\D T| \ll T \equiv (T_1 + T_2)/2$.
Since a positive $\delta E$ decreases the free energy, this
correctly gives the direction of the heat flow from hot to cold
regions. Luttinger sets out to understand such a flow in a Hamiltonian
context. Naively, this is not possible, since there is no Hamiltonian
description of diffusive processes, but he came up with a clever
trick: Due to the equivalence principle, an energy density couples
to the gravitational potential by the term,
\be{luttcoup}
F_g = \int d \mbr\, \mathcal {H} (\mbr) \frac 1 {c^2} \Phi (\mbr)\; ,
\ee
where $\mathcal H$ is the energy density, and $\Phi(y)=gy$ the gravitational
potential, $g$ being the force per unit of mass.  
Lifting the energy $\delta E$ in the gravitational potential will give
an energy contribution to the free energy,
\be{deltag}
\delta F_g =
\frac 1 {c^2} \left( \Phi(y_1) - \Phi (y_2) \right) \delta E = \frac 1
{c^2} \Delta \Phi \, \delta E \, .
\ee
Luttinger's insight was that in the equilibrium situation,
where there is  no energy
flow, we must have $ \delta_T F + \delta_g F = 0$, so,
\be{luttrel} \frac
{\Delta T} T = - \frac 1 {c^2} \Delta \Phi  \qquad \Rightarrow \qquad
\frac {\partial_y T} T = - \frac 1 {c^2} \partial_y \Phi\; ,
\ee
where the implication holds for  small gradients in $T$
and $\Phi$. This allows us to calculate the heat flow by an
equivalent gravitational problem.

\subsection{Heat flow in quantum Hall liquids}
\label{sec:gravbulkedge}

The previous Luttinger argument is not immediately applicable to the
quantum Hall effect since the bulk is gapped and the heat should be
transported by the gapless edge excitations as shown in
Fig. \ref{fig:hallbar}.  That Luttinger's relation \eqref{luttrel}
nevertheless applies to this case was shown by Stone in
\cite{PhysRevB.85.184503}, and in the rest of this section we follow
this paper rather closely.  In particular, we will show that although
a constant gravitational field will not amount to a bulk heat flow,
there is nevertheless a gravitational bulk-edge connection related to
an anomaly inflow caused by ``tidal'' effects that occur for
non-constant fields.

\subsubsection{Edge transport in a gravitational field}
\label{sec:edgegravtrans}
 From Sect. \ref{sec:thermgrav} we recall the result for
the thermal current,
\be{thermalc}
J_T = \left(  c-\bar c\right) \frac \pi {12} k_B^2 T^2 \, ,
\ee
so, again referring to Fig. \ref{fig:hallbar}, this current is larger
on the lower horizontal edge than on the upper one,
$J_{T_2}>J_{T_1}$. But this leads to an apparent contradiction, since
they are connected by the vertical edges, and the heat flow should be
conserved.

 The solution to this puzzle lies in the different definitions of
 time at the upper and lower edges due to the gradient of $\Phi(y)$,
 and the associated gravitational red and blue shifts of the modes on
 the vertical edges.  To calculate these shifts, we note that
   the wave equation for the chiral edge modes, $\vf$, depend on the
   metric as, $(\sqrt{g_{00}}\, \partial_t \pm v_F \partial_x)\vf = 0$,
   and thus $\omega \propto 1/\sqrt {g_{00}}$. By Wien's displacement
   law, the Planck distribution peaks at a frequency
   $\omega \propto 1/T$, so for the frequency shift to match the
   temperature difference we must have,
\be{redshift}
\frac {\omega (y_1)} {\omega (y_2)} =
\sqrt { \frac {g_{00} (y_2) } { g_{00} (y_1) } }= \frac{T_1}{T_2}\, .
\ee
In the limit of weak fields, the Newtonian approximation to
the metric is,\footnote{See \cite{Hobson} for the precise derivation
  of the gravitational red shift \eqref{redshift} for the metric
  \eqref{weak-g}.} 
\be{weak-g}
ds^2=g_{00}dt^2 -\d_{ij}dx^i dx^j\ , \qquad\quad {\rm with} \qquad \quad
g_{00}(y)\approx 1+ \frac{2\Phi(y)}{c^2}\; .
\ee
Using this, and also assuming that the temperature gradient is small,
\eqref{redshift} reproduces the Luttinger relation \eqref{luttrel}.

The equilibrium is upheld since the edge waves on left vertical edge
are red shifted, and those on the right vertical edge are blue
shifted, ensuring that the energy that is flowing from one horizontal
edge is ``arriving'' at the temperature of the receiving edge. This
gives a qualitative explanation for why different currents flow
on the two horizontal edges, even though they are connected by the
vertical edges.
Thus we conclude, with Stone, that the Luttinger picture also holds in
the quantum Hall case, where the heat transport is only along the
edges \cite{PhysRevB.85.184503}. 

This results may seem at variance with the case of the bulk
Hall current, but in the following we shall nevertheless find a
bulk action whose variation adsorbs the anomaly at the edge, thus
extending the anomaly inflow to the gravitational case.

\subsubsection{Gravitational bulk-edge connection in the quantum Hall effect}

Having seen how a chiral anomaly on a quantum Hall edge is canceled
by a bulk flow due to the lack of gauge invariance of the Chern-Simons
action, it is natural to ask if there is a related way to cancel the
gravitational and trace anomalies. For this, we define two
gravitational Chern-Simons actions, expressed in terms of the (matrix
valued) Christoffel one-form,
$\G^\mu_{\ \nu}=\G^\mu_{\nu\sigma}dx^\sigma$, and the spin connection one-form,
$\w^a_{\ b}=\w^a_{b \mu}dx^\mu$, respectively:
\begin{eqnarray}
S_{CS}[\G] &=&  \frac {c} {96 \pi}  \int_{\cal M} \mathrm{Tr}
\left(\G d \G + \frac {2}{3} \G^3 \right)   \, ,
\label{gamma-cs}\\
S_{CS}[\omega] &=&  \frac {c} {96 \pi}  \int_{\cal M}\mathrm{Tr}
\left(\omega d \omega + \frac 2 3 \omega^3 \right)   \, ,
\label{omega-cs}
\end{eqnarray}
where the trace is over the matrix indices, and 
where we, for simplicity, take $\bar c=0$ from now on.
These actions include cubic terms as needed for non-Abelian gauge
field, and they are, for closed manifolds $\cal M$, invariant
under the pertinent gauge transformations, diffeomorphisms and local
Lorentz transformations respectively (see App. \ref{app:calculus}).

Since we are considering gravitational backgrounds without torsion,
 the variables $\G$ and $\w$ are closely related,
$\G_\mu= e^{-1}(\w_\mu +\de_\mu )e$, and one can show that 
 the two Chern-Simons actions \eqref{gamma-cs} and
\eqref{omega-cs} only differ by a boundary term.
It turns out, however, that the variation of the  actions
define two different 3D stress tensors,
\be{3dt-def}
\!\!\!\!
\d S_{CS}[\G] = -\frac{1}{2}\int \sqrt{g}\; {\bf T}_{\mu\nu}\;
\d g^{\mu\nu}\ ,\qquad\qquad
\d S_{CS}[\w] = -\int e \; \widetilde{\bf T}_\mu^a\;\d e^\mu_a \; ,
\ee
where $\mu,\nu,a=0,1,2$.
The first one is symmetric with respect to its indices, while the
second one is, in general, not. This can be corrected for but let us focus
on $S_{CS}[\G]$, which is traceless and covariantly conserved
(there is no anomaly in odd spacetime dimensions).

Consider the half space 
$x_2<0$, with a non-trivial metric at the boundary $x_2=0$,
\be{2d-g}
ds^2=(dx^2)^2+g_{\a\b}(x^0,x^1)dx^\a dx^\b\ ,\qquad\qquad\a,\b=0,1\; .
\ee
The flux of momentum at the edge is then,
\be{flow-grav}
\dot P^\a =\int dx_1 \sqrt{g}\; {\bf T}^{2\a} \, ,\qquad\qquad \ \ \a,\b=0,1,
\ee
which should be compared to the  the anomalous conservation law of
the two-dimensional tensor $T_{\a\b}$ introduced in Sect. \ref{sec:curr},
\be{angravcons}
\dot P^\a= \int dx_1 \sqrt{g} D_\b T^{\b\a}, \qquad\quad\ \  \a,\b=0,1.
\ee

We now evaluate the expressions \eqref{flow-grav} and
  \eqref{angravcons} using the expressions, respectively, from the bulk
  and boundary theories and then verify that they match.
At  the $x_2=0$ boundary, the 3D tensor  obtained from $S_{CS}[\G]$ reads,
\be{grav-3d}
{\bf T}^{2 \a}  = -i\frac {c}{96\pi} \frac 1 {\sqrt g} \epsilon^{2\a\b}
\partial_\b {\cal R} \, ,\qquad\qquad \a,\b=0,1,
\ee
where ${\cal R}$ is the two-dimensional curvature.\footnote{
  The $i$ in front of the anomaly comes from $iS_{CS}[\G]$, which is
  omitted in \eqref{gamma-cs}, but  is important here for
  later comparison with Euclidean conformal field theory expressions.}
Here you should recall that \eqref{xx} shows that the bulk flow of the
electric current proportional to the field strength. Thus we would
naively expect that the energy-momentum flow would by proportional to
the curvature $\mathcal R$, while \eqref{grav-3d} shows that it
depends on its derivative. Thus a constant curvature gives no
contribution; the flow is a tidal effect. Also note that the curvature
is zero for a constant gravitational field, so the heat flow discussed
in Sect. \ref{sec:edgegravtrans} cannot be explained as a bulk effect.

The gravitational anomaly of the edge theory requires some discussion:
we start from the general expressions \eqref{chi-grav} parameterized
by the central charges $(c,\bar c)$,
\be{chi-grav2}
{\bf (I):}\qquad
D^z T_{zz}= \frac{c}{48\pi}\de_z {\cal R}, \qquad\quad
D^{\bar z} T_{\bar z\bar z} =\frac{\bar c}{48\pi}\de_{\bar z} {\cal R}, 
\qquad\quad T_{z \bar z}=T_{\bar z z}=0\; .
\ee
Local counterterms, which are polynomials in the metric and its derivatives,
can be added to the effective action, changing
the value of the stress-tensor trace
$T_\a^\a=g^{z\bar z}(T_{z\bar z}+T_{\bar z z})$, while keeping it symmetric
$T_{z\bar z}=T_{\bar z z}$, as explained in Ref. \cite{Witten:GravAnom}.
They can only depend on  the scalar curvature ${\cal R}$, which is
the unique local, covariant quantity that can parameterize the trace anomaly
in two dimensions. We may take,
\be{grav-tr}
\qquad {\bf (IIa):}\qquad
T_\a^\a=- \frac{\l}{48\pi} {\cal R},\qquad i.e. \qquad
T_{z\bar z}=- \frac{\l}{96\pi} g_{z\bar z}{\cal R}, 
\qquad\quad (\l=c+\bar c),\qquad\qquad
\ee
where the value of the proportionality constant $\l$ will become clear
momentarily.
Using \eqref{grav-tr} and the symmetry property
$T_{z\bar z}=T_{\bar z z}$, the anomalous conservation laws in
\eqref{chi-grav2} can be transformed into,
\be{grav-cons}
\qquad\
D^z T_{zz}+D^{\bar z} T_{\bar z z}=\frac{c-\bar c}{96\pi}\de_z {\cal R},
\qquad\qquad
D^{\bar z}T_{\bar z\bar z}+D^z  T_{ z\bar z}=
- \frac{c-\bar c}{96\pi}\de_{\bar z} {\cal R}\; .
\ee
In the case of equal chiralities $c=\bar c$, the right-hand sides of these
equations vanish, \ie $D_\a T^{\a\b}=0$, and thus
the gravitational anomaly \eqref{chi-grav2} 
has been transformed into the trace anomaly \eqref{grav-tr},
a possibility  mentioned in Sect. \ref{sec:curr}. 

For $c\neq\bar c$ there is no choice of the proportionality constant
$\l$ in \eqref{grav-tr} or any further counterterms that would completely
eliminate the gravitational anomaly \cite{Witten:GravAnom}.
Choosing $\l=c+\bar c$ is nonetheless convenient, because
it makes the anomaly equations \eqref{grav-cons} to acquire the following
covariant form, 
\be{a-grav2}
{\bf (IIb):}\qquad
D_\b T^{\b \a}  =-i\frac {c-\bar c}{96\pi} \frac 1 {\sqrt g} \epsilon^{\a\g}
\partial_\g {\cal R} \ ,  \qquad\qquad \a,\b,\g=0,1\; .
\qquad\qquad\
\ee
The expressions {\bf (I)} and {\bf(II)} in \eqref{chi-grav2} – \eqref{a-grav2}
give two equivalent forms of the anomaly. As usual,
what really counts is the interplay between the two conservation laws.

We are now ready to compare the edge anomaly (for $\bar c=0$) with the
bulk flow given by the three-dimensional Chern-Simons action
$S_{CS}[\G]$ \eqref{gamma-cs}. Inserting the two-dimensional
\eqref{a-grav2} and three-dimensional \eqref{grav-3d} tensors in the
respective flow equations \eqref{flow-grav} and \eqref{angravcons},
gives identical expressions which proves the asserted bulk-edge
connection \cite{PhysRevB.85.184503}.

For edge theories with both chiralities, $\bar c\neq 0$, the coupling
constant of gravitational Chern-Simons action $S_{CS}[\G]$
\eqref{gamma-cs} should be changed to $(c-\bar c)$, according to the
bulk-boundary correspondence just discussed. Note that under time
reversal transformations, this effective action changes sign and $c$ is
exchanged with $\bar c$ as expected (see App. \ref{app:jain} for
further discussion of edges with both chiralities).

As already pointed out, the coupling to gravity is a useful tool to
characterize neutral boundary degrees of freedom in higher dimensional
topological states. An important example are the Majorana fermions
describing topological superconductors, and in Sect. \ref{sec:global}
we shall indeed classify these states by using global gravitational
anomalies.

\subsection{Strain and viscosity}
\label{sec:strain}

In elasticity theory \cite{LandauEl}, deformations are expressed in
terms of the displacement field $u_i(x_j)$ as follows. On a lattice,
we can think of discretized coordinates $x_i=a n_i$
representing the atom positions, with $a$ the lattice spacing and
$n_i\in \Z$.  After a deformation, their positions are displaced to
$x_i'= x_i + u_i(x)$. This extends to  continuous media by
letting $x_i$ to be a real number.

The deformation clearly changes the distances between atoms, as measured
in terms of the Euclidean metric $g_{ij}=\d_{ij}$. The infinitesimal form
of the new distance is given by,
\be{new-g}
dl^2=\frac{\de x_k'}{\de x_i}\frac{\de x_k'}{\de x_j}
dx_i dx_j=\left(\d_{ij}+\de_i u_j+ \de_j u_i+
  \de_i u_k \de_j u_k\right)dx_i dx_j =g_{ij} dx_i dx_j\; .
\ee
Neglecting the quadratic term in $u_j$ for small deformations, the
expression \eqref{new-g} can be interpreted as defining a non-trivial
metric for distances between the original points. Note that is has
the same form as a diffeomorphism of the flat metric: thus, the response
to mechanical strain can be described by coupling the system
to a nontrivial metric. However,
unlike general relativity, diffeomorphisms do change the physical system,
\ie forces are not only of tidal type.

The response of quantum Hall states to gravitational backgrounds has been
extensively investigated \cite{Fro-rev, Avron, Read2011, Haldane2009}. The Chern-Simons effective action has been extended by
coupling it to a metric, leading to the Wen-Zee terms
\cite{Fro, Wen-Zee, Abanov1}. Other methods have also been
developed, such as explicit wave-function constructions
\cite{grav-bulk1, grav-bulk2},
hydrodynamic theory \cite{Hydro1, Hydro2} and the $W_{\infty}$
symmetry \cite{CR}.

\subsection{The Wen-Zee term}
\label{sec:wenzee}

In quantum Hall liquids there are some effects that are neither purely
topological, nor depend on the details of the interactions, but only  on
the geometry of the surface on which
the liquid resides. The reason is that both electrons and quasiparticles
are characterized not only by a charge, but also by an intrinsic
``orbital spin'', which is an internal angular momentum associated with
the particles, and is due both to their cyclotron motion in the magnetic field,
and their interaction. When these spins move on a curved surface, they
pick up Berry phases due to the coupling to the spin connection of the
manifold.

The addition of geometrical terms to the Chern-Simons action
has been discussed by Fr\"{o}hlich and collaborators \cite{Fro, Fro-rev}, and
by Wen and Zee \cite{Wen-Zee}. In order to motivate them, we recall 
the canonical way to couple  fermions to gravity (see App. \ref{app:calculus}), 
\be{gr-ferm}
\D\mathcal{L}=\w_{\m}^{ab}\bar{\psi}\g^\m\frac{1}{4}\left[\g_a,\g_b\right]
\psi,\qquad\qquad a,b=0,1,2\, .
\ee
In the non relativistic limit we consider a spatial metric
$g_{ij}=g_{ij}\left(t,x^i\right)$, while the other components are
$g_{0i}=0, g_{00}=1$. This reduces the local Lorentz symmetry to the rotation
in the plane $O(2)$ with Abelian connection
$\w_\m=\eps_{012}\w_{\m}^{12}$, and the interaction
\eqref{gr-ferm} reduces to $\w_\m \bar{\psi}\g^\m \psi$,
corresponding to 
the current coupling to another Abelian background $\w_\mu$.
This can be introduced in the hydrodynamics effective theory
\eqref{Seff} by the substitution,
\be{shift-sub}
j^\m A_\m\rightarrow j^{\m}\left(A_\m+s\w_\m\right),
\ee
where the constant $s$ is the orbital spin of the excitation in question.
Using the notation of forms, \ie $a=a_\m dx^\m$ \etcc we get,
\be{hydro-s}
S[a,A,\w]=-\frac{1}{4\pi\nu}\int ada
+\frac{1}{2\pi}\int \left( A +s \w \right)d a \, ,
\ee
where $\nu=1/k$ is the filling fraction.
Integration over $a_\mu$ yields the response action,
\be{Seffgrav}
S_{\rm ind}[A,g] = \frac{\nu}{4\pi}\int
\left(AdA\ +\ 2 s\; \w dA\ +\ s^2\; \w d\w \right)\ +\
\frac{c}{96\pi}\int \Tr\left(
\G d\G + \frac{2}{3}\G^3 \right) \; .
\ee
The last term is the one canceling the gravitational anomaly on the
edge as described earlier, and does not follows by the naive
evaluation of the Gaussian integral over $a_\mu$.  It was first found
by an explicit calculation for the case of a $n$ filled Landau level
of non-interacting electrons \cite{Abanov1}, and later by taking into
account the so called framing anomaly \cite{gromov2015}. This rather
subtle effect is discussed in Sect. \ref{sec:framing}.

The first term in the response action \eqref{Seffgrav}
is the standard one for the Hall current;
the other terms are called, respectively, the Wen-Zee term, 
\be{Swz}
S_{WZ}[A,g]=\frac{\n s}{2\pi}\int \w dA \; ,
\ee
and the gravitational Wen-Zee term,
\be{Sgwz}
S_{gWZ}[g]=\frac{\n s^2}{4\pi}\int \w d\w \; .
\ee

The action \eqref{Seffgrav} is not the full story.  There are other,
non-geometrical, terms that are local and gauge invariant and depend
on details of the microscopic Hamiltonian \cite{Son, Abanov1}, but
these non-universal contribution will be of no concern here.

The concept of orbital spin $s$ is relevant not only for quantum Hall
liquids, but also for paired states like superconductors. In
fractional states, $s$ depends on the interactions but for completely filled
Landau levels it is particularly simple. In the $i$-th level it is,
\be{si}
s_{i}=\frac{2i+1}{2}, \qquad i=0,\dots,n-1\; .
\ee
This expression is obtained by computing the total 
angular momentum $M_{i}$ of the $i-th$ Landau level densely filled with
$N$ electrons, starting from the origin, and then using the formula,
\be{tot-moment}
M_{i}=\frac{N^2}{2}-N s_i \, ,
\ee
whose term linear in $N$ gives the ``intrinsic'' angular momentum.

\subsection{Hall viscosity}
\label{sec:visc}

We now review the physical consequences of the
Wen-Zee action \eqref{Swz} which involves three terms,
\be{WZ3}
S_{WZ}=\frac{\n s}{2\pi} \int A d \w=
\frac{\n s}{2\pi}\int d^3x \left(\frac{\sqrt{g}}{2} A_0 \mathcal{R} + 
 \e^{ij} \dot{A_i} \w_j + \sqrt{g}\, B\, \w_0 \right)\; ,
\ee
where the dot indicates the time derivative, $i,j=1,2$ are spatial indices,
${\cal R}$ is the curvature of the spatial metric and ${\cal B}$
the magnetic field.\footnote{For a detailed discussion, see
  \cite{CR}. Curved space expressions are explained in the
  Appendix of this work and also in App. \ref{app:calculus}.}

The variation of $S_{WZ}$ with respect to $A_0$ gives an additional
contribution to the density that is expressed in terms of the 
spatial curvature. The total number of electrons $N$ for a closed
spatial geometry, \eg the sphere, is obtained by
integrating the density, leading to the result,
\be{shift}
N=\int d^2x \sqrt{g} \r= \frac{\n}{2\pi}\int d^2x\ \sqrt{g}
\left( B + \frac{s}{2} \mathcal{R} \right)= \n N_{\phi} +  
\n s \c \; ,
\ee
where $N_{\phi}$ is the  number of magnetic 
fluxes through the surface and $\c$ 
is its Euler characteristic \eqref{chi}.

Equation \eqref{shift} shows that the usual relation,
$N=\n N_{\phi}$, between electrons and fluxes is modified
on a curved surface by a $O(1)$ correction, called the ``shift'',
$\mathcal{S}= s \c$.
For the sphere, this is $\mathcal{S}=2 s$, and its value is obtained
from the total angular momentum of the wavefunction in this geometry
using \eqref{tot-moment}; for example,
$s=p/2$ for $\nu=1/p$ \cite{Wen-Zee}.
The shift is an important quantity since it can be calculated
numerically and is used to distinguish between competing model wave
functions having the same filling fraction.

Next, varying  the action (\ref{WZ3}) with respect to the spatial gauge
background gives the Hall current with an additional term,
\be{grav-E}
J^i=\frac{1}{\sqrt{g}} \frac{\d S[A,g]}{\d A_i}=\frac{1}{\sqrt{g}} 
\frac{\n}{2\pi}\e^{ij}\left(E^j + s\, E_{(g)}^j \right),
\qquad\quad E_{(g)}^i=\de_i\w_0-\de_0\w^i,
\ee
which defines the ``gravi-electric'' field $E^i_{(g)}$.

We now compute the induced stress tensor by varying the Wen-Zee action with
respect to the metric. For small fluctuations around flat space,
$g_{ij}=\d_{ij} + \d g_{ij}$, we should relate the variation of the
metric to that of the zweibeins $\d e^a_j$.
First, we must choose a gauge for the local O(2) symmetry: the
zweibeins $e^a_i$, with $a,i=1,2$, can be taken to be symmetric
matrices, \ie $e^a_i= \delta^i_j$ to lowest order.  
It follows that the variation of the metric is,
$\d g_{ij}=\d e^a_j \d_{ai} +\d e^a_i \d_{aj} =2 \d e_{ij}$.  Using
formulas explained in \cite{CR}, the Wen-Zee action \eqref{WZ3} can be
rewritten to quadratic order in $\d g_{ij}$ as follows,
\be{WZmetric} 
S_{WZ}= \frac{\n s}{4\pi}\int d^3x \left(  \ A_0 \mathcal{R} +  
\ \e^{ij} \dot{A_i} \G_{j,kl} \eps^{kl}
-\frac{B_0}{4}\e^{ij} \d g_{ik}\d \dot{g}_{jk} \right) \, ,
\ee
where $B_0$ is the constant magnetic field.

\begin{figure}[h]
\begin{center}
\includegraphics[width=0.5\textwidth]{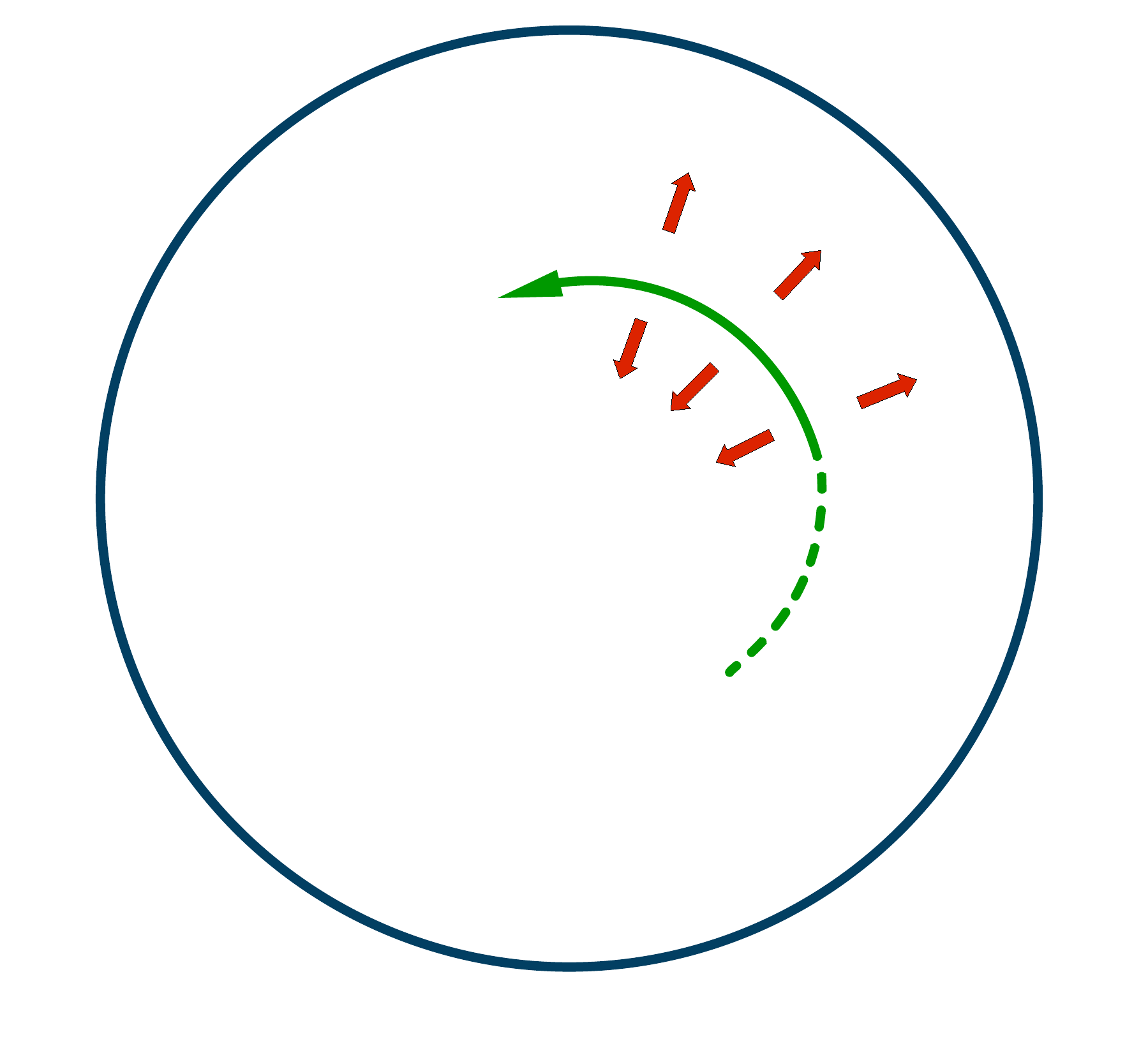}
\caption{Illustration of the Hall viscosity in a circular droplet of electrons:
a counter-clockwise stirring of the fluid in the bulk of the
droplet causes an orthogonal force (red arrows) (figure from \cite{CR}).}
\label{fig:visco}
\end{center}
\end{figure}

The induced stress tensor to leading order in the metric is therefore,
\be{stress-tensor}
T_{ij}=-\frac{2}{\sqrt{g}}\frac{\d S}{\d g^{ij}}= -
\frac{\h_H}{2}\left(\e_{ik}\dot{g}_{kj} + \e_{jk}\dot{g}_{ki} \right),
\ee
with
\be{hallvisco}
\h_H=\frac{\r_0\, s}{2}=\frac{\n\, s\, B_0}{4\pi}\; .
\ee
The quantity $\eta_H$ is a transport coefficient, called Hall
viscosity, which parameterizes the response of the fluid to a time-dependent
shear. It is only present in
two dimensional systems that violate time reversal symmetry, and
describes a non-dissipative viscous force orthogonal to the fluid
velocity (see Fig. \ref{fig:visco}), as first discussed by Avron, Seiler
and Zograf \cite{Avron, Read2009}.\footnote{
The relation \eqref{hallvisco} between the Hall
viscosity and the orbital spin $s$ is actually valid in general
\cite{Read2011}. For a non-trivial
  example at $\nu=2/5$ see \cite{Fremling}.}

Since $s$ is the coupling constant of an action of Chern-Simons type,
it describes a geometric response of the system that is independent of the
local dynamics. Furthermore, in compact geometries it parameterizes a
topological quantity (cf. \eqref{shift}), although it cannot be associated
to a anomaly of the edge theory. In Ref. \cite{CM2018}, it was shown
that $s$ amounts to a shift in the momentum of edge excitations, that is
a kind of Casimir effect, or chemical potential change. This quantity
can be unambiguously measured in edge transport experiments.

We remark that the stress tensor (\ref{stress-tensor}) is  first
order in time derivatives, and thus  corresponds to a non-covariant
force.  This is expected, since the Wen-Zee action \eqref{WZ3} is only invariant
under time-independent coordinate reparameterization.
Let us choose the conformal gauge for the metric at $t=0$,
$g_{ij}(0,x)=\sqrt{g}\,\d_{ij}$, and represent the deformations of the
fluid by time-dependent coordinate changes, $\d x^i=u^i(t,x)$.
These can be divided into: (i) Conformal transformations, for which
$\de g_{ij}=\de_i u_j+\de_j u_i=\d_{ij}\, \de_ku^k $
as explained in Sect. \ref{sec:curr};
(ii) Area-preserving diffeomorphisms, obeying $\de_k u^k=0$ and thus keeping
the determinant of the metric $g$ constant.

Upon substitution into (\ref{stress-tensor}), one finds
that the stress tensor is expressed in terms of area-preserving
diffeomorphisms only, causing the shear. These can
be parameterized by a scalar function $w(t,x)$, with the result,
\be{T-winf}
T_{ij}=\h_H\left(2\de_i \de_j- \d_{ij}\de^2 \right) \dot{w},
\qquad\qquad 
\d x^i=u^i=\eps^{ij}\,\de_j w(t,x)\, .
\ee
In conclusion, the orthogonal force in Fig. \ref{fig:visco}
is due to the deformations of the fluid given by
time-dependent area-preserving diffeomorphisms.

\section{Effective theories of symmetry protected
    topological states }
\label{sec:spt}

The last twenty years has seen an enormous progress in our
understanding of various topological states of matter.\footnote{
Or, equivalently, ``topological phases of matter''.}
We have already
discussed quantum Hall and quantum spin Hall states, as well as Weyl
semi-metals, but there are other systems, such as the topological
insulators and topological superconductors.  In the remaining parts of
these lectures we shall outline how anomalies have become an important
tool for classifying and understanding these states. To begin, we give
a brief introduction to symmetry protected topological states
of non-interacting fermionic systems (See the
first chapter in Ref. \cite{springerbook} for some complementary
material and Refs. \cite{Franz}, \cite{bernevigbook},
\cite{fradkinbook} for a thorough presentation).  We shall limit our
discussion to systems where the electronic excitations are
gapped.\footnote{ By only considering electronic excitations we
  disregard the gapless phonons that are present in all crystal
  materials.}

Next, we describe the effective theories of topological insulators and
superconductors, and explain the nature of their gapless
 boundary states. We also discuss the stability
argument by Fu, Kane and Mele that was instrumental to
understand topological insulators. Finally we mention the
possibility of having gapped boundary states that are themselves
topologically non-trivial.

In this and the following sections, you will gradually encounter the
different types of anomalies that are associated to these topological
states and learn how they are used to characterize them also in
presence of interactions.

\subsection{Introduction -- the ten-fold classification}
\label{sec:10fold}

Usual non-topological gapped systems are featureless, or ``frozen'',
when probed at energies below the gap. In topological states, there
are instead global effects, usually captured by a topological
effective theory, like massless boundary excitations, and non-trivial
responses to varying a gauge background or space geometry.  The
archetypical example is the quantum Hall effect, and the properties
captured by the Chern-Simons effective theory. In the following we
shall outline how this kind of physics extends to a wide range of
systems not necessarily involving a strong external magnetic field or
breaking of parity and time-reversal symmetries.

We have already emphasized that the integer Hall effect can be
understood in terms of free electrons, while interactions are crucial
in the case of the fractional effect.  The new topological states that
have been discovered and classified are close in spirit to the integer
case in the sense that free-fermion models capture their main aspects.
They are referred to as ``symmetry protected topological'' (SPT)
states and have gapless edge states, but no
fractionalization \cite{PhysRevB.80.155131, Wen2019}. 
They are different from the ``topologically
ordered'' (TO) states which are all fractionalized, and also have
degenerate ground states when defined on spaces with non-trivial
topology. The best known example is the $p$-fold degeneracy of the
$\nu = 1/p$ quantum Hall state on a torus, where the concept of
topological order was first introduced \cite{Wenbook}.  Yet another
feature of the TO states is that they are ``long-range entangled'',
which is manifested in a universal term in the ground-state
entanglement entropy \cite{PhysRevB.80.155131}. An SPT state is
instead ``short-range entangled'' \cite{Wen-long-range}.

There has been several proposals for fractional, and thus TO, versions
of the new topological states. These are based both on
effective field theories and explicit band theory models, but so far
they have not yet been observed experimentally
\cite{Levin-Stern,Neupert}.

The SPT states can be described by band models of non-interacting
fermions characterized by the presence/absence of the symmetries of time
reversal $(T)$ and charge conjugation $(C)$ (or particle-hole, symmetry).
The integer quantum Hall liquids are SPT states with no
symmetries.\footnote{It is fair to ask why a state without any
  symmetry should be called SPT. Some argue that it is protected by
  charge conservation. We consider this a matter of semantics.}
By definition, two states have the same TO if one can continuously
transform the Hamiltonian, and thus the ground-state wave function, of
one to that of the other without closing the gap. For the SPT states
the same holds but with the restriction that the Hamiltonian must
respect some symmetries during the whole transformation. Breaking a
symmetry can bring them into different SPT phases or trivial
insulators. 

\begin{figure}[!htb] \centering
\includegraphics[width=4.5in]{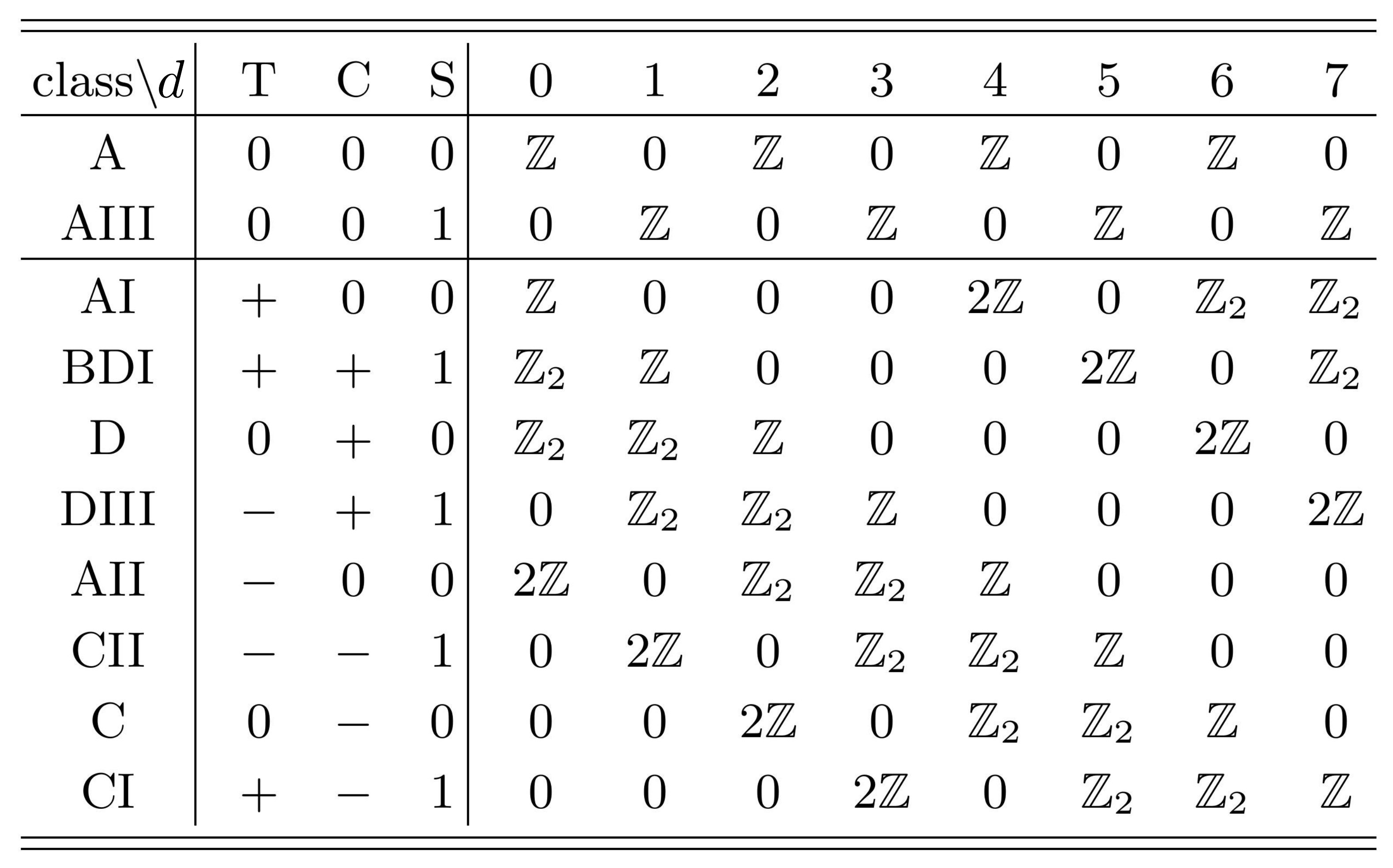}
\caption{The tenfold classification of symmetry protected topological
  phases of non-interacting fermions. See the text for explanation of
axis labels and symbols.}
\label{fig:classes}
\end{figure}

The classification of SPT states in any space dimension $d$ is shown
in Fig. \ref{fig:classes}.  We first explain the content of this table
and then discuss how it is derived.  Columns $T,C,S$ indicate the
symmetries of each class, where $S=TC$ is
called ``chiral'' or reflection symmetry of the energy spectrum with
respect to positive and negative values.
Both time reversal  and charge conjugation 
are antiunitary maps that square to $T^2=\pm 1$, and $C^2=\pm 1$. The
alternatives are denoted as $(+)$ and $(-)$ in the table, and the
third option $(0)$ means that there is no symmetry. This would imply
nine classes, but when neither $T$ nor $C$ is present there are
nevertheless two alternatives for the third entry $S$.
The labels in the first column refer to the classification of
maximally symmetric spaces due to the mathematician E. Cartan. These
space are similar to spheres in the sense that all point are
equivalent.  The correspondence between the two classifications is by
the theory of localization, to be discussed later.

The integer Hall effect has no symmetries and appears in the first row
in the table, labeled by (A); here $\Z$ denotes the possible values of
the Hall conductance in units of $e^2/h$.  As explained earlier, this
integer is the first Chern class as calculated from the Berry phase
(or alternatively, the monopole charge in the spectral flow of edge
states). Correspondingly, the integer sets listed in the other cases
give the possible values of other characteristic topological
invariants (with $(0)$ indicating no topology).  In more physical
terms, these integers count the number of branches of boundary excitations.

The topological insulators, already briefly introduced in
  Sect. \ref{sec:SQHE}, belong to class AII, are protected by TR
symmetry that squares to minus one, and have index $\Z_2$ in both two
and three dimensions, meaning that their boundary excitations have one
branch at most -- later we shall explain how this comes about.  An
interesting feature of the classification is that the same pattern
repeats itself in any dimension, shifted one step down, with period
two for the first two rows and eight for the remaining ones. This
feature is related to the properties of fermions in different
dimensions. 

The ten-fold classification was originally obtained using
three approaches:\footnote{For a review, see \cite{Ludwig-rev}.}
 \begin{itemize}
 \item
   The first, referred to as ``topological band theory'', studies
   general translation invariant lattice models using the mathematics
   of K-theory \cite{kitaev2009}.
 \item
   The second analyzes Anderson localization in non-interacting electron
   systems, and finds a relation  to the
   classification of random matrices and diffusion equations on
   maximally symmetric manifolds \cite{PhysRevB.78.195125}.
 \item
The third considers the representations of the $T$, $C$ and
   $S$ symmetries acting on fermion spinors.  In fact,
   Ref. \cite{Ryu_2010} showed that for any SPT state, one can find a
   ``representative'' massive Dirac theory in the continuum with the
   same topological properties as those of more realistic microscopic
   lattice Hamiltonians.
\end{itemize}

In these lectures, we do not explain these results, but describe
another approach that uses effective field theories 
and anomalies (initiated in Ref. \cite{PhysRevB.85.045104}).
Using the continuum theories associated to SPT states by the
third method above, we identify the associated massless boundary excitations
and use their characteristic anomalies to detect and classify them.

Actually the use of anomalies has major advantages: They are exactly
known even in the presence of interactions, and moreover are related to
topological quantities that are insensitive to small changes in the
Hamiltonian. Therefore, a classification based on anomalies also
applies to interacting systems. This is obviously very important since
the electrons in all real material are interacting.  In later
sections, we shall provide examples where the original ten-fold
classification breaks down by introducing cleverly chosen
interactions, which turn the SPT state into a trivial insulator
without closing the gap.  In these cases, anomalies give the correct
answer for the stable topological phases with interactions. We shall
come back to this in Sect. \ref{sec:intspt}.

\subsubsection{Some  historical remarks}

Here we describe some hallmarks in the development that led to the
discovery of topological states of matter beyond the quantum Hall
effect. For comprehensive reviews see \eg \cite{RevModPhys.82.3045}
and \cite{RevModPhys.83.1057}.

\medskip
\emph{The Chern Insulator.}

Haldane constructed a model of two-dimensional lattice fermions that
has gapless edge excitations and a quantized Hall conductance, in
absence of an external magnetic field \cite{PhysRevLett.61.2015}.  The
lattice is hexagonal, as in graphene, but the Hamiltonian has
additional next-to-nearest neighbor couplings with nontrivial phase
factors that break TR invariance. One can think of them as staggered
fluxes that average to zero over a unit cell.  Haldane found that the
Berry curvature is non-vanishing over a range of coupling constants,
and integrates to the Chern invariant and thus a quantized Hall
conductance $\s_H=\pm e^2/h$ \cite{tknn}.

In more physical terms, the effect of the external magnetic field is
traded for phase dependent couplings in the Hamiltonian, resulting
into a topologically non-trivial Berry connection, a gauge connection
in parameter space instead of real space. 

Regarding the effective field theory description, the low-energy
spectrum of the Haldane model includes two Dirac fermions in $(2+1)$
dimensions, that break TR symmetry when their masses obey
$m_1+m_2\neq 0$. In this system, there is a ``global'' anomaly, a new
kind of anomaly affecting the discrete symmetries of parity and time
reversal, to be discussed in Sect. \ref{subsec:parityanom}. It implies
a low-energy  Chern-Simons effective theory  with
coupling $k=\pm 1$.  This field theory result confirms the presence of
the Hall current and edge excitations.

\medskip
\emph{The quantum spin Hall effect.}

This system was already introduced in Sect. \ref{sec:SQHE} in a rather
formal way by considering a pair of quantum Hall system whose edge
excitations have opposite chiralities and spins, so as to obtain a TR
invariant system. It can be actually realized in some materials in
presence of strong spin-orbit coupling, of the form
$({\bf E} \times {\bf p}) \cdot {\bf S}$.\footnote{An electric field
  is not necessary to generate this state. The intrinsic spin-orbit
  coupling in heavy atoms, and the presence of a
  substrate can also lead to the quantum spin Hall effect. Actually,
  the interplay of these factors has led to the observation of conductive
  edge states even at room temperature \cite{reis2017bismuthene}.}
When the electric field $\bf E$ is confined to a plane
and linear in the $(x,y)$ coordinates, one obtains the TR invariant
Hamiltonian,
\be{sqhe}
H=\frac{1}{2m}\left( {\bf p} - {\bf A}\, \s_z\right)^2 , \qquad\qquad
{\bf A} = \frac B 2 \left(-y, x, 0\right)\; ,
\ee
which actually represents the two opposite Hall systems, respectively
for spin-up and spin-down electrons \cite{Bernevig-Zhang}.

The quantum spin Hall effect is interesting because TR symmetry makes
it rather different from the quantum Hall effect. The Hall
conductivity as well as the Berry phase vanish in this case, but there
is a non-vanishing spin Hall conductivity given by the difference of
currents of the two kinds of electrons. 

\medskip
\emph{ Topological insulators.}

One would naively think that edge excitations of opposite chirality
can interact and create a gap, thus making the quantum spin Hall
system a trivial insulator.  However, such interactions are not
possible without breaking time-reversal symmetry, because only half of
the possible scattering channels are present, given that the signs of
velocity and spin are paired.\footnote{For a more detailed explanation
  of this, see Sect. II C of \cite{RevModPhys.83.1057}.}
Still, there can be TR invariant interactions that causes
spin-flips and violate spin current  conservation. 

The question of stability of topological insulators,
\ie realistic TR invariant gapped systems with topological features
(class AII),
was addressed and solved by Kane, Mele and Fu in a series of papers
\cite{PhysRevLett.95.226801,PhysRevLett.95.146802,Fu-Kane-Mele}.
They introduced a $\Z_2$-valued topological quantity which generalizes
the Berry phase, and  measures the spin parity
$(-1)^{2S}=(-1)^F$ (\ie the fermion number parity) of the boundary modes.
This sign is the remnant of the $U(1)_S$ symmetry of the spin Hall
effect, once spin flip transitions are introduced while preserving TR
invariance.  These authors also generalized the Laughlin flux
insertion argument to explain how this index characterizes the TR
invariant topological phase in the presence of TR invariant
interaction.  In Sect. \ref{sec:tistab} we shall discuss these results
further.

\medskip
\emph{Experimental observation of topological insulators and the tenfold
  classification.}

Topological insulators were observed in two and three space dimensions
in 2007 and 2008, respectively, and their gapless boundary excitations
have been observed by measuring responses and photoelectric emission
\cite{RevModPhys.82.3045}.  At about the same time, the classification
of the SPT phases was achieved, using the methods outlined
above\cite{RevModPhys.83.1057}.

\subsection{Effective response action for 4D topological insulators
} \label{sec:2nspt}

In this section we analyze SPT states with an unbroken electromagnetic
$U(1)$ symmetry.  We shall derive the bulk topological effective
theories and the associated boundary dynamics. As seen already, the
response of a system is coded in the effective response action
$\Gamma[A]$, and we show that in 4D (as well as in any even
dimension), this contains a piece which encodes an ``anomalous
response'' to external fields, and that it implies the existence of
``protected'' surface modes.  As for the physical significance of this
result, we already mentioned that the Dirac continuum theory is just a
representative theory for extracting the topological properties of
more realistic models \cite{Hosur}.

\subsubsection{The $\theta$ term and the Chern-Simons boundary action }
\label{subsec:topthetaact}

We will now apply the techniques we used to derive chiral and axial
anomalies in even space-time dimensions to calculate possible
topological terms in the electromagnetic response action. The starting
point is the representative massive Dirac action,
\be{repact}
S_E =   \int dx\, \psib \left( i\slashed  \partial +
\slashed A  - M_0(\cos \theta + i \sin\theta  \gamma^{D+1} \right) \psi \; ,
\ee
and you can find the derivation of this action from topological
band theory in \eg \cite{Ryu_2010}.
In \eqref{repact} $\theta$ is a constant angle, and we note that
changing it from
0 to $\pi$ amounts to switching sign of the mass $M_0$.  Under the axial
symmetry transformation
$\psi \rightarrow \exp(-i\frac \xi 2 \gamma^{D+1})\psi$, theta changes
as $\theta \rightarrow \theta - \xi$.  We can now calculate the change
in the effective action $\Gamma_\theta[A]$ as we vary $\theta$. But
this is precisely the calculation that we did in
Sect. \ref{sec:anpath} to get the anomaly,
\be{effanom}
\delta \Gamma_\theta[A] = 2i \int dx\, \frac {\d\theta} 2 \mathcal {A}\; ,
\ee
where $\mathcal A$ is the anomaly given by \eqref{acutoff}. This can
immediately be integrated to give the ``theta term'' in the effective
action, which in 4D becomes,
\be{4Dtheta}
\Gamma_\theta[A]- \Gamma_0[A]= i {\theta} \frac 1 {16
  \pi^2} \int_{\cal M} dx\, F \tilde F = i\theta P= i \theta\,
\mathrm{ind} (\slashed D )\; .
\ee
Since for closed 4D manifolds $\cal M$ the index in \eqref{4Dtheta}
is an integer, the partition function $\exp( i \Gamma_\th[A])$ is 
a $2\pi$-periodic function of $\theta$.   The
lesson we get from the above calculation is that SPT
states in even dimensions can have a theta term in the effective
action.  Also note that symmetries will impose constraints on
$\theta$. Under time reversal, $i\Gamma$ changes sign, so for TR invariant
states, the only allowed values are $\theta\in \{0,\pi\}$, modulo $2\pi$ .  
If we further assume that $\theta = 0$
describes a trivial insulator equivalent to the vacuum, we also have a
non-trivial, TR invariant,  SPT phase for
$\theta = \pi$ which is the 3d topological insulator.\footnote{
Note that the theta term \eqref{4Dtheta} is purely imaginary both in
  the Euclidean formulation of Section \ref{sec:anpath} and in the
  Minkowskian version used for physical application, being subjected
  to TR invariance.  Furthermore, this symmetry can also be directly
  formulated in Euclidean space, as described in Sect. \ref{sec:globalindex}.}

An important point about $ \Gamma_\theta[A] $ is that
it is a topological invariant: Its variation with respect to $A$
vanishes, so it does not give any current response,
and also no contribution to Lagrangian equations of motion. This is,
however, true only if the theory is defined on a closed manifold.
In reality, any topological insulator has boundaries, and then
$\Gamma_\theta[A]$ is no longer a constant. To see this we write,
\be{thetacs}
\Gamma_\theta[A] &=&  i  \frac{\theta}   {8 \pi^2} \int_{\mathcal M} dx\,
\partial_\mu \epsilon^{\mu\nu\sigma\lambda} A_\nu \partial_\sigma A_\lambda 
\nl
&=&  i   \frac {\theta} {8 \pi^2}  \int_{\partial\mathcal M}
dx\, \epsilon^{\nu\sigma\lambda} A_\nu \partial_\sigma A_\lambda \; .
 \ee
and in the last equality  we recognize the 3D Chern-Simons action with
level $k=\th/2\pi$. Taking $\theta =\pm \pi$,
for topological insulators, we have $k=\pm 1/2$.
We recall from Sect. \ref{sec:CS} that this is effective action for
a quantum Hall state with Hall conductivity corresponding to
$\nu=\pm 1/2$.

\begin{figure}[t]
\begin{center}
\includegraphics[width=8cm]{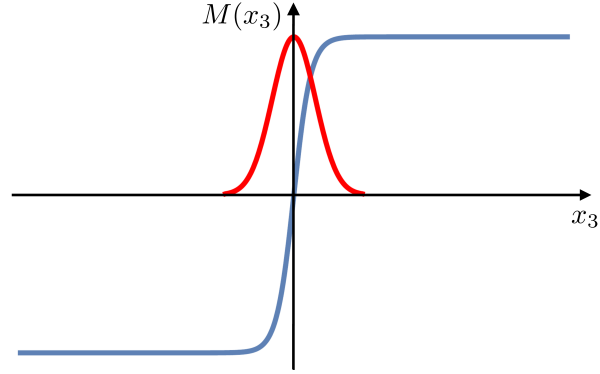}
\caption{Mass profile $M=M(x_3)$ (blue line) at the boundary of a topological
  insulator and wavefunction of the Dirac zero mode (red line).}
\label{fig:kink}
\end{center}
\end{figure}

This result raises two questions. First, a Hall state on the
boundary would seem to violate the TR symmetry. Second, the
sign of the Chern-Simons term differ depending on whether $\theta$ is
$\pi$ or $-\pi$ although $\theta$ is required to be defined only
modulo $2\pi$.  To resolve these issues, we will consider a
regularized version of the interface, and model it by taking the mass
in \eqref{repact} to have a kink profile $M(x_3)$ such that
$\lim_{x\rightarrow\pm\infty} M(x_3) = \pm M_0 $ (see Fig. \ref{fig:kink}).
This amounts to taking
$\theta = \pi$ in the left half space and $\theta = 0$ in the right
half space, with the boundary at $x_3 = 0$. Furthermore, we shall
explicitly break TR invariance by adding a small Pauli coupling with
magnetic moment $\mu$. The reason for this is twofold. It corresponds
to an interesting physical situation when a TI is subjected to a weak
magnetic field, and it also provides an infrared cutoff that allows
for a well defined perturbative calculation. The Pauli term for a
magnetic field in the $x^3$ direction, \ie perpendicular to the
surface, is,
\be{pauli} 
H_p = - \mu B \psi^\dagger \Sigma^3 \psi =
\mu B \bar\psi \gamma^5\gamma^3 \psi\; ,
\ee
where we used the relation
$\Sigma^i = i\eps_{ijk}[\g^j,\g^k]/4=\gamma^5\gamma^0\gamma^i$
for the Dirac spin vector \cite{IZ}. 

We now look for solutions to the Dirac equations that are localized
at the boundary. We first note that  the Dirac equation \eqref{repact}
admits a static (zero energy) solution localized at $x_3 = 0$
(see Fig. \ref{fig:kink}),
\be{transdirac}
[i\gamma^3\partial_3 - M(x_3)] \psi(x_3) = 0\; ,
\ee
as originally shown by  Jackiw and Rebbi in $(1+1)$-dimensions
\cite{Jackiw-Rebbi}.
This allows for the dimensional reduction of the 4D fermion to the
$3D$ boundary, as follows.

We use the following basis for the $\g$ matrices,
\begin{equation}
\label{rep2}
\gamma^0= \left(\begin{matrix}
0 & \sigma_3 \\
\sigma_3 & 0
\end{matrix}\right), \ \ 
\gamma^1= i\left(\begin{matrix}
0 & \sigma_1 \\
\sigma_1 & 0
\end{matrix}\right),  \ \ 
\gamma^2= i\left(\begin{matrix}
0 & \sigma_2 \\
\sigma_2
 & 0 \end{matrix}\right),  \ \ 
\gamma^3= i\left(\begin{matrix}
1 & 0 \\
0 & -1
\end{matrix}\right),
\end{equation}
and seek for a general solution with separation of variables,
\be{normsol}
\psi=\chi(x_\a)\phi(x_3), \qquad\qquad
\phi = N 
\begin{pmatrix} \chi \\ 0 \end{pmatrix}
\exp\left(-\int^{x_3} dx' M(x') \right)\, ,
\ee
where $\a=0,1,2$.
Upon substituting it in the Dirac equation \eqref{repact},
putting $A_\mu = 0$ and adding the Pauli term \eqref{pauli}, we get,
\be{edgediraclag}
{\mathcal L} = \begin{pmatrix} 0, &  \chi^\dagger \s_3  \end{pmatrix}
\begin{pmatrix}
  0 & i\s_3 \de_0 -\s_1 \de_1 -\s_2\de_2 -m \\
   i\s_3 \de_0 -\s_1 \de_1 -\s_2\de_2 -m & 0 \end{pmatrix} 
\begin{pmatrix} \chi \\ 0 \end{pmatrix} ,
\ee
where $m = -\mu B$.
Identifying the $3D$ gamma matrices,
$\gamma^\alpha= (\sigma_3, i \sigma_1, i\sigma_2)$, $\alpha = 0,1,2$, 
and $\bar\chi = \chi^\dagger \sigma_3$ we get the $3D$ Dirac equation,
\be{finboundary}
{\mathcal L} =\bar\chi \left(i\slashed \partial + \slashed A - m\right)
\chi\; ,
\ee
where we reintroduced the electromagnetic field in the gauge
$A_3 = 0$. Note that the TR breaking Pauli term appear as a mass in
the (2+1)D Dirac boundary theory. At this point we make a slight
digression and derive the effective response action for this theory.

\subsubsection{Chern-Simons action from Dirac fermions
  in 2+1 dimensions } \label{subsec:parityanom}

The most direct way to extract the electromagnetic effective action
$\G[A]$ is to start from the Lagrangian \eqref{finboundary} and 
simply calculate the Feynman diagram in Fig. \ref{fig:polten} with a
suitable regularization as was originally done in
Ref.\cite{redlich83}, and we will return to this in Sect. \ref{sec:global}. 
As an illustrative alternative, we shall here use
the same idea as in the calculation of the axial anomaly in
Sect. \ref{sec:3dhotel}, \ie solve the Dirac equation exactly in a
constant magnetic and then invoke gauge and Lorentz invariance to find
the general result.

We take the following $(2+1)$-dimensional Dirac $\alpha$-matrices,
\begin{align}
  (\beta,\alpha^{x},\alpha^{y})=(\sigma^{3},-\sigma^{2},\sigma^{1})
  \label{diral} \, ,
\end{align}
use the complex coordinates,
\begin{equation}
z=\sqrt{\frac{eB}{2}}(x+iy) \, ,
\end{equation}
and the notation $\partial=\partial_{z}$ and
$\bar{\partial}=\partial_{\bar{z}}$,  as  in Sect.~\ref{sec:LL}.
In the symmetric gauge, ${\bf A}=\frac{B}{2}(y,-x)$, where $B$ is
a constant magnetic field in the negative $z$-direction,
the Hamiltonian for the relativistic Landau problem becomes, 
\be{dirham}
H & =&\vec{\alpha}\cdot(\vec{p}+ e\vec{A})  + \beta m
\nl
& =& \sqrt{eB}\left(
  \begin{array}{cc}
\mu & \frac{1}{\sqrt{2}}(\partial-\bar{z})\ \\
-\frac{1}{\sqrt{2}}(\bar{\partial}+z) & -\mu
  \end{array}\right)=
\sqrt{eB}\left(\begin{array}{cc}
\mu & a^{\dagger}\\
a & -\mu \end{array}\right)\; ,
\ee
where  $[a,a^{\dagger}]=1$, and we defined $\mu = m/\sqrt{eB}$.
In terms of the number operator states $a^{\dagger}a\ket n=n\ket n$,
we easily find the following Landau level spectrum for the $m=0$ case,
\begin{align}
  \ket{\Psi_{0}}
&   =\left(\begin{array}{c} \ket 0\\ 0 \end{array}\right);
& & E_{0}=0 \; ,
\label{spect}\\
  \ket{\Psi_{n\pm}}
& =\frac{1}{\sqrt{2}}\left(\begin{array}{c} \ket n\\
   \pm\ket{n-1} \end{array}\right);
&  & E_{n\pm}=\pm\sqrt{neB}\ ,\qquad n>0\; .
\end{align}
Thus there are infinitely many positive and negative levels which are
paired by the charge-conjugation (particle-hole) symmetry. The
exception is the zero-mode which is self-conjugate.  The solution for
$m\neq0$ is easily found by observing that $H^2$ is a diagonal
matrix. The $\ket{\Psi_{0}}$ remains an eigenstate with the energy
$E_{0}=m$, while the rest of the spectrum has energies
$E_{n\pm}=\pm \sqrt{neB+m^{2}}$, for $n>0$.
We should also remember that, just as explained in
Sect. \ref{sec:LL}, all the Landau levels are highly degenerate.

The current operator $j^{\mu}= e :\bar{\psi}\gamma^{\mu}\psi:\, $       
includes the normal ordering to subtract
the contribution of the filled Dirac sea, while respecting charge conjugation
symmetry. In a static magnetic field the spatial components of the current
vanish: recalling that $\gamma^{0}=\beta$, we get the following expression
for the charge density,
\be{chargden}
\langle j^{0}\rangle=\frac{e}{2}\langle\W|\left(
\psi^{\dagger}\psi - \psi\psi^{\dagger} \right)|\W\rangle.
\ee
This expectation value is charge conjugation symmetric and receives
contributions from the states in the $E_0=m$ level only.
Setting the Fermi energy to zero, we find that
for $m<0$ all states in this level are all filled, and
the expectation value comes entirely from the first term
in \eqref{chargden}, while for $m>0$ only the
second term  contributes.
Recalling from Sect. \ref{sec:LL} that the density of a Landau level
is $\rho_0 = eB/2\pi$, we get the ground state charge density,
\begin{align}
  \av{j^0}= - \frac{m}{|m|} \half \frac{e^{2}}{2\pi} B\,\label{expchfin} \, .
\end{align}
To understand this result you should recall (or prove!) that a Dirac
mass term in (2+1)d violates both TR invariance and parity.\footnote{
  Parity in 2d is a reflection \eg $(x,y) \rightarrow (-x,y)$ since the
transformation $(x,y) \rightarrow (-x, -y)$ amounts to a $\pi$ rotation.}
A way to realize this is to first note that in 2d the
group of spatial rotations is just O(2), so the spin of a
fermion is a pseudoscalar number fixed to $S=\pm 1/2$.
Contrary to the 4D case the mass cannot change sign
by a chiral rotation. Furthermore
the sign of the spin and mass are related by
$S=\mathrm{sign}(m)/2$, and the antiparticles will have the opposite
sign.\footnote{To prove this, you can \eg work out the total angular
  momentum operator for 3d Dirac particles or derive the Zeeman term
  in a non-relativistic limit\cite{hansson1991anyons}. }
Since the TR operation changes the sign of angular momentum,
the mass term breaks TR invariance. 

Just as in the arguments based on the infinite hotel, we can now
invoke  Lorentz and gauge invariance to conclude, 
\be{3d-curr}
  \av{j^{\mu}}  =\frac{1}{i}\frac{\delta}  {\delta A_{\mu}} \Gamma[A] =
  - \frac{m}{|m|}\frac{e^{2}}{4\pi}\varepsilon^{\mu\nu\sigma}
  \partial_{\nu}A_{\sigma}\, ,\qquad\qquad
  \left( F_{\mu\nu}= {\rm const.}\right)\; ,
\ee
for constant electric and magnetic fields.  Integrating this
expression we get the Chern-Simons response action,
\be{3d-anom}
\G[A]=\frac{m}{|m|}\frac{e^2}{8\pi}\int d^3x
\eps^{\mu\nu\r}A_\mu\de_\nu A_\r \; ,
\ee
with coupling $k= {\rm sign}(m)/2$. You should note that even though
we had to introduce a mass to make this calculation well defined, the
final result only depends on its sign.  

Finally note that \eqref{3d-anom} does not include Maxwell terms, nor
higher derivatives and powers of $F_{\mu\nu}$ that do not
contribute to the current \eqref{3d-curr} for
$F^{\mu\nu}= {\rm const.}$ \cite{redlich84}. These additional terms
are non-anomalous and do not enter in the following discussions.

\subsubsection{Boundary action for the 3d topological insulator}
\label{subsubsec:fulltiaction}

We are now prepared to revisit the questions raised in
\ref{subsec:topthetaact} about the nature of the boundary states.
First we note that, not unexpectedly,
regularizing with a small Pauli term $\sim B$ determines the chirality
of the boundary, or equivalently, the sign of the Chern-Simons
action. But we are left with another serious problem.  Both the
argument leading to \eqref{thetacs} and the effective action
\eqref{3d-anom} give a Chern-Simons term with a half integer level
number, in contradiction with the expected integer value for a theory of free
fermions. So the problem is now whether these results should be added
to give an integer level number, or are they two ways to calculate the
same thing, in which case the level number is indeed half
integer. This question was addressed in a paper by Mulligan and
Burnell \cite{mb}, who did the full calculation of the two point
function taking into account both bulk and boundary effects. They used
the special mass profile,
\be{massprofile}
M(x) = M_0 \tanh (M_0 x )\; ,
\ee
where the full fermion propagator can be found analytically. The
calculations are cumbersome, and we only present the result for the
effective boundary action which is $\Gamma [A] = k \,S_{CS}[A]$ where
the level number is,
\be{mbexact}
k = \frac {m} {2|m |} - \frac 1 \pi \arctan \left( \frac {m}{M_0} \right)\; .
\ee
In the physical limit $m \ll M_0$ we retain the result from
\ref{3d-anom} while in general, the level number is not
quantized.\footnote{ The opposite limit $M_0\ll m$ where $k\rightarrow 0$
  is unphysical since the bulk gap vanish and there is no distinction
  between the topological and normal insulator.}  Although this
calculation was done using the special profile \eqref{massprofile},
there is no reason to believe that the result would be qualitatively
different for a more general profile. 

In summary we found that there is indeed a $\nu=1/2$ quantum Hall
effect on the boundary, with the chirality set by the weak magnetic
field. One should not add the contributions from the $\theta$-term in
the bulk action to the one originating from the boundary fermion, but
consider them as alternative derivations of the same effect.

With this said, there are still problems left. How come that we can
get a $\nu = 1/2$ state starting from free fermions, and is there any
way to handle the boundary theory for the case when $m$ is strictly
zero, and where we would not expect any TR non-invariant terms? To
address these questions we must learn about the parity anomaly, which
is an example of a global anomaly. This will be the subject of 
Sect. \ref{sec:global}.

Not surprisingly, one can also use anomalies to extract the thermal
response of 3d TIs by coupling to gravity and using
\eqref{gamma-cs}. We will not pursue this here, but refer to
\cite{Vozmediano} for a good explanation with references to the
original literature.

\subsection{Stability of topological insulators --
  the flux insertion argument}
\label{sec:tistab}

We have already stressed that the strength of the anomaly approach to
classifying SPT states is that it is topological and should thus be
robust against interactions as long as they respect the symmetry.
In the case of topological insulators there is, as we will now discuss,
a set of arguments for the stability which predates the analysis based on
anomalies but is very close in spirit and eventually 
  fits into it, as will become clear later.

The question, originally analyzed by Kane and Mele, is whether TR
symmetric interactions added to the boundary fermions can gap them
leading to the decay into trivial insulators.  We have already
mentioned that a TR symmetric mass term can be introduced for a pair
of fermions, while a single mode is anomalous and protected.  A
general answer beyond quadratic Hamiltonians and band theory can be
given using a flux insertion argument that is closely related to the
spectral flow we have have discussed
earlier\cite{PhysRevLett.95.226801,PhysRevLett.95.146802,Fu-Kane-Mele}.
The outcome of this analysis is that the $\Z_2$ characterization of
topological band insulators in two and three space dimensions (class
AII in Fig. \ref{fig:classes}) remains valid in the presence of
interactions.

\vskip 2mm \noi
\subsubsection{The Kramers theorem in time reversal invariant systems}

The action of the TR symmetry operation on states is represented by
the antiunitary operator ${\cal T}=UK$, where $U$ is unitary and $K$
is the complex conjugation, that also maps bra into kets. On a state
$|S\rangle$ of spin $S$, the square of the transformation gives,
\be{t-square}
{\cal T}^2|S\rangle = (-1)^{2S} |S\rangle =(-1)^F |S\rangle\; ,
\ee
where the sign can be called spin parity or fermion parity.
For fermionic states (\ie $2S$ odd integer), $\cal T$ squares to minus one:
in the case of Dirac fermions, using Euclidean gamma matrices
$\gamma^\mu = (\sigma^1,\sigma^2,\sigma^3)$,  the time-reversal
operator is ${\mathcal T} =i \sigma^2 K$,
that satisfies ${\mathcal T}^2 = -1$.

Let us consider a system with TR invariant Hamiltonian,
${\cal T H T}^{-1}={\cal H}$. Kramers theorem \cite{Griffiths}
uses antiunitarity of ${\cal T}$ and ${\mathcal T}^2 = -1$ to
prove that the following half-integer spin states,
\be{K-pair}
|S\rangle \qquad {\rm and}\qquad |S'\rangle= {\cal T}|S\rangle,
\qquad\qquad\quad ( 2S \ {\rm odd}),
\ee
called a Kramers pair, are orthogonal,
\be{K-pair2}
\langle S|S'\rangle=\langle S|{\cal T}|S\rangle=0 \; .
\ee
Furthermore, the following relations holds
for any Hermitian TR invariant operator $\cal O$, \ie obeying
${\cal T O T}^{-1}={\cal O}$,
\be{K-ref}
\langle S'|{\cal O}|S\rangle=0\, , \qquad\quad
\langle S'|{\cal O}|S'\rangle=\langle S|{\cal O}|S\rangle, \qquad\qquad
(2S\ {\rm odd})\, .
\ee
These imply that the degeneracy of the Kramers pair cannot be
lifted by any TR invariant interaction (Kramers degeneracy).

Let us now discuss the breaking of this symmetry by the addition of
a magnetic field. We consider geometries that allow flux insertions,
like the two-dimensional annulus in Sect. \ref{sec:hallcurr}.
The  Hamiltonian has the following properties 
 in the presence of fluxes,\footnote{For any compact space
 direction that allows flux insertion.}
\be{TR-H}
{\cal T}\;H[\Phi]\; {\cal T}^{-1}=H[-\Phi]\; ,\qquad\qquad
H[\Phi+\Phi_0]=H[\Phi]\; ,
\ee
where the second relation follows from the Byers-Yang theorem, which
implies that the spectrum returns to itself after adding one flux quantum.
From these relations, it follows that there are discrete flux values,
$\Phi=n\Phi_0/2$, $n\in\Z$, for which the system is TR invariant also
in the presence of a magnetic field.

\begin{figure}   \centering
\includegraphics[width=7cm]{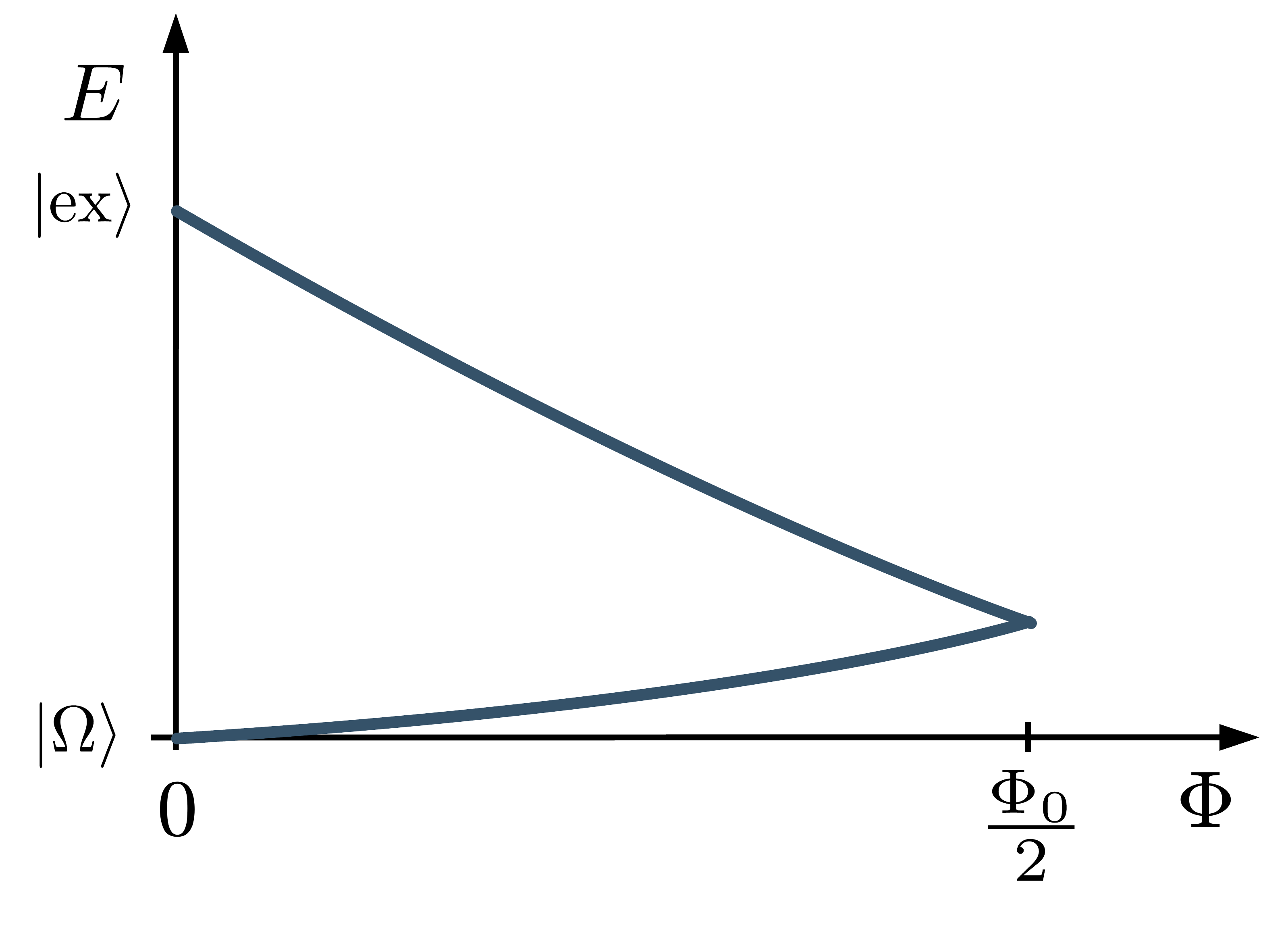}  
\caption{Energy levels of the ground state and an excited state as a
  function of the flux inserted, showing the Kramers degeneracy at
  half flux quantum.}
  \label{fig:stab}
\end{figure}

\vskip 2mm \noi
\subsubsection{The flux insertion argument and the $\Z_2$ invariant}

We first consider the quantum spin Hall effect in two dimensions
introduced in Sect. \ref{sec:SQHE}. This is a rather idealized
model of a topological insulator, because $S_z$ conservation is in
general broken to $(-1)^{2S}=(-1)^F$ by spin-flip transitions. However, it is
simple, and actually sufficient for the sake of the following argument
that only appeals to fermion parity conservation.

In Sect. \ref{sec:SQHE} we discussed the spectral flow in the quantum spin Hall
effect and showed that the addition of a flux quantum causes positive and
negative spin electrons to move in opposite directions, leading to
a spin Hall current with $\nu=2$. On a the outer edge of the annulus
(see Fig. \ref{fig:ti}), an edge excitation,
$|1\rangle=\Th[\Phi_0]|\W\rangle$,  is created, where $\Th[\Phi]$ is
the flux-insertion operator and $\ket\W$ the ground state.
The quantum numbers of $|1\rangle$ obtained by the flows are
$\D Q=Q^{\uparrow}-Q^{\downarrow}=1-1=0$ and 
$\D S_z=1/2- (-1/2)=1$ (a similar excitation appears in the inner edge with
opposite spin). 

Let us now consider what happens at half flux insertion.
This  creates a neutral $S_z=1/2$ excitation,
$|1/2\rangle=\Th[\Phi_0/2]|\W\rangle$,
on the outer edge. Since $\Phi_0/2$ is a TR invariant flux value, we
can use the Kramers theorem\footnote{A local version of the Kramers theorem
  at each edge is necessary as described in \cite{Stern}; here we
  follow closely their presentation.}
and conclude that this state is degenerate with another state of opposite spin
$|-1/2\rangle={\cal T}|1/2\rangle$, as shown in Fig. \ref{fig:stab}.
The two states are easily found in the massless free fermion spectrum,
for example, by recalling that a half flux changes the standard antiperiodic
boundary conditions into periodic ones, but of course, the result of the
flux insertion is the same in interacting theories.

We can now follow the state $|-1/2\rangle$ backward to
vanishing flux and show that this is an excited state $|{\rm ex}\rangle$,
because it is orthogonal to $|\W\rangle$ (see Fig. \ref{fig:stab}),
\be{kramers}
\langle \W|{\rm ex}\rangle=
\langle \W|\Th[-\Phi_0/2]|-1/2\rangle=
\langle \W|\Th[\Phi_0/2]^\dag\; {\cal T}\;
\Th[\Phi_0/2]|\W\rangle=0\; ,
\ee
where we used \eqref{K-pair2} for the last equality.
Since the work done on the system by inserting one flux goes to zero
as $O(1/R)$ in the limit of large radius $R$ of the annulus, the
energy of  $|{\rm ex}\rangle$ at $\Phi=\Phi_0/2$ and $\Phi=0$ is also
$E_{\rm ex}=O(1/R)$. We conclude that there is no  gap for
edge excitations  in the thermodynamic limit $R\to\infty$.

Therefore, we have proven that TR symmetry keeps a single edge fermion
gapless.  In the case of two fermion modes at the edge, the state
created by half flux has spin one and Kramers theorem cannot be
invoked. Thus an even number of fermions can acquire a gap.

Note that the argument does not change in the presence of edge
interactions that respects TR symmetry. Furthermore, it does not
require $S_z$ to be a good quantum number, but only charge and fermion
parity conservation.  We conclude that the $\Z_2$ characterization of
stable topological insulators is valid for general interacting
systems \cite{PhysRevLett.95.146802}.

The above argument extends to three-dimensional topological insulators by
considering the $(2+1)$-dimensional boundary fermions, in the geometry
${\cal M}_3=\mathbb{R}\times S^1\times S^1$ and inserting half flux
in each hole of the two-torus \cite{Fu-Kane-Mele}.

In the next chapter, we shall see that this flux insertion argument,
which is a discrete variant of the Laughlin proof of the integer Hall
conductivity (Sect. \ref{sec:hallcurr}) nicely relates to the spectral
flow of global anomalies.

\subsection{Electromagnetic response of topological
  superconductors} \label{sec:ts}

In this section we shall derive the electromagnetic response action of
3d topological superconductors, which are states where the electrons
form spin-triplet pairs. As of today, there is no confirmed
experimental candidates, but the B-phase of $^3$He is an example of a
triplet pairing state of neutral fermions. The time-reversal invariant
superconductors have very interesting properties: They have a
relativistic dispersion relation and gapless 2d Majorana fermions on
their surfaces. For a recent review with a comprehensive list of
references, see \cite{sato}.  Because of the Meissner effect the
electromagnetic fields are screened in a superconductor, so the most
natural would be to study thermal response by coupling to gravity and
using the index theorem.  We shall not cover this topic but refer to
the review \cite{Vozmediano}.

One can nonetheless derive an effective theory for the electromagnetic
field \cite{qwz}. This was originally done in  \cite{qwz} using an extension to
a (4+1)D theory and later in \cite{PhysRevB.103.235427} using the
representative Dirac theory, 
\be{lagdirac}
\mathcal{L} =  \bar \Psi_\pm  \left( i\slashed\partial \pm
\slashed A \gamma^5 - \Delta e^{\pm i\theta_\pm\gamma^5} \right) \Psi_\pm \, .
\ee
which can be derived from an appropriate non-relativistic
Bogoliubov-de Gennes (BdG) Hamiltonian for a triplet-paired
superconductor \cite{sato}. Here $\Psi$ is a Majorana fermion, and
$\theta_\pm$ are the phases of the superconducting order parameter
corresponding to the two helicities.  Note that in the representative
theory, the electromagnetic potential couples to the axial current.

As for the TI, the requirement of TR invariance is that the
angles $\theta_\pm$ can only take the values 0 or $\pi$ modulo
$2\pi$. Superficially we have a description similar to that of a TI,
and we would expect that integrating out the fermions gives a
topological $\theta$-term, $\sim \theta \int F \tilde F$, in the
response action. A crucial difference, however, is that in the TI case,
$\theta$ is a material parameter that is determined by the band
structure, while in the TSC it is the dynamical phase of the order
parameter, and the statement that $\theta$ takes a certain value
should be interpreted as referring to the expectation value.
The full action is then found by integrating the fermions
\cite{qwz,PhysRevB.103.235427},
\be{lowtheory}
S_{\text{eff}} &=&  \int d^4x \, \left[ - \frac 1 4 F_{\mu\nu}F^{\mu\nu}
  - \frac 1 {192\pi^2}  \left(\theta_+ - \theta_-\right)
  F_{\mu\nu} F_{\sigma\lambda} \right. \nonumber
\\
&&\qquad\qquad +
\left. \frac {\rho_+} 2 \left(\partial_\mu \theta_+ - 2e A_\mu\right)^2 +
  \frac {\rho_-} 2 \left(\partial_\mu \theta_- - 2e A_\mu\right)^2  \right]  ,
\ee
where, in addition to the topological and Maxwell terms, there is also
the usual kinetic terms $\sim |\vec D\, \theta_\pm|^2$ that give the
Meissner effect. Also note that in order to have a topological term,
you need the difference $\theta_+ - \theta_-$ to be an odd integer
times $\pi$. As compared with the TI case, the $\theta$-term has an
extra factor 1/2 because of the Majorana condition, and another factor
of 1/3 since the $A_\mu$ couples to the axial current so  the relevant
Feynman diagrams have three $\gamma_5$ vortices (this is  explained in App.
\ref{app:gaugeinvcurr}).\footnote{The factor of 1/3 is missing in \cite{qwz}.}

Because of the Meissner effect, we only expect interesting
electromagnetic phenomena on boundaries and vortex lines (the
superconductivity is type II). The simplest geometry to study is that
of a planar interface between a trivial and topological
superconductor, and here we encounter a potential contradiction -- by
the same argument that led to \eqref{thetacs}, it looks like there
would be a $\nu = 1/12$ quantum Hall effect on the boundary.  This,
however, cannot be correct since the Majorana particles in the
boundary Majorana theory are neutral \cite{sato}. To resolve this puzzle we
can again to do a full 4D calculation of the two-point function using
the same boundary profile \eqref{massprofile} as in the TI case, and
we also regularize the infrared divergences in the same way by adding
a small term $m \bar\Psi i\gamma^5 \Psi $, and thus take as a
starting point the (Euclidean) Lagrangian,
\be{reglagdirac}
\mathcal{L} =  \bar \Psi_\pm \left[ \gamma^\mu \partial_\mu +M(x)+
  i\gamma^5 m \right] \Psi_\pm \, ,
\ee
where $m$ is a chirally twisted  Majorana mass. Comparing with
\eqref{lagdirac}, the gap function is $\Delta = \sqrt{M^2(x) + m^2}$
and the twist angle $\theta = {\rm sign}(m) \tan^{-1} (m/M(x))$.

The explicit calculation in \cite{PhysRevB.103.235427} does give a
Chern-Simons term but with a level number, $k \sim m/M_0$, which,
contrary to the TI result \eqref{mbexact}, vanish in the limit
$m\rightarrow 0$.  The lesson here is that one must be careful with
applying results that are valid for infinite, or compactified, spaces
to systems with boundaries. In the case of the $\theta$-term, the
naive procedure leading to \eqref{thetacs} turned out to be correct in
the TI case, but not for the topological superconductor.

\subsection{Higher dimensional topological insulators  and
  descent relations } \label{subsec:descentrel}

In Sect. \ref{sec:10fold} we learned that SPT states exist in
different dimensions, and that the classification table in
Fig. \ref{fig:classes} shows regular patterns. Some of these
regularities are due to the relations existing between anomalies in
different dimensions.

The formulas for the topological insulators in
Sect. \ref{subsec:topthetaact} were derived for the special case of
4D, but it should be clear that, \emph{mutatis mutandis}, everything
generalizes to an arbitrary even dimensional space-time D.
Actually, anomalies and effective actions in different dimensions are
related by the following set of equations, called ``descent relations''
\cite{Bertlmann}. We start from the integrated anomaly
$\Omega_{D}(A) $ in even $D$ \eqref{asing} and use the Stokes
theorem\footnote{  See App. \ref{app:calculus}.}
to relate it to a Chern-Simons action $\Omega_{D-1}^{(0)} (A)$
generalizing \eqref{thetacs},
\be{descent1}
\int_{\mathcal {M}_{D}} {\Omega}_{D}(A) =
\int_{\partial \mathcal {M}_{D}} \Omega_{D-1}^{(0)}  (A)\; .
\ee
Next, the infinitesimal gauge transformation $\delta_\lambda$
of the latter,
\be{descent2}
\delta_\lambda\int_{\mathcal {M}_{D-1}} \Omega_{D-1}^{(0)}(A)
= \int_{\de \mathcal {M}_{D-1}} \Omega_{D-2}^{(1)} (A,\l)\; ,
\ee
defines the Wess-Zumino-Witten action $\Omega_{D-2}^{(1)} (A,\l)$, 
which is the integrated form of  the (D-2)-dimensional anomaly,
as described in App. \ref{app:WZW}.

\subsection{Symmetry enriched topological phases}

We have learned that the existence of a
Chern-Simons theory with half-integer level is the characteristic feature
for the boundary theory of topological insulators.
It turns out that this low-energy response can also be realized by
massive $(2+1)$-dimensional theories that are themselves topological, as
experience with quantum Hall states tell us.
In such cases, called ``symmetry enriched'' topological phases (SET),
the boundary fermions are strongly interacting and other degrees of
freedom may be present. \cite{Metlitski,Bonderson,Chen}

Several theories have been proposed that realize such close relatives
of $\nu=1/2$ Hall states; note, however, that
TR symmetry is not explicitly broken by the mass gap,
and the Chern-Simons response is due to the anomaly.
For example, anyon excitations may appear in TR invariant pairs,
having opposite charges.
These states have not yet been observed experimentally, but are are
not purely academic: in particular, the TR invariant version of the
quantum Hall ``Pfaffian state'' \cite{Pfaffian}, dubbed T-Pfaffian,
is a candidate for realizing anyons with non-Abelian fractional
statistics, a possible platform for decoherent-free quantum
computations \cite{TQC}.

\section{Global anomalies}  \label{sec:global}

The anomalies we studied so far manifested themselves as non-conserved
currents, corresponding to gauge non-invariance of the partition
function in the presence of background fields.  These anomalies can be
exactly determined from the response to weak fields calculated
in a low order perturbative expansion.  We showed that the anomalies
in 2D signal the existence of topological states in 3D, the connection
being the symmetry restoration due to the flow of a conserved
quantity, charge or energy-momentum, from bulk to edge.

It is natural to ask if there is a similar connection between a
  topological state in 4D and an anomaly in 3D, but from what we
  learned so far, this does not seem likely, since the anomalies we
  discussed are present only in even D. It turns out, however, that
there is a different class of anomalies that do appear in odd
dimensions.  These are the global, or nonperturbative, anomalies, that
imply that in the quantum theory one cannot preserve all
classical discrete symmetries and, at the same time, invariance under
large, or global, gauge transformations.  In this case, the
bulk-boundary compensation cannot be explained in terms of flow of a
current, but there is an analogous phenomenon in that symmetries are
restored in the partition function for the whole system.

In this section, we first discuss the inconsistencies in a 3D massless 
fermionic theory when considered in isolation, and the origin of the
$\Z_2$ parity anomaly (being also a TR anomaly).
More precisely we show that the partition function cannot be
regularized in a way that is both globally gauge invariant and
invariant under TR.  We start by rederiving the effective Chen-Simons
action \eqref{3d-anom} using a method that can also deal with the $m=0$
case.

Next we show that the partition function of a Chern-Simons
action with half integer level number is not globally gauge invariant,
and then that the same is true for a theory with an odd number of massless
fermions.

These problems are then resolved by attaching the 4D topological
bulk.  We derive the general form of the bulk-boundary partition
function in the presence of both gauge and gravitational backgrounds by
using the Atiyah-Patodi-Singer index theorem. Later we also treat the
case of topological superconductors that have Majorana fermions on
the boundary.

\subsection{The parity anomaly in 2+1 dimensions }
\label{sec:parityanom2}
\newcommand{\cald}{\mathcal D}

\subsubsection{Anomalous effective action}

Massive and massless 3D Dirac theory is of broad relevance
for condensed matter physics, since it describes both genuine
two-dimensional crystals such as graphene, and boundary layers in
semiconductor heterostructures that support quantum Hall
liquids. 
In Sect. \ref{subsec:parityanom} we derived the Chern-Simons effective
action \eqref{3d-anom} by studying the boundary fermionic modes of
topological insulators: We referred to it as anomalous, since it
breaks the rule that the Hall conductance of free fermions must be
integer in units of $e^2/h$.
Its violation of time reversal and parity was found to persist 
in the massless limit where the 3D Dirac action is symmetric,
signaling a clash between classical and quantum symmetries,
although the case of strict vanishing mass could not be described.

We now rederive the Chern-Simons action for the
isolated 3D theory and explain the origin of the anomaly \cite{redlich84}.
The effective action can be obtained by expanding the fermionic determinant 
to quadratic order in the gauge field as in \eqref{2deffact}, leading to, 
\be{2indA}
S_{\rm eff}[A]=\frac{1}{2} \int \frac{d^3k}{(2\p)^3} 
\ A_{\m}(k)\P_{\m\n}(k,m)A_{\n}(-k)\ + \ O\left(A^3 \right)\ .
\ee

The corresponding Feynman diagram  (see Fig. \ref{fig:polten}) has a
linear ultraviolet divergence in three dimensions (cf. \eqref{polten}),
that can be regularized by the Pauli-Villars method.\footnote{
  The Pauli-Villars regularization is  explained in more detail
  \eg in \cite{Zeebook}.}
This amounts to adding another spinor field 
with mass $M$ to the theory\footnote{Not to be confused with the 4D mass
  of previous section.}, but with Bosonic statistics,
that contributes to \eqref{2indA} by an analogous expression
with a minus sign, subtracting the ultraviolet infinities.
In the limit $M\to\infty$ the additional
field decouples leaving a finite result for
the polarization tensor $\P_{\mu\nu}$, 
\be{PV}
\P_{\m\n}(k,m) \to\P^{reg}_{\m\n}(k,m)=\lim_{M\to\infty} \left(
  \P_{\m\n}(k,m) - \P_{\m\n}(k,M) \right)\, .
\ee
The calculation of the regularized  (Euclidean) Feynman diagram
gives,
\be{Ploop1}
\P_{\m\n}(k,m) 
&=& \frac{1}{4\p} k_\a \e^{\a\m\n}\left(
     \frac{m}{|m|} \frac{\arctan(x)}{x} -\frac{M}{|M|} \right) 
\nl
&& - \left(k^2 \d_{\m\n} -k_\m k_\n\right) \frac{1}{8\pi |k|}
     \left(\frac{1}{x}- \frac{1-x^2}{x^2}\arctan(x) \right), 
\qquad
    x=\frac{|k|}{2|m|} \,.\qquad \ 
\ee
Note that the Pauli-Villars method \eqref{PV}
employs a gauge-symmetry breaking momentum cutoff
at an intermediate stage, but this cancels out in the final result, that 
is (locally) gauge invariant since $k^\mu\P_{\mu\nu}(k,m)=0$.
The first term in the expression \eqref{Ploop1}
is odd in momentum and breaks parity and time reversal
symmetries. Technically it originates from the non-vanishing gamma
matrix trace, $\Tr[\g^\a\g^\b\g^\d]=2i\eps^{\a\b\d}$ \, .

In the limit $|m|\to\infty$, the expression \eqref{Ploop1} becomes,
\be{Ploop2}
     \P_{\m\n}(k,m)=
\frac{1}{4\p} k_\a \e^{\a\m\n}\left(
\frac{m}{|m|}-\frac{M}{|M|} \right) 
  - \left(k^2 \d_{\m\n} -k_\m k_\n\right) \frac{1}{12\pi |m|} \ ,
\qquad (|k|\ll |m|).\ \ 
\ee
The result depends on the choice of sign for the Pauli-Villars mass $M$.
A natural choice for the isolated 3D Dirac theory is to require
a vanishing effective action in the limit $m\to\infty$ 
which fixes ${\rm sgn}(M)={\rm sgn}(m)$, but, and more importantly,
for either choice, the level number of the Chern-Simons term is integer.

The result for the massless theory is found by taking
$x\to\infty$ in \eqref{Ploop1},\footnote{
 The first term can in fact be easily be obtained by expanding
    the Feynman integral to leading order in $k$ without calculating
    the full $\Pi_{\mu\nu}$. You might, and rightly so, worry about
    infrared divergences in these perturbative expressions. A more
    complete analysis in \cite{redlich84} addresses this problem by
    summing all one loop graphs in the presence of a constant
    background field. The result for the Chern-Simons term is the same
    as in our simplified treatment, but the non topological part of the
    action has a non-analytic contribution $\sim |^\star F|^{3/2}$
    where $^\star F^\mu = \half \epsilon^{\m\n\s} F_{\n\s}$. }

\be{Ploop3}
\P_{\m\n}(k,0)=
-\frac{1}{4\p} k_\a \e^{\a\m\n} \frac{M}{|M|}
- \left(k^2 \d_{\m\n} -k_\m k_\n\right) \frac{1}{16 |k|},
 \qquad\qquad (m=0) \, ,
\ee
and the first term in \eqref{Ploop3} gives the Chern-Simons action,
\be{ind-CS}
S_{CS}[A] = \pm \frac{i}{8\pi} \int d^3x\, \eps^{\mu\nu\r}
A_\mu\de_\nu A_\r \, ,
\ee
so the parity and time reversal symmetries are broken at the quantum
level even though the classical theory preserves these symmetries --
this is the parity anomaly of $(2+1)$-dimensional fermions
\cite{redlich83,niemi}.

The analysis of the isolated 3D theory can be related to the
discussion of the topological insulator boundary in
Sect. \ref{sec:2nspt}, where the same theory appeared as an effective
low energy description for $|k|\ll M_0$, where $M_0$ is the bulk
gap. In that realization, $M_0$ played the role of a 3D ultraviolet
cutoff, but we also saw that a different regularization can be used in
the full 4D theory \cite{mb}, such that the Pauli-Villars
subtraction is not needed in \eqref{Ploop2} for $m\neq 0$.  Actually,
the topological Chern-Simons theory with half-integer level is the
correct low energy theory in this setting with a physically motivated
breaking of TR invariance.

An equivalent regularization choice, which is available in the full 4D
bulk-boundary system, is to include the 3D Pauli-Villars regulator in
\eqref{Ploop2}, but at the same time cancel it by adding a theta term
\eqref{4Dtheta} with $\theta=\pi$. This approach extends to the $m=0$
case, and is appealing since it implies a bulk-boundary anomaly
cancellation. However, more aspects need to be discussed before
reaching any definite conclusion about the fate of the boundary
anomaly in the full theory, and this will be clarified in
Sect. \ref{sec:z2bulkedge}.

\subsubsection{Global gauge non-invariance}\label{sec:globalg}

Above we learned, that
integrating out a massless Dirac fermion gives a Chern Simons action
with level number $\pm 1/2$, which violates parity. But imagine that
we did not know about the free fermion origin of this effective
theory, is there any fundamental reason for not accepting such
an effective action? It turns out that there is such a reason
-- the lack of invariance of the theory under large gauge transformations.
We now explain this other aspect of the parity anomaly .

We shall calculate the change of the effective action $\G[A]$ under a
general gauge transformation, and to do this we introduce an
additional parameter $\t$ varying in the interval
$\t\in [-{\cal T},{\cal T}]$, such that the gauge field is
adiabatically transformed as follows,
\be{3d-gauge}
A_\mu(x,\t ): \qquad A_\mu(x,-{\cal T})\ \to\ 
A_\mu(x,{\cal T})=U^{-1}(x)\left( A_\mu(x,-{\cal T}) +\de_\mu
  \right)U(x)\ ,
\qquad
\ee
where $x\equiv x^\mu$ and $ U(x)=\exp(i\phi(x))$ 
is the large $U(1)$ gauge transformation.\footnote{
  Note that the interpolating field $A_\mu(x,\t )$ at
  any intermediate $\t$ value is not a gauge transform of
  $A_\mu(x,-{\cal T})$.}
The change in the effective action \eqref{2indA} comes entirely from
the Chern-Simons term \eqref{ind-CS}. 
We can consider $\t$ as an additional coordinate and
assume the corresponding field $A_4(x)=0$, so we can write, using
Minkowski metric,
\be{deltaCS}
S_{CS}[A({\cal T})]-S_{CS}[A(-{\cal T})]
&=& \int_{-\cal T}^{\cal T}d\t \frac{k}{4\pi}\int_{{\cal M}_3}
\!\! \!\! d^3x\, \de_\t \left(\eps^{4\mu\nu\r}A_\mu\de_\nu A_\r \right)
\nl
&=& \frac{k}{16\pi} \int_{{\cal M}_4} \!\! \!\! d^4x\, \eps^{\mu\nu\r\s}
F_{\mu\nu}F_{\r\s} \qquad \qquad\ \ (\mu,\nu,\r,\s=1,2,3,4)
\nl
&=& 2\pi k P =2\pi k n, \qquad\qquad\qquad n\in \Z \, ,
\ee
where we recalled the definition \eqref{genasth4} of the
instanton number $P$, that takes integer values for compact
manifolds. That ${\cal M}_4$ is compact 
can be motivated by first taking ${\cal M}_3$ to be
compact, say the three-torus $S^1\times S^1\times S^1$, and then
${\cal M}_4=[-{\cal T},{\cal T}]\times {\cal M}_3$ can also be
considered compact by identifying the extrema
$\t=\pm {\cal T}$,  where for ${\cal T}\to\infty$
the field strength vanishes and the $A_\mu(x,\pm {\cal T})$
  differ by a large gauge transformation, as discussed
  in Sect. \ref{sec:antop} for the monopole field in 2D.
This compactified generalized cylinder is called a ``mapping torus''
\cite{RevModPhys.88.035001}.

From \eqref{deltaCS}, it follows that for $k=\pm 1/2$, and $P$ odd,
$\D \G[A]=\pm\pi$, so $Z=e^{i\G[A]}$ is not invariant but changes sign.
This discrete non-invariance is as unacceptable as 
the infinitesimal change associated with anomalies of continuous symmetries. 
Note that this argument relates the 3D global anomaly to the integrated 4D
chiral anomaly (cf. Sect. \ref{sec:3dhotel}, \eqref{4dquant}).

\subsubsection{Spectral flow analysis and the Euclidean
  hotel} \label{sec:globalhotel}

The sign change under large gauge transformations can be directly understood 
from the behavior of the Dirac spectrum and the related determinant
which we express as,
\be{fermpath}
Z[A] = \int d\mu \, e^{-\int dx \,  \psib i\slashed D \psi} =
\det( \cald )= \prod_i \lambda_i \; ,
\ee
where $d\mu$ is the fermionic path integral measure \eqref{defmes} and
$\cald = i\slashed D$ is an Hermitian operator with real eigenvalues
$\lambda_i$.  This would seem to imply that $Z$ is real, 
 in accordance with TR invariance, but since the determinant requires
a regularization, this conclusion is to naive.

To    proceed,     take    two    copies    of     the    theory    so
$ Z^2 = \prod_i \lambda_i^2  $.  From Sect. \ref{subsec:parityanom} we
recall that  although just  adding a  mass breaks  TRI, by  taking two
copies with opposite spin, \ie mass, one can couple them
so as  to preserve this  symmetry.\footnote{ For a discussion  of what
  kind of  perturbation in a  physical system  that will give  rise to
  such a mass term see \eg \cite{RevModPhys.83.1057}.}
Now, since we have a invariant mass term, we can regularize the theory
using the Pauli-Villars method, as we did earlier. Therefore, the renormalized
$Z^2$ is finite, and both TR and gauge invariant.  This means that the
only possible ambiguity in $Z$ is a sign, and  if this ``anomalous sign'' is
present, the quantum theory cannot be defined in a way that respects both
symmetries.

Indeed, we will find that some $\l_i$ in \eqref{fermpath} change
from being positive
to negative  (or vice versa) under an adiabatic variation of the background 
\eqref{3d-gauge}, thus changing the sign of the determinant.\footnote{
Clearly this cannot happen if the
spectrum is fully gapped, so there cannot be any problem in the massive
theory.}
Note that the Dirac spectrum is not symmetric with respect
to $\l_i \to -\l_i$  since there is no  chiral symmetry in 3D, otherwise all
sign changes would occur in pairs leaving the determinant invariant.

We now investigate the spectral flow of the Euclidean Dirac operator,
in close analogy with that of the Hamiltonians in the 1d and 3d
infinite hotels in Sects. \ref{sec:infinitehotel} and
\ref{sec:3dhotel}.  We shall first obtain the flow analytically and
then give a qualitative explanation.

It will be useful to introduce  the following auxiliary evolution
equations associated to the Dirac operator \cite{redlich84}
(using $\g^\mu=\s^\mu$ in 3D Euclidean space),
\be{3d-floweq}
\de_\t \psi_\pm(x,\t) =\pm i\slashed{D}(\t) \psi_\pm(x,\t)
= \pm i\s^\mu\left(\de_\mu-iA_\mu(\t)\right) \psi_\pm(x,\t)\; ,
\ee
which can be solved in an adiabatic approximation by taking
$\psi_\pm=f_\pm(\t) \phi_\l(x,\t)$, where $f_\pm(\t)$ are functions,
and $\phi_\l(x,\t)$ are bicomponent spinors diagonalizing
the Dirac equation $i \slashed{D}(\t)\phi_\l(x,\t)=\l(\t)\phi_\l(x,\t)$
in the $\tau$-varying gauge  background $A_\mu(\tau, x^i)$. The solutions are, 
\be{3d-zero}
\de_\t f_\pm =\pm \l(\t) f_\pm(\t)\ ,\qquad\qquad
f_\pm(\t)=f_\pm(-{\cal T})
\exp\left(\pm \int_{-{\cal T}}^\t d\t'\l(\t')\right) \, ,
\ee
and you should notice the analogy with the Jackiw-Rebbi
localized surface state  discussed
in Sect. \ref{subsec:topthetaact}. Thus, the solutions are normalizable
if and only if $\l(\t)$ change sign during the adiabatic evolution:
$f_+(\t)$ is normalizable for $\l(\t)$ going from positive to negative
values and vice versa for $f_-(\t)$. 
The next step is to recast the equations \eqref{3d-floweq} as an eigenvalue 
 equation for an auxiliary four-dimensional Euclidean Dirac operator,
\be{4D-D}
\slashed{D}_4\Psi=
\left[\g^4\de_\t +\g^\mu\left(\de_\mu-i A_\mu\right)\right]\Psi=0\ ,
\ee
where we identify $\tau \equiv x^4$, use  the gauge $A_4 = 0$ and take,
\be{4D-spin}
\Psi=\begin{pmatrix}\psi_+\\\psi_-\end{pmatrix}\ ,\qquad
\g_i=\begin{pmatrix}0 & i\s_i\\-i\s_i &0 \end{pmatrix}\ ,\qquad
\g_4=\begin{pmatrix}0 & 1\\1 &0 \end{pmatrix}\ ,\qquad
\g_5=\begin{pmatrix}1 & 0\\0 &-1 \end{pmatrix}\ ,\qquad
\ee
Note that in this formulation, $\psi_\pm$ are the chiral components of
the 4D theory.
So now we have a way to tell when one of the eigenvalues in
\eqref{fermpath} changes sign; it happens every time there is a zero
mode for the 4D auxiliary theory, and, taking signs properly into
account, this is nothing but the index
$\mathrm{ind}(\slashed D)= n_+ - n_-$. 
Finally, we can use the index theorem of Sect. \ref{sec:indexth}
for a flat space, stating that, $\mathrm{ind}(\slashed D) = P$,  to 
infer that $Z$ changes sign between
two 3D gauge configurations which interpolate by a 4D instanton. 
This is precisely the result obtained in the previous section for the
Chern-Simons action with a half integer level, and confirms that the
partition function is not invariant under 
large gauge transformations that are ``realized'' by an 
odd number of instantons.  

An intuitive picture of this spectral flow follows from
3d infinite hotel analysis in Sect. \ref{sec:3dhotel} (see
\eqref{weylgamma} and \eqref{hwsquare}). The
Hamiltonian described there is block diagonal, each block being precisely
 the 3D Euclidean Dirac operator $\pm i\slashed{D}$ in \eqref{3d-floweq}.
 Thus, that energy spectrum is the double of the present Dirac
 spectrum,\footnote{Note that the Dirac spectrum is not itself symmetric,
   as stressed earlier.}
obtained by reflecting it with respect to $\l=0$. 
The time evolution of the spectrum in Sect. \ref{sec:3dhotel}
corresponds to the $\t$ variation considered here.  It is apparent
from Fig. \ref{fig:hotel} that the flow amounts to two levels
switching sign. One of these levels could correspond to an eigenvalue
of $i\slashed{D}$ and the other of $-i\slashed{D}$, and in this case,
$Z=\det (i\slashed{D})$ changes sign. Another possibility is
  that they both belong to the spectrum of either $i\slashed{D}$ or
  $-i\slashed{D}$ leaving the determinant unchanged.  Since there is
  no chiral symmetry, this second option would, however, require an
  accidental degeneracy. The first option is indeed confirmed since
  the instanton number is computed to be 1 in \eqref{4dquant}, in
  agreement with the above result from the index theorem.\footnote{
  Another explicit description of this flow can be found in the
  appendix of Ref. \cite{Poly}. }

In conclusion we have shown the sign change of the 3D partition function
under large gauge transformations, both studying the effective action and
the Dirac determinant. Note that we made use of an auxiliary 4D
compact space and 4D Dirac equation.

{\bf Historical note:} The first occurrence of a global anomaly was
discovered by Witten in non-Abelian gauge theory with $SU(2)$
symmetry, leading to an inconsistent theory with an odd number of
chiral fermions carrying the fundamental (doublet) representation of
the gauge group \cite{Witten:su2}. The same phenomenon was later
observed in 3D Abelian gauge theory with odd number of Dirac
fermions as described here \cite{niemi} \cite{redlich84}.
It is rather remarkable that another global anomaly in non-Abelian
gauge theories has been discovered only recently \cite{Witten:su2new},
following the renewed interest in this subject in relation to
topological matter studies.

\subsection{Global anomalies, bulk-boundary cancellation
     and index theorems } \label{sec:globalindex}

In this section, we find the expression for the partition
function for the bulk-boundary combined system, and see how this
lets us understand how TR and gauge invariance are recovered.  The
analysis makes use of the index theorem introduced in
Sect. \ref{sec:indexth}, but extended to systems with boundaries.
We first consider the topological insulators, and later the
topological superconductors.  Our presentation is a shortened
version of that in \cite{RevModPhys.88.035001} and
\cite{Seiberg-Witten}, and we use essentially the same notation.

\subsubsection{The Atiyah-Patodi-Singer $\upeta$ invariant}
\label{sec:eta}

The study of the 3D Dirac determinant will not only
incorporate the gauge transformation properties found earlier, but
also give an explicit formula for the (regularized) partition function
in terms of the geometry and topology of the manifold and the
background $U(1)$ field.

We start from the path integral \eqref{fermpath}, coupling the
fermions not only to the $U(1)$ gauge field, but also to the metric
$g_{\mu\nu}$:
\be{fermpath2}
Z_\psi = \int d\mu \, e^{-\int dx \, \sqrt{g}\,  \psib i\slashed D \psi} =
 \mathrm{det} \cald = \prod_i \lambda_i \; ,
\ee
where $g = \det (g_{\mu\nu})$ and the
Dirac operator is given in \eqref{gendirac}.

We now let both the gauge field and the metric depend on a parameter
$\t$, $(A_\mu(\t), g_{\mu\nu} (\t))$, which interpolates between an
initial and final configuration as discussed above.
When $\t$ changes, so do the eigenvalues
$\lambda_i$, and in particular a number of them can pass through zero.

We regularize $Z$ by adding a Pauli-Villars fermion
with mass $M>0$ and opposite statistics that cancels
  the ultraviolet divergences \cite{Zeebook}.
Recalling from \eqref{sec:perturb} that the mass is
imaginary in the Euclidean Dirac equation, the partition function for
the regulator field is $Z_{PV} = 1/\onam{det} (\cald +iM )$, giving
the regularized expression,
\be{regz}
Z_\psi \to \frac{Z_\psi}{Z_{PV}} =
\prod_i  \frac {\lambda_i} {\lambda_i + iM} \, .  
\ee
For $M\to \infty$, 
we factor out a large positive mass-dependent factor from the determinant
and are left with a product of eigenvalues times the phase
$i$ or $-i$ depending on the sign of $\lambda_i$. We get,
\be{regzeta}
Z_\psi = |Z_\psi| \exp\left( -i \frac {\pi} 2 \sum_i \operatorname{sign}
  (\lambda_i) \right) =
|Z_\psi| \exp\left(-i \frac {\pi} 2  \upeta \right)\; ,
\ee
where,
\be{etadef}
\upeta = \lim_{s\rightarrow 0}   \sum_i \mathrm{sign}
(\lambda_i) |\lambda_i|^{-s}\; ,
\ee
defines the so-called Atiyah-Patodi-Singer (APS) $\upeta$
invariant.\footnote{This infinite sum can be
  made convergent by putting a power law suppression
  factor; this is usually called a zeta-function regularization
  \cite{nairbook}. Alternatively one can use a Gaussian cutoff as in
  the path integral calculation of the axial anomaly in
  Sect. \ref{sec:anpath}.}

The partition function \eqref{regzeta} is finite and well-defined but
suffers from two problems. First, it is not gauge invariant for large gauge
transformation: whenever an eigenvalue $\lambda_i$
changes sign under the spectral flow, $\upeta$ will change with
$ \pm 2 $ giving the phase change $e^{\pm i\pi}$ in $Z_\psi$ found earlier.
Secondly, $Z$ is not real, because the regularized value
of $\upeta$ \eqref{etadef} is not integer in general.
This violates TR symmetry, that acts by complex conjugation on $Z_\psi$.
Note also that, had we chosen a Pauli-Villars field with negative mass
$M$, we would have obtained the conjugate
expression $Z_\psi=|Z_\psi|\exp(i\pi\upeta/2)$.
We recover the same sign ambiguity of the effective action \eqref{3d-anom},
with the mass being that of the regulator.
Note also that in the case of the square of the partition function,
regularized by taking two TR invariant opposite-sign masses,
the phases would cancel out, leading to the harmless $Z_\psi^2=|Z_\psi|^2$,
as commented on earlier.

\subsubsection{Bulk-boundary anomaly cancellation in topological insulators}
\label{sec:z2bulkedge}

We have already pointed out that the effective action $\Gamma_\pi[A]$
in \eqref{4Dtheta} is TR invariant when the integral is over a compact 4D
space. In reality, topological insulators have a boundary, and
$\Gamma_\pi[A]$ is not a multiple of $2\pi$: in this case, TR
invariance is restored by the anomalous contribution coming from the
surface fermions (see \eqref{thetacs} and \eqref{3d-anom}).

In Euclidean quantum field theory, anomalous
terms are always given by the imaginary part of the effective action
\cite{Witten:GravAnom}, so we should focus on the phase of the
partition function. The bulk part $Z_b$ is obtained by substituting
the expression \eqref{genasth4} for the index of $\slashed D$
in the presence of both $(A_\mu,g_{\mu\nu})$ into the
effective action \eqref{4Dtheta},
\be{gravpart}
Z_b = |Z_b |  \exp\left(\pm i\pi({\mathcal P} - \hat{\mathcal A}) \right) \, ,
\ee
where $ \mathcal P$ is the Abelian theta term
and $\hat {\mathcal A}$ is the corresponding gravitational expression.
They differ from $P$ and $\hat A$ in \eqref{genasth4} because
they are not integrated over all spacetime due to the presence of
a boundary.
We now combine this with the regularized surface partition function,
\eqref{regzeta}, to get the total expression,\footnote{ The
  gravitational contribution is included in these equations,
  but is not relevant in the discussion of this section.}
\be{ztri}
Z_{TRI}  =  |Z_b ||Z_\psi|\exp\left[\pm i\pi\left(
    {\mathcal P} - \hat{\mathcal A} - \frac {\upeta}{2}\right)\right]
=  |Z_{TRI}|  (-1)^{\mathcal J}    \, .
\ee
  
In the last term of this equation, we used the Atiyah-Patodi-Singer theorem
to express the phase in the square brackets as the index of the Dirac operator
$\mathcal J$ on non-compact 4D manifolds with specific boundary
conditions.\footnote{
  For a relatively simple derivation of this theorem, see the
  appendix of Ref. \cite{LAG}.}
Since this is an integer, the total partition function \eqref{ztri}
is real and does preserve TR invariance.
We have thus demonstrated the  bulk-boundary cancellation of the anomaly.

This is a both  remarkable and comforting result. Comforting, since,
absent external magnetic fields, or magnetic impurities, the
microscopic physics is TR invariant, and we have shown, in great
  generality, that the effective theory, properly defined, preserve
this symmetry. That the proof goes via the use of deep mathematical
results is the remarkable part.

The lack of invariance of $Z_\psi$ under large gauge transformations
is also resolved within the full partition function $Z_{TRI}$:
this point needs some further explanation which is based on a thought
experiment laid out in Sect. 2.A.5 of \cite{RevModPhys.88.035001}.

\begin{figure}   \centering
\includegraphics[width=15cm]{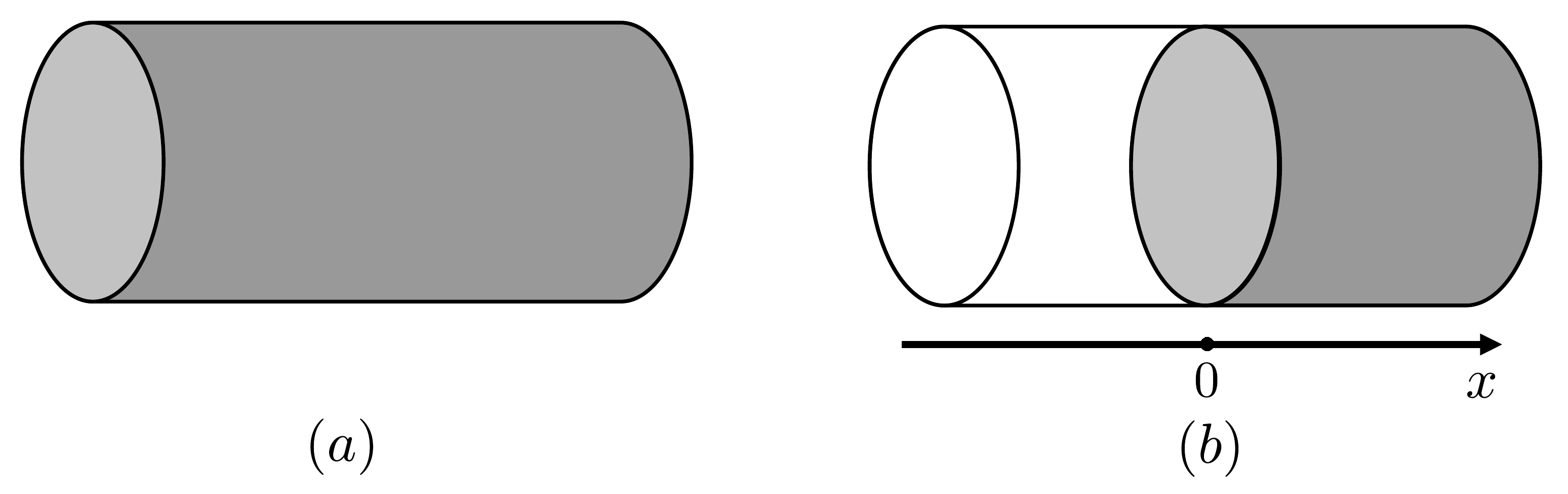}  
\caption{Simplified picture of a four-dimensional spacetime geometry
  with one unbounded coordinate $-\infty <x<\infty$ and three compact ones:
  $(a)$ The topological insulators is filling the whole space; $(b)$
  It is located in $x>0$ and separated from the empty space in $x<0$
  by a boundary at $x=0$.}
  \label{fig:gi}
\end{figure}

Let us consider the geometry in Fig. \ref{fig:gi} which is that of a
generalized cylinder ${\cal M}_4 = \RR \times {\cal M}_3$, where
$x\in\RR$ is a spatial coordinate and ${\cal M}_3$ is compact, \eg a
three-torus (only one circle of ${\cal M}_3$ is shown in the figure).
Panel $(a)$ show a space-filling topological insulator, where we
  identify the boundaries at $x=\pm\infty$.  This is an example of the
  so-called mapping torus discussed already in Sect. \ref{sec:globalg}, which
  is a compact Euclidean spacetime.  We can then, just as above, use
the Atiyah-Singer index $J={\rm ind}(\slashed{D}_4)$ for the 4D
Dirac operator on compact manifolds \eqref{asth4} discussed in
Sect. \ref{sec:indexth} to obtain the partition function,
\be{ti-comp}
Z_b=\exp\left(\pm i\left(P- \hat A\right)\right) = (-1)^J \; .
\ee
This quantity is clearly gauge invariant and its sign measures the
parity of the number of instantons inside the system. 

Panel $(b)$ in Fig. \ref{fig:gi} shows the physical situation of
having a topological insulator filling the $x>0$ space, and supporting
massless fermions at the $x=0$ boundary.  The other boundary at
$x=+\infty$ is disregarded assuming that the gauge field vanish there.
The partition function is now $Z_{TRI}$ in \eqref{ztri}, expressed in
terms of the APS index ${\mathcal J}$.  This is rather different from
the index $J$, since it involves the terms $\mathcal P$ and
$\hat{\mathcal A}$, which are integrated in the half spacetime $x>0$
only; in addition, there is the boundary part $\upeta$ that plays the
role of a generalized Chern-Simons term.

Contrary to $J$, the quantity ${\mathcal J}$ is not a topological
invariant, because its value depends on where the instanton background
is ``mostly'' located. Upon tuning the gauge field, this can be
squeezed in the region $x<0$, so that ${\mathcal J}$ vanishes;
conversely, it can be moved to $x>0$ inside the topological insulator,
leading to ${\mathcal J}=1$ which amounts to a   sign change of $Z_{TRI}$.
The transition between these two discrete values of ${\mathcal J}$
can be understood as  as follows: The $\upeta$ function depends smoothly on the
gauge field but it has $\pm 2$ discontinuities when one Dirac eigenvalue
switch sign, as discussed earlier. The $\mathcal P$ and $\hat{\mathcal A}$
terms actually cancel the smooth part of $\upeta/2$,
and one finds \cite{LAG, Nakahara:Geometry},
\be{eta-disc}
\frac{1}{2} {\rm disc}(\upeta[A(s)])_{s_1} =  {\mathcal J}, 
\ee
where $s$ is a parameter moving the instanton along $x$, and $s_1$
the value at which an eigenvalue vanishes and the discontinuity takes
place.\footnote{In case of several singularities 
  they should be summed over in \eqref{eta-disc}.}
Note that the partition function $Z_{TRI}$ is nonetheless continuous because
the determinant in $|Z_\psi|$ vanishes at the point $s_1$ due to the zero mode
precisely at the point where the eigenvalue changes sign.

We now discuss this jump of $\upeta$, \ie of ${\mathcal J}$, at the level
of effective actions. Using earlier results in
\ref{sec:globalhotel}, we know that the eigenvalue sign switch
occurs when the $3D$ gauge background at $x=0$
changes by a large gauge transformation.  Let us verify that this indeed
takes place under the displacement of the instanton from $x<0$ to $x>0$.

We start by observing that the instanton number $P$ is a total
derivative (cf. \eqref{thetacs}) and thus non-vanishing only when its
boundary Chern-Simons terms do not cancel each other.
In Fig. \ref{fig:gi}(a), the value $P=1$ is due to the backgrounds
$A_\mu(\pm)$ at $x=\pm\infty$ being different by a large gauge
transformation $U$ (see \eqref{deltaCS}),
\be{P-quant}
\pi P = \frac{1}{2}\left( S_{CS}[A(+)] - S_{CS}[A(-)] \right)
 = \frac{1}{2}\left( S_{CS}[A(+)] - S_{CS}[A^U(+)] \right) =\pi\ ,
\ee
where $A^U_\mu$ is the gauge transformed of $A_\mu$.  As in the case
of the monopole in Sect. \ref{sec:antop},
the demand that the gauge transformation $U$ is well defined, implies
that $P$ is quantized as an integer.

In Fig. \ref{fig:gi} (b),  the quantity ${\mathcal P}$ is
defined as the integral over the region
$x>0$ and its complement  $\overline{\mathcal P}$ for $x<0$, so that
$P= {\cal P} + \ov{\cal P}$.  They can be  expressed in terms of 
the Chern-Simons action  at $x=0$, $S_{CS}[A(0)]$ as,
\be{calp-quant}
\pi {\mathcal P} = \frac{1}{2}\left( S_{CS}[A(+)] - S_{CS}[A(0)] \right)
\; ,\qquad\quad
\pi\ov{\mathcal P} = \frac{1}{2}\left( S_{CS}[A(0)] - S_{CS}[A(-)] \right)
\; .
\ee
When the instanton is placed at $x\ll 0$, one has $\mathcal P=0$
and $\ov{\mathcal P}=1$; thus $A_\mu(0)=A_\mu(+)$, that vanishes by our
earlier assumption. As the instanton moves to the $x>0$ region, the gauge field
at $x=0$ grows and eventually accumulate a large gauge transformation,
$A_\mu(0)\to A_\mu(-)=A_\mu^U(+)$, leading to $\mathcal P=1$ and a
corresponding spectral flow.

Therefore, the large gauge transformation at $x=0$ is responsible for
the non-vanishing instanton number ${\mathcal P}$ inside the
topological insulator.  We conclude that what was seen as an
unphysical gauge-dependent sign of the 3D boundary partition
function $Z_\psi$ becomes a gauge-invariant physical effect for
$Z_{TRI}$ describing the full system \eqref{ztri}: This partition
function is just counting the number of instantons inside the TI
modulo two.  This completes the proof of gauge invariance for this
expression.

You may have noted that during the instanton displacement,
$\mathcal P$ changes smoothly while ${\cal J}$ jumps.
We can elaborate on this, by observing that
\eqref{ztri} and \eqref{eta-disc} imply (disregarding $\hat A (R)$),
\be{CS-jump}
\pi {\mathcal P} - \frac{1}{2} S_{CS}= \pi {\cal J}\; .
\ee
Said differently, the Chern-Simons action $S_{CS}[A]$ is
the smooth part of the  function $\upeta[A]$ \cite{RevModPhys.88.035001}.

\subsection{Topological superconductors and 
  Majorana boundary modes} \label{sec:tscmajo}

In a superconductor the gauge field acquires a gap, and magnetic
fields are expelled due due to the Meissner effect. Also, as
explained in \ref{sec:ts}, the boundary modes are Majorana fermions
and carry no $U(1)$ charge \cite{Read2009}.\footnote{ They do have a
  conserved fermion parity $(-1)^F$.}  We now analyze their TR anomaly
and the related bulk-boundary cancellation, paralleling the case of
topological insulators.  The charge neutrality means that the
global anomaly is manifested by symmetry violations related to certain
background geometries.

A Majorana fermion should be thought of as a real fermion, meaning
that $\psi$ and $\psib$ are not independent two-component spinors,
but (in Euclidean space) are related by
$\psib_\alpha = \psi^\beta \varepsilon_{\beta\alpha} $, where $\a$ and
$\b$ are spinor indices.  The Euclidean action for a 3D Majorana
fermion is,
\be{majact}
S_{Mj} = \int dx\, \psib i\slashed D \psi
=\psi^\gamma \varepsilon_{\gamma\alpha} i\slashed D^\alpha_\beta
\psi^\beta = \psi^\gamma \mathrm{D}_{\gamma\beta} \psi^\beta \, .
\ee
We used the notation of \cite{RevModPhys.88.035001} where spinor
indices are raised and lowered using the $\varepsilon$
tensor.\footnote{ In this reference you can find more details about
  the properties of Majorana fermions in Euclidean space-time, and a
  recent thorough discussion of this topic is \cite{mstone2020}. }

Before continuing, we briefly recall how to evaluate path integrals
for real, as opposed to complex, fermions. A general quadratic action
is $S = \sum_{\alpha\beta} \psi^\alpha \L_{\alpha\beta} \psi^\beta$
where $\psi^\alpha$ are real Grassmann numbers, and $\L_{\alpha\beta} $
an antisymmetric $2n\times 2n$  matrix. Using the rules of Grassmann
variable integration, we get:
\be{realpathint}
\int \prod_{\alpha=1}^{2n} d\psi_\a\exp(S_{MJ}) = \frac{1}{2^n n!}
\sum_{\{i_k,j_k\}}\eps_{i_1j_1i_2j_2 \dots i_n j_n}\;
\L_{i_{1}j_{1}}\L_{i_{2}j_{2}}\cdots \L_{i_{n}j_{n}} \equiv
\onam {Pf} (\L) \, ,
\ee
where the last equality defines the Pfaffian of the antisymmetric matrix $\L$.
To understand the logic of this formula you should
notice that in the expansion of the exponential you get a string of
elements like $\psi^1\L_{13}\psi^3 \psi^2 \L_{27} \psi^7 \dots$,
where each $\psi^i$ should occur once, and only once, for the integral
not to vanish. Thus we get all possible such strings, and by a few
examples you can convince yourself that the normalization and the
sign in \eqref{realpathint} is correct.\footnote{
  You might have encountered the Pfaffian in
  the theory of superconductivity, where it gives the BCS wave
  function for a fixed number of particles. There the Pfaffian, which
  is a sum of strings of pairs, $A_{13}A_{27}\dots$, tells that
  the wave function is built from Cooper pairs.}
Note the identity $\operatorname {Pf}(\L) ^2 = \det \L$ for antisymmetric
matrices $\L$.

With our conventions for the Euclidean gamma matrices
$\gamma^\mu = (\sigma^1,\sigma^2,\sigma^3)$, the time-reversal
operator is ${\mathcal T} =i \sigma^2 K$, where $K$ denotes
complex conjugation. It satisfies ${\mathcal T}^2 = -1$, and
$[ {\mathcal T}, i\slashed D ] = 0$. Thus the spectrum of
$ i\slashed D$ is doubly degenerate with eigenvalues
$\lambda_n$ and pairs of eigenfunctions which we combine in
the spinor $(\chi_\lambda, {\mathcal T}\chi_\lambda)^{\onam
  T}$. Acting on these spinors, the operator
$\onam D_{\a\b}=\eps_{\a\g}{\cal D}^\g_\b$ becomes,
\be{dmat}
\onam D = 
\begin{pmatrix}
0 & -\lambda \\ \lambda & 0
\end{pmatrix}\; ,
\ee
and the Pfaffian for this single mode is just $\lambda$.
Including the contributions from all modes we get, 
\be{pfafd}
\onam {Pf} (\onam D) = \widetilde\prod_i  \lambda_i \ ,
\ee
where the  product  $\widetilde\prod$ is defined to only pick one
$\lambda_i$ from each degenerate pair.

In the case of Dirac fermions in Sect. \ref{sec:globalhotel}, we showed
how a large gauge transformation resulted in a sign change of the
determinant. One can similarly study how $\onam {Pf} (\onam D)$
changes under a large diffeomorphism (here there is no gauge field),
that interpolate between two different metrics
$g_{\mu\nu}(0) \rightarrow g_{\mu\nu}(1)$, as some parameter $s$ is
changed from 0 to 1. The spectral flow gives
$\onam {Pf} (\onam D) \rightarrow \onam (-1)^{I/2} \onam{Pf} (\onam
D)$, where $I$ is the index taking all modes, \ie both $\gamma_i$ and
$-\gamma_i$ into account.  $I$ is related to the index for
Dirac fermions on a compact manifold introduced in \eqref{ti-comp} by
$I=2 J$ when both are defined, and is always even in 4D. Its value
is the gravitational analog of the instanton number and is given by
the corresponding theta term $\hat A(R)$ for compact manifolds with
$\theta=\pi$.

It turns out that $I=0$ for the geometry of the mapping torus relevant
for the spectral flow  discussed in Sect. \ref{sec:globalhotel};
thus, there is no anomaly for large gauge transformations
(diffeomorphism invariance for gravity) of the 3D Majorana partition
function \cite{RevModPhys.88.035001}.

Nonetheless, there is a violation of TR invariance as in the Dirac case.
To se this, we again regularize $\onam {Pf} (\onam D) $ in a
gauge-invariant way using the Pauli-Villars subtraction method to get,
\be{pfafreg}
Z_\psi^{Mj}\to \frac{Z_\psi^{Mj}}{Z_{PV}}=
  \onam{Pf} (\onam D) =   \widetilde\prod
\frac {\lambda_i} {\lambda_i \pm iM} \, .
\ee
We can now take over the expression \eqref{regzeta} if we
define $\upeta$ using the sum $\widetilde \sum$ which again picks only one
eigenvalue in each degenerate pair. We have,
\be{regzetamaj}
Z_\psi^{Mj} = |Z^{Mj}_\psi|\exp\left(
  \mp i \frac {\pi} 2 \widetilde\sum_i
    \operatorname{sign} (\lambda_i) \right) =
|Z^{Mj}_\psi|\exp\left(\mp i \frac {\pi} 2  \upeta \right) =
|Z^{Mj}_\psi|\exp\left(\mp i \frac {\pi} 4  \eta \right) \, ,
\ee
where the last expression is in terms of the quantity $\eta =2 \upeta$, whose
sum runs over all eigenvalues. Since $\eta$ is a non-trivial complex number
in general, the boundary partition function violates TR invariance.
How this function depends on the geometry of the manifold is subtle,
and will be discussed in the next section.

We now proceed to show that, as in the TI case, the TR invariance
is restored when the boundary is combined with the bulk.  
For this we need the Majorana version of the bulk partition function
$Z_b$ \eqref{gravpart}.
There is an extra factor of a half due to integration of
Majorana instead of Dirac fermions, and no gauge coupling, so we get,
\be{bulkmaj}
Z_b^{Mj} = |Z_b^{Mj}| \exp\left(  \pm i \frac {\pi}{ 2} \hat A(R)\right) \, .
\ee

Again combining boundary and bulk, we rely on the APS index theorem for
a manifold with boundary to obtain,
\be{ztrimaj}
Z_{TRI}^{Mj}  = |Z^{Mj}_\psi ||Z^{Mj}_b |
\exp\left[\pm i \frac {\pi}{2} \left(\hat A (R) -\half \eta \right)\right]  =
|Z^{Mj}_{TRI} | (-1)^{ {\mathcal I} / 2}
\, .
\ee
where $\mathcal I$ is the index of the Dirac operator on a manifold
with boundary, taking all modes into account.
Since $\mathcal I$ is integer, the expression \eqref{ztrimaj} shows the
cancellation between bulk and
boundary TR breaking terms, leading again to a real total partition
function.

We already mentioned that ${\mathcal I} =0$ for the mapping torus, but 
${\mathcal I} /2$ can be odd in general on a compact 4D ``spin manifold''.
We elaborate on this in Sect. \ref{sec:sptanom}.

\subsection{ Anomalies and interacting topological
  superconductors}  \label{sec:intspt}

At the end of Sect. \ref{sec:10fold} we emphasized that the
classification of SPT states based on anomalies is robust, and that
there are cases where it differs from the original 10-fold
classification based of free fermions. In this section we will expand
on this with some concrete examples.
  
Our first example is the 4D TR invariant topological superconductor
that we just discussed in some detail.  It belongs to class DIII (see
Fig. \ref{fig:classes}) and is characterized by an integer topological
number $\nu\in \Z$, meaning that there can be any number of boundary
gapless Majorana modes. Rather surprisingly, it was shown in
\cite{kitaevz16} that a symmetry preserving interaction can gap
$16$ of them, thus causing a reduction
$\mathbb{Z} \rightarrow \mathbb{Z}_{16}$.  An analogous breakdown of
the 10-fold classification was found earlier in the 2D TSC, the so
called Kitaev chain (class BDI in Fig. \ref{fig:classes}), where
Fidkowski and Kitaev showed that the $\mathbb{Z}$ index is reduced to
$\mathbb{Z}_{8}$ by a suitable interaction
\cite{intkitaevchain1,intkitaevchain2}.

Next we show how the global gravitational anomaly of
Majorana fermions in 3D, discussed in the previous section, 
also leads to the $Z_{16}$ symmetry, and
 we conclude by discussing the interacting Kitaev model.

\subsubsection{$\mathbb{Z}_{16}$ -- the anomaly perspective}

Fermions in 3D do not have any notion of chirality, but nevertheless
are of two kinds corresponding to non-equivalent representations
of the gamma matrices. In the massive case, we already saw that fermions
with positive and negative masses cannot be mapped into each other.
In the massless case this is reflected in that  there are two different
ways the  TR transformation can act \cite{Seiberg-Witten}.
In terms of Majorana fermions $\psi_\pm$, these are
${\cal T} \psi_+ = \gamma_0 \psi_+$ and
${\cal T} \psi_- = -\gamma_0 \psi_-$.  Since these can be gapped in
pairs, the number of massless modes at the boundary of TSC is given by
$\nu = n_+ - n_-\in \Z$.

In the interacting case, however, $16$ fermions can be gapped while
respecting TR symmetry. This  suggests the existence of a an anomaly
for $\nu$ mod $16$, that manifests itself as an obstruction to define
a TR symmetric 3D partition function.
We already saw that the phase of the boundary $Z^{Mj}_\psi$
\eqref{regzetamaj} is in general a complex number, but to find out
what values are actually possible, we have to discuss a class of
background geometries that describe unorientable manifolds.

A well-known unorientable surface is the M\"obius strip, where one cannot in
a consistent way define the orientation of a curve, or equivalently
the direction of the normal vector.\footnote{The generalization of the
  concept to higher dimensional spaces is less intuitive
  \cite{Nakahara:Geometry}.}.
Since time reversal transformations change the
orientation, a TR symmetric theory does not bother about orientation
and can thus be defined also on unorientable manifolds.
  
Using a Pauli-Villars regulator for the bulk (gapped) 4D TSC, similar
to the one used for the 3D Majorana theory in \eqref{pfafreg}, one can
compute the partition function as $Z_{4D} = \exp(-i\nu \eta/2)$, where
the $\eta$-function is defined on the compact 4D manifold. In the
special case of an orientable manifold this result equals
$(-1)^{ {\mathcal I} / 2}$, but in general one must use the formula
involving $\eta/2$.\footnote{ For a complete discussion, see sections
  IV.C, IV.E and App. A.3 in \cite{RevModPhys.88.035001}.}
 
Now mathematicians have proved that for the general, not necessarily
orientable, four-dimensional manifolds of interest, $\exp(i \eta/2)$
is a 16$^{th}$ root of unity, which implies that the partition
function is periodic in $\nu$ with a period 16. Or put differently,
the topological number characterizing a 4D TSC is an element of
$\mathbb Z_{16}$ rather than of $\mathbb Z$.

It is fair to ask whether this analysis has any relevance for physics. After
all, there are not a lot of unorientable materials around, at lest not
in 3d. But before rejecting the whole exercise as purely academic, one
should recall that similar excursions into physically non-realizable
manifolds can be quite fruitful. A well-known example is the integer
QHE on a torus. A real QH sample always have boundaries since our labs
have no supply of the magnetic monopoles that would be needed to get a
net flux out of a closed 2d surface. However, there is no fundamental
physical principle that prevent us from having magnetic monopoles, so
a microscopic theory of electrons on a torus enclosing monopoles is
perfectly consistent. This implies that any correct low energy
theory should also be such that it \emph{can} be consistently defined
on a torus, and this in turn implies that the level number $k$ in front
of the CS action \eqref{CSaction} must be integer, meaning that the Hall
conductance is quantized.  In the same vein, there are no principles
that forbid us from studying TIs and TSCs on unorientable surfaces. In
the case of the TIs which are described by $\Z_2$ topological numbers,
this does not add anything, but for TSCs we might hope that these 
rather technical properties of the partition function may be captured by
some observables in more mundane geometries.
A final remark is that another interesting physical relation exists between
non-trivial geometries and properties of the entanglement spectrum
\cite{top-entang}.

\subsubsection{$\mathbb{Z} \rightarrow   \mathbb{Z}_{8} $ in
  the interacting TR invariant Kitaev chain}
The above argument is certainly rather formal, so to illustrate that it
does capture an important aspect of the physics of TSCs, we now
outline the concrete 1d example given in Ref. \cite{intkitaevchain1}
for how the symmetry class reduction
$\mathbb Z \rightarrow \mathbb Z_{8}$ is caused by interactions. We
use the notation from this paper that you should consult for further
details.

\begin{figure}   \centering
\includegraphics[width=10cm]{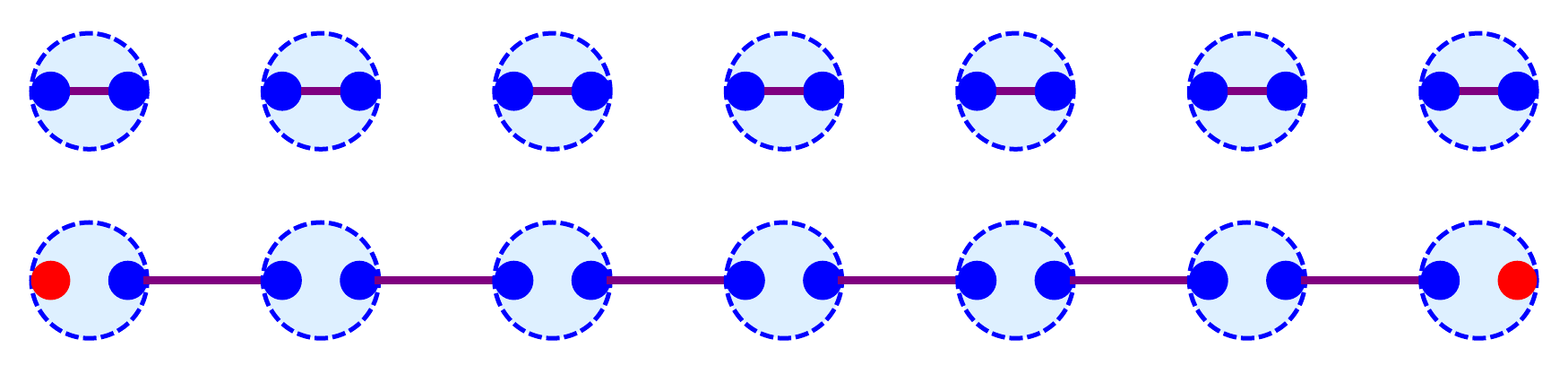}  
\caption{ Top panel: Ground state of $H_{\rm triv}$. Bottom panel:
  Ground state of $H_{\rm top}$. The small blue circles are the bounded
  Majorana fermions and the small red circles the free ones.}
  \label{fig:kitaev}
\end{figure}

First we give a short summary of the relevant properties of the Kitaev
chain, following Kitaev's original paper \cite{kitaevchain}. The model
is a 1d BdG lattice superconductor with a
Hamiltonian,
\be{kitham}
H_K = \sum_{j=1}^N \left[ -w(a_j^\dagger a_{j+1} + a^\dagger_{j+1} a_j)
  - \mu\left(a_j^\dagger  a_j - \half \right) + \Delta a_j a_{j+1} +
  \Delta^\star a^\dagger_{j+1} a^\dagger_j \right]\; ,
\ee
where $\Delta$ is the complex (proximity) superconducting order
parameter, and for simplicity we shall take it to be real. The $N$
annihilation and creation operators satisfy
$\{ a_i, a^\dagger_j\} = \delta_{ij}$, and can be rewritten in terms
of $2N$ real Majorana operators,
\be{majdef}
c_{2j-1} = a_j + a_j^\dagger \ , \qquad\qquad
c_{2j} = -i(a_j - a_j^\dagger)\; ,
\ee
satisfying $c_i = c_i^\dagger$ and $\{c_i , c_j\} = 2\delta_{ijj} $.
Two points in the parameter space $(w,\mu,\Delta)$ are particularly
easy to analyze. The first is $\Delta = w =0$, $\mu > 0$, where $H_K$
becomes,
\be{trivham}
H_{\rm triv} = -\mu  \sum_{j=1}^N \left(a_j^\dagger  a_j - \half \right) =
-\frac \mu 2  \sum_{j=1}^N i c_{2j-1} c_{2j} \; ,
\ee
so the ground state $\ket{\Psi_0}_{\rm triv}$ is clearly the one where
all the sites are occupied, as illustrated in the top panel of
Fig.~\ref{fig:kitaev}. The second parameter set is $\Delta=w >0$ and
$\mu = 0$ which gives the Hamiltonian,
\be{topham}
H_{\rm top} = 2w \sum_{j=1}^{N-1} \left(\tilde a_j^\dagger \tilde  a_j -
  \half \right) = w  \sum_{j=1}^{N-1} i c_{2j} c_{2j+1} \, ,
\ee
were a new set of fermion operators, with standard properties, are defined by:
\be{altferm}
\tilde a_j = \half(c_{2j} + i c_{2j + 1}) \ , \qquad\qquad
\tilde a^\dagger_j = \half(c_{2j} - i c_{2j + 1}),
\qquad\qquad j = 1 , \dots , N-1   \, .
\ee
The corresponding ground state, $\ket{\Psi_0}_{\rm top}$, is
annihilated by the $N-1$ fermion operators $\tilde a_j$ but note that
$H_{\rm top}$ does not depend on the two operators $c_1$ and $c_{2N}$,
which can be combined to the single fermion $b = ( c_1 + i c_{2N})/2$.
Thus the ground state is doubly degenerate, since the
fermion state corresponding to $b$ can be filled or empty. In the
bottom panel of Fig.~\ref{fig:kitaev}, we illustrate this ground state
with the two Majoranas edge modes. Such a degenerate edge state is
characteristic of a topological state, as seen earlier,  so we shall
refer to the Hamiltonians $H_{\rm triv}$ and $H_{\rm top}$ and the related
ground states as trivial and topological, respectively. In Kitaev's
paper it is shown that the two phases correspond to extended regions
of parameter space, these two special points being
the easiest to analyze.
 
Next we generalize the Kitaev chain in two ways. First we shall extend
to a multicomponent model with $n$ identical copies of the TR invariant chain,
and label the fermions by $c_i^\alpha$, where $\alpha = 1, \dots ,
n$. Notice that if we have two copies in the topological state, by
adding the terms $i c_1^1 c_1^2$ and $i c_{2N}^1 c_{2N}^2$ to the
Hamiltonian, we can gap out the end Majorana modes and get a trivial
model. The same can be done for any even number of copies. Thus, we have
obtained the $ \mathbb{Z}_{2} $ classification of 1d models
in absence of TR symmetry (class D in Fig. \ref{fig:classes}).

The second generalization introduce TR invariance, by demanding
that $\mathcal T c_j \mathcal T^{-1} = c_j$ on even sites and
$\mathcal T c_j \mathcal T^{-1} = - c_j$ on odd sites. Thus both
$ i c_{2j-1} c_{2j}$ and $i c_{2j} c_{2j+1}$ are TR invariant, and in
general this is true for any quadratic term $M_{ij} i c_jc_k$ where
the sites $j$ and $k$ have different parity and $M_{ij}$ is real and
symmetric.

For this TR invariant model we cannot gap out the edge Majoranas by adding
terms like $i c_1^1 c_1^2$, since they are not TR invariant, but
Fidkowski and Kitaev showed that this is possible if one allows for
other interactions. Their strategy was to consider the Hamiltonian,
\be{extham}
H_{\rm ext} = H_{\rm triv}  + H_{\rm top} + W \; ,
\ee  
where $W$ is interaction term that has no matrix elements between
$\ket{\Psi_0}_{\rm triv}$ and $\ket{\Psi_0}_{\rm top}$.

Again take $\Delta = w$ and consider the following adiabatic
transformation in the $(\mu, w)$ space,
$(\mu, 0) \rightarrow (0,0) \rightarrow (0,w)$, where $w$ is kept zero
in the first part and $\mu$ in the second part. Absent $W$, the system
remain gapped during this full evolution, except at $w = \mu = 0$
where the Hamiltonian is identically zero and the gap vanish,
signaling a change of phase. The crucial insight in
Ref. \cite{intkitaevchain1} is now to consider interaction terms of
the form $c_i^{\alpha_1}c_i^{\alpha_2}c_i^{\alpha_3}c_i^{\alpha_4}$,
which are Hermitian and cannot mix the states $\ket{\Psi_0}_{triv}$
and $\ket{\Psi_0}_{top}$. The latter follows since $c_i$, by
\eqref{majdef} and \eqref{altferm}, is a linear combination of the
fermion creation and annihilation operators, and the two ground states
are eigenstates of the corresponding number operators.

The question is now if we can choose $W$ so that the gap is maintained
under the evolution through the point $w=\mu = 0$. If so, for some
number $N$ of copies, the states with $n=1$ and $n=N$ are
topologically the same. The obvious first try is $n=4$, where there is
only one possible choice for the interaction on each site, namely
$W_i = A_i c_i^1 c_i^2 c_i^3 c_i^4$. It is not hard to show that this
term alone has a doubly degenerate state independent of the sign of
$A_i$, meaning that at the point $w = \mu = 0$, where the kinetic term
vanish, the system has a large ground state degeneracy and no gap. By
a group theoretical analysis, it was proven in
\cite{intkitaevchain1} that there is a way to continuously connect
$\ket{\Psi_0}_{\rm triv}$ and $\ket{\Psi_0}_{\rm top}$ without closing a gap
for $n=8$, meaning that the free fermion $\mathbb Z $ classification
is reduced to $\mathbb Z_{8}$. We shall not here try to explain the
analysis, which is rather involved, but Fidkowski and Kitaev also give
the following heuristic argument in favor of $\mathbb Z_{8}$: Divide
the system in two parts with 4 chains each, and add in both groups the
above interaction that will give doubly degenerate ground states. This
system can now be viewed as two spin 1/2 particles, so if we could
construct an interaction that favors the singlet, and has a gap to the
triplet, we would expect the gap to remain even when the kinetic term
vanish. For how to construct such a term we refer to
Ref. \cite{intkitaevchain1}, which also contains a derivation of the
result using conformal field theory. In a later paper
\cite{intkitaevchain2}, the same authors used matrix product state
techniques to extended their work, both by a more detailed analysis of
the Majorana chain, and by providing a classification of all
interacting 1d fermion systems.

\subsection{Global anomalies and the classification of SPT states}
\label{sec:sptanom}

In previous sections, we  found a general pattern by which
quantum field theory anomalies are related to topological phases;
a boundary (gapless) quantum field theory displays an anomaly
and a bulk massive theory ``adsorbs'' it,  restoring the symmetry.
For systems in even d, such as quantum Hall states,
the anomalies are perturbative; in odd d, 
as for topological insulators, they are global and the bulk
action is a purely topological theta term.

In a sense, the former, perturbative, anomalies are ``simple'', since
the same pattern repeats in any even dimension, involving generalized
bulk Chern-Simons theories and non-conserved boundary currents, as
discussed in Sect. \ref{subsec:descentrel}.
The study of global anomalies in odd d is 
more difficult, since one would need a complete list of topological actions
(theta terms) and topological partition functions, including those
related to discrete symmetries and  the associated gauge
fields.\footnote{
  Gauge symmetry can be generalized to
  discrete groups by using finite transformations as for ordinary gauge
  theories on a lattice.}
Actually, also in even d, a further global anomaly could be
included, since it does not hamper the perturbative pattern.

Thus, the classification of interacting topological phases
can be recast as the purely mathematical problem of finding 
topological quantum field theories  (TQFT) in any dimension.
A complete solution of this problem is still missing, but
different approaches have been proposed for attacking it.
In this section we  present some basic elements of the approach based
on ``cobordism invariance'' \cite{Kapustin},
which recovers known topological theories in low dimension
and encompasses results obtained by other methods.

We shall discuss TQFTs defined on  compact manifolds, assuming that the 
bulk-boundary correspondence can be understood afterwards.
Topological theories are characterized by partition
functions  that are just phases, $Z_{\cal M}[A, g]=\exp(i\phi_{\cal M}[A,g])$,
as in the earlier examples \eqref{ti-comp} and \eqref{bulkmaj}
of TIs and TSCs respectively. These partition functions obey a composition
rule when a manifold ${\cal M}$ is obtained by gluing together
two other manifolds ${\cal M}_1$ and ${\cal M}_2$. This ``connected sum''
is obtained by cutting a hole in each of them, and then gluing them along
the created boundaries \cite{Yonekura}. The composition rule is,
\be{compos}
Z_{\cal M}=Z_{{\cal M}_1} Z_{{\cal M}_2}\; ,
\ee
corresponding to just summing the respective effective actions.
Actually, we already used this property  for the instanton number
in Sect. \ref{sec:z2bulkedge}, when we wrote $P={\cal P} +\hat{\cal P}$ (\cf
\eqref{P-quant} and \eqref{calp-quant}).

Note that \eqref{compos} may not hold in the presence of ground
state degeneracy as in the fractional quantum Hall effect.  The topological
theories we are considering here have a single ground state on a compact
manifold, and they are referred to as  ``invertible'' TQFTs. 
Here you should recall the discussion of TO contra SPT phases in Sect.
\ref{sec:10fold}; only the latter will be discussed here. 

Partition functions of TQFTs depend on the manifold ${\cal M}$
and its properties, but possess some invariances in the sense that
they are the same for all manifolds that can be  related by certain
rules, which thus  define equivalence classes.
We can characterize manifolds by the following properties:
\begin{itemize}
\item In cases of fermionic systems, the possibility to unambiguously define
  spinors, \ie to consistently assign the minus sign associated to
  $2\pi$ rotations. This is called a ``spin structure'' and the manifold
  a ``spin manifold''.
\item For TR invariant theories, we should also consider unorientable manifolds,
that possess generalized spin structures 
  called Pin${}^\pm$ structures, depending on how  TR invariance is realized:
  ${\cal T}^2=1$ (Pin${}^-$ structure) and ${\cal T}^2=(-1)^F$
  (Pin${}^+$ structure).
\item A smooth gravitational background should be defined and,
  for charged states, a $U(1)$ gauge field.
\end{itemize}

\begin{figure}
  \centering
  \includegraphics[width=2.5in]{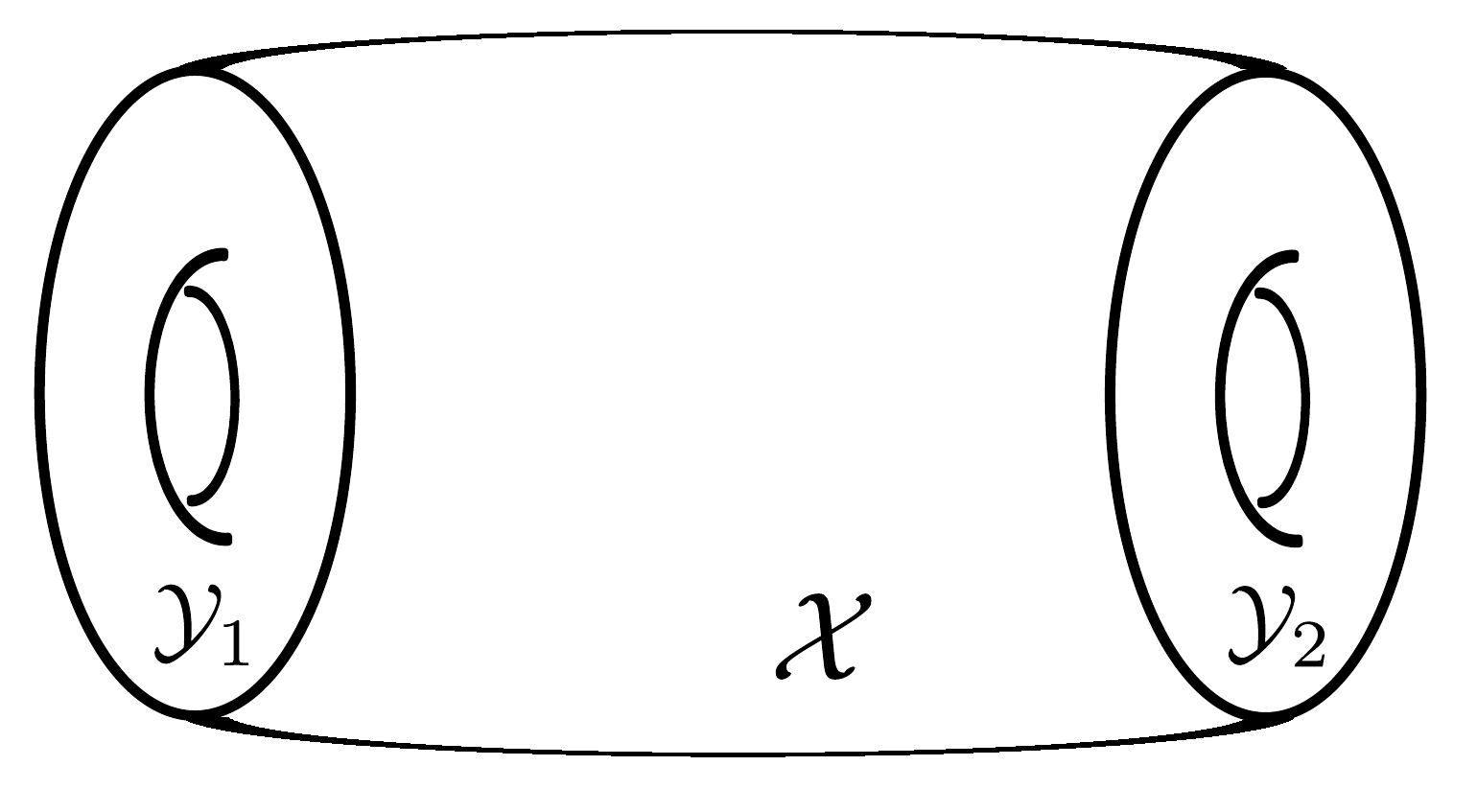}
  \caption{Two-dimensional compact manifolds ${\cal Y}_1$ and ${\cal Y}_1$
    are boundaries of the  3D manifold ${\cal X}$.}
 \label{fig:cobord}
\end{figure}

Manifolds with these properties can be divided into equivalence
classes according to the so-called cobordism invariance: two
D-dimensional compact manifolds ${\cal Y}_1$ and ${\cal Y}_2$ are
bordism equivalent, or cobordant, if there exist a (D+1)-dimensional
manifold ${\cal X}$ that joins them, \ie the two manifolds are
disjoint boundaries of it (see Fig. \ref{fig:cobord}).  The structures
listed above, defined on ${\cal Y}_1$ and ${\cal Y}_2$, are supposed
to extend smoothly to ${\cal X}$.
These equivalence classes are elements of the so-called bordism groups
that are known in the mathematical literature.

\begin{figure}
  \centering
\includegraphics[width=2.5in]{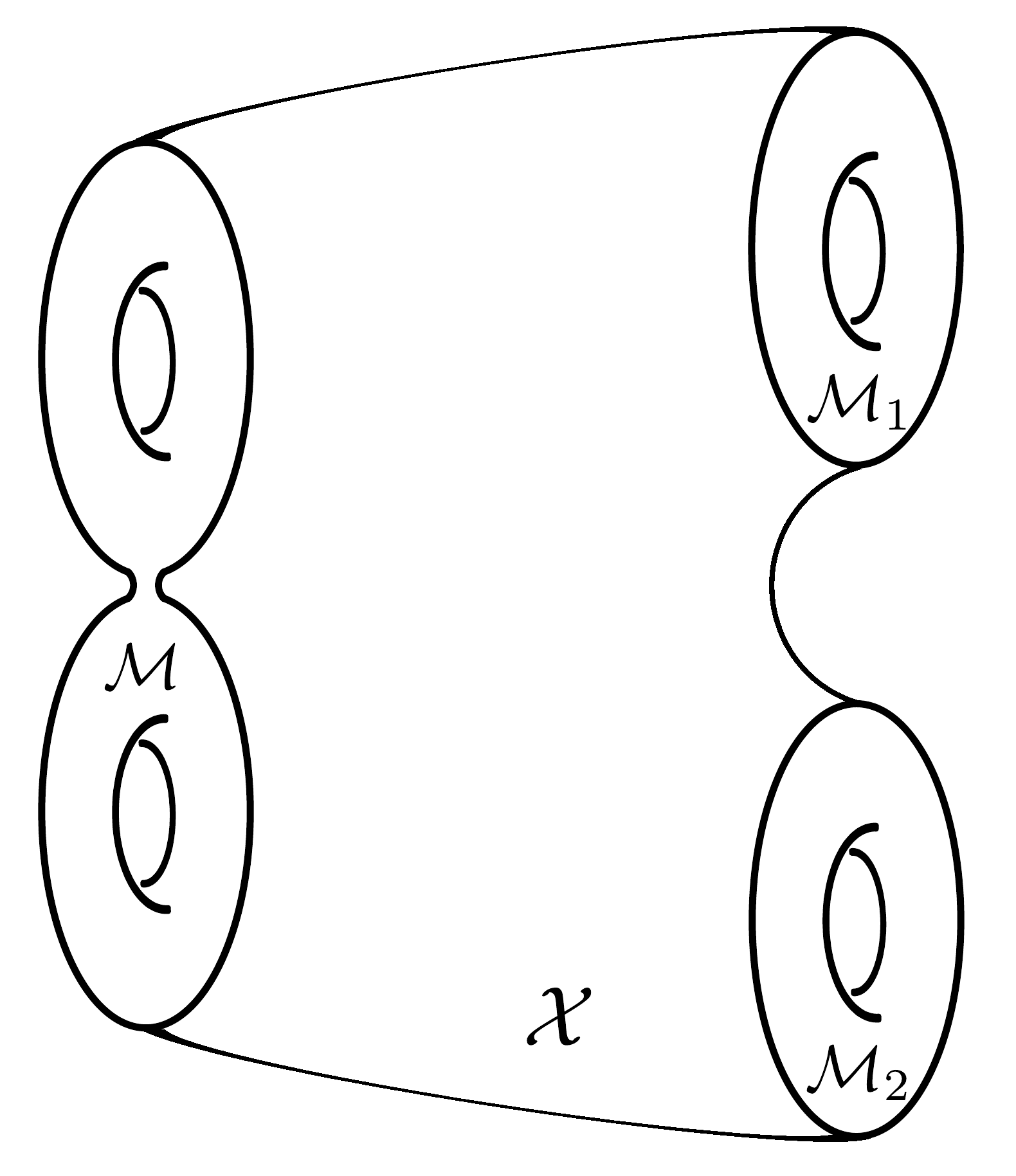}
\caption{The 3D manifold ${\cal X}$ with boundaries ${\cal M}_1$, ${\cal M}_2$
and their connected sum ${\cal M}$.}
 \label{fig:pants}
\end{figure}

TQFT partition functions are naturally invariant under cobordism, in
the sense that $Z_{{\cal Y}_1}=Z_{{\cal Y}_2}$ if ${\cal Y}_1$ and
${\cal Y}_2$ are bordism equivalent. This property was
first conjectured in \cite{Kapustin} and then proved in \cite{Yonekura}.

The above equivalence can be restated in the following way.
Let $\cal W$ be a $D$-dimensional compact manifold that is the boundary of a
$(D+1)$-dimensional manifold ${\cal X}$, to which geometric structures extend
without obstructions. The partition function $Z_{\cal W}$ is then said to be
a cobordism invariant if it is trivial, \ie if
\be{cob-inv}
Z_{\cal W}=1\, , \qquad\quad {\rm for}\qquad\quad   {\cal W}=\de {\cal X},
\qquad\qquad ({\rm cobordism\ invariance}).
\ee
In the previous case of two boundaries, we take
${\cal W}={\cal Y}_1\cup{\cal Y}_2$ (disjoint union), and apply
\eqref{cob-inv} to obtain $Z_{{\cal Y}_1}^{-1}Z_{{\cal Y}_2}=1$, as
anticipated.  The inversion relating one boundary to the other can be
understood as due to orientation, but it is actually more general. It
follows from unitarity of the TQFT, as one can understand by
thinking of the extra dimension in Fig. \ref{fig:cobord} as time and
the whole picture as a kind of quantum mechanical amplitude
\cite{RevModPhys.88.035001}.

Let us verify that the relation \eqref{cob-inv} is fulfilled by
the partition function of a a 3d topological insulator.
For this, consider the instanton number on the 4D manifold ${\cal W}$,
smoothly extend the gauge field to the 5D space ${\cal X}$,
and use  Stokes theorem to find,
\be{cob-inst}
\int_{\cal W}  F^2 = \int_{\cal X} d \, F^2 =0 \; .
\ee
The last result holds since the anomaly density
$\W_4\propto F^2$ is a closed form.
Thus, for this simple TQFT, cobordism invariance is just a 
consequence of Stokes theorem.

The composition rule of partition functions \eqref{compos} also
follows from cobordism invariance: Consider the $(D+1)$-dimensional
manifold ${\cal X}$ in the form of a ``pair of pants'', with boundary
${\cal Y}$ given by three disconnected parts, ${\cal M}_1$,
${\cal M}_2$ and their connected sum ${\cal M}$ (see
Fig. \ref{fig:pants}). In this case, taking
into account the inversion rule, \eqref{cob-inv} implies,
$Z_{\cal Y}=Z_{\cal M}^{-1}Z_{{\cal M}_1} Z_{{\cal M}_2}=1$, which is equivalent to
\eqref{compos}.

With this background, it is hopefully clear that the identification of
topological partition functions as cobordism invariants allows one to
associate them to elements of the bordism groups of manifolds, and
classify them by borrowing results from the mathematical literature
\cite{Kapustin}.

\begin{figure}
  \begin{center}
{\large
\be{eq-cob}\nonumber
  \begin{array}{c|ccc|cccccccc}
      \hline\hline
  {\rm class}\backslash d & {\rm T}& {\rm C}& {\rm S} &0 &1 &2 &3 &4 &5 &6 &7
\\ \hline
{\rm A}  & 0& 0& 0& \Z &0 &\Z &0 &\Z &0 &\Z &0
\\ 
& & & & {\CR\Z^2_2} &{\CR\Z^2_2}  &{\CR\Z\times\Z_8}  &{\CR 0} &
        {\CR 0} &{\CR 0}&{\CR\Z^2\times\Z_{16}} &{\CR 0}
\\ \hline
{\rm BDI}  & +& +& 1& \Z_2 &\Z &0 &0 &0 &2\Z &0 &\Z_2
\\ 
& & & & {\CR\Z^2} &{\CR\Z_8}  &{\CR 0}  &{\CR 0} &
        {\CR 0} &{\CR \Z_{16}}&{\CR 0} &{\CR \Z_2^2}
\\ \hline
{\rm D}  & 0& +& 0& \Z_2 &\Z_2 &\Z &0 &0 &0 &2\Z &0
\\ 
& & & & {\CR\Z_2} &{\CR\Z_2}  &{\CR\Z}  &{\CR 0} &
        {\CR 0} &{\CR 0}&{\CR\Z^2} &{\CR 0}
\\ \hline
{\rm DIII}  &- & + &1 & 0& \Z_2& \Z_2& \Z &0 &0 &0 &2\Z 
\\ 
& & & & {\CR 0} &{\CR\Z_2}  &{\CR\Z_2}  &{\CR \Z_{16}} &
        {\CR 0} &{\CR 0}&{\CR 0} &{\CR\Z_2\times\Z_{32}}
\\ \hline\hline
\end{array}
\ee}
\caption{Topological classes of the free-fermion tenfold
  classification (black, see Fig. \ref{fig:classes}) compared with the
  results of cobordism invariance (red) for topological theories of
  interacting fermions.}
\label{fig:cob}
\end{center}
\end{figure}

The results of this analysis are summarized in Fig. \ref{fig:cob}:
Four of the ten  classes of the free fermion classification
in Fig. \ref{fig:classes} are
displayed, with the corresponding groups written in black, indicating
the possible numbers of massless boundary modes. Below each class, the bordism
groups of manifolds are given in red. 
Very importantly, we see that symmetry reductions $\Z\to\Z_8$ for the
Kitaev chain in 1d (class BDI), and $\Z\to\Z_{16}$ for 3d
topological insulators (class DIII), are captured by the cobordism
analysis, which clearly is powerful enough to correctly classify the
interacting phases.  It also shows that there are symmetry enhancements,
indicating topological phases of interacting fermions completely different
from band systems.  Also note that the periodicity in space dimension
modulo eight is lost.

A virtue of the cobordism approach is that it reproduces results
obtained by other methods, such as ``group cohomology'' \cite{Gu-Wen},
that will not be described here. A drawback is that the cobordism method is
not constructive, in that it does not give explicit expressions form
of the partition function/effective action for the various bordism
classes.\footnote{Of course, known low-dimensional theories have
  been shown to fit the scheme \cite{Kapustin}.}
The group cohomology method, on the other hand, is
constructive and can analyze topological phases in the presence of
additional, non-generic symmetries. We hope that this brief description
has given at least a flavor of this mathematically advanced
subject, which reveals deep relations between geometrical properties
and physical phenomena.

\section{  Chern-Simons theory and the
  framing anomaly} \label{sec:framing}

In our study of quantum Hall liquids we encountered Chern-Simons terms
in two related, but distinct, appearances. In \eqref{S-ind} as an
effective action $S[A]$ for the electromagnetic field that encodes the Hall
response, and in \eqref{Seff} as the kinetic term in the effective
action \eqref{Seff} for the hydrodynamical field $a_\mu$ that
parameterizes the current. We also showed how $S[A]$ could be
obtained by integrating out the hydrodynamic field, but deferred the
calculation of the gravitational part of the full Wen-Zee action
\eqref{Seffgrav} to this section. It is worth noticing that there is
yet another condensed matter context where dynamical Chern-Simons
terms appear, namely as actions for the ``statistical gauge fields''
that are employed in microscopic mean field theories for the quantum
Hall effect based on composite bosons \cite{zhang1992rev} or composite
fermions \cite{jainbook}. In these theories, both electrons and
quasiparticles are charge-flux composite, and during braiding they
will pick up Aharonov-Bohm type phases that endows them with the
correct integer or fractional statistics.

We now proceed to the derivation of gravitational Chern-Simons
action by evaluating the path-integral of the hydrodynamic field.
We consider the action \eqref{Seff}, which has no explicit coupling
to the metric, and write it as
\be{qcsact}
S_{CS}[A,\tilde A] &=& -\frac k {4\pi}\int d^3x\,
\epsilon^{\mu\nu\sigma}  A_\mu \partial_\nu  A_\sigma +
\frac {k}{2\pi} \int d^3x\,
\epsilon^{\mu\nu\sigma} A_\mu \partial_\nu \tilde A_\sigma \\
&\equiv& -\frac k {4\pi} \int  A \wedge d A +
\frac {k} {2\pi} \int A\wedge d\tilde A \, . \nonumber
\ee
To simplify comparison with  \cite{wittenjones}, which we will
follow rather closely, we shall in this section change notation and
use $ A_\mu$ for the hydrodynamic Chern-Simons field,
$-\frac{e}{k}\tilde A_\mu$ for
the background electromagnetic field, and also write the wedge
products explicitly in the differential forms (\cf App.
\ref{app:calculus}). Furthermore, we will also assume some basic
knowledge of the Faddeev-Popov method to quantize gauge theories
\cite{nairbook}.

From \eqref{qcsact} follows the classical equations of motion
$d A = d\tilde A$, and the effective action $S[\tilde A]$ is obtained by
integrating out the $A_\mu$ field.  We define $Q$ as the quantum field
fluctuating  around the mean field, 
\be{qdef}
Q_\mu =  A_\mu - \tilde A_\mu\; ,
\ee
and write the path integral as,
\be{cspath}
Z[\tilde A] =  e^{iS[\tilde A]} =
\exp\left(i \frac {k} {4\pi} \int  \tilde A\wedge d\tilde A \right)
\int \mathcal {D} Q\, \exp\left(i\frac {k} {4\pi} \int Q\wedge dQ \right) \, .
\ee
Naively, we might think that the problem is now solved, since
the path integral is independent of $ A$, but there is more to
the story. For the theory to
be topological, it should not depend on the
metric. Superficially this looks obvious, since the Lagrangian
\eqref{qcsact} does not, but for this to be true also   for
\eqref{cspath}, we must
 establish that the path integral is also independent of the
metric. And here comes the problem: there is no way to regularize the
theory without introducing a metric. This is intuitively clear, since
any regularization involves changing the short distance behavior, and
this does not make any sense without a way to measure lengths.

For a TQFT to be interesting, it is not enough to have a metric independent
partition function, one must also be able to calculate correlation functions
of metric independent operators. In gauge theories these operators are the
Wilson loops,
\be{wilsonloop}
W[C] = \mathrm {Tr}\left[ {\mathcal P} \exp\left(i\oint dx^i \, A_i\right)
  \right],
\ee
where $\mathcal P$ denotes path ordering, and the average is defined
with respect to the action in \eqref{qcsact}. 
 
Below we show that both the partition
function, and the correlation functions $\av{W[C_1] \dots \\W[C_n] } $
do depend on the  metric, but  that this dependence can  be compensated for by adding a gravitational Chern-Simons counter term to
the action. Still, however, there remains a subtle 
dependence on the geometry that is referred to as the ``framing anomaly''.

How to properly quantize the path integral in \eqref{cspath} was shown
in the classic paper by Witten \cite{wittenjones}, and we shall now
summarize some of its key results. Another useful technical
reference is the later extension in  \cite{BarNatan} and also
chapter 5 in the lecture notes \cite{Hu}.

Following these papers, we generalize \eqref{cspath} to the 
 case of a compact simple non-Abelian gauge group,  \eg 
$SU(N)$, and use the corresponding Chern-Simons action,
\be{nacs}
S[ A]=k I[ A] &=& \frac k {4\pi} \int_{\cal M}
\mathrm {Tr} \left(
   A \wedge dA +
  \frac {2}{ 3} A \wedge A \right)   \\
&=& k I[\tilde A]  + \frac k {4\pi} \int_{\cal M}  \mathrm {Tr} \left(
  Q\wedge DQ \right)  \; , \nonumber
\ee
where, as in Sect. \ref{sec:indexth}, $A$, $Q$ and
$\tilde A$ are Lie algebra valued one forms, $A= \tilde A+Q$ and
${\cal M}$ is a closed manifold.   $D$ is the covariant derivative
with respect to the $\tilde A$ background which is an 
extremal point of the CS action. 
In \cite{desjactem}, a homotopy argument was used to
show that for the gauge groups $SU(N)$, the level number $k$ must be
an integer in order for \eqref{nacs} to be well defined.

In \cite{wittenjones} the CS theory is quantized in two ways. The
most general method uses a canonical framework, but here we will
describe the other method based on a saddle point evaluation of the
path integral \eqref{nacs}. This calculation, which is valid for
large $k$, amounts to calculating a functional determinant somewhat
similarly to the ones we have already encountered. At the end we comment
on the relationship between the two methods.

As is always the case for gauge theories, the naive path integral
\eqref{nacs} includes gauge copies, so to be properly defined one
must pick a gauge for the quantum field $Q$.  A natural choice is the
covariant background field gauge
$D_\mu Q^\mu = [\partial_\mu +\tilde  A_\mu, g^{\mu\nu} Q_\nu]=0$
which explicitly depends on the metric. The essence of the
Faddeev-Popov method is to integrate the gauge condition over
the gauge volume, and include
the appropriate Jacobian as a path integral over ghost fields. The
appropriate Lagrangian that implement this is,
\be{gaugfix}
{\mathcal L}_{\rm gf} =\frac k {4\pi} \int_{\cal M}
\mathrm {Tr} \left( \phi D_\mu Q^\mu + \bar c D_\mu D^\mu c \right)\, ,
\ee
where $\phi$ is a scalar
multiplier field implementing the gauge condition, and $c$ and
$\bar c$ are the Grassmann valued ghost fields. With this we can write
the path integral as,
\be{fppath}
Z[\tilde  A] &=& e^{i k I[\tilde  A]}
\int {\mathcal {D} }Q\, {\mathcal {D} }\phi \, {\mathcal {D} } c\,
{\mathcal {D} }\bar c\,
\exp\left(\frac {ik}{4\pi} \int_{\cal M} \mathrm {Tr} \left( Q\wedge DQ
    + \phi \star D\star Q + \bar c \, \star D \star D c \right)\right)
  \nl
&\equiv & e^{i k I[\tilde  A]}\; \mu[\tilde  A] \, ,
\ee
where $\phi$ is a 3-form and $\star$ is the (metric dependent) Hodge
operator that maps $k$ forms to $3-k$ forms.\footnote{
    See App. \ref{app:calculus} for the properties of differential
    forms used in this analysis.}

It is a non-trivial mathematical task to evaluate this path integral,
and we shall just outline how to proceed and then state the
result. The first observation is that (after a rescaling with
$\sqrt{k/2\pi}$, detailed
in \cite{BarNatan}), the bosonic part of the action can be expressed
as,
\be{bosact}
(H\cdot L_- H ) =  \mathrm {Tr}
\left(  Q \wedge DQ + \phi \star D \star Q \right) \; ,
\ee
where the inner product  is defined by,
\be{inner}
(\phi \cdot \phi' ) = \int_{\cal M} \mathrm {Tr} \left( \phi\wedge\star\,
  \phi'\right) \, ,
\ee
$H = (Q,\phi)$, and $L_- = \star D + D\star$ is a self adjoint
operator which exists in 3 dimensions and maps odd forms into odd
forms. The expression \eqref{bosact} is thus rewritten
$(H \cdot L_-H)= Q\cdot L_- Q+ \\ Q\cdot L_-\phi+\phi \cdot L_- Q+
\phi \cdot L_- \phi$. In the fermionic part we recognize
$ \star D \star D = \triangle$ as the Laplacian. We can now formally
write the result of the Gaussian integrations in terms of determinants
which can be computed by the methods discussed in Sects.\ref{sec:eta},
\ie by finding the eigenvalues of $\triangle$ and $L_-$ and
regularizing the relevant products. Doing so one gets,

\be{inres}
\mu[\tilde A]  = \frac {  \operatorname {Det }   \triangle  }
{ \sqrt{ \operatorname {Det}  L_- }   } = \sqrt \tau\,
\exp\left(- \frac {i\pi}4 \eta(\tilde A) \right)\ .
\ee
The quantity $\tau$, appearing in the absolute value, is the
Ray-Singer analytic torsion which is a topological invariant. The proof
of the second identity requires mathematics beyond this exposition,
but it can be traced from the references given above.  The phase is
proportional to $\eta(\tilde  A) \equiv \eta(L_-(\tilde  A))$ where the
$\eta$ invariant was introduced in Sect. \ref{sec:eta}.
Using the APS index theorem and specializing to the gauge group
$SU(N)$ gives the relation,
\be{etarel}
\eta(\tilde A) =  \eta(0) +  \frac {2N} \pi  I[\tilde  A] \, ,
\ee
so the second term in the RHS will just renormalize the coefficient of
$I[\tilde A]$ in \eqref{fppath}.  The first term is more troublesome;
it is not topologically invariant, but potentially depends on the
metric. For $ A = 0$ the theory is just $d_G$ copies of the purely
gravitational theory where $d_G$ is the dimension of the gauge group $G$, so
writing $\eta(0) = d_G\, \eta_g$ the interesting phase factor is
$\exp(id_G\pi \eta_g /4)$. One can now again refer to the APS index
theorem that in this case states that the combination,
 \be{apsres}
-\frac 1 4  \eta_g + \frac 1 {24\pi}  I[g] \; ,
\ee
is a topological invariant. Here $I[g]$ is  the gravitational Chern-Simons
term \eqref{omega-cs},
\be{gravcs}
I[g] = \frac 1 {4\pi} \int_M \mathrm {Tr} \left(
  \omega d\omega + \frac 2 3 \omega\wedge\omega\wedge\omega\right) \, ,
\ee
which is normalized to be $2\pi \Z$ for a closed manifold.
It is now clear how one can make $Z[\tilde A]$ independent
of the metric, namely by the substitution,
\be{topsubst}
\exp\left(-i d_G\frac \pi 4 \eta_g\right)  \rightarrow \exp\left(i d_G \left(
    -\frac \pi 4 \eta_g +  \frac 1 {24} \frac 1 {2\pi} I[g] \right)\right) ,
\ee
which amounts to adding a metric dependent counter term. 

It turns out, however, that although the above expression does not
depend on the metric, $I[g]$ will depend on another property of the 
manifold, namely how one globally define the coordinate system. 
Mathematically this is related to how one ``trivializes'' the
tangent bundle, or equivalently the frame bundle. In fact, $I[g]$ will 
change by a multiple of $2\pi$
under a global ``twist'' just as the $U(1)$ CS term can be changed by a
non-trivial global gauge transformation as discussed in
Sect. \ref{sec:globalhotel}. This remaining dependence on the
geometry,
\be{framanom}
Z \rightarrow Z\exp\left(i s \frac{2\pi d_G} {24} \right)\; ,
\ee
under a unit $s$ twist, is the framing anomaly. Note, however, that
for a given framing, all correlation functions are topological
invariants.\footnote{ For this to be true the Wilson loops
  \eqref{wilsonloop} must be modified by a ``framing'' which turns it
  into a ribbon, as discussed in \cite{wittenjones}.  }

As already mentioned, the above analysis was based on a saddle point
approximation, which in general is appropriate only for large $k$. For
the special case of an Abelian theory, \ie a collection of $U(1)$
fields it is exact since the integral is Gaussian. In
Ref. \cite{wittenjones} Witten also treats the general non-Abelian case
using canonical quantization. The result of that analysis is a formula
very similar to \eqref{framanom} but with $d_G$ replaced by the central
charge $c(k)=kd_G/(k+ 2N)$ of the two dimensional
current algebra of the group $G=SU(N)$ at level $k$.

The gravitational Chern-Simons term \eqref{gravcs} obtained by functional
integration of the hydrodynamic field completes the derivation of the
Wen-Zee effective action of quantum Hall fluids discussed in Sect.
\ref{sec:wenzee} (see \eqref{Seffgrav}).
Its coupling constant clearly becomes the conformal central charge
in the case of interacting fermions.
In this way the effective theory directly generalizes to
general Abelian quantum Hall liquids, but also to non-Abelian ones,
(for details on this you should read \cite{gromov2015}).

\section{To conclude -- some important points}
We conclude by stressing a number of important points made in these lectures. 

\begin{enumerate}
\item Anomalies occur when a quantum field
  theory cannot be quantized so that all the classical symmetries are
  preserved.
\item In a perturbative anomaly, a local gauge invariance is violated,
  and the related current is not locally conserved.
\item In a global anomaly, the partition function $Z$ is not invariant
  under non trivial global gauge transformations, and typically some
  discrete symmetries, such as parity and time reversal, are violated.
\item The perturbative anomalies occur in even
  space-time dimensions, and can be calculated by various methods, the
  most important being:
\begin{itemize}
\item Analysis of spectral flow
\item Path integrals -- the Fujikawa method
\item Perturbation theory
\end{itemize}
\item Currents coupled to dynamical gauge fields, be they fundamental
  or emergent, cannot be anomalous, since this makes the theory
  inconsistent. This puts important constraints on theories, an
  important example being the particle content of the Standard Model
  of fundamental interactions.
\item Anomalies have a Janus face. In the infrared they appear as
  symmetry violating spectral flows, and in the ultraviolet as
  symmetry violating regularization procedures.
\item An anomaly in a fundamental theory, will also be present in a
  corresponding effective low energy theory, that can be either
  bosonic or fermionic, and these ``'t Hooft anomaly matching conditions''
  put important restrictions on model building.
\item A theory that can be formulated on a lattice is anomaly free. An
  important consequence is the phenomenon of fermion doubling where
  anomalies cancel between different Fermi points (or, more generally,
  Fermi surfaces).
  Cancellations also occur between bulk and boundary and between
    different boundaries of the system.
\item In condensed matter physics, anomalies provide a powerful tool for
  classifying symmetry protected topological states of matter
  by relating them to topological invariant
    exact quantities that are not changed by interactions.
\item Even in condensed matter physics, it is often useful to consider 
  theories  on curved manifolds. The related gravitational anomalies
  can be used to classify states of neutral fermions, as well as provide a tool to exactly compute
  thermal response functions.
\end{enumerate}

\section*{Acknowledgements}
AC would like to thank Alexander Abanov, Domenico Seminara and Paul
Wiegmann for interesting discussions on anomalies, and Nigel Cooper
for reading and commenting the draft.  THH thanks Jens
Bardarson and Parameswaran Nair for valuable comments on parts of
the manuscript. AC thanks Nordita, Stockholm, for hospitality and the
G. Galilei Institute for Theoretical Physics, Arcetri, for running the
2021 program ``Topological properties of gauge theories and their
applications to high-energy and condensed-matter physics''. RA and THH
thank Cristiane Morais Smith and the Institute for Theoretical Physics
at Utrecht University, where a shorter version of these lectures was
given. RA acknowledge funding from CAPES, NWO, CNPq, and the Knut and
Alice Wallenberg Foundation.

\begin{appendix}

\section{Notations and conventions}  \label{app:notations}

Coordinates in Minkowski space are denoted by
$x = (t, \vec{r})= (x^0,x^i)$ and their metric is
$\eta_{\mu\nu} = diag (1, -1, \dots,-1)$. Most of the time we will put
$c = \hbar =1$, and use $\sigma^i$, $i=1,2,3$ for the Pauli
matrices. Space-time dimension is denoted by D, and space dimension by
d, so D=d+1. For Euclidean spacetime we write \eg 4D, while for
Minkowski (3+1)d etc.
 
The $(1+1)$-dimensional gamma matrices are given by,
 \be{2dgam}
 \gamma^0 = \begin{pmatrix} 0 & -i\\ i & 0 \end{pmatrix} = \sigma^2 \ ,
 \qquad\qquad
\gamma^1 = \begin{pmatrix} 0 & i\\ i & 0 \end{pmatrix} =i\sigma^1\ ,
\ee
and satisfy $\{\gamma^\mu,\gamma^\nu\} = 2\eta^{\mu\nu}$,
so $\gamma^3 = \gamma^0\gamma^1 = \sigma^3$.
In Euclidean space, the metric is $\delta^{\mu\nu}$ and the gamma matrices
are Hermitian, satisfying, $\{\g_E^\mu , \g_E^\nu \} = 2 \delta^{\mu\nu}$,
with $\mu,\nu=1,\dots,D$.

In 2D Euclidean space, we take $\gamma_E^\mu = \sigma^\mu$, $\mu=1,2$
and we introduce
(omitting the subscript $E$ whenever there is no risk of confusion),
\be{2Dgam}
\gamma^3 = - i\gamma^1\gamma^2 = \sigma^3   \qquad\quad  \mathrm {implying}
\qquad\quad \Tr [\gamma^3\gamma^\mu\gamma^\nu]= 2i\epsilon^{\mu\nu} \, ,
\ee
and we also have the useful relation, 
\be{specrel}
\gamma^3\gamma^\mu = i \epsilon^{\mu\nu} \gamma_\nu\ ,
\ee
In 3D space, we similarly take $\g^\mu=\s^\mu, \mu=1,2,3$.

In (3+1) dimensions, we consider the following Weyl representation
of gamma matrices,
\be{4D-weyl}
\g_i=\begin{pmatrix}0 & i\s_i\\-i\s_i &0 \end{pmatrix}\ ,\qquad
\g_4=\begin{pmatrix}0 & 1\\1 &0 \end{pmatrix}\ ,\qquad
\g_5=\begin{pmatrix}1 & 0\\0 &-1 \end{pmatrix}\ .\qquad
\ee
The matrix $\g^5$ obeys,
\be{4Dgam}
\gamma^5 = -\gamma^1\gamma^2\gamma^3\gamma^4\; , \qquad\qquad
(\g^5)^2=1 \qquad {\rm and}\quad
\Tr \left[\gamma^5\gamma^\mu\gamma^\nu \gamma^\sigma\gamma^\lambda\right]=
- 4\epsilon^{\mu\nu\sigma\lambda} \; ,
\ee
and anticommutes with all the other $\gamma^\mu$. It defines
spinors of positive or negative chirality depending on its eigenvalue
$\pm 1$. In odd dimension this is not such matrix and we
cannot define chirality of spinors and a axial current.

The Euclidean Dirac action is written in terms of the Hermitian operator,
$i\slashed D_E = \gamma^\mu_E (i\partial_\mu   + A_\mu) $, as follows,
\be{euclact2}
S_E =  \int dx_E \, \bar\psi (i\slashed D_E  + im ) \psi  \, ,
\ee
where $dx_E$ is the D-dimensional measure.

We use $A$ to denote the vector gauge potentials and $A_5$ to denote
the axial gauge potentials. All quantities with a index $5$ are the
axial ones, regardless of the spacetime dimension, and with a $+$
($-$) index are the ones with positive (negative) chirality. Using
this convention, the axial current will be denoted by $J_5$ and a
gauge field by $A_-$ for negative chirality, for example.

Currents are obtained by functional derivative of the effective action
with respect to the backgrounds, $J^\mu=\d S [A, A_5]/\d A_\mu$ and
similarly for $J^\mu_5$. The relation between the Euclidean $(E)$ and
Minkowskian $(M)$ actions, $S^{(M)}=i S^{(E)}$, gives a quick way to
map the corresponding currents, e.g. $J^{(M)}_\mu=i J^{(E)}_\mu$.

\section{Multicomponent Chern-Simons theory of
  edge excitations} \label{app:jain}

In this appendix we show how the chiral edge theory can be extended to
edges with several modes of both chiralities. To motivate the
construction we start with a brief introduction to the physics of the
quantum Hall states in the so called Jain series.

\subsection{Effective bulk theory of Jain states}

Besides the Laughlin series $\nu=1/(2q+1)$, the most stable plateaus are
observed at filling fractions:
\be{Jain-series2}
\nu=\frac{n}{2p n\pm 1}, \qquad\qquad\qquad n,p=1,2,\dots \; .
\ee
The theory for these
Hall states has been developed by Jain, generalizing Laughlin
approach \cite{jainbook}.
The basic physical picture involves a correspondence
between integer and fractional Hall effects. Let us analyze the
inverse filling fraction, that gives the number of flux quanta per particle:
\be{inverse-nu}
\frac{1}{\nu}= 2q \pm \frac{1}{n}= 2q \pm \frac{1}{\nu^*}\; .
\ee
These values correspond to integer fillings $\nu^*=n$, if 
$2q$ fluxes per particles could be removed, leaving a residual magnetic
field $B^*=B-2q\Phi_0 \r_o$.
The Jain theory assumes that the $2q$ fluxes disappear
because they are bound to each electron forming a composite state,
called composite fermion.
Therefore, the experimentally observed filling fractions
\eqref{Jain-series} can be interpreted
as integer Hall states of composite fermions, that
completely fill effective Landau levels for the magnetic field $B^*$.
Note that the fermionic statistics is not hampered by adding an even number of
fluxes.

Flux attachment is precisely implemented in wave-function constructions for
the ground state and excitations, and finds remarkable numerical and
experimental confirmations \cite{jainbook}.
For example, low-energy local
excitations have been observed with the expected properties of the
composite fermion, \ie  weakly interacting and moving as if they were
in the field $B^*$. Furthermore, effective 
field theories have been formulated that implement Jain flux attachment
\cite{DasSarma-Pinczuk}.

Jain as well as Laughlin theories have not yet been derived from a
study of the microscopic dynamics, but several analytic and numerical
results confirm their predictions.  The extensive phenomenology based
on the composite fermion picture let us argue that the structure of
several branches is correct \cite{jainbook}.

\subsection{Multi-component Chern-Simons theory} 

The Chern-Simons effective theory introduced in Sect. \ref{sec:fracQHE} cannot
describe the Jain states, because its Hall conductivities $\nu=1/p$ do
not match the values \eqref{Jain-series2} for $n>1$.  Since these
states can been interpreted as integer Hall states of composite
fermions, $\nu^*=n$, it is rather natural to generalize the effective
low-energy theory
by including several field components \cite{Wenbook}.

The current of matter excitations is written
\eqref{CScurrent}:
\be{multi-j}
j^\mu=\frac{1}{2\pi}\eps^{\mu\nu\r} \de_\nu \; \sum_{i=1}^n t_i a^i_\r ,
\ee
where the Abelian gauge fields $a^i_\mu$
are labeled by $i=1,\dots,n$ and $t_i$ are their
charges. This kind of generalization is rather natural for extending the
earlier analysis to several non-interacting filled Landau levels, \ie
$\nu=n$, but can also describe fractional cases as follows.
The multicomponent Abelian Chern-Simons action is:
\be{CS-multi}
 S_{\rm eff}=\frac{1}{4\p}\int d^3x\,
 K_{ij}\eps^{\m\n\r}a^i_\mu\de_\n a^j_\r+
 \frac{1}{2\pi}\int
 d^3x\, t_i\eps^{\m\n\r}A_{\m}\de_\n a^i_\r.
 \ee
 In this expression, the summation over the components
 $i,j=1,\dots ,n$ is assumed, and they are mixed through
the symmetric matrix $K_{ij}$ of couplings.
It turns out that both $t_i$ and $K_{ij}$ should take integer values.
Upon repeating the analysis of the equations of motion done in
Sect. \ref{sec:CS}, one finds the values of the Hall conductivity
and the spectrum of charge and statistics of anyons in this theory
(excitations are specified by a set of integers
$l_i$, $i=1,\dots,n$):
\be{nuKt}
\n= t_i K^{-1}_{ij} t_j,\qquad
Q= t_i K^{-1}_{ij} l_j, \qquad
\frac{\th}{\pi}=l_i K^{-1}_{ij} l_j.
\ee

The multicomponent theory is able to describe many filling fractions and
spectra. The Jain series \eqref{Jain-series2}
are obtained by choosing the form of $K_{ij}$ and $t_i$
that is completely symmetric over the permutation of the $n$ fluids,
namely \cite{Wenbook}:
\be{JainK}
\ t_i=1, \qquad\quad K_{ii}= 2q \pm 1, \qquad\quad K_{ij}=2q,
\qquad\quad i\neq j=1,\ldots,n \; .
\ee
Working out the inverse matrix $K^{-1}$, and inserting it in the first
of \eqref{nuKt}, one verifies that the
filling fractions \eqref{Jain-series2} are obtained.

The corresponding theory at the edge is given by the
multicomponent generalization of the chiral boson described
in Sect. \ref{sec:chilut}.
The action involves the scalar fields $\vf^i$, $i=1,\dots,n$, and reads
(cf. \eqref{S-edge}):
\be{Sn-edge}
S_{\rm edge}=-\frac{1}{4\pi}\int_{\cal C}d^2x\;K_{ij}\de_x\vf^i\de_t\vf^j
+V_{ij} \de_x\vf^i\de_x\vf^j,
\ee
where $K$ is the earlier matrix and $V$ is a positive-definite
symmetric matrix of velocities, not encoding any universal feature.

Let us remark that the simplest one-component effective theory (bulk
Chern-Simons and edge boson) uniquely identifies the Laughlin Hall
states. However, its multicomponent generalization predicts
far more Hall states than the Jain series, upon changing the
integer parameters in the symmetric matrix $K$ \cite{Frohlich:1995}.
These states are not observed experimentally.
A solution to this puzzle has been found in the works \cite{CTZ}
\cite{Sakita1} \cite{Sakita2}, by relying on a property of
the two-dimensional incompressible fluids made by Laughlin
and Jain states. These are characterized by constant bulk density
$\r_0$ and fixed particle number $N=\r \cal A$, thus constant area
$\cal A$. It follows that excitations are generated by deformations
of the fluid that keep the area constant: these are the area-preserving
transformations of spatial coordinates,
obeying the so-called $W_\infty$ symmetry.  It was shown that
the Jain states are characterized by special `minimal' realizations of
this symmetry \cite{Cappelli:MinimW}, involving reduced
multiplicities, as \eg a single
electron excitation instead of $n$ of them as given by the spectrum
\eqref{nuKt}, \eqref{JainK}.
These minimal theories are associated to non-trivial representations
of the $W_\infty$ algebra.

\section{Detailed derivation of 3D infinite hotel}
\label{app:infinitehotel}
  
The free Dirac Hamiltonian is $H=\vec{\alpha}\cdot\vec{p}$, so in the
presence of a potential it becomes
$H=\vec{\alpha}\cdot\left(\vec{p}+\vec{A}\right)$.
Using the $\gamma$ matrices defined in \eqref{weylgamma}, we get:
\be{alphai}
\alpha_i\equiv \gamma^0\gamma^i=
\left(\begin{matrix}\sigma_i&0\\0&-\sigma_i\end{matrix}\right),
\ee
so that the Hamiltonian in matrix form becomes
\be{Hamil3dhotel}
H=\vec{\alpha}\cdot\left(\vec{p}+\vec{A}\right)
=\left(\begin{matrix}\vec{\sigma}\cdot\left(\vec{p}+\vec{A}\right)&0
    \\
    0&-\vec{\sigma}\cdot\left(\vec{p}+\vec{A}\right)\end{matrix}\right).
\ee

In the Weyl representation, each block describes a specific helicity
of the wavefunctions, such we can write the Hamiltonian as the direct
sum $H_{+}\oplus H_-$, such that the upper block of the Hamiltonian is
related to the states with positive helicity
($\vec{\sigma}\cdot\hat{p}=1$) and the lower one with negative
helicity ($\vec{\sigma}\cdot\hat{p}=-1$).

To diagonalize this Hamiltonian, we square it and use the relations
$\sigma_i \sigma_j=\delta_{ij}+\imath \epsilon^{ijk}\sigma_k$, to get,
\be{magterm} \sigma_i\sigma_j\left(p_i+ {A}_i\right)\left(p_j+
  {A}_j\right)=p^2+2 {\vec{A}}\cdot \vec{p}+{A}^2 +
\left(\vec{\nabla}\times
  {\vec{A}}\right)_z\sigma_z=p_z^2+\left(\vec{q}+
  {\vec{A}}\right)^2- B\sigma_z \; ,
\ee
where we used that,
\be{off-diag}
\epsilon^{ijk}
\left(A_ip_j+p_iA_j\right)=\epsilon^{ijk}\left[A_i,p_j\right]=
i\epsilon^{ijk}\partial_j
A_i=-i B^k=i B \, \delta^{k, z}.
\ee
Thus,
\be{Hsquaresimp}
H^2&=\left[p_z^2+\left(\vec{q}+ \vec{A}\right)^2- B\sigma_z\right]\otimes 1_2,
	\ee
where $1_2$ is the $2\times2$ identity .
This Hamiltonian is a sum of three different terms:
\begin{description}
\item[$\mathbf{p_z^2}$]{Momentum along $z$ is a good quantum
    number. This part of the
    wavefunction will be plane waves along $z$, $e^{\pm i p_z z}$.}
\item[$\mathbf{\left(q+A\right)^2}$]{This will lead to Landau
    levels, as discussed in Sect. \ref{sec:LL}. The energies will be
    given by the harmonic oscillator levels $\left(2n+1\right)B$,
    $n\in \mathbb{N}$, and the wavefunctions are highly degenerated.}
\item[$\mathbf{B\sigma_z}$]{The energies will be $B s$, where
    $s=\pm 1$ is the eigenvalue of $\sigma_z$. The wavefunctions are
    spinors $\chi_{1}=\left(\begin{matrix}1\\0\end{matrix}\right)$ and
    $\chi_{-1}=\left(\begin{matrix}0\\1\end{matrix}\right)$.}
	\end{description}
	
Then the eigenvalues of $H^2$ are given by
$\left(E_{p_z, n, s}\right)^2=p_z^2+\left(2n+1\right)B+s B$,
so that the energies becomes
$E_{p_z, n,s}=\pm \sqrt{p_z^2+\left(2n+1\right)B+ s B}$ and
they are shown for some values of $n$ in Fig.~\ref{fig:hotel}.
For $n=0$ and  $s=-1$, the dispersion relation is linear.

\section{Covariant and consistent mixed anomalies}
\label{app:mixed}

We have at several occasions referred to the freedom of adding local
counterterms to the effective action, and in this appendix we extend
the discussion
about this. It turns out that it is also possible to add local terms to the
currents themselves, independently of the action,
according to definite rules that have geometric and physical meaning.
There are in fact two kinds of currents, called ``consistent'' and
``covariant'', and correspondingly two types of  anomalies.
The consistent/covariant distinction is important in the case
of mixed Abelian anomalies,
and more prominently for non-Abelian gauge anomalies. 

Recall that we used three methods to calculate anomalies: the spectral
flow and path integral calculations gave us the currents written as
$J^\mu$ and $J_5^\mu$ in the main text -- we shall see that they are gauge
invariant. In the non-Abelian case they transform covariantly
under gauge transformations which motivates the name ``covariant''
which is used also in the Abelian case.
The currents evaluated using perturbation theory are in general
different, and will be denoted by $j^\mu$ and $j_5^\mu$ and 
named ``consistent'', because they are obtained from an effective
action $\G[A,A_5]$ and satisfy certain consistency conditions.
We first analyze the simpler 2D case
before turning to the 4D case which is of relevance for Weyl
semi-metals in Sect. \ref{sec:WSManom}.

\subsection{The 2D case}

The path integral method gives the covariant anomalies \eqref{covanom}
when both vector and axial vector gauge fields are present (mixed anomalies),
\be{covanom2}
\de_\mu J^\mu=\frac{1}{\pi}F_{5}\, , \qquad\qquad
\partial_\mu J_5^\mu = \frac 1 {\pi}F \; ,
\ee
where we used
the notation $F = \frac 1 2 \eps^{\mu\nu}F_{\mu\nu}$ and similarly for
$F_5$. The spectral flow method yields the same results, see Sect.
\ref{sec:SQHE}.

In Sect. \ref{sec:pert}, we discussed the perturbative calculation of the
effective action $\G[A,A_5]$ \eqref{2deffact}, leading by differentiation
to the following currents,\footnote{
  Recalling that in Minkowski space the currents
  acquire a factor of $(i)$.}
\be{curr2}
j^\mu &=&
\Pi^{\mu\nu} (i A_\nu - \eps_{\nu\a} A_5^\a)  
=\left(c\eta^{\mu\nu}-\frac{q^\mu q^\nu}{q^2}\right)
(i A_\nu - \eps_{\nu\a} A_5^\a)  \ ,
\\
j_5^\mu &=& \eps_{\mu\a}\Pi^{\a\nu} ( A_\nu +i \eps_{\nu\a} A_5^\a) \; .
\label{curr25}
\ee

As explained in the text, the value of the
constant $c$ depends on the renormalization
conditions we choose for the theory. 
 
From the above currents, we get the anomalies,
\be{2dpertan}
\partial_\mu j^\mu =  \frac {c-1} \pi (-F_5 + i\partial\cdot A ) \, , \qquad
\qquad \partial_\mu j_5^\mu =  \frac c {\pi}  (F -i \partial\cdot A_5)  \ ,
\ee
so for the particular values of $c$  picked in
Sect. \ref{sec:perturb} we reproduce the results \eqref{2dpertanom0}
and \eqref{2dpertanom1} when either $A$ or $A_5$ vanishes.
In the presence of both backgrounds, the expressions \eqref{2dpertan} are
not invariant under neither vector nor axial gauge transformations as
opposed to the covariant anomalies \eqref{covanom2} obtained from
spectral flow or path integrals.

To understand what is going on we revisit the freedom to add
counterterms to $\G[A,A_5]$ \cite{nairbook}.
These are local polynomials in fields and derivatives
and in 2D we only have three possible terms with dimension 
equal to two that are compatible with Lorenz invariance:
$A_\mu A^\mu$, $A_{5\mu} A_5^\mu$ and $ \eps_{\mu\nu}A^\mu A_5^\nu $. The
first term has in fact already been used, and correspond to fixing the
value of $c$. If we choose to maintain the vector gauge invariance we
take $c=1$ (corresponding to having a transverse
$\Pi^{\mu\nu}$); we can furthermore eliminate the term
$ \partial_\mu A_5^\mu $ in \eqref{2dpertan} by adding the counterterm
$i A_5^\nu A_{5\mu}/(2\pi)$, with the result,
\be{2dpertanom2}
\partial_\mu j^\mu = 0 \, , \qquad\qquad
\partial_\mu j_5^\mu =  \frac 1 {\pi}  F  \, . 
\ee
The only remaining freedom by the term
$\frac {c'} \pi \eps^{\mu\nu}A_\mu A_{5\nu}$ leads to the
anomalies,
\be{2dpertanom3}
\partial_\mu j^\mu = \frac {c'} \pi F_5 \, , \qquad\qquad
\partial_\mu j_5^\mu =  \frac {1-c'} {\pi}  F   \, .
\ee

Putting $c' = 1/2$ almost reproduces \eqref{covanom2}, but differs by
a significant factor of 2. It is clear that no choice of counterterms
will allow us to reproduce the path integral result. Therefore, in the
mixed case, the ``consistent'' currents $j^\mu$ and $j_5^\mu$ that are
obtained by the effective action are definitely different from the
covariant currents $J^\mu$ and $J_5^\mu$.

\subsection{Gauge invariance of currents and Bardeen-Zumino terms}
\label{app:BZ}

How to reconcile the results \eqref{covanom2}, and
\eqref{2dpertanom2}? The first answer to this problem was given in the
context of four-dimensional non-Abelian gauge theories by Bardeen and
Zumino \cite{BARDEEN1984421}. They noticed that the \emph{currents}
(not the action!) can be redefined by adding local terms. Let us begin
by showing how this can be done and then explain the rationale behind
the modifications.

The consistent vector current in \eqref{2dpertanom2} can be changed by
adding a ``Bardeen-Zumino term'' $\D j_\mu=\eps_{\mu\nu}A_5^\nu/\pi$
while keeping the axial current unchanged, as follows:
\begin{eqnarray}
&&  \tilde j^\mu =  j^\mu + \frac 1 {\pi} \epsilon^{\mu\nu} A_{5\nu}\ ,
\qquad {\mathrm {implying}}  \qquad  \partial_\mu \tilde j^\mu =
   \frac 1 {\pi}  \,  F_5 \ ,
   \label{2dcondef}\\ 
&&  \tilde j_5^\mu =
j_5^\mu\ ,  \qquad\qquad\qquad\quad {\mathrm {implying}} \qquad
   \partial_\mu \tilde j_5^\mu =   \frac 1 {\pi}  \,  F \; .
   \label{2dconde}
\end{eqnarray}
The new currents $\tilde j$ and $\tilde j_5$ have the same 
anomalies \eqref{covanom2} that were obtained using spectral flow
or path integrals
and are identified with the covariant expressions , $\tilde j^\mu=J^\mu$ and
$\tilde j^\mu_5=J^\mu_5$.
One can convince oneself that the modifications \eqref{2dcondef}
\eqref{2dconde} cannot be
obtained by adding counterterms to the effective action $\Gamma[A,A_5]$.

Let us now explicitly check the gauge invariance of the
currents $j^\mu$ and $j^\mu_5$.
The expressions \eqref{curr2}, \eqref{curr25}
with the choice of counterterms leading to
the anomalies \eqref{2dpertanom2} become:
\be{currents3}
j^\mu &=&\frac{1}{\pi}\left(\eta^{\mu\nu}-\frac{q^\mu q^\nu}{q^2}\right)
(i A_\nu - \eps_{\nu\a} A_5^\a) \ ,
\\
j_5^\mu &=& \frac{1}{\pi} \eps^{\mu\nu} A_\nu
-\frac{1}{\pi}\eps_{\mu\a}\frac{q^\mu q^\nu}{q^2}
( A_\nu +i \eps_{\nu\a} A_5^\a)  \ .
\ee
Under gauge transformations $\d_\l A_\mu=\de_\mu \l$ and
$\d_\xi A_{5\mu}=\de_\mu\xi$, their variations are:
\be{cov-cons}
\d_\l j^\mu=\d_\l j_5^\mu=\d_\xi j_5^\mu=0,
\qquad\qquad \d_\xi j^\mu= -\frac{1}{\pi} \eps^{\mu\a}\de_\a \xi \neq 0 \; .
\ee
Thus, the consistent current $j_\mu$ is not invariant under axial
gauge transformations, but  note that the Bardeen-Zumino term
\eqref{2dcondef} is precisely the correction needed to restore the invariance.
We have then shown that the so-called covariant
currents $J_\mu$ and $J_{5\mu}$ are indeed gauge invariant, as advertised
earlier.

We now discuss gauge invariance from the point of view of the
effective action $\Gamma[A,A_5]$.
The functional variations are defined by,
\be{gtrans}
\delta_\lambda = \int dx\, \partial_\mu\lambda \frac{\delta}{\delta A_\mu}
\ , \qquad\qquad 
\delta_\xi = \int dx\, \partial_\mu\xi \frac{\delta}{\delta A_{5\mu}} \ ,
\ee
leading to,
\be{gvar}
G(\l)\equiv \delta_\lambda \Gamma = \int dx\, \partial_\mu\lambda\,
\frac{\delta \Gamma}{\delta A_\mu}
= - \int dx\, \lambda\, \partial_\mu j^\mu \, ,
\qquad  
G(\xi) \equiv\delta_\xi \Gamma = - \int dx\, \xi\, \partial_\mu j_5^\mu \, ,
\ \ \
\ee
where we used the notation $G$ as in \cite{nairbook}, and the currents
$j^\mu$ and $j^\mu_5$ are those obtained from the perturbative calculation.
Of course, in the presence of anomalies, the effective action is not invariant,
but its variations  obey the  consistency conditions,
\be{wzcond}
\d_{\l_1}\d_{\l_2}\G -\d_{\l_2}\d_{\l_1}\G=\d_{[\l_1,\l_2]}\G \; ,
\ee
since they must obey the composition rules of the gauge group.
Specializing to the axial case,  the definitions \eqref{gvar}
of the anomalies, directly yields,
\be{WZcond2}
\d_{\l_1}G(\l_2) -\d_{\l_2}G(\l_1) = G([\l_1,\l_2])\, ,
\ee
which  is the celebrated Wess-Zumino consistency condition \cite{Wess-Zumino}.

In our Abelian case, the right-hand side of \eqref{WZcond2}
vanishes, and the same holds for its axial counterpart.
The left side also vanish since  the anomalies
\eqref{2dpertanom2} are expressed in terms of field strengths only, so the
Wess-Zumino conditions are satisfied for both symmetries.  
Given that the anomalies obey the consistency conditions,
the corresponding currents $j_\mu,j_{5\mu}$ are also called consistent.

We now reconsider the gauge invariance of the currents themselves. We have:
\be{cov-curr}
\delta_\lambda j^\mu(x) =\d_\l \frac{\d}{\d A_\mu(x)}\G=
\frac{\d}{\d A_\mu(x)}\d_\l \G \, , \qquad\quad
\delta_\xi j^\mu(x) = \frac{\d}{\d A_\mu(x)}\d_\xi \G \, ,
\ee
where we used that the two functional derivatives commute.
Analogous expressions hold for the axial current.
Upon evaluating the variation \eqref{cov-curr} using \eqref{gvar}
together with the consistent anomalies \eqref{2dpertanom2}, one reobtains
the result \eqref{cov-cons} from the explicit calculations, showing that
$j_\mu$ is not gauge invariant.
This second argument for the gauge transformation of currents will be
useful later in the 4D case. Note that this derivation was only
possible since the consistent anomalies were derived
from an effective action; it could not have been used for
the covariant anomalies.

Summarizing the discussion so far, we have shown that
there are two kinds of currents and anomalies in the mixed Abelian case:
the covariant currents obey gauge invariance but cannot be
obtained from an effective action, while the opposite holds for the
consistent currents. 
These same features are found in non-Abelian gauge theories even in
presence of a single background gauge  field \cite{BARDEEN1984421}.

Below, we shall give another, geometrical,  explanation for the
difference between consistent and covariant currents and the origin of
Bardeen-Zumino terms using the so-called
Wess-Zumino-Witten action. This will also clarify why the covariant
currents absorbs the anomaly inflow, while the consistent ones do not.

\subsection{The 4D case}
We now extend the previous analysis to the 4D case. Recall the perturbative
result \eqref{gtensor} for the three-point function,
\be{gtensor2}
q_\mu \Pi^{\mu\nu\lambda}  &=&   \frac{-i}
{8\pi^2} (2c) \epsilon^{\nu  \lambda \alpha \beta} k_\alpha p_\beta\ ,
\nonumber\\
k_\nu \Pi^{\mu\nu\lambda} &=&  \frac{-i}
{8\pi^2}  (1-c) \epsilon^{\lambda \mu \alpha \beta} p_\alpha q_\beta\ ,
\nonumber  \\ 
p_\lambda\Pi^{\mu\nu\lambda}  &=&   \frac{-i}
{8\pi^2} (1-c)\epsilon^{ \mu \nu  \alpha \beta}
q_\alpha k_\beta \, ,
\ee
where the momentum $q^\mu$ flows into the vertex with the $\gamma^5$,
and $c$ is the free parameter. In the main text, we took $c=1$, since this 
enforces vector gauge invariance.
Another possibility is $c = 1/3$ which treats the vertices symmetrically.
For chiral fermions coupled  to independent gauge fields
$A_+$ and $ A_-$, this choice  is \emph{required} by Bose symmetry
for the currents to be obtained by taking functional derivatives of an
effective action.
Evaluating all the diagrams in Fig. \ref{fig:triangles}, and recalling
that $A_\pm = A \pm A_5$ so $J_\pm = \half (J \pm J_5)$, the chiral
4D consistent anomalies become,
\be{cons-anom}
\partial_\mu j_\pm^\mu = \pm\frac {1} {48\pi^2}  F_\pm\tilde F_\pm \, ,
\ee
with the notation $F \tilde F = F_{\mu\nu} \tilde F^{\mu\nu}$, or  
equivalently, 
\be{aconanom}
 \partial_\mu j ^\mu =   \frac 1 {12\pi^2}  F_5\tilde F \, , \qquad\qquad
\partial_\mu j_5 ^\mu = \frac 1 {24\pi^2} ( F\tilde F + F_5\tilde F_5)  \, .
\ee
As in the 2D case, we can modify anomalies by adding local
terms to the effective action. Let us consider the expression,
\be{4dwzterm}
\D\Gamma  =  \frac 1 {6\pi^2} \int dx\, \epsilon^{\mu\nu\sigma\lambda}
A_\mu A_{5\nu} \partial_\sigma A_\lambda\; ,
\ee
that leads to the form,
\be{conanom2}
\partial_\mu j_5 ^\mu = \frac 1 {8\pi^2} (  F\tilde F +
\frac 1 3  F_5\tilde F_5 ) \ , \qquad\qquad     \partial_\mu j ^\mu = 0\; ,
\ee
so that the vector current is conserved.

\subsubsection{Relation between consistent and covariant anomalies in 4D}
\label{app:gaugeinvcurr}

The expressions \eqref{cons-anom} and  \eqref{aconanom} should
be compared with the  covariant anomalies derived by the  path-integral method,
\be{4covchir}
\de_\mu J_\pm^\mu=\pm  \frac {1} {16\pi^2}  F_\pm\tilde F_\pm \, ,
\ee
or in terms of vector and axial currents,
\be{4covan}
\partial_\mu J^\mu =   \frac 1 {4\pi^2}  F_5\tilde F \, ,
\qquad \qquad
\partial_\mu J_5 ^\mu = \frac 1 {8\pi^2} ( F\tilde F + F_5\tilde F_5 ) \, .
\ee
The 4D consistent and covariant anomalies differ by a factor of $3$
which is the counterpart of the factor of 2 that we
encountered in the 2D case. This again shows that the covariant
anomalies cannot be obtained by an effective action. For vanishing
$A_5$, however, the equations \eqref{4covan} and \eqref{conanom2} are
identical, so the difference is only present in the mixed case, just
as in 2D.

We now derive the gauge transformations of the
consistent currents \eqref{conanom2}, using, \emph{mutatis mutandis}
the same method as in the  2D case.
Again, the gauge transformations of the currents are obtained from
the variations \eqref{cov-curr} with \eqref{gvar} evaluated with the
consistent anomalies \eqref{conanom2}. We find the following violations
of gauge invariance,
\be{4dg-var}
&& \d_\l j_\mu = \d_\l j_{5\mu}=0 \ , \nonumber
\\
&& \d_\xi j^\mu = -\frac 1 {4\pi^2} \eps^{\mu\nu\r\s} \de_\nu\xi F_{\r\s} \ ,
\\
&& \d_\xi j_5^\mu = -\frac 1 {12\pi^2} \eps^{\mu\nu\r\s} \de_\nu\xi F_{5\r\s}\;
. \nonumber
\ee
From this  it is not hard to see that we can make
the currents gauge invariant by adding Bardeen-Zumino terms as,
\be{4dcorresp1}
\tilde j^\mu &=& j^\mu +
\frac 1 {4\pi^2} \eps^{\mu\nu\r\s} A_{5\nu} F_{\r\s} = J^\mu\ ,
\\
\tilde j^\mu_5 &=& j^\mu_5 +
\frac 1 {12\pi^2} \eps^{\mu\nu\r\s} A_{5\nu} F_{5\r\s} =J^\mu_5\; .
\label{4dcorresp2}
\ee
The anomalies of these currents agree with those of the currents
$J_\mu$ and $J_{5\mu}$ in \eqref{4covan}, so up to terms with zero
divergence, we can identify $J_\mu \equiv \tilde j_\mu$ and
$J_{5\mu}\equiv \tilde j_{5\mu}$, which proves that $J_{5\mu}$ and
$J_\mu $ are indeed covariant.  The relation \eqref{4dcorresp1}
between the conserved consistent current $j_\mu $ and the
non-conserved covariant one $J_\mu$ is used in the discussion of Weyl
semi-metals in Sect. \ref{sec:WSMresp}.

\subsection{Anomaly inflow and the Wess-Zumino-Witten action}
\label{app:WZW}

In this section we  discuss the relation between covariant
and consistent currents and anomalies from a geometric perspective,
closely following the exposition in  Ref. \cite{PhysRevB.85.184503}.

The first observation is that the anomaly inflow provides
a definition for the covariant anomalies.
We first explain this point in the simpler case of a 
chiral two-dimensional theory at the edge of a quantum Hall liquid.
In the main text, we showed
that charge non-conservation in the chiral theory is compensated by a
current flowing from an extra dimension, describing the attached bulk.
The three-dimensional inflow current is
obtained by varying the Chern-Simons action  $S_{CS}[A]$ \eqref{S-ind}
which gives,
\be{covdef}
&& J^\r_{(3D)}= \frac{\d}{\d A_\r} S_{CS}[A]=
\frac{1}{2\pi} \eps^{\r\mu\nu}\de_\mu A_\nu \, ,
\qquad\qquad \mu,\nu,\r=1,2,3,
\\\label{covdef2}
&& J^3_{(3D)}= \de_\a J_{(2D)}^\a=
\frac{1}{2\pi} \eps^{3\mu\nu} \de_\mu A_\nu, \qquad\qquad\qquad \a=1,2 \; .
\ee
where $x^3$ is the coordinate perpendicular to the edge.

The equation \eqref{covdef2} can be considered as a {\it definition}
of the covariant current $J^\a_{(2D)}$ and its anomaly: since the bulk
current $J^\r_{(3D)}$, is itself invariant covariant, so is the
corresponding edge current.
We have thus shown that the covariant anomaly can also be derived from an
effective action, albeit one that is defined in  one extra dimension. 
This inflow argument extends to mixed anomalies in $3D$ topological
insulators, as discussed in Sect. \ref{sec:SQHE}. 

We now turn to the consistent anomaly. As stressed earlier, it can be
derived from the effective action $\G[A]$ defined in the ambient
dimension which here is two.  This action can be computed, \eg by a
perturbative expansion, but we now show that it can also be partially
reconstructed from the knowledge of the anomaly using a universal
method.

The idea is to  integrate the infinitesimal gauge variation
$\d_\l\G[A]$ in \eqref{gvar}  over a finite range as follows:
\be{WZ}
S_{WZW}[A,\l] = \int_0^1 dt\int dx\; \frac{d\bar\l}{dt}\;
\de_\mu j^\mu[A^{\bar\l(t)}]= \G[A]-\G[A^\l] \, ,
\ee
where, $\bar\l(t)=\bar\l(x,t)$ for $t\in [0,1]$ is a parametric
variation of the gauge transformation from $\bar\l(0)=0$ to a
finite value $\l(x)=\bar\l(x,1)$, and the consistent anomaly in the
integrand is evaluated on the gauge transformed background $A^{\bar\l}$
which equals  $A_\mu+\de_\mu \bar\l$ in the Abelian case. 

The quantity $S_{WZW}[A,\l]$ is called the Wess-Zumino-Witten action; it is
not itself the effective action but its gauge variation, so it
depends on two fields. It is clear that the anomaly only allows a
partial reconstruction of $\G[A]$, whose full expression is usually
unknown.  Nonetheless, the part given by the Wess-Zumino-Witten action is exact
and has a very general form, possessing a number of
remarkable properties \cite{Wess-Zumino} \cite{WITTEN1983422}:
\begin{itemize}
\item
   $S_{WZW}[A,\l]$ obeys the Wess-Zumino consistency conditions
  and its variation gives, by definition, the consistent anomaly.
\item
  $S_{WZW}[A,\l]$ can be written as a local expression in one more dimension,
  upon interpreting
  the parameter $t$ as the extra coordinate; however, any variation with
  respect to its variables $A$ and $\l$ turns out to be a total derivative
  and is thus a  local expression in the original dimension.
\item
  Owing to general properties of anomalies, the form of $S_{WZW}$
  is known and given by,
  \be{WZ-CS}
  S_{WZW}[A,\l]=S_{CS}[A^\l]-S_{CS}[A] \ .
  \ee
In this relation, the Chern-Simons action is defined on a
  three-dimensional space with a boundary, on which the
  two-dimensional anomalous theory is defined.
\end{itemize}
In particle physics applications, the addition of one dimension is a
technical trick, and the ambiguities in this extension are shown to be
harmless due to the integer quantization of the Chern-Simons coupling
constant \cite{WITTEN1983422}. In condensed matter systems such as the
quantum Hall effect, the extra dimension is physical and the
Chern-Simons action is implementing the anomaly inflow discussed
earlier, so that the relation \eqref{WZ-CS} simply expresses the
overall gauge invariance.

Summarizing, we have established that the Chern-Simons theory gives
the covariant anomaly and the Wess-Zumino-Witten action the consistent
one, and we also found the relation \eqref{WZ-CS} between the two
actions. We then can explain the geometrical origin of the
Bardeen-Zumino terms, relating the two currents.  Before carrying a
detailed analysis, let us anticipate the result: It turns out
that while varying the Chern-Simons action to obtain the bulk current,
there is a (often disregarded) boundary term.  Once this is taken
into account, it gives a correction to the boundary (consistent)
current that exactly equals the Bardeen-Zumino term, thus matching the
inflowing (covariant) current.

\subsubsection{The 2D case}

We recall from Sect. \ref{sec:SQHE} the Chern-Simons action for the anomaly
inflow in the mixed case,
\be{BF2}
S_{CS}[A_+,A_-]=\frac{1}{4\pi}\int_{\cal M} A_+ dA_+ -A_-dA_- \; ,
\ee
where we use chiral backgrounds $A_+$ and $A_-$, whose gauge variations
are $A_\pm\to A_\pm +d\l_\pm$, and $\cal M$ is the three-dimensional
geometry with the two-dimensional boundary $\de\cal M$.
According to \eqref{WZ-CS}, the corresponding Wess-Zumino-Witten action is:
\be{WZ2}
S_{WZW}[A_\pm,\l_\pm]=\frac{1}{4\pi}\int_{\cal M} d\l_+dA_+ -
d\l_-dA_-=\frac{1}{4\pi}\int_{\de\cal M}\l_+dA_+ -\l_-dA_- \, .
\ee
Note that this action is explicitly two-dimensional, a
simplification occurring in the Abelian case. This expression
should be compared with the full effective action $\G[A,A_5]$ in
\eqref{2deffact} and \eqref{polten2}, which is  nonlocal and
only approximately known. 

The gauge variation of $S_{WZW}$ gives the consistent anomalies,
\be{WZ2-cons}
\de_\mu j^\mu_{(2D)\pm}=\pm\frac{1}{4\pi} \eps^{\mu\nu}\de_\mu A_{\pm \nu} \, ,
\ee
in agreement with earlier findings (cf. \eqref{2dpertanom3} for 
$c'=1/2$), and differ by a factor of 2 compared to the covariant
anomalies \eqref{covanom2}.

Next we compute the bulk current by varying the Chern-Simons action
\eqref{BF2}, being careful with the boundary terms,
\be{BF2var}
\d S_{CS}[A_+,A_-]&=&\frac{1}{4\pi}\int_{\cal M} \d A_+ dA_+ 
+A_+ d\left(\d A_+\right)\ - \ \left( A_+ \leftrightarrow A_-\right)
\\\label{BF2var2}
&=& \frac{1}{2\pi}\int_{\cal M} \d A_+ dA_+
+\frac{1}{4\pi}\int_{\de\cal M} \d A_+ A_+\ -\
\left( A_+ \leftrightarrow A_-\right)\; ,
\ee
where the corresponding terms with $A_+$ replaced by $A_-$ are not written.
 From the variation of the first and second term, we read the quantities,
\be{cov2}
&&J^3_{(3D)\pm}= \pm \frac{1}{2\pi} \eps^{3\mu\nu}\de_\mu A_{\pm \nu}=
\de_\a J^\a_{(2D)\pm} \; ,
\\\label{cov22}
  &&\D j^\a_{(2D)\pm}=\pm \frac{1}{4\pi} \eps^{\a\b}\de_\a A_{\pm \b} \; ,
\ee
defined on ${\cal M}$ and ${\de\cal M}$, respectively.

The expression \eqref{cov2} shows the inflow of the bulk current giving
the expected form of covariant currents \eqref{covanom2}.
The quantity \eqref{cov22} coming from the boundary part of \eqref{BF2var2}
reproduces the Bardeen-Zumino terms that
relate the consistent and covariant currents, namely
$j^\mu_{(2D)\pm} +\D j^\mu_{(2D)\pm}= J^\mu_{(2D)\pm}$, providing the
missing factor of $2$ between the corresponding anomalies,
respectively \eqref{WZ2-cons} (\ie \eqref{2dpertanom3} with $c'=1/2$)
and \eqref{covanom2}.

We conclude that covariant anomalies are relevant when the
two-dimensional theory is part of a higher dimensional system, as for
edge states in topological systems, the covariant currents being
unambiguously defined by the bulk theory.

For isolated two-dimensional theories there is an unresolved ambiguity
in which kind of currents to use, and different methods (\eg
perturbative versus path-integral) may yield one kind of anomaly or
the other, sometimes ``adsorbing'' the boundary term \eqref{cov22}
(Bardeen-Zumino) coming from the ``outside''.  Further physical input
is necessary for choosing the proper currents.

\subsubsection{The 4D case}

The previous analysis of Chern-Simons and Wess-Zumino-Witten actions
directly extends to four dimensions. We  first write the
five-dimensional action that realizes the anomaly inflow as,
\be{CS4}
S_{CS}[A_+,A_-]=\frac{1}{24\pi^2}\int_{\cal M} A_+ dA_+dA_+ -A_- dA_-dA_- \, ,
\ee
where $\cal M$ is now a five-dimensional
manifold with a four-dimensional boundary $\de\cal M$,
and varying this action gives,
\be{BF4var}
\d S_{CS}[A_+,A_-]&=&\frac{1}{24\pi^2}\int_{\cal M} \d A_+ dA_+ dA_+ 
+2 A_+ dA_+ d\left(\d A_+\right)\ -
\ \left( A_+ \leftrightarrow A_-\right)\\
&=& \frac{1}{8\pi^2}\int_{\cal M} \d A_+ dA_+dA_+
+\frac{1}{12\pi^2}\int_{\de\cal M} \d A_+ A_+ d A_+\ -
\ \left( A_+ \leftrightarrow A_-\right)\, .\qquad 
\ee
The first term in this equation gives the bulk currents,
\be{cov4}
&&J^5_{(5D)\pm}=
\pm \frac{1}{8\pi^2} \eps^{5\mu\nu\r\s}
\de_\mu A_{\pm \nu}\de_\r A_{\pm \s}=\de_\a J^\a_{(4D)\pm},
\ee
whose inflow match the covariant anomalies \eqref{4covchir}.
The second, boundary part gives the following Bardeen-Zumino terms,
\be{BZ4}
\D j^\mu_{(4D)\pm}=\pm \frac{1}{12\pi^2}
\eps^{\mu\nu\r\s}A_{\pm \nu}\de_\r A_{\pm \s} \; .
\ee
The Wess-Zumino-Witten action is obtained by the gauge variation of
\eqref{CS4}, according to the relation \eqref{WZ-CS}, and reads, 
\be{WZ4}
S_{WZW}[A_\pm,\l_\pm]=\frac{1}{24\pi^2}\int_{\de\cal M}
\l_+dA_+ dA_+ -\l_-dA_- dA_- \, .
\ee
This action in turns gives the consistent anomalies \eqref{cons-anom},
\be{WZ4-cons}
\de_\mu j^\mu_{(4D)\pm}=\pm\frac{1}{24\pi}
\eps^{\mu\nu\r\s}
\de_\mu A_{\pm \nu}\de_\r A_{\pm \s} \; ,
\ee
that are off by a factor of three with respect to the covariant ones
\eqref{cov4}, in agreement with earlier analyses.
The addition of the Bardeen-Zumino terms \eqref{BZ4} establishes
the map between the two kinds of currents,
$j^\mu_{(4D)\pm} + \D j^\mu_{(4D)\pm} =  J^\mu_{(4D)\pm}$,  and similarly for the
axial ones, thus connecting the anomalies  \eqref{cov4} and \eqref{WZ4-cons}.
Note that the earlier Bardeen-Zumino terms \eqref{4dcorresp1},
\eqref{4dcorresp2} are different from those in \eqref{BZ4}, because
they related the covariant anomalies \eqref{cov4} to the other form
\eqref{conanom2} of the consistent anomalies, equivalent up to
counterterms in $\G[A,A_5]$.

In conclusion, the relation \eqref{WZ-CS} between Wess-Zumino-Witten
and Chern-Simons actions clarifies the geometrical meaning of
the two types of anomalous currents and provides a general method
for deriving Bardeen-Zumino terms.

\section{Elements of differential geometry and curved
  space calculus } \label{app:calculus}

This appendix give a summary of the tools used for describing
topological and geometrical properties of gauge and
gravitational backgrounds. This is just a guide to more
comprehensive and rigorous presentations that can be found in
\cite{Nakahara:Geometry, Zeebook, GSW}. We start by
introducing manifolds and differential forms, then discuss the metric
structure, gauge fields and fermions on curved backgrounds.
In the final part we derive some consequences of the
trace anomaly in two-dimensional conformal field theory.

\subsection{Manifolds and differential forms}\label{subsec_man}

A manifold ${\cal M}$ is basically a $n$-dimensional surface that
\emph{locally} looks like flat space, namely $\mathbb{R}^n$.
Around each point there is a well-defined neighborhood
of other points, that can be described by an open set $U_{(i)}$
called patch, and mapped it into a open subset of $\mathbb{R}^n$. The
map associates an element $P$ in the manifold to an element
$\left(x^1(P), x^2(P), ..., x^n(P)\right)$ of $\mathbb{R}^n$ called
the coordinates of $P$. As every neighborhood  can be mapped
into $\mathbb{R}^n$, we can carry over known tools of calculus.

Several open subsets are needed to cover ${\cal M}$ and they
should be consistent on overlapping
regions. Two patches $U_{(1)}, U_{(2)}$  give equivalent descriptions
on their intersection if there exists a transition function $f_{(1,2)}$
between $x^\mu_{(1)}(P)$ and $x^\mu_{(2)}(P)$, \ie
$x^\mu_{(1)}=f_{(1,2)}(x^\mu_{(2)})$, corresponding to a change of
coordinates on $\mathbb{R}^n$.
The properties of these transitions functions specify the structures
that are build on ${\cal M}$.
The first one is the differentiable structure, the possibility to
draw smooth curves on the manifold, define derivatives and
the tangent space at each point $P$, as we shall see momentarily.
For this to work, the transition functions $f_{(1,2)}$ should be
differentiable functions.  More structures will be needed later.

Derivatives on the manifold are defined through the
tangent vector to a curve, specified by a function $x^\mu(\lambda)$
parameterized by $\lambda$. Therefore, a function $f(x^\mu)$ will depend
implicitly on $\lambda$ along the curve and we can define the
directional derivative,
\be{dir_der}
\frac{d}{d\lambda}=\frac{\partial
  x^\mu}{\partial \lambda}\partial_\mu.
\ee
In this expression, ${d}/{d\lambda}$ is the natural definition of a vector
on the manifold with basis spanned by
$\partial_\mu$ and components $(\partial x^\mu/\partial \lambda)$
\cite{Nakahara:Geometry} .

We write a vector as $V=V^{\mu}\partial_\mu$, where derivatives are
the basis elements and $V^\mu$ its components;
the differential $df=\partial_\mu f dx^\mu$ is instead an element of
the dual space with basis $dx^\mu$.
Then, the directional derivative of a function $f$, 
$V[f]=V^\mu \partial_\mu f$, can be considered as an inner product.
Vectors and dual vectors satisfy,
\be{basis_form}
\left( dx^\mu , \de_\nu \right) =\frac{\de x^\mu}{\de x^\nu}=\d^{\mu}_\nu.
\ee

The differential $df=\partial_\mu f dx^\mu$ is
also called a differential form of degree one, or 1-form.
A generic form can be written as $\omega=\omega_\mu dx^\mu$.
From the basis of vectors and forms one can build the $(n, p)$-rank tensors,
\be{tensor}
T=T^{\mu_1\dots\mu_n}_{\nu_1\dots\nu_p}\; \partial_{\mu_1}\cdots
\partial_{\mu_n}dx^{\nu_1}\cdots dx^{\nu_p} .
\ee
The tangent space and its dual at one point
are invariant under linear coordinate transformations of the group $GL(n,\RR)$.
The tensors \eqref{tensor} carry representations of this group and
in this context their lower (upper) indices are called covariant
(contravariant).

A special class is given by the totally antisymmetric $(0,n)$-tensors that are
associated to higher differential forms. Let us start with the definition of
the exterior (wedge) product $\wedge$ of $n$ dual vectors as their totally
antisymmetric tensor product $\otimes$,
\be{def_wedge_n}
dx^{\mu_1} \wedge dx^{\mu_2} \wedge\cdots \wedge dx^{\mu_n}=
\sum_{\s \in {\cal S}_n} \text{sign}(\s)\,
dx^{\mu_{\s(1)}}\otimes dx^{\mu_{\s(2)}}\otimes\cdots\otimes dx^{\mu_{\s(n)}},
\ee    
where the sum extends to permutations of $n$ elements and includes
their sign.  This product of differentials is just the $n$-dimensional
volume element.

An important property of differential forms is that they
can be used without introducing any metric on ${\cal M}$
and thus are suited for describing topological quantities that
are metric independent, as encountered in the text.
Examples will be analyzed later.

We define the $2$-forms and $p$-forms as follows,
\be{def_pform}
F=\frac{1}{2!}F_{\mu\nu}\, dx^\mu \wedge dx^\nu,\qquad\qquad
H=\frac{1}{p!}H_{\mu_1\cdots \mu_p}\, dx^{\mu_1}
\wedge \cdots \wedge dx^{\mu_p}.
\ee
In a $D$-dimensional manifold, the form of degree $D$ has only one
component, because the antisymmetric basis vector is unique up to
permutations; this is called the top form. Forms of higher degree vanish.

The exterior product of forms follows from the properties of the basis
\eqref{def_wedge_n}: it is written $A\wedge B$, or simply  $AB$.
The components of the product of a $p$-form $A$ and a $q$-form $B$
are given by,
\be{form-prod}
\frac{1}{p!q!}(A\wedge B)_{\mu_1\cdots\mu_{p+q}}=
\frac{1}{(p+q)!}\left(A_{\mu_1\cdots\mu_{p}} B_{\mu_{p+1}\cdots\mu_{p+q}}
  \pm {\rm permutations}\right)\; .
\ee
Note the property,
\be{prod-sign}
    A\wedge B=(-1)^{p+q} B\wedge A\; .
\ee

The exterior derivative $d$ of a $p$-form is defined by generalizing
the earlier derivative of a function ($0$-form) $df=\partial_i f dx^i$.
The operator $d$ is a map from a $p$-form $A$ to a $(p+1)$-form $dA$,
defined by,
\be{d-form}
dA=\frac{1}{p!}\left(\partial_\nu\; A_{\mu_1\cdots\mu_p}\right)
    dx^\nu \wedge dx^{\mu_1}\cdots \wedge  dx^{\mu_p}.
\ee
The distribution rule for the exterior derivative is,
\be{d-prod}
    d\left(A\wedge B\right)=dA\wedge B+(-1)^p A\wedge dB\; ,
\ee
where $A$ is a $p$ form, that can be memorized by
saying the $d$ is a $1$-form to be carried close to the quantity on which
it is acting.

\subsection{Metric, connection and curvature}

We now define the notion of distance on the manifold. Given a
point $x^\mu$ on ${\cal M}$, the distance to a close point
$x^\mu+dx^\mu$ is given by $ds$, where,
\be{inter}
ds^2=g_{\mu\nu}(x)dx^\mu dx^\nu,
\ee
with $g_{\mu\nu}(x)$ the metric tensor.
As discussed several times in the text, metric backgrounds are not only
relevant for general relativity but also for studying the
response of quantum systems to strain, dislocations and disorder, to
model thermal gradients and expose the effects of gravitational anomalies.

The metric provides the inner product (scalar product) of vectors
\be{inn}
(U \cdot  V )=g_{\mu\nu} U^\mu V^\nu,
\ee 
and, consequently, a way to upper and lower indices of components of
vectors and tensors.  If the metric has only
positive eigenvalues it is called a Riemannian metric, while a
pseudo-Riemannian metric has both positive and negative eigenvalues,
as in the case of Lorentzian signature.
The inner product \eqref{inn} is positive definite for Riemannian metrics.

The tangent vectors to the curves on ${\cal M}$ passing through the
point $x^\mu(P)$ form a vector space
that is called the tangent space $T(P)$. In order to define
derivatives of vectors, one needs a way to relate
the tangent spaces at $x^\mu$ and $x^\mu+dx^\mu$, and
transport vectors between them.
This is specified by the connection $\G_{\mu\l}^\nu$,
that gives the variation $\d_\G V^\nu = u^\mu \G_{\mu\l}^\nu V^\l$
of the vector $V^\nu$ under the transport along a curve of tangent $u^\mu$
connecting the two points.
This should be added to the flat-space variation to obtain the derivative
on the manifold,
\be{cov_v}
D_\mu V^\nu=\partial_\mu V^\nu+\Gamma^{\nu}_{\mu\lambda} V^\lambda \; .
\ee

A natural choice of connection $\G$ is that realizing the parallel transport,
under which
the angle between the tangent to the curve $u$ and $V$ remains constant.
Otherwise said, the parallel transport of the tangent vector $u^\a$ vanishes:
this condition is equivalent to the geodesic equation for the metric,
\be{geod}
u^\a D_\a u^\mu=0, \qquad\qquad
\frac{d u^\mu}{d\t} + \G^\mu_{\nu\l} u^\nu u^\l=0\; ,
\ee
where $u^\a=d\xi^\a/d\t$ is the tangent to the curve $\xi^\a(\t)$.

The parallel transport identifies the Levi-Civita connection: the
corresponding derivative is called the covariant derivative, that is
compatible with the metric because it obeys,
\be{g-comp}
D_\mu g_{\nu\l}(x)=0\; .
\ee
This condition ensures that the parallel transport of the scalar product
of two vectors vanishes, so that the covariant derivative reduces to
the flat-space one,
\be{prod-comp}
D_\mu(A^\a g_{\a\b}B^\b)=\de_\mu (A^\a g_{\a\b} B^\b)\; .
\ee
In order to prove this statement, one should use the relation
$D_\mu(A^\a B_\a)=\de_\mu (A^\a B_\a)$ to obtain the other covariant
derivative $D_\mu V_\nu=\partial_\mu V_\nu-\G^\l_{\mu\nu} V_\l$,
apply it to \eqref{g-comp} and solve for $\Gamma$
under the assumption of symmetric lower indices,
$\G^\mu_{\nu\l}=\G^\mu_{\l\nu}$, as in \eqref{geod}.
One obtains the explicit expression of $\G$ in terms of the metric,
the so-called Christoffel symbol,
\be{christ}
\Gamma^{\nu}_{\mu\lambda}=\frac{1}{2}g^{\nu\rho}\left(\partial_\mu
  g_{\lambda\rho}+\partial_\lambda g_{\mu\rho}-\partial_\rho
  g_{\mu\lambda}\right) \; .
\ee
This expression checks \eqref{prod-comp}.
Note also that covariant derivatives obey the Leibniz rule,
$D_\mu(A^\a B_\a)=(D_\mu A^\a)B_\a+A^\a (D_\mu B_\a)$.

We now consider the parallel transport on a sphere along 
parallels and meridians: starting from a point and moving in the
two directions on the sides of a quadrangle reaches
the opposite point with two different results.
The infinitesimal difference of parallel transports on the two paths
is given by the commutator of covariant derivatives,
\be{comm_der}
\left[D_\mu, D_\nu\right]V^\alpha=
R_{\mu\nu,}\,^\alpha_{\ \beta} V^{\beta}\; .
\ee
that defines the curvature, or Riemann tensor $R_{\mu\nu,\a\b}$.
Such difference is
$\d_R V^\a= \d \S^{\mu\nu} R_{\mu\nu,}\,^\a_{\ \b} V^\b$,
where $\d\S^{\mu\nu}$ is the infinitesimal square. One can show that $\d_R V^\a$
is a rotation, thus the Riemann tensor is antisymmetric on the second pair
of indices too.

We remark that the connection and curvature can be associated
to matrix-valued differential forms, as follows,
\be{gamma-f}
\G^\nu_{\ \l} =\G^\nu_{\mu\l}dx^\mu, \qquad\qquad
R^\a_{\ \b}= \frac{1}{2} R_{\mu\nu,}{}^\a_{\ \b}\; dx^\mu\wedge dx^\nu
=\left( d\G +\G^2 \right)^\a_{\ \b} \ .
\ee
having used \eqref{comm_der} to write the last expression for $R^\a_{\ \b}$.

\subsection{Gauge fields}

The wavefunction $\Psi(x)$ introduce an additional complex vector space
at each point of the manifold. The same occurs when considering a
charged quantum field $\phi$. This new vector space is 
called a ``fiber'' on the manifold and the collection of fibers is
a fiber bundle.\footnote{
  For a precise definition of fiber bundles, see \cite{Nakahara:Geometry}.}
The connection relating such vector spaces at different points is the gauge
field $A_\mu(x)$ and the covariant derivative is defined according
to the usual expressions. The correspondence with the gravitational
connection \eqref{cov_v}
is better seen in the case of a non-Abelian gauge group $G$:
the field $\phi^i$ transforms under a representation
${\cal R}(G)$ of the group,  with $i=1,\dots,{\rm dim}{\cal R}(G)$,
and the covariant derivative is,
\be{non-a}
(D_\mu\phi)^i=\de_\mu\phi^i+iA_\mu^a (T_a)^i_{\ j} \phi^j,
\qquad \qquad a=1,\dots,{\rm dim}G,
\ee
where $(T_a)^i_{\ j}$ are generators obeying the Lie algebra of the group
in that representation.
These generators are usually omitted, to write
$A_\mu^a (T_a)^i_{\ j}\equiv (A_\mu)^i_j$.

The curvature associate to the gauge connection is clearly the
field strength, $F_{\mu\nu}=[D_\mu,D_\nu]/i$, in close analogy with
\eqref{comm_der}. Note that
this formula applies both to Abelian and non-Abelian fields.

\subsection{Reparameterization invariance and integrals
of forms}

The presence of a metric \eqref{inter} introduces an invariance under
general coordinate transformations, $x'^\mu=x'^\mu(x^\a)$, that is
also called general covariance or
diffeomorphism invariance. Gravitational forces possess this
symmetry which can be viewed as a kind of gauge invariance.
Effective field theories can naturally be generalized to curved space
where they are written in covariant form, 
unless specified differently, as in the case of the
elasticity effects discussed in Sect. \ref{sec:visc}.

The length element is invariant under reparameterizations,
\be{diff-g}
ds^2=g_{\mu\nu}(x)dx^\mu dx^\nu=
\left(g_{\mu\nu}(x)\frac{dx^\mu}{dx'^{\mu'}}\frac{dx^\nu}{dx'^{\nu'}}
\right)dx'^{\mu'}dx'^{\nu'}=g'_{\mu'\nu'}(x')dx'^{\mu'}dx'^{\nu'},
\ee
and vectors transform accordingly, \eg
$V'^{\mu'}(x')=(dx'^{\mu'}/dx^\mu)\; V^\mu(x)$.

Reparameterization invariant actions involve covariant
scalar quantities made up from the fields, the metric and
their covariant derivatives. The
integration measure is given by the invariant volume element,
\be{vol_element}
dV=\sqrt{\left|g\right|}dx^1\wedge dx^2\wedge\dots \wedge dx^D\; ,
\ee
where $g=\det(g_{\mu\nu})$, and under reparameterizations
\eqref{diff-g}, the transformation of the determinant cancels 
the Jacobian factor from the volume. Another form of the same relation is
given by the covariant totally antisymmetric tensor.
For example in 2D,
\be{eps-cov}
\frac{1}{\sqrt{|g|}}\eps^{\mu\nu}A_\mu B_\nu=
\frac{1}{\sqrt{|g'|}}\eps^{\mu'\nu'}A'_{\mu'}B'_{\nu'}\; .
\ee

Using these results we can write for example the Maxwell action on a
curved manifold as,
\be{YM}
S_{YM}= -\frac{1}{4}\int_{\cal M} d^D x\; \sqrt{|g|}F_{\mu\nu}
F_{\a\b}g^{\mu\a}g^{\nu\b}\; .
\ee
In the case of the Chern-Simons action, we similarly have
\be{CS-M}
S_{CS} =\frac{k}{4\pi} \int_{\cal M} d^3 x\; \eps^{\mu\nu\r}
A_\mu \de_\nu A_\r \; ,
\ee
and remark that the $\sqrt{|g|}$ factors from the measure and the
epsilon tensor cancel out.

This indicates that the Chern-Simons
action can be naturally rewritten in terms of differential forms
that are independent of the metric structure.
The non-Abelian case can be discussed without much more effort.
The gauge connection and field strength are associated to the following forms,
\be{ym-forms}
A=A_\mu dx^\mu, \qquad\qquad F=dA +A^2=
\frac{1}{2}F_{\mu\nu}dx^\mu\wedge dx^\nu\; ,
\ee
that are matrix valued  in a representation ${\cal R}(G)$, as in \eqref{non-a}.

The Chern-Simons form is,
\be{CS-form}
\w_{CS}=AdA +\frac{2}{3} A^3, 
\ee
and it obeys $ d\Tr\, \w_{CS}=\Tr\, F^2 $ \cite{Nakahara:Geometry}.

The integral of a differential form of maximal degree (top form) is
given by the simple integration of its unique independent
component. For the $3$-form $\th$ in three dimensions, for example,
\be{form-int} \int_{\cal M}\th =\int_{\cal M} \frac{1}{3!}  \th_{ijk}
dx^i\wedge dx^j \wedge dx^k= \int_{\cal M} \frac{1}{3!} \eps^{ijk}
\th_{ijk} dx^1 dx^2 dx^3 \; .  \ee In the case of the Chern-Simons
form \eqref{CS-form}, the integral is, \be{CS-int} S_{CS}=
\frac{k}{4\pi}\int \Tr\left(AdA+\frac{2}{3} A^3\right) \; .  \ee The
gauge indices of the form are resolved by taking the trace over the
gauge indices, obeying ${\rm Tr}(T^a T^b)=C({\cal R}) \d^{ab}$, where
the normalization factor\footnote{ The normalization of
  \eqref{CS-int} is such that the action is gauge invariant for
  integer $k$ values \cite{desjactem}; for example, $C({\cal R})=1/2$
  for the fundamental representation of $SU(N)$.}
$C({\cal R})$ depends on the representation.  Note that \eqref{CS-int}
reproduces \eqref{CS-M} in the Abelian case, where $A^3$ vanishes.

In $D$-dimensional space, the integration of a $k$-form on a
surfaces with $k<D$, such as a two-sphere in $\RR^3$,
can be defined by first rewriting $\w$ as a top form
in $k$ dimensions using the parameterization of the surface
and then applying the earlier definition
\eqref{form-int} \cite{Nakahara:Geometry}.
In the case of the sphere, the parameterization $x^\mu(z^\a)$,
where $\mu=1,2,3$ and $\a=1,2$, can be given by
the polar coordinates $(\vf,\th)$.
The two form $F$ can then be rewritten in term of surface differentials
$(dz^1=d\vf, dz^2=d\th)$ as follow,
\be{form-ref}
F=\frac{1}{2}F_{\mu\nu}\; dx^\mu\wedge dx^\nu=
\frac{1}{2}\left(F_{\mu\nu}\frac{dx^\mu}{dz^\a}
  \frac{d^\nu}{dz^\b}\right) dz^\a\wedge dz^\b=
\frac{1}{2}\hat F_{\a\b}\; dz^\a\wedge dz^\b \; .
\ee
In the last expression, $\hat F$ is a top form on the sphere and can
be integrated using \eqref{form-int}. Note that the metric is again
not necessary.

It turns out that the Maxwell action can also be written in terms of
differential forms, although the metric dependence is unavoidable.
We introduce the Hodge dual of a  $p$-form $\omega$ as the following
$(D-p)$-form ${}^*\w$,
\be{hodge-d}
{}^*\w=\frac{\sqrt{|g|}}{p!(D-p)!}\w_{\mu_1\dots\mu_p}
\eps^{\mu_1\dots\mu_p}{}_{\mu_{p+1}\dots\mu_d} dx^{\mu_{p+1}}\wedge\cdots
\wedge dx^{\mu_D} \; ,
\ee
where indices in epsilon tensor are raised with the help of the metric.

If two forms $A$ and $B$ have equal degree $p$, then
$A\wedge{}^*B$ is a top form that can be integrated in $D$ dimensions.
The integral defines the inner product of forms,
\be{inner2}
(A \cdot B)\equiv \int A\wedge{}^*B=\frac{1}{p!}
\int A_{\mu_1\cdots\mu_p}B^{\mu_1 \cdots \mu_p}
\sqrt{\left|g\right|}d^D x,
\ee
that is symmetric and positive-define for Riemannian manifolds.
In particular, the Yang-Mills action can be written,
\be{YM-f}
S_{YM}=-\frac{1}{2}\int\Tr[ F\wedge {}^*F] \; .
\ee

The inner product \eqref{inner2} can be used to define the adjoint
derivative $d^\dagger$ by $(A, dB)=(d^\dagger A, B)$; one finds the
equivalent expression when acting on a $p$ form,
\be{adj}
    d^\dagger=(-1)^{np+n+\alpha}\; {}^* d^*,
\ee
where $\alpha$ is equal to $1 (0)$ for Riemannian (Lorentzian) manifolds.

From $d$ and $d^\dagger$, we define the Laplacian acting on $p$ forms,
\be{lap}
\Delta=\left(d+d^\dagger\right)^2=dd^\dagger+d^\dagger d=
(-1)^{np+n+\alpha}\left(d^*d^* +{}^*d^*d\right).
\ee

\subsection{The first Chern class and the monopole charge}

A differential form $\w$ is said closed if $d\w=0$, and exact if
$\w=d\L$. Stokes theorem is particularly simple once
expressed in terms of forms, where it reads,
\be{stok}
\int_{\cal M} \w=\int_{\de{\cal M}} \L, \qquad\qquad {\rm for}\qquad \w=d\L\; .
\ee
Familiar expressions in $\RR^3$ can be recovered by working out examples
and using earlier definitions.

We now discuss some further mathematical aspects of topological
invariant local functionals of the gauge field, like the first Chern class.
This is the integral of the field strength two-form $F$
on a closed two-dimensional surface ${\cal M}$,
\be{mon-form}
I[A]=\frac{1}{2\pi}\int_{\cal M} F=
\frac{1}{4\pi}\int_{\cal M} \eps^{\mu\nu}F_{\mu\nu}d^2x=
\frac{1}{2\pi}\int \vec{B}\cdot d\vec{S}= n \in \Z,
\ee
where $\vec{B}$ is the magnetic field orthogonal to the surface.
Using $F=dA$ and the Stokes theorem \eqref{stok}, one would conclude that this
integral vanishes since the surface has no boundary.
However, the monopole field is
singular, due to the Dirac string emanating from
inside the surface and piercing it at one point. It follow that
the relation $F=dA$ is not well defined on the whole surface and
\eqref{stok} cannot be applied immediately.

We consider the sphere and cover it with two patches,
corresponding to the Northern and Southern hemispheres, respectively
$D^{(+)}$ and $D^{(-)}$, that overlap at the equator
$S^1=D^{(+)}\cap D^{(-)}$.
Let us consider two vector potentials for the monopole:
$A^{(+)}$ is well defined in the upper hemisphere because the string
is sent downward, and $A^{(-)}$ is regular in the
lower patch, having moved the string upward. We then have well defined
expressions: $F=dA^{(+)}$ in $D^{(+)}$ and $F=dA^{(-)}$ in $D^{(-)}$.
At the equator, the two forms should give the same $F$, thus
the two potentials differ by a gauge transformation,
$A^{(+)}- A^{(-)}=d\chi$, as one can check from the explicit expressions
\cite{Nakahara:Geometry}.

The Stokes theorem \eqref{stok} applied to $F$ defined in two
patches gives,
\be{mono-stok}
\int_{\cal M} F &=&\int_{D^{(+)}} dA^{(+)} +\int_{D^{(-)}} dA^{(-)}
= \int_{S^1} A^{(+)}-A^{(-)}
\nl
&=&\int_{S^1} d\chi=\chi(2\pi)-\chi(0) =2\pi n \; ,
\ee
having paid attention to boundary orientations.  The result is
nonvanishing because the gauge transformation $\L(\vf)=e^{i\chi(\vf)}$
is nontrivial: it is a map from the equator, $\vf \in S^1$, into the
group, $\L(\vf)\in U(1)\equiv S^1$, that winds around that
circle. Such maps fall into equivalence classes under smooth
deformations that form the additive homotopy group
$\Pi_1(U(1))=\Z$. The representative element of the $n$-th class is
the gauge transformation $U=e^{in\vf}$ leading to monopole charge $n$
in \eqref{mono-stok}.

In conclusion, the quantization of the monopole
charge follows from the periodicity of the large $U(1)$ gauge
transformation on the equator of the sphere: this is a well-defined
transition function for the gauge field at patch overlaps.
A similar argument was used in Sect. \ref{sec:antop} for the
monopole inside the cylinder $S^1\times[-T,T]$: the field
values at $\pm T$ where also found to differ by a large gauge transformation.

\subsection{Vielbeins, spin connection and fermions}

The definition of spinors on manifolds requires additional elements
to be introduced because they carry a double-valued representation of
Lorentz transformations: a $2\pi$ rotation returns
a minus sign. This feature is not present in tensors \eqref{tensor},
that have integer spin only, thus one should introduce a new
``basis of vectors'' on which Lorentz
transformations can act, and an associated connection for
their parallel transport.\footnote{
  In this section, we follow closely the discussion of Sect. 12 in
  \cite{GSW}.}
In so doing, we introduce additional degrees of freedom and a new gauge
symmetry, that of local Lorentz transformations, as we shall see momentarily.

The new vectors $\hat{e}_a$ are linear combinations of the tangent
vectors introduced at the beginning of this appendix,
\be{v-def}
\hat{e}_a=e_a^{\ \mu}(x)\de_\mu \; ,
\ee
where $\det(e_a^\mu)>0$ for keeping orientation and $a=1,\dots,D$
is the index counting these vectors.
We require that the new basis is orthonormal at any point $x^\mu$,
as it were flat Euclidean (resp. Minkowskian) space,
\be{riem_viel}
(\hat{e}_a\cdot \hat {e}_b )=
g_{\mu\nu}e_{a}^{\ \mu}e_{b}^{\ \nu}=\delta_{ab}\; , \qquad\qquad
({\rm resp.}\ =\eta_{ab}) .
\ee

It follows that  the Lorentz group acts by,
$e^\mu_a(x)\to e^\mu_b(x) \L^b_{\ a}(x)$, where $\L^b_{\ a}(x)$ is the
local Lorentz transformation. The index $a$ is called a Lorentz index
and denoted by Latin letters, while we keep Greek symbols for spacetime
indices.

The matrices $e_{a}^{\mu}$ are called vielbeins, where \textit{viel}
comes from the German word for many; depending on dimension, they
are also named zweibein for 2D and vierbein for 4D. Further names
for $e^\mu_a$ are local frame vectors and, in 4D, tetrads. Their indices can be
lowered and raised using the corresponding metrics.
Note the relations,
\be{viel-rel}
e^a_{\ \nu}e_b^{\ \nu}=\d^a_b\; , \qquad\qquad
  g_{\mu\nu}=e^a_{\ \mu}e^b_{\ \nu} \d_{ab}\; ,
\ee
showing that $e^a_{\ \mu}$ is the inverse\footnote{Sometimes vielbein and
  its inverse are given different names to avoid confusion,
  as \eg $(e^a_\mu,e_a^\mu)\to (e^a_\mu,E_a^\mu$).}
of $e_a^{\ \mu}$ and parameterizes the
dual vectors, $\hat{\th}^a=e^a_\mu dx^\mu $.
Note that all tensors can be transformed into the Lorentz basis, \eg
$V^a=e^a_\mu V^\mu$.

The parallel transport of Lorentz vectors is specified by the
spin connection $\omega^a_{\mu b}$ by,
\be{cov_w}
D_\mu V^a=\partial_\mu V^a+\omega^{a}_{\mu b} V^b \; .
\ee
The spin connection causes an infinitesimal Lorentz transformation on
the index $a$, so it obeys $\omega_{\mu,ab}=-\omega_{\mu, ba}$.
The corresponding one-form is $\omega^{a}_b=\omega^{a}_{\mu b} dx^\mu$.

The compatibility of the parallel transport with the metric is
specified by the following condition,
\be{cov_basis}
D_\mu e^a_\nu= \partial_\mu e^a_\nu-\Gamma_{\mu\nu}^{\lambda}e_{\lambda}^a+
\omega^{a}_{\mu b} e^{b}_\nu=0 \; ,
\ee
that extends the earlier condition \eqref{g-comp}.

The parallel transport of a Lorentz vector on a infinitesimal square
determines again the Riemann curvature,
\be{Lor-R}
[D_\mu,D_\nu] V^a =R_{\mu\nu,}{}^a_{\ b}V^b \; ,
\ee
that is now expressed in terms of the spin connection only,
\be{cartan}
R_{\mu\nu,}{}^a_{\ b}= \de_\mu \w^a_{\nu b} + \w^a_{\mu c}\w^c_{\nu b}
- (\mu \leftrightarrow \nu) \; .
\ee
In analogy with \eqref{gamma-f}, the curvature can be written as the
Lorentz-valued two-form,  $R^a_{\ b}=(d \w +\w^2)^a_{\ b}$.

Having established a local frame analogous to flat space, we can consider
Dirac fields in it. The usual gamma matrices obey\footnote{
  We are assuming that the manifold is Minkowskian.}
$\left\{\gamma^a,\gamma^b\right\}=2\eta^{ab}$,
where $a,b$ are the local Lorentz indices. The covariant derivative
acting on spinors $\psi$ should be consistent with Lorentz transformations,
thus it takes the form,
\be{cov_spin}
D_a\psi=e_a^{\mu}\left(\partial_\mu-i\omega^{bc}_{\mu}\mathcal{\s}_{bc}\right),
\ee
where $\s_{ab}=i\left[\gamma_a, \gamma_b\right]/4$ is the generator
of Lorentz transformations on spinors \cite{nairbook}.

Thus the expression of the Dirac action in curved space is given by,
\be{Dir-m}
S[\psi,e^a_\mu]=\int_{\cal M} d^Dx\; e\; \bar\psi\left(i\g^a D_a -m\right)\psi,
\ee
where $e=\det(e^a_\mu)$.

A few words of explanation are needed at this point.  We started by
explaining that the connection $\G$ and the metric $g$ are independent
concepts, but eventually the requirement that they are
  compatible among themselves implied the condition \eqref{g-comp}.
  This in turn lead to the expressions \eqref{christ} and
  \eqref{gamma-f} respectively, for the connection $\G=\G(g)$ and the
  curvature $R(g)$ in terms of the metric.  In this subsection, we
introduced vielbeins $e$ and spin connection $\w$, that are also
independent quantities in general; similarly, the compatibility
condition \eqref{cov_basis} can be shown to be sufficient to determine
$\w(e)$ and $R(e)$, given that $\G(e)$ and $g(e)$ follow from
\eqref{christ} and \eqref{viel-rel} \cite{GSW}. In particular, it is
found that
$\G_{\mu\nu}^{\r}=e^{\r}_a\de_\mu e^{b}_\nu +e^{\r}_a\omega^{a}_{\mu
  b} e^{b}_\nu $.

It turns out that the curved space calculus in the $(e,\w)$ variables,
called Einstein-Cartan or first-order formulation, is more general
 and closer to gauge theory than the $(g,\G)$
formulation. The introduction of
Dirac fermions and corresponding localized spin sources also provides an
extension of general relativity, leading to torsion
$T^\l_{\mu\nu}=\G^\l_{\mu\nu}-\G^\l_{\nu\mu}\neq 0$, and effects not
completely accounted by the metric. Actually, the transport of a
vector around an infinitesimal square \eqref{comm_der}
results into a rotation and a translation,
$\d V^\a=\d \S^{\mu\nu}(R_{\mu\nu,}{}^\a_{\ \b}V^\b + T^\a_{\mu\nu})$,
the latter being given by the torsion. This is the continuum
formulation of a dislocation on the lattice. In these lectures, we never
discuss the effects of torsion and consider the metric and vielbein
formulations as equivalent descriptions.

In absence of torsion, the Levi-Civita connection $\G(g)$ is given by
\eqref{christ} and the spin connection $\w (e)$ follows by solving
\eqref{cov_basis} and the relation between $\Gamma$ and $\omega$
\cite{GSW},
\be{lc_conn_omega}
\omega^{ab}_\mu=\frac{1}{2}
  e^{\nu [a} \partial_{[\mu} e^{b]}_{\nu]} -\frac{1}{2}
  e^{\rho [a}e^{\sigma b]} e_{\mu c}\partial_\rho e_{\sigma}^ c \; ,
\ee 
where square brackets means antisymmetrization of indices, \ie
$A_{[\a} B_{\b]}= A_\a B_\b -A_\b B_\a$.

\subsection{The example of the sphere}

To make the previous discussion more concrete, we derive the geometric
properties of a familiar manifold. Consider a sphere in $\RR^3$: its
metric can be obtained by the parameterization of the surface given by
the polar coordinates,
$x^i(z^\a)=(r\sin\theta \cos\phi, r\sin\theta \sin\phi,
r\cos\theta)$,  where $i=1,2,3$, $\a=1,2$ and $z^\a=(\th,\phi)$, as follows,
\be{ind}
ds^2 &=& \d_{ij}dx^i dx^j=
\d_{ij} \frac{dx^i}{dz^\a}\frac{dx^j}{dz^\beta} dz^\a dz^\b
\nl
&=& g_{\a\b} (\th,\phi)\; dz^\a dz^\b
=r^2 d\theta^2 +r^2\sin^2\theta d\phi^2\; ,\qquad\qquad \a,\b=\th,\phi\; .
\ee
Using \eqref{riem_viel} and \eqref{viel-rel}, we can obtain the
following expression of the zweibeins, after choosing a gauge condition
for the Lorentz frame, $a,b=1,2$,
\be{zwei}
&& g_{\a\b}=e_\a^a \d_{ab} e_\b^b\; , 
  \nl
&& e^1_\th=e_{\th 1}= \left(e^{\th 1}\right)^{-1}=r\ ,
\qquad\qquad
e^{2}_{\phi}=  \left( e^{\phi2} \right)^{-1} =r\sin\theta\; ,
\ee
with the other components being equal to zero.

Using \eqref{lc_conn_omega} and
$\partial_\gamma e^a_\beta= r\cos\theta \delta^{a2}
\delta_{\gamma\theta} \delta_{\beta \phi}$, we obtain that
$\omega^{ab}_\gamma=-\epsilon^{ab} \cos\theta
\delta_{\gamma\phi}$. From \eqref{cartan} follows the Riemann tensor,\footnote{
  Notice that the antisymmetric symbol is
  $\epsilon_{\gamma\beta}=\delta_{\gamma\theta}\delta_{\beta\phi}-
  \delta_{\gamma\phi}\delta_{\beta\theta}$.}
$R_{\gamma\beta,ab}=\sin\theta \epsilon_{\gamma\beta}\epsilon_{a b}$.
Finally, one finds the scalar curvature of the sphere
\be{curv_sphere}
\mathcal{R}=R_{\gamma\beta,ab} e^{\gamma a} e^{\beta b}=
\frac{2}{r^2}.
\ee

\subsection{Trace anomaly in conformal field theory}

The constant $c$ appearing in the trace anomaly,
\be{a-conf2}
D^\mu T_{\mu\nu}=0\ , \qquad \qquad
T_\mu^\mu=-\frac{c}{24\pi} {\cal R} \; ,
\ee
is very important in conformal theory because it parameterizes
several physically relevant quantities. Since
the anomaly manifests itself on curved metric backgrounds, it is convenient
to relate it to observables that have a non-vanishing
flat space limit, $g_{\mu\nu}\to\d_{\mu\nu}$.
Let us write the path-integral expression of an observable $\cal O$,
\be{o-grav}
\langle {\cal O}(x)\rangle_{g}=\frac{1}{Z}
\int {\cal D}\phi e^{-S[\phi,g]}\; {\cal O} \; ,
\ee
where the integration is carried out over some fields $\phi$ in the presence of
the background denoted by $g$, and where $Z$ is the path integral
without the insertion of ${\cal O}$.
Under an infinitesimal change of the metric, we have the identity:
\be{wi-grav}
\frac{\d}{\d g^{\mu\nu}(y)}\langle {\cal O}(x)\rangle_{g}=
\frac{\sqrt{g}}{2}  \langle  {\cal O}(x) T_{\mu\nu}(y)\rangle_g \; ,
\ee
where the insertion of the stress tensor follows from the variation of
the action inside the path integral using the definition \eqref{t-def}.

We consider \eqref{wi-grav} for ${\cal O}=T_\mu^\mu$ and evaluate the
metric variation of the trace anomaly \eqref{a-conf2} on the left-hand side
by using the formula 
$\d {\cal R} =\left(g_{\mu\nu} D^2- D_\mu D_\nu\right)\d g^{\mu\nu}$
(See \cite{Cappelli-Coste}, appendix B).
Taking the limit to flat space, we get:
\be{ttheta}
\langle  T_\a^\a(x) T_{\mu\nu}(y)\rangle=
-\frac{c}{12\pi}\left(\de_\mu\de_\nu -\d_{\mu\nu} \de_\r^2\right)
\d^{(2)}(x-y)\; ,
\ee
where $\de_\a=\de/\de x^\a$ and $\d^{(2)}(x)$ is the two-dimensional Dirac
delta function. The trace anomaly partially
determines the stress-tensor two-point function in Euclidean space.
Note also that the flat-space limit is non-vanishing because 
the scalar curvature $\cal R$ is approximately linear in the
metric. In four dimensions the trace anomaly is quadratic
in the curvature for dimensional reasons, and one would
need two derivatives in \eqref{wi-grav} for getting a non-vanishing
limit, leading to informations on the stress-tensor
three-point function \cite{Cappelli-Guida-Magnoli}.

Two-dimensional conformal invariance
determines the stress tensor two-point function up to a constant
\cite{Cappelli-Friedan-Latorre} by,
\be{tt-conf}
\langle  T_{\a\b}(x) T_{\mu\nu}(y)\rangle=
\l \left(\de_\a\de_\b -\d_{\a\b} \de_\r^2\right)
\left(\de_\mu\de_\nu -\d_{\mu\nu} \de_\r^2\right) G(x-y) \; ,
\ee
where $G(x-y)=-\left[\log(x-y)^2\right]/(4\pi)$ is the two-dimensional massless
scalar Green function obeying $\de_\r^2G(x-y)=\d^{(2)}(x-y)$.
Note that \eqref{tt-conf} has the correct scale dimension $4$ and the index
structure is appropriate for enforcing transversality due to $T_{\mu\nu}$
conservation \eqref{a-conf2}.
The comparison between \eqref{tt-conf} and \eqref{ttheta} gives
$\l=c/(48\pi)$.

The correlator \eqref{tt-conf} is only non-vanishing for $x\neq y$ for
the traceless components of the stress tensor, that are given by
$T_{zz}$ and $T_{\bar z\bar z}$, using the complex notation introduced
in Sect. \ref{sec:gravbulkedge}. One finds,
\be{tzz}
\langle  T_{zz}(z) T_{zz}(0)\rangle=- \frac {c}{12}
\de_z^4 \log(z\bar z) = \frac{c}{2 z^4}\; ,
\qquad\qquad
\langle  T_{\bar z\bar z}(z) T_{\bar z\bar z}(0)\rangle=\frac{c}{2 \bar z^4}
\; ,
\ee
where we also rescaled $T_{\mu\nu}\to 2\pi T_{\mu\nu}$ as is
customary in the CFT literature.

We can now use the  result \eqref{tzz} for checking 
 the Virasoro algebra \eqref{Vir-alg2}.
We need the expression of the Virasoro generators $L_n$ as Fourier
modes of the stress tensor on the unit circle $z=\exp(i\theta)$
\cite{Ginsparg}, as well as the ground-state conditions \eqref{hws},
respectively given by:
\be{Vir-def}
T_{zz}(z)=\sum_{-\infty}^{\infty}\frac{L_n}{z^{n+2}}\ ,
\qquad\qquad L_n\vert \W\rangle =\langle \W\vert L_{-n}=0,
\quad n=-1,0,1,\dots \ .
\ee
Note that the $L_n$ generators were introduced earlier in \eqref{L-def} and
\eqref{L-bose} as the Fourier modes of the Hamiltonian density
${\cal H}=T_{zz}+T_{\bar z \bar z}$, for chiral theories where actually
$T_{\bar z \bar z}$ vanished.

Another derivation of the stress tensor two-point function is obtained
by using the Virasoro algebra \eqref{Vir-alg2} together with the
properties \eqref{Vir-def}, as follows,
\be{tt-alg}
\langle\W\vert T_{zz}(z)T_{zz}(0)\vert\W\rangle=
\lim_{w\to 0} \sum_{n,m=2}^\infty\langle\W\vert[L_n,L_{-m}]\vert\W\rangle
\frac{w^{m-2}}{z^{n+2}}=\langle\W\vert[L_2,L_{-2}]\vert\W\rangle
\frac{1}{z^4}=\frac{c}{2z^4} \; .
\ee
This expression matches the form \eqref{tzz} given by the trace anomaly.
This completes the proof that the Virasoro central charge $c$
and the coefficient of the trace anomaly are equal, as claimed
in Sect. \ref{sec:curr}

Another interesting property parameterized by the central charge
is the anomalous transformation law of the stress tensor under
conformal transformations. This result is used in the derivation of the
thermal Hall current in Sect. \ref{sec:thermgrav}.
Let us briefly discuss this point.

As mentioned in Sect. \ref{sec:curr}, analytic coordinate transformations
in two dimensions change the metric by an overall (conformal) factor,
\be{gzz2}
z=z(w), \qquad \qquad ds^2= dz d\bar z=
\left\vert\frac{dz}{dw}\right\vert^2 dw d{\bar w}
=2g_{w\bar w}\; dw d{\bar w},
\ee
namely $g_{\mu\nu}(w)=|dz/dw|^2g_{\mu\nu}(z)$.

The transformation of the traceless part of stress tensor found to be
\be{t-trans}
T_{ww}(w)\; dw^2 = T_{zz}(z)\; dz^2 + \frac{c}{12}\left[
  \frac{z'''(w)}{z'(w)}-
  \frac{3}{2}\left(\frac{z''}{z'}\right)^2\right] dw^2 \; ,
\ee
(a corresponding expression for $T_{\bar z \bar z}$ involves $\bar c$).
One recognizes the usual covariant transformation of a two-index
tensor and the additional term given by the central charge times
the quantity between square brackets called Schwarzian
derivative $\{z,w\}$. The derivation of \eqref{t-trans} is rather
technical and can be found in Sect. 4 of Ref. \cite{Cappelli-Coste}:
in short, it is based on the calculation of the Wess-Zumino-Witten action
by integration of the trace anomaly \eqref{a-conf2}, using the analogue
of \eqref{WZ}. Upon a metric variation, this action then gives the full
expression of the stress tensor, not only its trace part, 
more precisely its variation between the two geometries \eqref{t-trans}
related by the conformal map $z=z(w)$.

The result \eqref{t-trans} is used in Sect. \ref{sec:thermgrav} to obtain the
stress tensor in the cylinder geometry with periodic time
representing the thermal state. As described in \cite{heat2}, the map
between the complex $z$-plane and the $w$-cylinder is:
\be{therm-map}
z(w)=\exp\left[\frac{i2\pi w}{v\b}\right]\ , \qquad\qquad
w=v\t+ix \; ,
\ee
where the Euclidean time $\t$ has period $\b=1/k_BT$, 
related to temperature $T$, and $v$ is the Fermi velocity.
Insertion of the map \eqref{therm-map} into \eqref{t-trans}
yields the  ground-state value, 
\be{t-therm}
\langle T_{ww}(w)\rangle_{\rm cyl} =\langle T_{zz}(z)\rangle_{\rm plane} 
+ \frac{\pi^2 c}{6v^2\b^2} = \frac{\pi^2 c}{6v^2\b^2} \; ,
\ee
being normal-ordered to zero in the plane. This result is used to
compute the thermal current and energy \eqref{E-P} given in the text
(after restating the $1/2\pi$ factor and suitable powers of the
Fermi velocity).

\end{appendix}

\bibliography{review}

\nolinenumbers

\end{document}